\pdfoutput=1
\documentclass[nofootinbib,aps,prd,a4paper,superscriptaddress,eqsecnum]{revtex4}
\usepackage{amsmath}
\usepackage{amsfonts}
\usepackage{booktabs}
\usepackage{amssymb}
\usepackage{multirow}
\usepackage{siunitx} 
\usepackage{adjustbox}

\usepackage[T1]{fontenc}
\usepackage[greek,english]{babel}
\usepackage[linktocpage]{hyperref}

\usepackage{graphicx}
\usepackage{color}
\usepackage{braket}
\usepackage{dcolumn}
\usepackage{bm,url}
\usepackage{subfigure}
\usepackage[dvipsnames,svgnames]{xcolor}  
\definecolor{oxfordblue}{rgb}{0.0, 0.13, 0.28}
\definecolor{burgundy}{rgb}{0.5, 0.0, 0.13}
\definecolor{darkolivegreen}{rgb}{0.33, 0.42, 0.18}
\definecolor{darkblue}{rgb}{0,0,0.5}
\definecolor{richcarmine}{rgb}{0.84, 0.0, 0.25}
\definecolor{darkblue}{rgb}{0,0,0.5}
\definecolor{bluer}{rgb}{0.00,0.50,0.75}{}
\hypersetup{colorlinks=true, citecolor=red, linkcolor=blue,
 urlcolor = magenta, filecolor=magenta}

\providecommand{\U}[1]{\protect\rule{.1in}{.1in}}
 
 \DeclareUnicodeCharacter{0327}{\c{\ }}
\newcommand{\newc}{\newcommand}

\newc{\be}{\begin{equation}}
\newc{\ee}{\end{equation}}

\newc{\ba}{\begin{eqnarray}}
\newc{\ea}{\end{eqnarray}}
\newc{\bea}{\begin{eqnarray*}}
\newc{\eea}{\end{eqnarray*}}
\newc{\D}{\partial}
 
\newc{\eg}{{\it e.g.} }
\newc{\etc}{{\it etc.} }
\newc{\etal}{{\it et al.}}
\newc{\lcdm}{$\Lambda$CDM }

\newc{\ra}{\Rightarrow}

\allowdisplaybreaks

\begin{document}

\title{Entropy applications in cosmology: spacetime thermodynamics, 
holographic dark energy, entropic gravity and beyond - a review}

\author{Emmanuel N. Saridakis}
\email{msaridak@noa.gr}\affiliation{Institute for Astronomy, Astrophysics, 
Space 
Applications and 
Remote Sensing, National Observatory of Athens, 15236 Penteli, Greece}
\affiliation{CAS Key Laboratory for Researches in Galaxies and Cosmology, 
Department of Astronomy, University of Science and Technology of China, Hefei, 
Anhui 230026, P.R. China}
\affiliation{Departamento de Matem\'{a}ticas, Universidad Cat\'{o}lica del 
Norte, 
Avda.
Angamos 0610, Casilla 1280 Antofagasta, Chile}

\author{Giuseppe Gaetano Luciano}
\email{giuseppegaetano.luciano@udl.cat}
\affiliation{Department of Chemistry, Physics and Environmental and Soil 
Sciences, Polytechnic School, University of Lleida, Av. Jaume II, 69, 25001 
Lleida, Spain
}

\author{Andreas Lymperis}\email{alymperis@upatras.gr} 
\affiliation{Department of Physics, University of Patras, 26500 Patras, Greece}

\author{Eirini C. Telali} 
\email{etelali@perimeterinstitute.ca}
\affiliation{Perimeter Institute for Theoretical Physics,
31 Caroline Street North, Waterloo, Ontario, Canada N2L 2Y5}
\affiliation{Department of Physics and Astronomy, University of Waterloo,
200 University Avenue West, Waterloo, ON, N2L 3G1, Canada}

\begin{abstract}
Entropy has emerged as a key concept at the intersection of gravity, quantum
theory, and cosmology, revealing deep connections between geometry, information,
and thermodynamics. The realization that black holes are thermodynamic systems,
characterized by an entropy proportional to their horizon area, has provided
strong evidence that spacetime itself may possess an underlying microscopic
structure. Since then, entropy has acquired a central role across quantum field
theory, holography, quantum gravity, and cosmology.
In this review, we present a unified account of entropy in gravitational and
cosmological contexts.
We first survey entropy in gravitational systems, focusing on black-hole
thermodynamics, entropy bounds, the holographic principle, entanglement entropy,
and the area law, together with the main corrections and generalizations arising
from quantum field theory, generalized uncertainty principles, modified
statistics, and extended theories of gravity.
We then discuss three conceptually distinct ways in which entropy is employed in
cosmology. The first is through the spacetime-thermodynamics conjecture, in 
which
gravitational dynamics emerge from thermodynamic relations applied to
cosmological horizons, and   generalized entropy forms consequently lead to
modified cosmological evolution. The second is the holographic approach, in
which entropy bounds constrain the vacuum energy of the Universe, giving rise to
the holographic dark energy scenario and its various extensions. The third is
entropic gravity, where gravity is interpreted as an emergent phenomenon of
entropic origin, leading likewise to interesting cosmological phenomenology.
By reviewing the theoretical foundations and cosmological implications of these
entropic frameworks, we aim to clarify the role of entropy, discuss open issues
and outline directions toward a deeper thermodynamic and informational
understanding of gravity and cosmology.
\end{abstract}

\maketitle
 
\tableofcontents

\section{Introduction}

Entropy is among the most profound  concepts in  
physics, incorporating within a single quantity ideas as diverse as
irreversibility, information loss, coarse graining, quantum correlations, and
the emergence of macroscopic behaviour from microscopic laws.
The term \emph{entropy} was introduced by Rudolf Clausius in 
1865~\cite{Clausius1865}, who deliberately adopted a Greek-derived terminology 
in 
analogy with the word ``energy''. The term \emph{\textgreek{ἐντροπία}} 
(entropy) 
stems from the Greek words 
\emph{\textgreek{ἐν}} (``in'') and \emph{\textgreek{τροπή}} 
(``transformation'' or ``change'') and 
was intended to characterize the transformation and irreversible dispersal of 
energy within physical processes. Clausius was inspired by the Aristotelian 
term \emph{\textgreek{ἐνέργεια}} (energy), derived from 
\emph{\textgreek{ἐν}}  (``in'') and \emph{\textgreek{ἔργον}}
(``work'' or ``activity''), which in ancient Greek philosophy denoted 
``actuality'' or ``being-in-action''.
 
 Originally introduced within classical thermodynamics as a quantity governing
the direction of physical processes, entropy was initially regarded as a purely
phenomenological concept, disconnected from microscopic physics.
This perspective changed dramatically with the development of statistical
mechanics, where entropy was reinterpreted as a measure of the number of
microscopic configurations compatible with a given macroscopic state.
In this framework, entropy acquired a fundamental informational character,
connecting thermodynamics with probability theory and laying the foundations for
its later incorporation into quantum theory.

The quantum-mechanical generalization through the von Neumann entropy elevated
entropy to a central role in quantum information theory.
In quantum systems, entropy quantifies entanglement, loss of information due to
coarse graining, and the degree of correlation between subsystems.
Hence, entropy is no longer merely a thermodynamic quantity associated with
heat and disorder, but a universal descriptor of physical states across
classical, statistical, and quantum domains. 

Over the past decades, the conceptual significance of entropy has expanded even
further with the realization that it plays a fundamental role in gravitational
systems.
Unlike ordinary matter systems, gravity intertwines geometry, dynamics, causal
structure, and information.
This feature reached a decisive turning point with the discovery that black
holes are thermodynamic objects, endowed with both a temperature and an entropy
proportional to the area of their event horizon.
The Bekenstein-Hawking entropy,
\begin{equation}
S_{BH}=\frac{A}{4G},
\end{equation}
established a deep     connection between geometry,
thermodynamics, and quantum 
theory~\cite{Bekenstein:1973ur,Hawking:1974rv,Hawking:1975vcx}. Moreover, this 
result 
overturned the conventional expectation that entropy should scale
with volume, instead revealing a fundamentally non-extensive character of
gravitational degrees of freedom.

The area scaling of black-hole entropy, together with the generalized second
law of thermodynamics, has provided one of the strongest indications that
gravity is intimately related to information theory.
It suggests that spacetime itself may possess an underlying microscopic
structure whose degrees of freedom are encoded not in the bulk volume, but on
lower-dimensional boundaries.
The robustness of the area law across a broad variety of classical and
semiclassical settings strongly indicates that it is not an accidental feature
of specific solutions, but rather a universal property of gravitational
systems.
At the same time, the semiclassical nature of the Bekenstein-Hawking formula
makes it evident that it represents only the leading contribution in a more
general entropy functional, which should incorporate quantum, statistical, and
dynamical corrections.

These developments have motivated an enormous body of work aimed at
understanding the microscopic origin, universality, and possible modifications
of gravitational entropy.
Corrections to the Bekenstein-Hawking relation arise in a wide variety of
frameworks, including quantum field theory in curved spacetime,
entanglement entropy,
string theory,
loop quantum gravity,
generalized uncertainty principles,
non-extensive statistical mechanics,
and modified theories of 
gravity~\cite{Bombelli:1986rw,Srednicki:1993im,
Strominger:1996sh,Ashtekar:1997yu,
Kaul:2000kf,Solodukhin:2011gn}.
Remarkably, despite their diverse physical origins and mathematical
implementations, many of these approaches preserve the area law at leading
order while introducing subleading logarithmic, power-law, fractal,
non-extensive, or exponential corrections.
The repeated emergence of similar structures across otherwise unrelated
frameworks strongly suggests that entropy is not merely a derived or auxiliary
concept, but rather a fundamental ingredient of gravitational dynamics and
quantum spacetime.

Beyond black holes, entropy bounds and holographic considerations have further
reinforced this viewpoint. The Bekenstein bound, the covariant entropy bound
and the holographic principle all suggest that the maximum entropy contained
within a spacetime region is determined by the area of its boundary rather than
by its 
volume~\cite{Bekenstein:1980jp,Bousso:1999xy,tHooft:1993dmi,Susskind:1994vu}.
This remarkable departure from the conventional extensive behavior of entropy
provides strong evidence that the fundamental degrees of freedom of
gravitational systems are organized in a fundamentally non-local manner.

These ideas culminated in the holographic paradigm, according to which the
complete physical information contained in a gravitational system can be
encoded on a lower-dimensional hypersurface. Their most concrete realization
is provided by the AdS/CFT correspondence~\cite{Maldacena:1997re}, which
established a profound duality between gravity in the bulk and a quantum field
theory defined on its boundary, thereby revealing deep connections among
gravity, quantum field theory, geometry and quantum entanglement. In this
framework, spacetime geometry itself appears to emerge from patterns of quantum
correlations, elevating entropy and entanglement from thermodynamic concepts
to fundamental building blocks of spacetime. These developments have profoundly
influenced modern approaches to quantum gravity and provide a powerful
conceptual foundation for many of the entropy-based cosmological models
reviewed in the following sections.

The deep connection between gravity and entropy naturally raises the question
of whether entropy can play an equally fundamental role in cosmology.
Cosmology provides a unique arena in which gravitational physics,
thermodynamics,
quantum theory,
and observational astrophysics coexist at the largest observable scales.
The Universe   possesses causal horizons, such as apparent and event
horizons, which share many thermodynamic properties with black-hole horizons
and admit an associated entropy and temperature.
At the same time, cosmology faces some of the deepest open problems
in fundamental physics, including the nature of dark energy,
the origin of cosmic acceleration, the inflationary paradigm,
the thermodynamic arrow of time, 
and the possible quantum origin of spacetime itself.
Within this broader context, entropy naturally emerges as a powerful  
concept in modern cosmological research.

Indeed, over the last two decades, the application of entropy-related ideas in
cosmology has generated a vast and rapidly expanding literature.
These studies investigate how entropy can influence cosmic dynamics,
constrain the energy content of the Universe,
modify gravitational field equations,
or even provide a microscopic interpretation of spacetime geometry itself.
However, despite the abundance of results, it is important to emphasize that
entropy enters cosmology in several conceptually distinct ways.
Failure to distinguish between these different roles can lead to conceptual
confusion and misinterpretation of physical results. Hence, clarifying these 
distinctions is   one of the principal motivations of
the present review.

The use of entropy in cosmology can be classified into three
main categories. 
The first category  starts from the conjecture of \emph{spacetime 
thermodynamics}, in
which gravitational dynamics emerge from thermodynamic relations applied to
spacetime horizons.
A seminal result in this direction was obtained by Jacobson, who demonstrated
that the Einstein field equations can be derived from the Clausius relation
\begin{equation}
\label{Clre}
\delta Q = T dS\,,
\end{equation}
applied to local Rindler horizons, provided that entropy is proportional to
horizon area~\cite{Jacobson:1995ab}.
This remarkable result suggests that gravity itself may not be fundamental, but
rather an emergent phenomenon arising from deeper thermodynamic or statistical
principles.
Subsequent developments extended this perspective to cosmological spacetimes,
where the Friedmann equations can be recovered from the first law of
thermodynamics applied to the apparent horizon of a
Friedmann-Robertson-Walker (FRW) Universe~\cite{Cai:2005ra,Akbar:2006kj}.

Within this spacetime-thermodynamics framework, modifications of the
entropy-area relation naturally translate into modifications of cosmological
dynamics. In particular, generalized entropy forms inspired by non-extensive 
statistics,
quantum gravity,
minimal-length physics, modified gravitational theories, etc., lead to altered 
Friedmann equations and,
consequently, to novel cosmological 
scenarios~\cite{Tsallis:1987eu,Renyi:1961EEE,Barrow:2020tzx,Kaniadakis:2002zz}.
In this approach, entropy plays an active dynamical role, directly shaping the
gravitational field equations and the evolution of the Universe.

The second category of entropy applications in cosmology is conceptually
distinct and is based  on \emph{holographic considerations}.
Here, entropy does not modify gravitational dynamics directly, but instead
constrains the allowed energy content of the Universe.
The holographic dark energy proposal provides a prominent realization of this
idea.
Motivated by entropy bounds and the holographic principle, holographic dark
energy models posit that the vacuum energy density is limited by the condition
that the total entropy within a region does not exceed the entropy of a black
hole of the same size~\cite{Cohen:1998zx,Li:2004rb}.
This leads to a dark-energy density determined by an infrared cutoff scale,
typically associated with a cosmological horizon.
The holographic dark-energy framework has subsequently been extended in numerous
directions, including interacting models, generalized entropy forms, modified
gravity constructions, and alternative infrared cutoffs.

The third category is associated with the idea of \emph{entropic gravity},
according to which gravity itself emerges as an entropic force generated by the
statistical tendency of physical systems to maximize entropy.
In this perspective, gravitational attraction is not a fundamental interaction
but an emergent macroscopic phenomenon associated with information storage on
holographic screens~\cite{Verlinde:2010hp}.
This approach has stimulated extensive discussion concerning the microscopic
origin of gravity and has also led to interesting cosmological applications,
including modified Friedmann equations and alternative scenarios for inflation
and dark energy.

The purpose of the present review is to provide a comprehensive and unified
account of entropy in gravitational and cosmological physics, while carefully
distinguishing the conceptual foundations and domains of validity of the
different entropic approaches.
Our goal is not merely to summarize existing results, but also to place them
within a coherent theoretical framework that reveals the common ground,
conceptual connections, open questions, and future directions. Furthermore, we 
will give particular emphasis   on the role of generalized entropy
expressions, the emergence of modified cosmological dynamics from
thermodynamic considerations, and the interplay between entropy, holography,
and dark energy.

The  manuscript is organized as follows.
In Sec.~\ref{Entropydefinitions}, we review entropy in gravitational
physics, focusing on black-hole thermodynamics, entropy bounds,
the holographic principle, entanglement entropy, quantum-gravity approaches,
and generalized entropy expressions arising from quantum,
statistical,
and geometrical considerations.
In Sec.~\ref{Spacetimethermodynamics} we discuss the spacetime
thermodynamics conjecture and the derivation of gravitational field equations
from thermodynamic principles, while in
Sec.~\ref{SpacetimethermodynamicsApplications} we investigate the
cosmological implications of generalized entropy-area relations and the
resulting modified cosmological scenarios.
In Sec.~\ref{Holographicdarkenergy} we present the holographic dark-energy
framework, discussing its theoretical foundations, phenomenological
implications, and observational aspects.
Sec.~\ref{ExtendedHolographicdarkenergy} is devoted to extended models of
holographic dark energy, including interacting scenarios and models based on
alternative infrared cutoffs or modified gravitational theories.
In Sec.~\ref{HolographicdarkenergyModified} we review holographic
dark-energy models arising from generalized entropy forms, such as Tsallis,
R\'enyi, Sharma-Mittal, Kaniadakis, Barrow, logarithmic-corrected,
power-law-corrected, and Luciano-Saridakis entropies.
In Sec.~\ref{Entropicgravity} we discuss entropic gravity and its
cosmological applications.
Finally, in Sec.~\ref{Conclusions} we summarize the main results and 
we present
our concluding remarks.
For ease of comparison, throughout this review we generally retain the notation, 
conventions and system of units adopted in the original references whenever 
appropriate.

\section{Entropy: definitions, origins, corrections, extensions and conceptual 
developments}
\label{Entropydefinitions}

  Historically, the concept of entropy
emerged during the nineteenth century in the context of classical 
thermodynamics, primarily through the works of Clausius, Kelvin and Boltzmann, 
as an attempt to quantify irreversibility and the directionality of physical 
processes. 
 In its original thermodynamic formulation, entropy 
was introduced by 
Clausius as a state function satisfying
\begin{equation}
dS=\frac{\delta Q_{\rm rev}}{T},
\end{equation}
where $\delta Q_{\rm rev}$ denotes the reversible heat transfer and $T$ the 
temperature of the system~\cite{Clausius1865}. Within this framework, entropy 
measures the degradation of usable energy and constitutes the cornerstone of 
the second law of thermodynamics, according to which the entropy of an isolated 
system never decreases.

A decisive conceptual step was achieved by Boltzmann, who provided the first 
microscopic interpretation of entropy through statistical mechanics. In this 
picture, entropy no longer characterizes merely macroscopic irreversibility, 
but instead counts the number of microscopic configurations compatible with a 
given macroscopic state. This interpretation is embedded in the celebrated 
Boltzmann relation
\begin{equation}
S=k_B \ln W,
\end{equation}
where $W$ denotes the number of accessible microstates and $k_B$ is Boltzmann’s 
constant~\cite{Boltzmann1877}. Subsequently, Gibbs generalized this framework 
to statistical ensembles and probability distributions, leading to the 
Boltzmann-Gibbs entropy
\begin{equation}
S_{BG}=-k_B\sum_i p_i\ln p_i,
\end{equation}
where $p_i$ is the probability associated with the $i$-th microscopic 
state~\cite{Gibbs1902}. In parallel, Shannon later recognized that the same 
mathematical structure naturally quantifies missing information in 
communication 
theory~\cite{Shannon:1948dpw}. Entropy therefore acquired a dual statistical 
and 
informational interpretation, a perspective that would eventually become 
central in quantum theory and gravitational physics.

For many decades entropy was regarded primarily as a property of matter and 
statistical systems. However, this perspective changed dramatically with the 
development of black-hole thermodynamics. The realization that black holes 
possess a temperature and an entropy proportional to the area of their horizon 
established a profound connection between gravity, quantum theory, geometry and 
thermodynamics~\cite{Bekenstein:1973ur,Hawking:1975vcx}. Unlike conventional 
systems, whose 
entropy typically scales with volume, gravitational entropy obeys an area law, 
suggesting that the fundamental degrees of freedom of spacetime behave in a 
highly non-trivial and intrinsically holographic manner. This discovery not 
only transformed the conceptual status of entropy, but also provided one of the 
strongest indications that gravity itself may emerge from microscopic 
thermodynamic or informational principles.

Entropy therefore plays a central role in modern gravitational and cosmological
physics, establishing deep connections between geometry, quantum theory and
thermodynamics. In cosmological applications, entropy does not merely quantify
disorder or information, but often acts as a fundamental ingredient that
governs the dynamical laws of spacetime itself. This perspective has given rise
to a vast body of literature in which entropy enters cosmology either as a
microscopic quantity encoding the underlying quantum degrees of freedom or as
an effective macroscopic functional that captures aspects of quantum gravity
beyond our present understanding. Before addressing these applications, it is
therefore essential to review the various notions of entropy that emerge in
gravitational systems, together with their physical origins, theoretical
motivations and possible generalizations.

Accordingly, in this section we provide a unified overview of the entropy
concepts most relevant to cosmology, starting from black-hole thermodynamics
and horizon entropy and progressively extending to quantum, statistical and
gravitational generalizations. We emphasize both the remarkable universality of
the Bekenstein-Hawking area law and the systematic ways in which it can be
modified through quantum corrections, generalized statistical frameworks, or
extensions of the underlying theory of gravity. Our objective is not to
advocate a particular entropy functional, but rather to equip the reader with
the conceptual and technical background needed to understand the origin,
physical interpretation and cosmological implications of the different entropy
proposals discussed throughout this review. This framework will provide the
foundation for the following sections, where generalized entropy plays a key
role in spacetime thermodynamics, modified cosmological dynamics, holographic
dark-energy models and related approaches to gravitational and cosmological
physics.

\subsection{Black-hole entropy and horizon thermodynamics}
\label{EESectionBekensteinHawking}

Black holes are not merely mathematical solutions of the gravitational field 
equations but genuine thermodynamic systems. Any consistent physical 
description of a black hole must therefore endow it with well-defined 
thermodynamic properties, foremost among them entropy. For more than five 
decades, this entropy has been identified with the celebrated 
\textit{Bekenstein-Hawking entropy}, given by Eq.~\eqref{EEBekensteinHawking} 
below. 
Its most striking and conceptually profound property is that it scales with the 
\emph{area} of the event horizon rather than with the enclosed volume, a 
feature 
that sharply distinguishes gravitational systems from ordinary thermodynamic 
ones and that will play a frequent role throughout this review.

A key conceptual step towards understanding black-hole entropy is provided by 
the \textit{no-hair theorem}~\cite{Israel:1967wq,Carter:1971zc,Hawking:1971vc}, 
according to which any 
stationary black-hole solution of Einstein-Maxwell theory is completely 
characterised by only three macroscopic parameters: mass, electric charge and 
angular momentum. All other information about the matter that formed the black 
hole becomes permanently hidden behind the horizon. From a thermodynamic 
viewpoint, this dramatic loss of accessible information strongly suggests the 
existence of an entropy associated with microscopic degrees of freedom that are 
no longer observable.

This intuition was placed on firmer footing by Hawking’s \textit{area 
theorem}~\cite{Hawking:1971vc}, which states that, under classical evolution 
and 
assuming 
the null energy condition, the area of a black-hole horizon cannot decrease,  
namely
$dA \geq 0$. Moreover, during black-hole mergers the final horizon area exceeds 
the sum of the initial ones. This monotonic behaviour closely mirrors the 
second 
law of thermodynamics and naturally identifies the horizon area as the 
gravitational analogue of entropy.

The explicit connection between entropy and horizon area was first argued by 
Bekenstein in 1973~\cite{Bekenstein:1973ur}. Interpreting black-hole entropy as 
a 
measure of the information inaccessible to an external observer, similar to the 
fact that 
thermodynamic entropy quantifies missing microscopic information, Bekenstein 
concluded, on general physical grounds, that black-hole entropy must be 
proportional to the horizon area measured in Planck units.

The essence of Bekenstein’s reasoning can be summarised as follows. Let the 
black-hole entropy be a function of the horizon area,
\begin{equation}
S = f(A),
\end{equation}
with $f(A)$ a monotonically increasing function. The smallest possible entropy 
increase corresponds to the absorption of a single bit of information, 
$\Delta S_{\min}=\ln 2$. Consider a particle of rest mass $\mu$ and proper 
radius 
$b$ falling into the horizon. Then the corresponding increase of the horizon 
area is 
bounded from below by $(\Delta A)_{\min}=2\mu b$, independently of the black 
hole’s global parameters. One then obtains
\begin{equation}
\ln 2 = (\Delta A)_{\min}\frac{df}{dA},
\end{equation}
which, upon integration, yields an entropy proportional to the area,
\begin{equation}
S_{BH}=\gamma A,
\end{equation}
with $\gamma$ a dimensionless constant of order unity. Although the argument is 
formulated assuming spherical symmetry, encoded in the estimate of 
$(\Delta A)_{\min}$, its validity extends more generally, provided the minimal 
area increase remains independent of the detailed black-hole parameters.

Additionally, Bekenstein proposed the \textit{generalised second law} 
(GSL)~\cite{Bekenstein:1972tm,Bekenstein:1973ur,Bekenstein:1974ax}, according 
to 
which 
the sum of black-hole entropy and the entropy of matter and radiation outside 
the horizon never decreases, namely
\begin{equation}
dS_{\text{total}} = dS_{BH} + dS_{\text{matter}} \geq 0.
\end{equation}
The GSL plays a central role in establishing the thermodynamic consistency of 
black holes, especially in the presence of quantum processes.

The precise proportionality constant $\gamma$ was determined by Hawking in 
1975~\cite{Hawking:1975vcx}, following the discovery that black holes emit 
thermal 
radiation with a characteristic temperature, now known as the 
\textit{Hawking temperature}. This led to the definitive expression for the 
maximal entropy of a black hole, i.e. 
\begin{equation}\label{EEBekensteinHawking}
S_{BH}=\frac{c^{3}A}{4\hbar G}=\frac{A}{4l_{P}^{2}},
\end{equation}
where $A$ denotes the horizon area, $G$ is Newton’s constant, and 
$l_{P}=(\hbar G/c^{3})^{1/(D-2)}$ is the Planck length in $D$ dimensions 
(although originally derived for four-dimensional black holes within Einstein
gravity, the Bekenstein-Hawking relation admits a natural extension to
arbitrary spacetime dimensions and more general horizon geometries). In 
natural units, this reduces to the compact form $S_{BH}=A/4$, which we shall 
adopt in the remainder of this section. Relation~\eqref{EEBekensteinHawking} is 
commonly referred to as the \textit{area law}, although area-scaling entropies 
emerge in a much broader context beyond black-hole physics.

Finally, it is important to emphasise that the Bekenstein-Hawking entropy is a 
\emph{semiclassical} result, relying on the validity of Einstein gravity, weak 
curvature, and horizon regularity. Deviations from these assumptions, arising 
from quantum effects, modified gravitational dynamics, or non-trivial 
microscopic structure, are expected to induce corrections to the entropy. 
Remarkably, however, in a wide range of physical settings the leading area 
dependence persists, a fact that will resurface in the following subsections 
and 
will prove particularly significant in cosmological applications.

\subsection{Entropy bounds and holographic principle}
\label{EESectionEntropyBounds}

The recognition of black holes as thermodynamic systems naturally prompts the 
question of whether entropy plays a comparable role in more general 
gravitational 
settings. In particular, when extending the discussion beyond black holes to 
arbitrary matter configurations or to cosmological horizons, the 
 Bekenstein-Hawking entropy  acquires a broader interpretation: rather 
than characterising the entropy of a specific object, it acts as an upper bound 
on the entropy that any physical system can contain within a given spacetime 
region.

The first explicit formulation of such a bound was proposed by Bekenstein in 
1980~\cite{Bekenstein:1980jp}, building upon the 
 generalised second law  (GSL)~\cite{Bekenstein:1972tm}. He argued that 
any 
weakly gravitating, isolated matter system in asymptotically flat spacetime 
must 
satisfy
\begin{equation}
S \leq \frac{2\pi k E R}{\hbar c},
\end{equation}
where $E$ is the total mass-energy of the system and $R$ is the radius of the 
smallest sphere that can enclose it. This relation, known as the 
\textit{Bekenstein bound}, is remarkable in that it links entropy to purely 
macroscopic quantities, revealing an intimate connection between gravity, 
information, and thermodynamics.

Although the Bekenstein bound does not hold in complete 
generality~\cite{Unruh:1982ic,Unruh:1983ir}, it remains valid under additional 
physically 
reasonable assumptions, such as limited deviations from spherical 
symmetry~\cite{Wald:1999vt,Flanagan:1999jp}. In this restricted but physically 
relevant 
sense, it provides a robust guideline. Notably, the bound is saturated by a 
four-dimensional Schwarzschild black hole, for which $R=2E$ and the entropy 
reduces to the Bekenstein-Hawking value $S=A/4=\pi R^{2}$. Black holes thus 
emerge as the most entropic objects permitted by gravity.

A closely related, though conceptually distinct, proposal was introduced by 
Susskind in 1995~\cite{Susskind:1994vu}. The \textit{spherical entropy bound} 
states that the entropy of any matter system enclosed by a surface of area $A$ 
must satisfy
\begin{equation}\label{EEBound}
S_{\text{matter}} \leq \frac{A}{4}.
\end{equation}
The physical reasoning behind this bound is gravitational collapse: allowing 
more than $A/4$ independent degrees of freedom within a region of area $A$ 
would 
inevitably result in black-hole formation, whose entropy is itself given by 
$A/4$. From this perspective, entropy bounds encode a fundamental gravitational 
limit on information storage.

However, when one considers spacetimes containing more general horizons, 
such as cosmological horizons, the spherical entropy bound proves 
insufficient. 
To address this limitation, Bousso formulated the 
\textit{covariant entropy bound}~\cite{Bousso:1999xy}, which asserts that the 
entropy passing through any light-sheet $L(B)$ generated by a surface $B$ is 
bounded by the area of that surface, namely
\begin{equation}
S[L(B)] \leq \frac{A(B)}{4}.
\end{equation}
Unlike earlier proposals, this bound is fully covariant and applicable to 
arbitrary spacetime geometries. In this sense, it represents the most general 
entropy bound currently known and is expected to hold for all physically 
admissible horizons. The technical construction of light-sheets will not be 
reviewed here.

It is important to stress that all entropy bounds discussed above were derived 
within the framework of Einstein gravity. Consequently, their validity in 
modified or extended theories of gravity is not guaranteed and must be examined 
anew in each case. This issue will be revisited in subsequent subsections.

The emergence of entropy bounds and the universal area scaling of black-hole 
entropy strongly suggest that gravitational degrees of freedom behave in a 
fundamentally different manner from those of conventional local field theories. 
This observation culminated in the formulation of the 
\textit{holographic principle} by ’t~Hooft in 1993~\cite{tHooft:1993dmi}. The 
principle posits that, in any consistent theory of quantum gravity, the number 
of independent quantum states contained in a spatial region is bounded by the 
area of its boundary rather than by its volume.

Equivalently, the holographic principle may be stated as the requirement that 
the maximal number of degrees of freedom associated with a region of boundary 
area $A$ does not exceed $A/4$, corresponding roughly to one bit of information 
per Planck area. This principle is supported by two complementary arguments. 
The 
first is gravitational collapse, i.e. a system endowed with more than $A/4$ 
degrees 
of freedom cannot avoid forming a black hole. The second is unitarity. If a 
region were to contain $\sim V$ degrees of freedom, proportional to its volume, 
black-hole formation would seemingly reduce this number to $\sim A/4$, implying 
an irreversible loss of information. Preserving unitarity therefore requires 
that the number of fundamental degrees of freedom be bounded by the area from 
the outset.

The idea that quantum gravity may admit a lower-dimensional description 
predates 
the holographic principle itself. In particular, Klebanov and 
Susskind~\cite{Klebanov:1988ba} showed that a phase of $(2+1)$-dimensional 
lattice gauge 
theory in the large-$N$ limit reproduces the dynamics of free fundamental 
strings. Related insights were developed by Thorn~\cite{Thorn:1991fv}, who 
suggested that strings could emerge as composite objects built from more 
elementary degrees of freedom in a lower-dimensional framework.

These developments strongly indicated that gravitational systems may admit a 
holographic description in terms of lower-dimensional degrees of freedom. 
Their first concrete and mathematically precise realization was eventually 
achieved through the AdS/CFT correspondence, which we briefly review in the 
following subsection.


 \subsection{Holography and the AdS/CFT correspondence}
\label{EESectionAdSCFT}

The   AdS/CFT correspondence was proposed by Maldacena in 
1997~\cite{Maldacena:1997re}. The correspondence asserts that string theory, or 
more 
generally quantum gravity, formulated in a $(D+1)$-dimensional asymptotically 
anti-de Sitter (AdS) spacetime is exactly dual to a conformal field theory 
(CFT) living on its $D$-dimensional boundary. This duality provides a 
non-perturbative definition of quantum gravity in AdS backgrounds and 
constitutes one of the deepest realizations of holography currently known.

The original and most studied example relates type-IIB string theory on
$AdS_5\times S^5$ to $\mathcal{N}=4$ supersymmetric Yang-Mills theory in four 
dimensions with gauge group $SU(N)$~\cite{Maldacena:1997re,Gubser:1998bc,
Witten:1998qj}. In the large-$N$ and strong-coupling limit, the gauge theory 
admits a classical gravitational description in the bulk, while finite-$N$ and 
finite-coupling effects correspond to quantum and stringy corrections on the 
gravity side. The correspondence is therefore a strong/weak duality, allowing 
strongly coupled quantum systems to be studied through semiclassical geometry.

A central element of the duality is the UV/IR 
correspondence~\cite{Susskind:1998dq,Peet:1998wn}, according to which 
ultraviolet scales in 
the boundary theory correspond to infrared scales in the bulk geometry. In this 
sense, the radial AdS coordinate acquires the interpretation of an emergent 
renormalization-group scale. The holographic dictionary identifies the boundary 
value of a bulk field $\phi$ with a source for a dual operator $\mathcal{O}$ in 
the CFT, through the relation
\begin{equation}
Z_{\mathrm{grav}}[\phi_0]
=
\left\langle
\exp\left(
\int d^Dx\,\phi_0 \mathcal{O}
\right)
\right\rangle_{\mathrm{CFT}},
\end{equation}
where $\phi_0$ denotes the asymptotic boundary value of the bulk field. This 
relation, formulated explicitly in~\cite{Gubser:1998bc,Witten:1998qj}, provides 
the operational basis of the correspondence.

Entropy plays a central role in the AdS/CFT framework. In particular, thermal 
states of the boundary CFT are dual to black-hole geometries in the bulk, with 
the thermal entropy of the field theory identified with the Bekenstein-Hawking 
entropy of the corresponding AdS black hole~\cite{Witten:1998zw,Hawking:1982dh}. 
Thus, black-hole thermodynamics emerges 
naturally as a thermodynamic manifestation of holography. The Hawking-Page 
phase transition~\cite{Hawking:1982dh}, describing a transition between thermal 
AdS space and an AdS black hole, acquires a dual interpretation as a 
confinement/deconfinement transition in the gauge theory~\cite{Witten:1998zw}.

An even deeper connection between entropy and geometry emerged with the 
Ryu-Takayanagi proposal~\cite{Ryu:2006bv,Ryu:2006ef}, according to which the 
entanglement entropy of a spatial subregion $A$ in the boundary CFT is given by
\begin{equation}\label{RTformula}
S_A
=
\frac{\mathrm{Area}(\gamma_A)}
{4G_N^{(D+1)}},
\end{equation}
where $\gamma_A$ is the codimension-2 minimal surface in the AdS bulk whose 
boundary coincides with     that of the region $A$. Equation~\eqref{RTformula} 
is formally 
identical to the Bekenstein-Hawking entropy, but now interpreted as a measure 
of quantum entanglement. This result provided the first direct and quantitative 
connection between spacetime geometry and quantum information.
The proposal was later generalized to time-dependent geometries by Hubeny, 
Rangamani and Takayanagi~\cite{Hubeny:2007xt}, replacing minimal surfaces by 
extremal ones.

Furthermore, quantum corrections to holographic entanglement 
entropy were derived in~\cite{Faulkner:2013ana}, leading schematically to
\begin{equation}
S_A
=
\frac{\mathrm{Area}(\gamma_A)}{4G_N}
+
S_{\mathrm{bulk}}
+
\cdots,
\end{equation}
where $S_{\mathrm{bulk}}$ denotes the bulk entanglement entropy across the 
extremal surface. These developments strongly suggest that spacetime geometry 
itself may emerge from patterns of quantum 
entanglement~\cite{VanRaamsdonk:2010pw,Lashkari:2013koa,Faulkner:2013ica}.

An especially important result in this direction was obtained by Ryu and 
Takayanagi~\cite{Ryu:2006ef}, Casini \textit{et al.}~\cite{Casini:2011kv}, and 
subsequently by Lashkari \textit{et al.} and Faulkner \textit{et 
al.}~\cite{Lashkari:2013koa,Faulkner:2013ica}, who showed that the linearized 
Einstein equations in the bulk can be derived from the first law of 
entanglement 
entropy in the boundary theory. This observation provides one of the strongest 
indications that gravitational dynamics may emerge from quantum-information 
principles rather than constituting fundamental interactions.

Despite its original formulation in anti-de Sitter spacetime, holography is 
widely believed to possess a much broader validity. Extensions to de Sitter 
space~\cite{Strominger:2001pn,Anninos:2011ui}, flat-space 
holography~\cite{Bagchi:2010zz,Barnich:2010eb}, cosmological 
spacetimes~\cite{McFadden:2009fg}, tensor-network 
realizations~\cite{Swingle:2009bg}, and quantum-error-correcting 
interpretations~\cite{Almheiri:2014lwa,Pastawski:2015qua}, have significantly 
expanded the 
scope 
of the holographic paradigm. Although many aspects remain conjectural, the 
AdS/CFT correspondence has profoundly reshaped modern understanding of entropy, 
gravity, quantum field theory, and spacetime itself, establishing entropy and 
entanglement as fundamental ingredients in the microscopic description of 
geometry.

\subsection{Area law and its physical interpretation}
\label{EESectionAreaLaw}

Entropy relations that scale proportionally with the area of a spatial or 
causal 
boundary are collectively referred to as obeying an \textit{area law}. The most 
celebrated and physically significant realization of such behaviour is the 
Bekenstein-Hawking entropy~\eqref{EEBekensteinHawking} itself. What upgrades 
this 
relation 
from a striking coincidence to a genuine ``law'' is its remarkable 
universality. Area-scaling entropies repeatedly arise in a broad range of 
physical settings, 
including classical and quantum gravity, string theory, loop quantum gravity, 
quantum field theory, condensed matter systems, and lattice models. In most of 
these cases the area law constitutes the leading contribution to the entropy, 
while subleading terms encode additional dynamical or microscopic information.

The ubiquity of the area law naturally raises the fundamental question of what 
physical 
mechanism underlies its emergence across such disparate theoretical frameworks. 
Although the interpretation depends on the context, two complementary 
viewpoints 
have been established. In gravitational systems, the area law is widely 
interpreted 
as 
a manifestation of holography, reflecting a fundamental limitation on the 
number 
of independent degrees of freedom that can be accommodated within a region of 
spacetime. By contrast, in non-gravitational quantum systems, such as quantum 
field 
theories and condensed matter models, the area law originates from the locality 
of 
interactions and the structure of short-range quantum correlations.

In quantum mechanics and quantum field theory, entropy is defined through the 
quantum generalization of the Boltzmann-Gibbs entropy, namely the 
\textit{von Neumann entropy}
\begin{equation}\label{EEVonNeumannDefinition}
S(\rho) = -\mathrm{Tr}(\rho \ln \rho),
\end{equation}
where $\rho$ denotes the density matrix of the system, or the reduced density 
matrix associated with a spatial subregion. When applied to ground states of 
extended systems, this quantity measures the entanglement between a subregion 
and 
its complement and is therefore commonly referred to as \textit{entanglement 
entropy}. Although additional contributions may arise in more general 
situations, 
this terminology will be adopted throughout the present review.

\paragraph{Entanglement entropy in quantum field theory}

In quantum field theory, entanglement entropy quantifies the information loss 
induced by the presence of a geometrical boundary across which correlations are 
no 
longer accessible~\cite{Callan:1994py}. For this reason it is sometimes also 
called 
\textit{geometrical entropy}. The first explicit demonstration of an area law 
in 
this context was provided by Srednicki~\cite{Srednicki:1993im}, who numerically 
computed the entanglement entropy of a spherical region for a free scalar field 
in 
its ground state and showed that it scales with the area of the boundary. 
Closely 
related results were obtained earlier by Bombelli \textit{et 
al.}~\cite{Bombelli:1986rw} 
for quantum fields in black-hole backgrounds.

The physical origin of the area law in quantum field theory can be traced to 
the 
locality of correlations. For instance, for a free massless scalar field in $D$ 
spacetime dimensions, the two-point function behaves 
as~\cite{Casini:2009sr,Solodukhin:2011gn}
\begin{equation}\label{EEMasslessSCalarCorrelator}
\langle \phi(x)\phi(y)\rangle = \frac{\Omega_d}{|x-y|^{d-2}},
\end{equation}
with $\Omega_d=\Gamma\!\left(\frac{d-2}{2}\right)/(4\pi^{d/2})$. While 
correlations 
decay at large separations, they diverge in the ultraviolet limit 
$|x-y|\to\epsilon$. This short-distance behaviour leads to the dominant 
contribution to the entanglement entropy, namely
\begin{equation}\label{EEAreLawScalarMasslessQFT}
S \sim \frac{A(\Sigma)}{\epsilon^{d-2}},
\end{equation}
where $A(\Sigma)$ is the area of the boundary separating the region from its 
complement. The precise coefficient depends on the field content, such 
as scalars~\cite{Casini:2009sr,Solodukhin:2011gn}, 
fermions~\cite{Larsen:1994yt,Larsen:1995ax,Kabat:1995eq}, or gauge 
fields~\cite{Kabat:1995eq}. Detailed analytical results for free theories can be 
found 
in~\cite{Casini:2009sr}.

It is important to stress that in quantum field theory the area law represents 
the leading term of an expansion rather than an exact relation. Subleading 
corrections, which will be discussed later, encode further physical 
information. 
Nevertheless, the area law has been rigorously shown to provide an upper bound 
on 
the entropy of a wide class of systems, including $D$-dimensional harmonic 
lattice 
models approximating scalar quantum field 
theories~\cite{Plenio:2004he,Cramer:2006EEE,Cramer:2005mx}.

\paragraph{Conformal field theories}

Exact computations of entanglement entropy in generic quantum field theories 
are 
often technically challenging, especially in the presence of massive degrees of 
freedom~\cite{Katsinis:2017qzh,Katsinis:2019lis}. A particularly tractable and 
physically important class is provided by conformal field theories (CFTs), 
which 
are invariant under conformal transformations~\cite{Qualls:2015qjb} and 
therefore 
describe massless excitations.

In $(1+1)$-dimensional CFTs, the entanglement entropy of an interval of length 
$L$ 
takes the universal 
form~\cite{Fiola:1994ir,Holzhey:1994we,Calabrese:2004eu,Callan:1994py}
\begin{equation}\label{EECFT2D}
S \sim c\,\ln (L/\epsilon) + \mathcal{O}(L^{0}),
\end{equation}
with $c=1/6$ for bosons and $c=1/12$ for fermions. In higher dimensions, the 
leading contribution again scales with the area of the 
boundary~\cite{Callan:1994py,Ryu:2006ef}, i.e.
\begin{equation}\label{EECFTDgr2}
S = \frac{V_{D-2}}{(2\sqrt{\pi})^{D-2}(D/2-1)}\,\epsilon^{2-D} + \cdots .
\end{equation}
Beyond their intrinsic interest, CFT techniques have proven valuable in 
describing 
near-horizon symmetries and deformations of black-hole 
horizons~\cite{Carlip:1998wz}, a topic to which we shall return.

\paragraph{Quantum fields in black-hole backgrounds}

A long-standing question is whether the Bekenstein-Hawking entropy itself can 
be 
interpreted as an entanglement entropy. While a definitive answer remains open, 
it is natural to consider the entanglement entropy of quantum fields outside a 
black-hole horizon, since interior degrees of freedom are causally inaccessible 
to external observers. In this sense, the horizon acts as a geometrical 
partition.

This idea was first explored by Bombelli \textit{et 
al.}~\cite{Bombelli:1986rw}, 
who showed that the entanglement entropy of a scalar field outside the horizon 
contributes to the total entropy, namely
\begin{equation}
S_{\text{tot}} = S_{BH} + S_{\text{ent}},
\end{equation}
in agreement with the generalized second law. Subsequent studies confirmed that 
the leading entanglement contribution near the horizon scales with the area, 
for 
scalar fields~\cite{tHooft:1984kcu,Frolov:1993ym} as well as for fermionic and 
gauge 
fields~\cite{Kabat:1995eq,Carlip:1998wz}. These analyses, however, typically 
rely 
on Einstein gravity and assumptions such as horizon regularity.

Taken together, these diverse realizations strongly suggest that the area law 
captures a fundamental property of quantum correlations in both gravitational 
and non-gravitational systems. Understanding how this behaviour persists and 
how 
it is modified by quantum, statistical, or dynamical effects, will be central 
to 
the discussions that follow.

 \subsection{Entanglement entropy and quantum field theoretic 
origins}
\label{EESectionEntanglementOrigins}

The interpretation of entropy as a manifestation of quantum entanglement and 
information loss provides one of the deepest modern explanations for the 
emergence of area-law behaviour in both gravitational and non-gravitational 
systems. From this perspective, entropy does not primarily count microscopic 
configurations in the conventional thermodynamic sense, but instead quantifies 
the information inaccessible to a given observer after tracing out part of the 
system. This viewpoint is particularly powerful in quantum field theory and 
many-body physics, where entanglement is an intrinsic property of vacuum and 
low-energy states. Remarkably, the same conceptual structure reappears in black 
hole physics, holography and quantum gravity, strongly suggesting that quantum 
entanglement might lie at the heart of spacetime thermodynamics itself.

\paragraph{Many-body systems}

Area-law scaling of entanglement entropy is a well-established feature of a 
wide class of quantum many-body systems, provided certain physically reasonable 
conditions are satisfied. In particular, locality of interactions and a 
sufficiently fast decay of correlations ensure that the entanglement entropy of 
a spatial region grows proportionally to the area of its boundary rather than 
to its volume. More precisely, the area law is expected to hold when 
(i) correlations decay rapidly with distance and 
(ii) the density of low-energy states does not increase exponentially with the 
system volume~\cite{Masanes:2009tg}.

Under these conditions, harmonic, bosonic and fermionic lattice systems away 
from criticality exhibit entropy scaling linear in the boundary area. While 
early investigations focused mainly on one-dimensional systems, where exact 
results are often available, the area law has also been shown to persist in 
higher dimensions, with subleading corrections encoding additional physical 
information. A comprehensive review of entanglement entropy in many-body 
systems can be found in~\cite{Eisert:2008ur}.

\paragraph{String theory}

From the viewpoint of quantum gravity, it is natural to expect that the 
Bekenstein-Hawking entropy should admit a microscopic derivation within string 
theory, since Einstein gravity emerges as its low-energy effective 
description. A first general argument supporting this expectation was advanced 
by Susskind and Uglum~\cite{Susskind:1994sm}, who argued that black-hole 
entropy should originate from fundamental string degrees of freedom. Related 
insights emerged from the connection between string theory and 
$\sigma$-models in quantum field theory~\cite{Kabat:1995jq,Kabat:1995eq}, 
suggesting that the effective infrared 
description naturally reproduces area-law scaling.

A decisive breakthrough occurred with the development of D-brane techniques, 
which enabled Str\"ominger and Vafa to exactly compute the entropy of a 
five-dimensional extremal black hole through microscopic counting of BPS bound 
states~\cite{Strominger:1996sh}. Their result precisely reproduced the 
Bekenstein-Hawking entropy and provided the first explicit microscopic 
derivation of the black-hole area law in string theory. This success was later 
extended to non-extremal black 
holes~\cite{Callan:1996dv,Horowitz:1996ay,Sfetsos:1997xs}, rotating 
configurations~\cite{Breckenridge:1996is,Breckenridge:1996sn}, and 
four-dimensional black 
holes~\cite{Maldacena:1996gb,Johnson:1996ga}.

Area-law entropy has also been reproduced in 
M-theory~\cite{Maldacena:1997de,Vafa:1997gr,Horowitz:2007xq,Castro:2008ne}, 
matrix 
models~\cite{Banks:1997hz,Banks:1997tn,Klebanov:1988ba,Horowitz:1997fr,
Kabat:1997im}, 
and twisted-sector constructions~\cite{Dabholkar:2001if}. In practice, 
however, direct microstate counting is often technically demanding, and many 
results are obtained indirectly through dual conformal field theories or 
holographic arguments. For detailed reviews we refer the reader 
to~\cite{Mohaupt:2000mj,Solodukhin:2011gn,Lust:2018cvp,DeHaro:2019gno}.

\paragraph{Holography}

A conceptually distinct but closely related understanding of entanglement 
entropy emerged with the advent of holography. As we mentioned above, within 
the AdS/CFT 
correspondence~\cite{Maldacena:1997re}, Ryu and Takayanagi proposed a 
geometric prescription for computing the entanglement entropy of a region $A$ 
in a $D$-dimensional conformal field theory, namely 
relation~\eqref{RTformula}~\cite{Ryu:2006bv}. Remarkably, this expression is 
formally identical to the Bekenstein-Hawking entropy, with the area computed 
in the bulk geometry.

The Ryu-Takayanagi proposal established a  connection between quantum 
information and spacetime geometry, enabling the computation of entanglement 
entropy in strongly coupled quantum field 
theories~\cite{Ryu:2006ef,Casini:2011kv}. In particular, Casini \textit{et 
al.}~\cite{Casini:2011kv} demonstrated that the entanglement entropy of a 
spherical 
region in a vacuum CFT can be mapped, through a conformal transformation, to 
the thermal entropy of a black hole in hyperbolic spacetime. Related 
developments were presented in~\cite{Jensen:2013lxa}.

An especially intriguing aspect of holographic entanglement entropy is its 
direct identification with black-hole entropy when the minimal surface 
coincides with a horizon, either in thermal CFT states~\cite{Ryu:2006bv} or in 
CFTs containing black holes~\cite{Emparan:2006ni}. A fully covariant extension 
was later formulated in~\cite{Hubeny:2007xt}, inspired by the covariant 
entropy bound~\cite{Bousso:1999xy,Bousso:2002ju}. Extensions to spacetimes 
with positive curvature, including de Sitter backgrounds, have also been 
explored in~\cite{Hawking:2000da}. For comprehensive reviews 
see~\cite{Nishioka:2009un,Rangamani:2016dms}.

Although holographic dualities and the Ryu-Takayanagi prescription have not 
been rigorously proven in full generality, compelling evidence strongly 
supports their validity. In particular, Lewkowycz and 
Maldacena~\cite{Lewkowycz:2013nqa} derived holographic entanglement entropy 
using 
Euclidean gravitational path integrals, placing the proposal on firmer 
theoretical grounds.

\paragraph{Generalized entropy, quantum extremal surfaces and the Page curve}

A major recent development in the interplay between entropy, holography and 
quantum gravity emerged from renewed investigations of the black-hole 
information paradox and the calculation of the Page curve for evaporating black 
holes. In semiclassical gravity, the relevant quantity is not simply the 
Bekenstein-Hawking entropy, but the \textit{generalized 
entropy}~\cite{Bekenstein:1974ax,Engelhardt:2014gca},
\begin{equation}
S_{\rm gen}
=
\frac{\mathrm{Area}(\partial I)}{4G_N}
+
S_{\rm matter}(R\cup I),
\end{equation}
where $R$ denotes the Hawking radiation region, $I$ an ``island'' region in 
the bulk, and $S_{\rm matter}$ the von Neumann entropy of quantum fields 
outside the corresponding quantum extremal surface. The entropy of Hawking 
radiation is then determined through the island 
prescription~\cite{Almheiri:2019psf,Almheiri:2019qdq,Penington:2019npb} 
\begin{equation}
S(R)
=
\min
\left\{
\mathrm{ext}_{I}
\left[
\frac{\mathrm{Area}(\partial I)}{4G_N}
+
S_{\rm matter}(R\cup I)
\right]
\right\},
\end{equation}
which successfully reproduces the unitary Page curve and resolves the apparent 
information-loss paradox at the semiclassical level.

These developments arise from replica wormhole contributions in the Euclidean 
gravitational path integral, and provide strong evidence that spacetime 
geometry, entropy and quantum entanglement are fundamentally interconnected. 
In particular, the appearance of area contributions within the generalized 
entropy once again highlights the central role of holography and horizon 
thermodynamics in the microscopic description of gravitational systems.

While holography and string theory suggest that spacetime geometry and entropy 
emerge from quantum entanglement and microscopic degrees of freedom, canonical 
approaches to quantum gravity provide an alternative perspective in which 
entropy arises directly from the quantization of geometry itself.

\paragraph{Canonical quantisation and loop quantum gravity}

Within loop quantum gravity (LQG), the area law extends beyond stationary black
hole horizons to a broader class of \textit{isolated horizons}, including
cosmological horizons. Early work by Rovelli~\cite{Rovelli:1996dv} and
Ashtekar~\cite{Ashtekar:1997yu} showed that the entropy of a large,
non-rotating black hole takes the form
\begin{equation}
\label{EELQGAreaLawBH}
S_{\mathrm{bh}}
=
\frac{\gamma_{0}}{4\ell_{P}^{2}\gamma}A_S,
\qquad
\gamma_{0}=\frac{\ln 2}{\pi\sqrt{3}},
\end{equation}
where $\gamma$ is the Immirzi parameter and $A_S$ denotes the horizon area.
Choosing $\gamma=\gamma_{0}$ reproduces the Bekenstein-Hawking
entropy~\eqref{EEBekensteinHawking}. Generalizations to isolated horizons in
four dimensions~\cite{Ashtekar:2000eq} and in arbitrary
dimensions~\cite{Bodendorfer:2013sja,Wang:2014cga} yield analogous results.
Moreover, for non-extremal black holes with near-horizon Rindler geometry, the
area law can be recovered without explicit dependence on the Immirzi
parameter~\cite{Bianchi:2012ui,Frodden:2011eb}. Collectively, these results
provide strong evidence that area-law entropy is a robust prediction of
canonical quantum gravity.

Beyond the standard statistical treatment of LQG horizon microstates,
non-extensive statistics can also be incorporated into the counting of
quantum-geometric degrees of freedom. This leads to a generalized LQG horizon
entropy of the form
\cite{Majhi:2017zao,Czinner:2015eyk,Mejrhit:2019oyi,Liu:2021dvj}
\begin{equation}
S_{\rm LQG}
=
\frac{1}{1-q}
\left[
\exp\!\left((1-q)\Lambda(\gamma)S_{\rm BH}\right)-1
\right],
\label{SPCG22}
\end{equation}
where $q$ is the non-extensive entropic index, while the dependence on the
Barbero-Immirzi parameter is encoded through
\begin{equation}
\Lambda(\gamma)
=
\frac{\ln 2}{\sqrt{3}\,\pi\,\gamma}.
\label{LQGLambda}
\end{equation}
Thus, this construction combines the microscopic quantum-geometric origin of
LQG entropy with a generalized statistical description through the additional
parameter $q$. In the limit $q\rightarrow1$, the entropy reduces to
$\Lambda(\gamma)S_{\rm BH}$, and for
$\gamma=\ln 2/(\sqrt{3}\pi)$ one recovers the standard
Bekenstein-Hawking entropy.

\subsection{Entropy from quantum gravity approaches}
\label{EESectionQGEntropy}

The Bekenstein-Hawking entropy~\eqref{EEBekensteinHawking} and the associated 
entropy bounds, as originally formulated by Bekenstein~\cite{Bekenstein:1980jp} 
and Bousso~\cite{Bousso:1999xy}, are derived within a semiclassical framework, 
namely 
gravity is treated classically, while matter fields are quantised. From the 
standpoint of a fundamental theory, however, this separation is only an 
effective approximation. Once gravity itself becomes a quantum system, the 
area law is not expected to disappear (its universality is precisely what 
makes 
it so remarkable) but rather to acquire subleading contributions encoding the 
microscopic structure of spacetime. Importantly, such corrections arise in 
essentially all approaches to quantum gravity, despite the fact that the 
underlying mechanisms differ substantially. In the following we briefly review 
how string theory, holographic corrections, and loop quantum gravity 
corrections 
modify the classical 
entropy formula, and we finally discuss the conceptually distinct proposal of 
Barrow entropy, which effectively incorporates quantum-gravity effects through 
a geometrical deformation of the horizon.

\paragraph{Higher-curvature and stringy corrections}

As a UV-complete framework, string theory provides a natural arena for 
investigating quantum corrections to black-hole entropy. As we mentioned in 
the previous subsection, in the early work of 
Susskind and Uglum~\cite{Susskind:1994sm}, the Bekenstein-Hawking entropy was 
interpreted as arising from string world-sheets intersecting the horizon, with 
the leading contribution originating from genus-zero configurations and 
reproducing the classical area law. In general, corrections arise through two 
independent expansions intrinsic to string theory. The first is the string 
coupling expansion in $g_s$, where higher-genus world-sheets encode genuine 
quantum effects. The second is the $\alpha'$ expansion, with 
$\alpha'=l_s^2$, which captures finite string-length effects and generates 
higher-derivative corrections in the low-energy effective gravitational 
action. Both mechanisms produce subleading contributions beyond the classical 
area term.

Explicit corrections can be computed in controlled settings, often with the 
help of dual descriptions. For example, in heterotic string theory, tree-level 
$\alpha'$ corrections modify the entropy of supersymmetric black holes 
according to
\begin{equation}
S_{\mathrm{W}} = 2\pi \sqrt{n w \tilde{N}}
\left(1+\frac{2}{\tilde{N}}\right),
\end{equation}
where $n$ denotes the momentum, $w$ the fundamental string winding number, and 
$\tilde{N}$ an appropriate combination of brane and monopole 
charges~\cite{Cano:2018qev,Cano:2021nzo,Elgood:2020xwu,DominisPrester:2008ynb}. 
Analogous expressions arise for non-supersymmetric configurations, typically 
with additional charge-dependent structure. More generally, once 
higher-curvature terms are included, entropy is no longer determined solely by 
the horizon area and must instead be computed using the Wald entropy 
prescription~\eqref{EEWaldEntropyDefinition}. Detailed discussions of 
higher-curvature and string-induced corrections can be found 
in~\cite{Mohaupt:2000mj,Polchinski:1998rr,DeHaro:2019gno}.

An important class of corrections arises from logarithmic contributions. 
Motivated by the correspondence between black holes and highly excited string 
states~\cite{Horowitz:1996nw}, Solodukhin showed in~\cite{Solodukhin:1997yy} 
that both free string theory and Schwarzschild black 
holes exhibit identical logarithmic corrections of the form
\begin{equation}
\delta S \sim \ln(l_s M_{BH}),
\label{auxiliary22}
\end{equation}
where $l_s$ is the string length and $M_{BH}$ the black-hole mass. Such terms 
suggest that black-hole entropy effectively captures the degeneracy of 
interacting string states. Similar logarithmic contributions arise from 
one-loop effects in supersymmetric black holes, from Cardy-type conformal field 
theory calculations~\cite{Carlip:2000nv}, and from matter multiplet 
contributions near the horizon in $\mathcal{N}=4$ 
supergravity~\cite{Banerjee:2010qc}.

String theory also predicts constant, topology-related contributions to the 
entropy. Such terms appear naturally in the study of extremal black holes in 
M-theory~\cite{Maldacena:1997de,Vafa:1997gr}, both from macroscopic and 
microscopic analyses, and are closely connected to topological sectors of the 
underlying theory.

Finally, an especially intriguing class of corrections originates from 
T-duality and the associated double-field-theory framework. Since string 
theory cannot resolve distances below the string scale, theories at inverse 
length scales become physically equivalent. This leads to genuinely 
non-perturbative modifications of black-hole 
entropy~\cite{Nicolini:2019irw,Pourhassan:2019luf},
\begin{equation}
S\left(r_{+}\right)=\frac{A_{+}}{4} 
\left[\left(1-\frac{8 \pi l_{0}^{2}}{A_{+}}\right) 
\sqrt{1+\frac{4 \pi l_{0}^{2}}{A_{+}}}\right.\\
+\left.\frac{12 \pi l_{0}^{2}}{A_{+}}
\left(\operatorname{arsinh} \sqrt{\frac{A_{+}}{4 \pi l_{0}^{2}}}
-\operatorname{arsinh} \sqrt{2}\right)\right],
\label{auxiliary11}
\end{equation}
where $l_{0}$ denotes the zero-point length and $A_{+}$ the horizon area. 
These corrections encode the existence of a fundamental minimal length and 
therefore provide a direct manifestation of genuinely stringy spacetime 
structure.

\paragraph{Quantum and holographic corrections}

For a spherical entangling region, the entanglement entropy generally takes the 
form
\begin{equation}
S = \frac{C}{4G_N^{(D+1)}} \sum_i c_i 
\left(\frac{r}{\epsilon}\right)^{D-2i}
+ c_* \ln\!\left(\frac{r}{\epsilon}\right),
\end{equation}
where logarithmic terms appear in even boundary dimensions and agree with the 
expectations from conformal field 
theory~\cite{Ryu:2006ef,Casini:2009sr,Solodukhin:2011gn}. In addition, constant 
terms 
associated with topological entanglement entropy may also 
arise~\cite{Kitaev:2005dm}. 

Bulk quantum effects further modify the entropy functional. In particular, 
Faulkner \textit{et al.}~\cite{Faulkner:2013ana} showed that the one-loop 
corrected holographic entropy takes the schematic form
\begin{equation}
S_q = S_{\text{bulk-ent}} + \frac{\delta A}{4G_N}
+ \langle \Delta S_{\text{Wald-like}} \rangle
+ S_{\text{counterterms}},
\end{equation}
where the various contributions encode bulk entanglement across the extremal 
surface, geometric backreaction, higher-curvature effects, and renormalisation 
counterterms. Extensions to de Sitter spacetimes have also been developed, 
yielding both divergent and finite corrections governed by the Hubble 
scale~\cite{Hawking:2000da,Alishahiha:2004md,Casini:2010kt}.

\paragraph{Discrete geometry and loop corrections}

Canonical approaches to quantum gravity, and in particular loop quantum gravity 
(LQG), provide a conceptually distinct mechanism for entropy corrections. In 
LQG, geometric operators possess discrete spectra, and entropy arises from 
counting the microscopic quantum-geometric configurations compatible with a 
given horizon. In the macroscopic limit, the leading contribution reproduces 
the Bekenstein-Hawking area law, but one generically expects a subleading 
structure of the form
\begin{equation}\label{EEEntropyGeneralCorrectionLQGclean}
S_H = S_{BH}
+ \alpha \sqrt{S_{BH}}
+ \beta \ln S_{BH}
+ C
+ \mathcal{O}\!\left(e^{-\delta S_{BH}}\right),
\end{equation}
where $S_{BH}=A/(4\ell_P^2)$ and $\alpha$, $\beta$, $C$, and $\delta$ are 
dimensionless 
coefficients~\cite{Perez:2017cmj,Rovelli:2008zza,Rovelli:2010bf,
Chatterjee:2020iuf}. These 
terms encode quantum fluctuations of geometry and, depending on the specific 
framework, contributions from matter and gauge degrees of freedom associated 
with the horizon. 
In particular, the  logarithmic corrections have attracted particular 
attention. Their coefficient 
depends on the statistical ensemble and on the treatment of horizon gauge 
symmetries and constraints, with values such as $\beta=-1/2$ or 
$\beta=-3/2$ commonly 
reported~\cite{Jing:2000yn,Domagala:2004jt,Kaul:2000kf,Das:2000bx,Kaul:2012pf}. 
Constant and exponentially suppressed terms have also been identified, although 
their phenomenological relevance becomes significant mainly for microscopic or 
near-Planckian horizons.

\paragraph{Barrow entropy}

A conceptually different modification of horizon entropy was proposed by 
Barrow~\cite{Barrow:2020tzx}, motivated by the possibility that quantum-gravity 
effects induce a fractal-like deformation of horizon geometry. In this picture, 
even mild geometric irregularities modify the effective scaling of entropy with 
area, leading to
\begin{equation}\label{EEBarrowEntropyClean}
\mathcal{S}_{\mathrm{B}}
= \left(\frac{A}{4G}\right)^{1+\Delta/2},
\end{equation}
where the deformation parameter $\Delta\in[0,1]$. The standard area law is 
recovered for $\Delta=0$, while larger values correspond to increasingly rough 
or intricate horizon geometries. Although the resulting functional form is 
formally related to Tsallis-type non-extensive entropies for suitable parameter 
choices, the underlying physical origin here is geometrical rather than 
statistical.

\subsection{ Corrected entropy expressions} 
\label{Correctedentropy1}

Although the Bekenstein-Hawking entropy successfully incorporates the leading 
thermodynamic behavior of gravitational horizons, it is now widely understood 
that it represents only the first term of a more general entropy expansion. In 
essentially all microscopic approaches to gravity, as well as in quantum field 
theory, entanglement entropy, holography, and generalized statistical 
frameworks, additional subleading contributions naturally emerge. These 
corrections become particularly relevant near the Planck scale, in regimes of 
strong curvature, or whenever the underlying microscopic structure of spacetime 
cannot be neglected.

As we discussed in the previous subsections,  entropy corrections do 
not arise from a single physical origin. 
Some are associated with ultraviolet divergences and quantum entanglement near 
horizons, others originate from topology, conformal symmetry, holography, or 
non-perturbative quantum effects, while additional modifications may emerge 
from   minimal-length physics, or modified gravity 
theories. Nevertheless, despite their different derivations, many of these 
approaches exhibit strikingly similar correction structures, most notably 
logarithmic, power-law, square-root, constant, and exponential terms. This 
remarkable universality strongly suggests that the entropy-area relation is not 
fundamental in itself, but rather the semiclassical limit of a deeper 
microscopic description.

In this subsection we review the main classes of corrected entropy expressions 
that appear in the literature, emphasizing both their physical origin and 
their relevance for gravitational and cosmological applications.

\subsubsection{Quantum field theoretic and entanglement corrections}

Various subleading corrections to the  Bekenstein-Hawking entropy arise once 
effects 
such as finite mass, excitations, interactions, curvature, topology, or finite 
temperature are taken into account. These corrections encode important 
information about the microscopic structure of quantum correlations and the 
short-distance behavior of spacetime.

In free quantum field theories, logarithmic corrections appear generically in 
even dimensions~\cite{Casini:2006hu,Casini:2009sr}. For instance, for a 
massive scalar field in four dimensions one obtains~\cite{Solodukhin:2011gn}
\begin{equation}
S_{m\neq 0}=
\frac{A(\Sigma)}{12(4\pi)}
\left(
\frac{1}{\epsilon^{2}}
+2m^{2}\ln\epsilon
+m^{2}
+O(1)
\right),
\end{equation}
where $m$ denotes the field mass and $\epsilon$ the ultraviolet cutoff. 
Additional constant contributions arise in inverse-mass 
expansions~\cite{Katsinis:2017qzh}, while finite-temperature corrections were 
studied in~\cite{Katsinis:2019lis}. Excited states further induce power-law 
corrections 
of the form~\cite{Das:2007mj,Das:2008sy}
\begin{equation}
S = c_{1}\frac{A}{\epsilon^{2}}
+ c_{2}\left(\frac{A}{\epsilon^{2}}\right)^{-\gamma},
\end{equation}
with $\gamma\in[0,1]$, while interaction terms generate both divergent and 
finite contributions~\cite{Solodukhin:2009sk}.

Quantum fields propagating on black-hole geometries also produce subleading 
entropy corrections through one-loop effects~\cite{Kabat:1995eq}. In even 
dimensions, logarithmic contributions again 
appear~\cite{Solodukhin:1994st,Cognola:1995km,Mann:1996bi,Sen:2012dw}, 
indicating that such terms are a   robust feature across different 
quantum-field-theoretic settings.

Conformal field theories provide particularly clean realizations of these 
structures. In four dimensions the entropy typically takes the form
\begin{equation}
S
=
c_{1}\frac{A}{\epsilon^{2}}
+c_{2}\ln\epsilon
+O(\epsilon^{0}),
\end{equation}
with the coefficients depending on the specific CFT under 
consideration~\cite{Solodukhin:2008dh,Solodukhin:2011gn,Ryu:2006ef}. At finite 
temperature, 
additional exponentially suppressed corrections 
arise~\cite{Cardy:2014jwa,Herzog:2014fra,Herzog:2014tfa,Herzog:2015cxa}, while 
alternative entanglement measures such as logarithmic negativity have also 
been extensively studied~\cite{Calabrese:2014yza}.

Another modification that arises from  
quantum-entanglement considerations is the power-law one.
In particular, when the quantum state of a 
field is not restricted to its ground state but includes excited-state 
contributions, the entanglement entropy receives power-law corrections to the 
area law~\cite{Das:2007mj}. The resulting entropy can be written schematically 
as
\begin{equation}
\label{powerlawentropy}
S=S_{BH}+c\,S_{BH}^{-\gamma},
\end{equation}
where $c$ is a constant and $\gamma>0$ characterizes the strength of the 
correction. Unlike logarithmic terms, which typically originate from quantum 
fluctuations around equilibrium configurations, power-law corrections become 
important when the entanglement between ground and excited states is 
significant. Since the correction term decreases with increasing horizon area, 
the standard Bekenstein-Hawking relation is naturally recovered in the 
large-area semiclassical limit.

Finally, related correction structures appear in holographic and lattice 
systems. In particular, conformal descriptions of black-hole horizons lead, 
through the Cardy formula~\cite{Cardy:1986ie}, to logarithmic corrections of 
the form $\delta 
S_{\log}=-\frac{3}{2}\ln(A/4G)$~\cite{Carlip:1998wz,Carlip:2000nv}, while 
higher-dimensional generalizations 
are described by the Cardy-Verlinde formula~\cite{Verlinde:2000wg}. Moreover, 
fermionic lattice systems exhibit logarithmic violations of the area law, with 
entanglement entropy satisfying bounds of the 
form~\cite{Barthel:2006ct,Gioev:2006zz,Wolf:2006zzb,Eisert:2008ur}
\begin{equation}
c_{1}L^{D-1}\ln L
\leq S \leq
c_{2}L^{D-1}(\ln L)^{2}.
\end{equation}
Collectively, these results demonstrate that although the area law is 
remarkably universal, its subleading corrections carry highly non-trivial 
information about quantum correlations, topology, field content, and the 
microscopic structure of spacetime.

\subsubsection{Logarithmic and topological corrections}

Among the various modifications to the Bekenstein-Hawking entropy, 
logarithmic and topological corrections are particularly important since they 
appear universally across a wide range of quantum-gravitational frameworks. 
Unlike model-dependent higher-order terms, these corrections arise repeatedly 
in string theory, conformal field theory, loop quantum gravity, Euclidean 
quantum gravity, entanglement entropy calculations, and holographic 
constructions, suggesting that they capture generic features of the microscopic 
structure of spacetime.

\paragraph{Logarithmic corrections.}

Logarithmic terms constitute the most common subleading modification to the 
area law and typically appear in the schematic form
\begin{equation}
S = S_{BH} + \alpha \ln S_{BH} + \cdots ,
\label{Logarithmic11}
\end{equation}
where $\alpha$ is a dimensionless coefficient depending on the underlying 
theory and ensemble. In string theory, such corrections emerge naturally from 
quantum fluctuations in the perturbative expansion in the string coupling 
$g_s$. Motivated by the correspondence between strings and black 
holes~\cite{Horowitz:1996nw}, it was showed in \cite{Solodukhin:1997yy} 
that 
free string theory and Schwarzschild black holes exhibit identical logarithmic 
corrections of the form given in
Eq.~(\ref{auxiliary22}).

Similar logarithmic structures arise in a variety of other approaches. In 
particular, one-loop corrections in $\mathcal{N}=4$ supergravity, conformal 
field theory methods based on the Cardy and Cardy-Verlinde 
formulas~\cite{Carlip:2000nv}, matter multiplets propagating near black-hole 
horizons~\cite{Banerjee:2010qc}, loop quantum gravity, Euclidean path-integral 
methods, 
and entanglement entropy calculations all produce analogous logarithmic terms. 
The recurrent appearance of these contributions strongly suggests that they 
represent universal quantum corrections to the classical area law.

\paragraph{Topological corrections.}

In addition to logarithmic terms, many quantum-gravity frameworks predict 
constant entropy contributions independent of the horizon area. Such terms are 
usually interpreted as topological corrections, since they depend on global 
properties of the spacetime or compactification manifold rather than on local 
geometric quantities. 
An especially important class of such corrections consists of constant, 
topological contributions of the form
\begin{equation}
S=c_{1}\frac{A}{\epsilon^{2}}+c_{2},
\end{equation}
which remain finite even when the entangling area vanishes. These terms are 
associated with the so-called topological entanglement 
entropy~\cite{Kitaev:2005dm,Hamma:2005EEE}, defined through combinations such as
\begin{equation}
S_{\text{top}}
=
S(A)+S(B)-S(A\cup B)-S(A\cap B),
\end{equation}
and depend only on the topology of the entangling surface rather than on local 
geometric details~\cite{Levin:2006zz,Donnelly:2011hn}. Similar constant terms 
appear in a variety of holographic and entanglement  
constructions~\cite{Ryu:2006bv,Dong:2008ft,Tsilioukas:2023tdw,
Petronikolou:2025mlm}.
 
Additionally, constant corrections were identified, for example, in extremal 
black holes in 
M-theory~\cite{Maldacena:1997de,Vafa:1997gr}, where they arise both from 
microscopic state counting and from macroscopic effective-action analyses. 
Related topological terms also appear in entanglement entropy and holographic 
constructions, where they are associated with topological entanglement 
entropy~\cite{Kitaev:2005dm,Levin:2006zz}. Since these contributions remain 
finite 
even when the horizon area becomes small, they encode genuinely global and 
non-local information about the microscopic quantum structure of spacetime.

\paragraph{T-duality and non-perturbative corrections.}

String theory further predicts genuinely non-perturbative entropy corrections
associated with T-duality, namely the invariance of the theory under the
exchange of large and small length scales. Since strings cannot probe
distances below the string scale, T-duality effectively introduces a
fundamental minimal length and modifies the entropy-area relation at short
distances.

In the double-field-theory framework, these effects lead to modified entropy
expressions of the form of (\ref{auxiliary11})
\cite{Nicolini:2019irw,Pourhassan:2019luf}, where $A_{+}$ denotes the horizon
area and $l_0\propto\sqrt{\alpha'}$ is the zero-point length associated with
the Regge slope $\alpha'$. These corrections become important near the string
scale, while the standard Bekenstein-Hawking entropy is recovered in the
semiclassical limit.

A related implementation of T-duality effects, particularly useful for
gravitational and cosmological applications, can be formulated through the
zero-point-length modification of the Newtonian potential. In units where
$G=1$, the corresponding potential takes the form
\begin{equation}
\Phi(r)=-\frac{M}{\sqrt{r^2+l_0^2}},
\label{Tdualpotential}
\end{equation}
which regularizes the short-distance gravitational interaction. The
associated entropy of a spherical horizon of radius $R$ then satisfies
\cite{Jusufi:2022mir,Luciano:2024mcn}
\begin{equation}
dS_h=2\pi R
\left(1+\frac{l_0^2}{R^2}\right)^{-3/2}dR .
\label{STDentropy}
\end{equation}
Hence, the zero-point length modifies the standard horizon entropy at small
scales, while in the limit $l_0\rightarrow0$ one recovers
$dS_h=2\pi R\,dR$, and therefore the usual Bekenstein-Hawking area law.

\subsubsection{Quantum-gravity  induced corrections}

Beyond logarithmic and topological terms, several approaches to quantum gravity 
predict additional subleading modifications to the Bekenstein-Hawking entropy, 
associated with genuinely quantum properties of spacetime and horizon 
microstructure. In contrast to perturbative higher-curvature corrections, these 
terms are typically connected to the existence of a minimal length or area 
scale, the discreteness of quantum geometry, or non-perturbative horizon 
degrees of freedom. Although their explicit form depends on the underlying 
framework, a number of characteristic correction structures repeatedly emerge 
in the literature.

\paragraph{Square-root corrections.}

A first class consists of square-root corrections of the   form
\begin{equation}
S = S_{BH}+\alpha \sqrt{S_{BH}}+\cdots ,
\end{equation}
with $\alpha$ a dimensionless coefficient. Such terms arise, for instance, in 
grand-canonical descriptions of horizon degrees of 
freedom~\cite{Ghosh:2013iwa,Perez:2017cmj}, where matter fields near the horizon 
are 
treated thermodynamically in the presence of an effective Unruh temperature. 
Square-root contributions are especially interesting because they indicate the 
onset of short-distance quantum effects and often signal the presence of a 
fundamental minimal length. Similar structures also appear in generalized 
uncertainty principle corrections, discussed below.

\paragraph{Exponential corrections.}

A qualitatively different behavior is associated with exponentially suppressed 
terms,
\begin{equation}
\delta S \sim e^{-\delta S_{BH}},
\end{equation}
where $\delta$ is a positive dimensionless constant~\cite{Chatterjee:2020iuf}. 
Such corrections typically originate from genuinely 
non-perturbative effects, including instanton contributions, wrapped branes, 
or quantum-isolated horizon 
configurations~\cite{Dabholkar:2014ema,Pourhassan:2020bzu}. Although negligible 
for large 
macroscopic horizons, they may become important when the horizon area 
approaches the Planck scale, where the continuum approximation of spacetime 
breaks down.

\paragraph{Minimal area and discreteness effects.}

A recurring prediction of loop quantum gravity and related quantum-geometric 
approaches is the existence of a minimal area eigenvalue. In this case the 
entropy-area relation is modified at small scales, with representative 
expressions of the form
\begin{equation}
S=\frac{\sqrt{A^{2}-a_{0}^{2}}}{4},
\end{equation}
where $a_0$ denotes the minimal area scale~\cite{Corichi:2006bs,Corichi:2006wn}. 
More generally, the discreteness of the 
area spectrum implies that entropy increases effectively in quantized steps, 
forming a microscopic ``ladder'' structure. Nevertheless, for horizons much 
larger than the Planck scale the standard Bekenstein-Hawking entropy is 
recovered with excellent accuracy~\cite{BarberoG:2011gvo}, ensuring consistency 
with semiclassical black-hole thermodynamics.

Collectively, these corrections illustrate how quantum-gravity effects can 
modify the entropy-area relation beyond the familiar logarithmic terms, while 
still preserving the area law as the leading semiclassical contribution.

\subsubsection{Minimal length and generalized uncertainty principle corrections}
\label{GUPss}

An especially economical way to encode quantum-gravity effects, without 
explicitly modifying the gravitational field equations, is to acknowledge the 
existence of a fundamental minimal length scale. Independent arguments from 
string theory~\cite{Veneziano:1986zf}, loop quantum gravity, and black-hole 
physics converge to the conclusion that distances below the Planck length 
$l_{P}$ cannot be operationally resolved. This expectation already arises at 
the level of Gedankenexperiment, such as the Heisenberg microscope once 
gravitational backreaction is taken into 
account~\cite{Mead:1964zz,Mead:1966zz,Scardigli:1999jh}, and was later refined 
by incorporating general relativistic effects~\cite{Adler:1999bu}.

The existence of a minimal observable length inevitably leads to a deformation 
of the standard Heisenberg uncertainty principle and gives rise to the 
so-called generalized uncertainty principle (GUP)~\cite{Bosso:2023aht}. 
Conceptually, the GUP reflects the fact that arbitrarily sharp localization of 
events is incompatible with quantum gravity. Instead, attempts to probe ever 
smaller distances unavoidably trigger gravitational effects that counteract 
further localization. Comprehensive reviews of minimal-length scenarios and 
the resulting GUP formulations can be found in~\cite{Hossenfelder:2012jw}.

A particularly transparent derivation of the GUP follows from treating black 
holes themselves as probes of spacetime geometry, as originally argued by 
Maggiore~\cite{Maggiore:1993rv}. Suppose one attempts to measure the radius of 
a black hole using its Hawking radiation. The typical wavelength of the emitted 
quanta depends on the black-hole radius, which in turn becomes subject to 
quantum uncertainty once a minimal length is admitted. Two independent sources 
of uncertainty then contribute to the measurement, namely the usual 
quantum-mechanical 
uncertainty in momentum and an additional gravitational uncertainty associated 
with fluctuations of the horizon size. Combining these effects leads to a 
generalized uncertainty relation of the form
\begin{equation}
\label{EEGUP1clean}
\Delta x \Delta p \ge 
\frac{\hbar}{2} + \frac{\lambda}{\hbar}(\Delta p)^2 ,
\end{equation}
where $\lambda \sim \mathcal{O}(l_P^2)$ parametrizes the minimal length scale.  
Once such a relation is imposed, the phase-space density of states is modified 
and the standard microstate counting must be revisited~\cite{Li:2002xb}. As a 
result, both black-hole entropy and the entropy of quantum fields receive 
corrections. Early applications of the GUP to black-hole thermodynamics showed 
that the area law can be recovered without ultraviolet divergences in the 
brick-wall model, provided the GUP is employed as a physical regulator. Related 
cosmological implications were discussed in~\cite{Chang:2001bm}.

A systematic and widely used implementation of a minimal length proceeds by 
deforming the canonical commutation relations between position and momentum 
operators. In this framework, the GUP arises directly from modified operator 
algebras. A particularly well-studied realization, motivated by string theory, 
black-hole physics, and doubly special relativity~\cite{Cortes:2004qn}, was 
proposed in~\cite{Ali:2009zq,Das:2010zf}. The resulting uncertainty relation 
can be written as
\begin{equation}
\label{EEGUP2clean}
\Delta x \Delta p \ge 
\frac{\hbar}{2}\left[
1 - 2\alpha \langle p \rangle
+ 4\alpha^2 \langle p^2 \rangle
\right],
\end{equation}
where $\alpha = \alpha_0 \ell_P / \hbar$ and $\alpha_0$ is a dimensionless 
parameter expected to be of order unity. This form explicitly encodes both 
linear and quadratic momentum corrections, reflecting different quantum-gravity 
contributions.

The modified uncertainty relations~\eqref{EEGUP1clean} 
and~\eqref{EEGUP2clean} induce characteristic corrections to black-hole 
entropy, 
whose precise structure depends on the details of the deformation. Explicit 
calculations for a variety of black-hole spacetimes have been performed 
in~\cite{Majumder:2011xg,Tawfik:2015kga}. Retaining only the quadratic term 
in~\eqref{EEGUP2clean}, the entropy of a Schwarzschild black hole acquires a 
leading logarithmic correction, i.e.
\begin{equation}
\label{EEGUPEntropyBHquadclean}
S_{\mathrm{GUP}}
= S + \frac{\alpha^2}{4}\pi \ln S
- \frac{(\alpha^2\pi)^2}{8S}
+ \cdots + C ,
\end{equation}
where $S=A/4$ denotes the standard Bekenstein-Hawking entropy. If the linear 
term in~\eqref{EEGUP2clean} is also included, an additional square-root 
contribution appears, namely
\begin{equation}
\label{EEGUPEntropyBHlinearclean}
S_{\mathrm{GUP}}
= S - 2\alpha\sqrt{\pi}\sqrt{S}
+ \alpha^2 \pi \ln S + C .
\end{equation}
Remarkably, this square-root structure coincides with corrections found in 
several other quantum-gravity-motivated approaches discussed earlier, strongly 
suggesting that the presence of a minimal length constitutes a common physical 
origin of such terms.

For cosmological-horizon applications, a particularly useful realization is 
provided by the standard quadratic GUP, characterized by a minimum measurable 
length. Within the thermodynamic description of gravity, the presence of such 
a minimum length naturally induces corrections to the entropy associated with 
gravitational horizons. In particular, applying the GUP to the standard 
derivation of horizon entropy leads to the modified entropy-area relation
\cite{Kouwn:2018rmp,Luciano:2025ezl}
\begin{equation}
\label{GUP6}
S_{\beta}(A)\simeq
\frac{A}{4l_p^2}
\left[
1-\frac{\beta\pi l_p^2}{4A}
\log\left(\frac{A}{4l_p^2}\right)
\right],
\end{equation}
where $\beta$ denotes the dimensionless GUP deformation parameter. Alternative
GUP-induced entropy relations have also been obtained in
Refs.~\cite{Adler:2001vs,Medved:2004yu,Setare:2004sr,Zhao:2006ri,
Nouicer:2007jg,Wang:2012ku,Ali:2011ap,Amelino-Camelia:2005zpp,
Majumder:2011xg,Anacleto:2015mma,CaboBizet:2022hpz,Hong:2025bae}.
As expected, the standard Bekenstein-Hawking area law is recovered in the
limit $\beta\to0$.

A further extension of the minimal-length framework is obtained by allowing 
for both a minimum and a maximum measurable length scale. A notable example is 
provided by the Generalized and Extended Uncertainty Principle (GEUP), whose 
modified uncertainty relation reads~\cite{KMM,Bojowald:2011jd}
\begin{equation}
\label{GEUPrelation}
\Delta x \Delta p \geq \frac{1}{2}
\left(
1+\frac{\alpha}{L_x^2}\Delta x^2+\beta l_p^2 \Delta p^2
\right),
\end{equation}
where $\alpha$ and $\beta$ are dimensionless deformation parameters, and $L_x$ 
denotes a characteristic infrared length scale. For negative values of 
$\alpha$, the above relation implies the existence of a maximum observable 
length in addition to the minimum length associated with the GUP sector. In 
this framework, the corresponding entropy-area relation is modified 
according to~\cite{Kouwn:2018rmp} as
\begin{equation}
\label{GEUPentropy}
S_{\alpha,\beta}(A)=\frac{A}{4l_p^2}
\left[
1-\frac{\alpha}{2\pi L_x^2}A
-\frac{\beta\pi l_p^2}{4A}
\log\left(\frac{A}{4l_p^2}\right)
\right].
\end{equation}
Thus, the GEUP introduces an additional correction controlled by $\alpha$, 
associated with the infrared or maximum-length sector, alongside the 
$\beta$-dependent correction associated with the minimum-length sector. The 
standard GUP expression~(\ref{GUP6}) is recovered for $\alpha\to0$, while the 
Bekenstein-Hawking area law is restored when both deformation parameters 
vanish.

Recently, the connection between the standard Bekenstein bound and generalized 
uncertainty relations has been explored in~\cite{Buoninfante:2020guu}, where 
the Bekenstein bound was shown to emerge from general thermodynamic arguments 
as a consequence of the Heisenberg uncertainty principle. By extending the 
latter to the GUP framework, a generalized entropy bound was obtained, 
providing a framework to investigate Planck-scale corrections to the 
information content of bounded physical systems.

The impact of the GUP is not restricted to black-hole thermodynamics. Modified 
state counting based on the GUP has also been employed to revisit entropic 
formulations in extended gravity~\cite{DAgostino:2025axy} and bounds in local 
quantum field theory. In four spacetime dimensions, this approach reproduces 
the standard holographic scaling at leading order, while generating subleading 
square-root corrections analogous to those 
in~\eqref{EEGUPEntropyBHlinearclean}~\cite{Wang:2012ku}. These results further 
support the view that minimal-length physics leaves a universal imprint on 
entropy, extending from black holes to quantum fields and cosmological systems.

Along these lines, further evidence for the deep interplay between generalized 
uncertainty relations, information-theoretic uncertainty measures, and 
generalized entropy formalisms has been provided 
in~\cite{Shababi:2020evc,Luciano:2021ndh,Jizba:2022icu,Jizba:2023ygi}, where 
the decoherence limit of quantum systems obeying the GUP was shown to be 
closely connected with Tsallis non-extensive thermostatistics, suggesting that 
quantum-gravitational modifications of the uncertainty principle may naturally 
induce generalized entropic structures at the quasi-classical level.

\subsubsection{Third-law and information-theoretic corrections}

A conceptually different modification of black-hole entropy has recently been
proposed from thermodynamic and information-theoretic considerations
\cite{Kehagias:2026yrn,Kehagias:2026wni}. The starting point is the 
observation
that, although the Bekenstein-Hawking entropy is consistent with the first and
second laws of black-hole thermodynamics, it does not satisfy the Nernst
formulation of the third law. For a Schwarzschild black hole,
$T\rightarrow0$ formally corresponds to $M\rightarrow\infty$, whereas
$S_{\rm BH}\propto M^2$ diverges instead of approaching a universal constant.
This motivates interpreting the Bekenstein-Hawking relation as an asymptotic
semiclassical expression rather than the complete black-hole entropy.

Imposing the Nernst condition as an additional thermodynamic requirement and
assuming a universal maximum black-hole mass $M_0$, at which the temperature
vanishes, leads to the modified entropy~\cite{Kehagias:2026yrn}
\begin{equation}
S_{\rm bh}(M)
=
4\pi M_0^2
\left[
\left(1-\frac{M}{M_0}\right)
\ln\left(1-\frac{M}{M_0}\right)
+
\left(1+\frac{M}{M_0}\right)
\ln\left(1+\frac{M}{M_0}\right)
\right],
\label{KehagiasEntropy}
\end{equation}
in units where $G=1$. The Bekenstein-Hawking result is recovered for
$M/M_0\ll1$, whereas for $M\rightarrow M_0$ the entropy approaches a finite
universal value. Introducing
\begin{equation}
N=\frac{M_0^2}{M_P^2},
\end{equation}
the entropy admits the large-$N$ expansion
\begin{equation}
S_{\rm bh}
=
\frac{A}{4}
\left[
1+\frac{A}{12N}
+\frac{A^2}{30N^2}
+\cdots
\right],
\label{KehagiasEntropyExpansion}
\end{equation}
so that the standard area law emerges as the leading term, while the
subleading contributions can be interpreted as finite-information
corrections.

The full entropy (\ref{KehagiasEntropy}) admits a direct
information-theoretic interpretation. For $N$ binary degrees of freedom with
mass-dependent probabilities
\begin{equation}
p_{\pm}
=
\frac{1}{2}
\left(
1\pm\frac{M}{M_0}
\right),
\end{equation}
$S_{\rm bh}$ can be written as the Kullback-Leibler divergence between the
corresponding biased Bernoulli ensemble and an unbiased maximally mixed
distribution. Thus, black-hole entropy measures an information deficit, or
statistical distinguishability from a maximally random reference state, with
the Bekenstein-Hawking area law arising as its leading large-$N$
approximation.

An accompanying analysis investigated the gravitational origin of this
entropy~\cite{Kehagias:2026wni}. It was shown perturbatively that its
temperature and entropy can arise from an infrared deformation of General
Relativity, whose Wald entropy agrees with the information-theoretic result to
the order considered. Since the corrections grow with positive powers of the
black-hole mass, the underlying modification involves infrared, including
nonanalytic or inverse-curvature, contributions rather than conventional
ultraviolet higher-curvature terms. This construction should be viewed as an
existence proof rather than a unique gravitational completion.

The same framework generates a positive effective cosmological constant,
\begin{equation}
\Lambda_{\rm eff}
=
\frac{1}{16G_N^2M_0^2}
\sim
\frac{M_P^2}{N},
\label{KehagiasLambda}
\end{equation}
thus relating the universal scale $M_0$ to the cosmological infrared scale.
Although this does not solve the cosmological constant problem, it provides an
interesting connection between information-theoretic black-hole entropy,
infrared gravity, and the cosmological constant.

 \subsection{Generalized entropy functionals}

Although the Bekenstein-Hawking relation provides the standard semiclassical 
description of gravitational entropy, many physical arguments suggest that 
the fundamental entropy functional may differ from the simple 
area law. In particular, long-range gravitational interactions, strong quantum 
correlations, non-equilibrium effects, holographic scaling, fractal horizon 
structures, and generalized microscopic statistics, may all modify the standard 
Boltzmann-Gibbs framework and lead to alternative entropy constructions.

In particular, generalized entropy functionals can arise in at 
least three conceptually distinct ways. Firstly, they may emerge from modified 
statistical descriptions, where the standard assumptions of extensivity, 
ergodicity, or separability are relaxed. Secondly, they may effectively encode 
microscopic quantum-gravity effects, such as spacetime discreteness, horizon 
deformations, or multifractal geometric structures. Thirdly, they may be viewed 
phenomenologically, namely as effective entropy parametrizations capable of 
capturing unknown ultraviolet corrections while preserving the semiclassical 
area law in the appropriate limit.

Remarkably, despite their diverse motivations, many generalized entropies share 
common features. Some preserve the area scaling at leading order, 
introducing additional power-law, logarithmic, non-extensive, or multi-scaling 
corrections, while others lead to different   functions of the area. 
Furthermore, several apparently different entropy proposals can be 
related to one another through suitable parameter choices or limiting cases, 
suggesting the possible existence of a deeper underlying statistical or 
geometrical framework.

In the following we review the main generalized entropy constructions that have 
been employed in gravitational and cosmological applications, beginning with 
entropies motivated by generalized statistics and non-extensive thermodynamics.

\subsubsection{Generalized entropy from modified statistics: Tsallis, R\'enyi,
Sharma-Mittal, Kaniadakis, Luciano-Saridakis, Kruglov, and Viaggiu entropies}
\label{entropymap}

\paragraph{Motivation and conceptual framework.}
A natural, yet often understated, question in the discussion of entropy, is
whether the underlying statistical framework itself is always appropriate.
In both quantum mechanics and thermodynamics, entropy is conventionally defined
through the von Neumann and Boltzmann-Gibbs expressions, respectively.
For a quantum system described by a density matrix $\rho$, the entanglement
entropy is given by
\begin{equation}
S=-\mathrm{Tr}(\rho\ln\rho),
\end{equation}
which, upon diagonalisation $\rho=\sum_i p_i |i\rangle\langle i|$, reduces to
\begin{equation}
S=-\sum_i p_i\ln p_i.
\end{equation}
Its thermodynamic counterpart is the Boltzmann-Gibbs entropy
\begin{equation}\label{EEBoltzmannGibbsEntropy}
S_{BG}=-k_B\sum_i p_i\ln p_i,
\end{equation}
which implicitly relies on Gaussian statistics, short-range correlations, and
ergodicity.

However, many systems of physical interest, including strongly correlated
quantum systems, gravitational configurations, and cosmological spacetimes, do
not satisfy these assumptions. Long-range interactions, memory effects, and
non-ergodic dynamics are intrinsic features of such systems, rendering the
Boltzmann-Gibbs framework conceptually incomplete. This observation naturally
motivates the consideration of generalized entropy functionals defined already
at the statistical level, rather than as perturbative corrections to a fixed
entropy formula.

\paragraph{Tsallis entropy and non-extensive statistics.}
A prominent generalization is provided by non-extensive statistical mechanics,
developed by Tsallis~\cite{Tsallis:2009zex,Tsallis:2017fhh,Tsallis:2019giw}. In 
this framework, the
number of microstates $W$ need not grow exponentially with the number of degrees
of freedom $N$, but may instead scale as a power law, $W\sim N^\tau$. The
corresponding entropy is non-additive and is defined as
\begin{equation}\label{EETsallisEntropyDef}
S_q=\frac{k_B}{q-1}\left(1-\sum_{i=1}^{W}p_i^q\right),
\end{equation}
which reduces to the Boltzmann-Gibbs entropy in the limit $q\to1$. For two
statistically independent subsystems $A$ and $B$, Tsallis entropy satisfies the
composition rule
\begin{equation}\label{EETsallisEntropyNonAdditive}
S_q(A,B)=S_q(A)+S_q(B)+\frac{1-q}{k_B}S_q(A)S_q(B),
\end{equation}
explicitly encoding its non-extensive character.

In the context of black-hole thermodynamics, Tsallis has argued that an
area-law entropy is thermodynamically inadmissible for systems governed by
long-range gravitational interactions, and that extensivity should instead be
restored by modifying the entropy functional 
itself~\cite{Tsallis:2017fhh,Tsallis:2019giw}. Applying non-extensive statistics 
to
horizons leads to the Tsallis-modified black-hole entropy
\begin{equation}\label{EETsallisBlackHoleDelta}
\mathcal{S}_{\mathrm{T}}
=\frac{A_0}{4G}\left(\frac{A}{A_0}\right)^{\delta},
\end{equation}
where $\delta>0$ and $A_0$ denotes a fundamental area scale. The
Bekenstein-Hawking entropy is recovered for $\delta=1$. Closely related
non-extensive entropy expressions have also been obtained within loop quantum
gravity and related 
approaches~\cite{Majhi:2017zao,Czinner:2015eyk,Liu:2021dvj,Mejrhit:2019oyi}.

\paragraph{ R\'enyi entropy.}
Another widely employed generalization is the  R\'enyi 
entropy~\cite{Renyi:1961EEE},
defined as
\begin{equation}\label{EERenyiEntropyDef}
S_\alpha=\frac{1}{1-\alpha}\ln\!\left(\sum_{i=1}^{W}p_i^\alpha\right),
\end{equation}
which reduces to the von Neumann entropy in the limit $\alpha\to1$. Although
originally introduced as a mathematical generalization,  R\'enyi entropy has 
proven
particularly useful in gravitational and cosmological applications, partly due 
to its additive composition law for statistically independent systems.

Importantly,  R\'enyi and Tsallis entropies are not independent. They are 
related
through
\begin{equation}\label{EERenyiTsallisRelation}
\mathcal{S}_{\mathrm{R}}
=\frac{1}{1-\alpha}
\ln\!\left[1+(1-\alpha)\mathcal{S}_{\mathrm{T}}\right],
\end{equation}
which allows results obtained in one framework to be translated into the other.
Inserting the Tsallis-modified black-hole entropy into this relation yields a
 R\'enyi-type generalization of horizon entropy, extensively used in 
cosmological
model building.

\paragraph{Sharma-Mittal entropy.}
The Sharma-Mittal entropy~\cite{Sharma:1975EEE,SayahianJahromi:2018irq}
provides a two-parameter generalization that interpolates between Tsallis and
 R\'enyi statistics. It is defined as
\begin{equation}\label{EESharmaMittalEntropyDefinition}
S_{\mathrm{SM}}
=\frac{1}{1-r}
\left[
\left(\sum_{i=1}^{W}p_i^{1-\delta}\right)^{\frac{1-r}{\delta}}-1
\right],
\end{equation}
and can be expressed compactly in terms of Tsallis entropy as
\begin{equation}\label{EESharmaMittalEntropyRenyiTsallisRelation}
\mathcal{S}_{\mathrm{SM}}
=\frac{1}{R}
\left[
\left(1+\delta\mathcal{S}_{\mathrm{T}}\right)^{R/\delta}-1
\right],
\end{equation}
where $R$ and $\delta$ are free parameters. The additional freedom provided by
this two-parameter structure has proven useful in phenomenological studies,
particularly in cosmology, where observational data may favor departures from
simpler entropy models.

\paragraph{Kaniadakis entropy.}
Kaniadakis statistics 
\cite{Kaniadakis:2001EEE,Kaniadakis:2002zz,Kaniadakis:2005zk}
emerge from the requirement of consistency with Lorentz symmetry and 
relativistic
kinematics. The associated entropy is defined as
\begin{equation}\label{EEKaniadakisEntropyDef}
S_K
=-k_B\sum_{i=1}^{W}p_i\ln_{\{K\}}p_i,
\end{equation}
where $\ln_{\{K\}}$ denotes the $K$-deformed logarithm and $|K|<1$. The
Boltzmann-Gibbs entropy is recovered smoothly in the limit $K\to0$. An
equivalent representation is
\begin{equation}\label{EEKaniadakisEntropyDef2}
S_K
=-k_B\sum_{i=1}^{W}
\frac{p_i^{1+K}-p_i^{1-K}}{2K}.
\end{equation}

When applied to black-hole horizons under the assumption of a uniform
distribution of microstates, Kaniadakis entropy leads to the modified horizon
entropy~\cite{Moradpour:2020dfm}
\begin{equation}\label{EEKaniadakisBekensteinHawkingEntropy}
S_K=\frac{1}{K}\sinh(KS_{BH}),
\end{equation}
which admits the perturbative expansion
\begin{equation}\label{EEKaniadakisBekensteinHawkingEntropyPerturbation}
S_K
= S_{BH}+\frac{K^2}{6}S_{BH}^3+\mathcal{O}(K^4).
\end{equation}
The deformation parameter $K$ has a clear physical interpretation related to the
relativistic composition of momenta and the existence of an invariant speed of
light~\cite{Kaniadakis:2002zz}. See also~\cite{Luciano:2026zke} for its 
relation to $\kappa$-entropic statistical paradigm and relativistic extensions 
of the Heisenberg uncertainty principle.

\paragraph{Unified parametric entropy.}

The various generalized entropies discussed above can be embedded
into a single parametric entropy functional with four free parameters, proposed
in~\cite{Nojiri:2022dkr},
\begin{equation}\label{EEParametricEntropyModifiedStatisticsAll}
S_{\mathrm{g}}
\!\left[\alpha_{+},\alpha_{-},\beta,\gamma\right]
=\frac{1}{\gamma}
\left[
\left(1+\frac{\alpha_{+}}{\beta}S\right)^{\beta}
-\left(1+\frac{\alpha_{-}}{\beta}S\right)^{-\beta}
\right],
\end{equation}
where $S$ denotes the Bekenstein-Hawking entropy. Appropriate choices of the
parameters reproduce Tsallis,  R\'enyi, Sharma-Mittal, Kaniadakis, as well as
Barrow entropy. This unifying phenomenological formulation is particularly 
appealing in
cosmological applications, where observational constraints may be used to
effectively reconstruct the underlying entropy functional and thereby gain
insight into the microscopic statistical structure of spacetime.

\paragraph{ Luciano-Saridakis entropy.}

While the above  
constructions   postulate modified area laws directly at the macroscopic 
level, Luciano-Saridakis entropy    arises from a well-defined microscopic 
entropic functional and an 
associated generalized microstate counting  
through a controlled violation of the separability 
axiom~\cite{Luciano:2026ufu}.
Since in the conventional holographic picture, the number of microscopic states 
grows 
as $
W=g(L)\,\xi^{L^2},
$
the authors of~\cite{Luciano:2026ufu} proposed the generalized entropic 
functional
\begin{equation}
S_{\delta,\epsilon}
=\eta_\delta \sum_i 
p_i\left(\log\frac{1}{p_i}\right)^\delta
+\eta_\epsilon \sum_i 
p_i\left(\log\frac{1}{p_i}\right)^\epsilon,
\end{equation}
where $\delta,\epsilon>0$ are two independent exponents and 
$\eta_\delta,\eta_\epsilon$ are constants. For equiprobable distributions this 
expression reduces to
\begin{equation}
S_{\delta,\epsilon}
=\eta_\delta(\log W)^\delta
+\eta_\epsilon(\log W)^\epsilon.
\end{equation}
The above entropy corresponds to the generalized microstate scaling
\begin{equation}
W_{\delta,\epsilon}
=g(L)\,\xi^{L^{2\delta}}
\,\tilde{\xi}^{L^{2\epsilon}},
\end{equation}
which can be interpreted as the simplest multi-scaling extension of the 
standard holographic growth of states, analogous to the multifractal behaviour 
encountered in complex systems. In the case of a gravitational system bounded 
by an area $A\sim L^2$, the resulting horizon entropy takes the 
holographic-like 
form~\cite{Luciano:2026ufu}  
\begin{equation}
S_{\delta,\epsilon}
=\gamma_\delta A^\delta
+\gamma_\epsilon A^\epsilon,
\label{LSentropy}
\end{equation}
where $\gamma_\delta$ and $\gamma_\epsilon$ are positive constants with the 
appropriate dimensions. 

  \paragraph{Kruglov entropy.}

A recent non-additive entropy proposal was introduced by Kruglov
\cite{Kruglov:2025kat}, starting from the statistical functional
$S_{\rm Kr}=-\sum_i p_i\ln p_i/(1-\gamma\ln p_i)$, where $\gamma$ is a
deformation parameter. For equiprobable microstates, identifying
$\ln W=S_{\rm BH}$ leads to the generalized horizon entropy
$S_{\rm Kr}=S_{\rm BH}/(1+\gamma S_{\rm BH})$. The standard
Bekenstein-Hawking entropy is recovered in the limit $\gamma\rightarrow0$,
while the entropy is non-additive for statistically independent systems.
For positive $\gamma$, $S_{\rm Kr}$ is positive and monotonically increasing,
but unlike $S_{\rm BH}$ it approaches the finite value $1/\gamma$ as
$S_{\rm BH}\rightarrow\infty$. This saturation removes the divergence of the
horizon entropy in the $H\rightarrow0$ limit of a spatially flat FRW
Universe and has motivated applications to apparent-horizon thermodynamics.

 \paragraph{Viaggiu entropy.}

A different generalization of horizon entropy was proposed by 
Viaggiu~\cite{Viaggiu:2014woa,Viaggiu:2015cra}, motivated by the observation 
that
the standard Bekenstein-Hawking entropy is derived for asymptotically flat
spacetimes and therefore does not account for the dynamical degrees of freedom
associated with an expanding Universe. Starting from theorems concerning the
formation of trapped surfaces in Friedmann geometries, Viaggiu argued that the
energy required to form a black hole in an expanding background exceeds the
corresponding static value. By combining the generalized trapped-surface
condition with the Bekenstein entropy bound, he obtained a modified entropy
expression containing an additional contribution proportional to the Hubble
expansion rate, namely
\begin{equation}
\label{Viaggiu1}
S_V
=
\frac{k_B A}{4L_P^2}
+
\frac{3k_B}{2cL_P^2}\,VH,
\end{equation}
where $A$ and $V$ are respectively the area and volume enclosed by the
horizon, and $H$ is the Hubble parameter.
The first term reproduces the standard Bekenstein-Hawking entropy, while the
second encodes the contribution of the cosmological expansion to the total
entropy budget. Contrary to many generalized entropies that originate from
modified statistics or quantum-gravity corrections, the Viaggiu entropy
emerges from gravitational and geometric considerations in non-static
spacetimes. It has subsequently found applications in horizon thermodynamics,
the generalized second law, and holographic dark-energy 
constructions~\cite{Saha:2019qnx,Saha:2026lnb,Halder:2026wvg}, providing an 
interesting
example of how cosmological dynamics can directly modify the entropy-area
relation.

\subsubsection{Generalized mass-to-horizon entropy}

In~\cite{Nojiri:2022sfd,Nojiri:2021czz} it was claimed that due to the first 
law 
of  thermodynamics, altering the entropy should require a change in energy or 
temperature    in order to obtain self-consistency. Similarly, 
in~\cite{Gohar:2023lta}  it was argued that in order to restore consistency 
with 
the Clausius relation while retaining the Hawking temperature, one 
should introduce a  generalized mass-to-horizon relation. In other words,  
extending a single thermodynamic quantity in isolation is generally 
insufficient, and self-consistency requires that the remaining thermodynamic 
relations should be appropriately adjusted.  

In conventional entropic cosmologies, this consistency is typically ensured by 
assuming a linear relation between the mass enclosed within the horizon and the 
horizon radius. However, it turns out that modifying the entropy alone is 
insufficient to generate non-trivial cosmological dynamics. Irrespective of the 
specific entropy or temperature adopted in the definition of the entropic 
force, 
the latter remains unchanged as long as the thermodynamic quantities on the 
horizon are treated consistently. In this sense, the functional form of the 
entropic force is primarily dictated by the assumed mass-to-horizon relation.

Motivated by these considerations, the authors of~\cite{Gohar:2023lta} proposed 
a generalized mass-to-horizon relation of the form
\begin{equation}
M=\gamma \frac{c^2}{G}L^{n},
\label{Masstohor}
\end{equation}
where $M$ denotes the mass associated with the cosmological horizon of radius 
$L$, $n$ is a non-negative constant, and $\gamma$ is a parameter with 
dimensions 
$[L]^{1-n}$ (in natural units $\hbar = k_{B} = c = 1$).

Substituting this relation into the Clausius law, $dE = c^{2} dM = T dS$, while 
assuming that the temperature retains its Hawking form  and 
identifying $L$ with the apparent horizon $r_a$, one obtains a generalized 
entropy expression given by~\cite{Gohar:2023lta}
\begin{equation}
S_{n}=\gamma \frac{2n}{n+1} r_a^{\,n-1} S_{BH},
\label{snentropy}
\end{equation}
where $S_{BH}$ is the standard Bekenstein-Hawking entropy. 
It is straightforward to verify that for $\gamma = n = 1$ one recovers both the 
usual linear mass-to-horizon relation and the standard Bekenstein-Hawking 
entropy. Finally, note that  in any 
diffeomorphism-invariant gravitational theory, the entropy associated with a 
horizon can, in principle, be derived from the gravitational Lagrangian via the 
Noether-charge construction~\cite{Wald:1993nt}. From this perspective, it is 
natural to expect that the generalized mass-to-horizon 
entropy~(\ref{snentropy}) 
may arise as the Noether entropy of an as-yet-unknown extension of General 
Relativity, in close analogy with the correspondence recently established for 
Tsallis entropy~\cite{DAgostino:2024sgm}.

\subsection{Entropy in modified gravity theories}
\label{ModGr}

A conceptually more radical modification of entropy arises when the underlying
gravitational dynamics themselves are altered \cite{CANTATA:2021asi}. In this 
case, deviations from
the Bekenstein-Hawking area law are not introduced phenomenologically, nor do
they originate from modified statistical frameworks, but instead emerge
directly from the structure of the gravitational action. Provided that the
modified theory admits Einstein gravity as an appropriate low-energy or 
weak-curvature limit, such deviations are expected to remain under control and 
to
preserve the leading area scaling at macroscopic scales.

\paragraph{Black-hole entropy and Wald's formula.}
In diffeomorphism-invariant theories of gravity, black-hole entropy is most
naturally defined in terms of the Noether charge associated with
diffeomorphism symmetry. This construction was developed by Wald and
collaborators~\cite{Wald:1993nt,Jacobson:1993vj,Iyer:1995kg}, who demonstrated
that, for stationary black holes, the entropy can be expressed as an integral
of the Noether charge over a spatial cross-section $\Sigma$ of the event
horizon, namely
\begin{equation}\label{EEWaldEntropyDefinition}
S = 2\pi \int_{\Sigma} Q ,
\end{equation}
up to the standard normalization conventions, where $Q$ is the Noether charge 
$(D-2)$-form associated with the
horizon-generating Killing vector.

For a general curvature-based modified gravity theory described by a 
diffeomorphism-invariant
Lagrangian of the form~\cite{Jacobson:1994qe}
\begin{equation}\label{EEModifiedGRAbstractLagrangianClean}
L = L\!\left(
\psi,\nabla_a\psi,g_{ab},R_{abcd},
\nabla_e R_{abcd},
\nabla_{(e_1}\nabla_{e_2)} R_{abcd},\ldots
\right),
\end{equation}
the corresponding entropy takes the general form
\begin{equation}\label{EEModifiedGRAbstractEntropyClean}
\mathcal{S}
= -2\pi \oint d^2x\,\sqrt{h}
\sum_{m=0}^n (-1)^m
\nabla_{(e_1}\!\cdots\nabla_{e_m)}
Z^{e_1\cdots e_m:abcd}
\,\epsilon_{ab}\epsilon_{cd},
\end{equation}
with
\begin{equation}
Z^{e_1\cdots e_m:abcd}
\equiv
\frac{\partial L}{\partial
\nabla_{(e_1}\!\cdots\nabla_{e_m)} R_{abcd}},
\label{Waldcurv1}
\end{equation}
where $h$ denotes the determinant of the induced metric on the horizon
cross-section and $\epsilon_{ab}$ is the binormal to $\Sigma$.

In the particular case of Einstein gravity, 
expression~\eqref{EEModifiedGRAbstractEntropyClean} reduces exactly to the
Bekenstein-Hawking entropy $S=A/(4G)$. Moreover, in metric $f(R)$ gravity, the 
entropy of
a Killing horizon retains an area-law structure, but with an effective
gravitational coupling~\cite{Cognola:2005de,Briscese:2007cd,Bamba:2010kf},
\begin{equation}
S_{BH}=\frac{A}{4G_{\mathrm{eff}}},
\end{equation}
or, equivalently,
\begin{equation}
S_{BH}=\frac{f'(R)\,A}{4G},
\end{equation}
where the factor $f'(R)$ modifies the standard area term. Detailed analyses of
entropy in $f(R)$ gravity can be found 
in~\cite{Faraoni:2010yi,Geng:2019wgd,Anand:2026byv}.

Recently, the inverse application of Wald's formalism, namely reconstructing the 
underlying gravitational action from a prescribed horizon entropy, has also been 
implemented for the Tsallis entropy. Within the framework of metric $f(R)$ 
gravity, this procedure leads to an effective modified gravitational Lagrangian 
of the form $f(R)\propto R^{1+\epsilon}$, with $\epsilon=\delta-1$ quantifying 
deviations from General Relativity and thus providing a gravitational origin for 
the Tsallis non-extensive entropy~\cite{DAgostino:2024sgm}. Interestingly, the 
same reconstruction strategy has very recently been extended to the generalized 
mass-to-horizon entropy~\cite{Mondal:2026mqb}. These results suggest that the 
inverse Wald procedure provides a powerful and rather universal framework for 
establishing a direct correspondence between generalized black-hole entropy 
functionals and the underlying modified theories of gravity.

The Wald construction can also be generalized to non-stationary horizons,
although additional subtleties arise and the Noether charge must be defined
with care~\cite{Iyer:1994ys}. In dilaton gravity, for instance, one obtains
entropy expressions containing curvature-dependent contributions in addition
to the area term, with the Einstein limit smoothly recovered when the extra
couplings vanish.

\paragraph{Holographic entropy in modified gravity.}
Modified gravity theories that admit holographic dual descriptions introduce
further conceptual and technical challenges in the definition of entanglement
entropy. In the original Ryu-Takayanagi prescription, holographic entanglement
entropy is identified with the area of a minimal surface in the bulk and
coincides with the Bekenstein-Hawking entropy. It is therefore natural to
expect that, in modified gravity, holographic entropy should be obtained by
evaluating Wald entropy on the corresponding extremal surface.

However, it has been shown that Wald’s formula alone is generally insufficient
to reproduce the correct boundary conformal field theory 
entropy~\cite{Hung:2011xb}. The discrepancy originates from contributions 
involving the
extrinsic curvature of the holographic surface, which are not captured by the
Noether charge construction. As a result, the holographic entanglement entropy
in higher-derivative gravity theories takes the   form
\begin{equation}
S_{\mathrm{HEE}}^{\mathcal{L}(R_{\mu\nu\rho\sigma})}
= S_{\mathrm{Wald}} + S_{\mathrm{anomaly}},
\end{equation}
where the additional anomaly  term accounts for extrinsic-curvature 
effects~\cite{Tsilioukas:2024seh,Anagnostopoulos:2025tax,Tsilioukas:2026gvy}.

Explicit entropy functionals have been derived for general $2n$-derivative
gravity~\cite{Dong:2013qoa}, as well as for quadratic and cubic curvature
theories~\cite{Camps:2013zua,Bueno:2020uxs}, including Gauss-Bonnet, Lovelock,
quasi-topological, and Einsteinian cubic gravity. We refrain from reproducing
these lengthy expressions here, as their explicit form is not essential for the
present discussion.

\paragraph{Remarks and limitations.}
While modified gravity offers a systematic and well-defined framework for
generalizing entropy, it also highlights important conceptual limitations. In
particular, the entropy relevant for black-hole thermodynamics and the entropy
governing holographic entanglement need not coincide once higher-curvature
terms are present. This distinction becomes especially significant in
cosmological applications, where horizons are typically dynamical and not
associated with exact Killing vectors. These issues will reappear in later
sections, when discussing spacetime thermodynamics and holographic dark
energy. Finally, note that in classes of modified gravity that are not based on 
curvature, such as torsional~\cite{Cai:2015emx} and 
non-metricity~\cite{Heisenberg:2023lru} 
theories,
one should go 
beyond~(\ref{EEModifiedGRAbstractLagrangianClean})-(\ref{Waldcurv1}). 
In these formulations, the corresponding horizon-entropy construction must 
be expressed directly in terms of the torsional or non-metricity variables 
and the derivatives of the respective gravitational Lagrangian ( see, e.g., 
Refs.~\cite{Bamba:2011pz,Hammad:2019oyb,Heisenberg:2022nvs,Rao:2024ncj}).

  More recently, horizon-entropy modifications have also been obtained in Cotton
gravity through the additional Codazzi sector. Since their explicit form
depends on the cosmological background and the apparent-horizon evolution, we
discuss them in subsection \ref{Cottongravitysub}.

\subsection{Equilibrium versus non-equilibrium horizon thermodynamics}
\label{noneq}

The modifications of horizon entropy discussed above naturally raise the
question of whether the standard thermodynamic description of spacetime
remains valid in extended theories of gravity. In Einstein gravity, the
derivation of the gravitational field equations from thermodynamic principles
relies on the assumption that local causal horizons are in thermodynamic
equilibrium, so that the Clausius relation~\eqref{Clre}
holds for every local Rindler horizon \cite{Jacobson:1995ab}.  

A natural extension of this construction was proposed by Eling, Guedens and
Jacobson \cite{Eling:2006aw}, who investigated whether the thermodynamic
derivation could be generalized to theories in which the horizon entropy
contains curvature corrections. Instead of assuming a constant entropy density,
they considered an entropy density proportional to a generic function $f(R)$ of
the Ricci scalar. The corresponding entropy variation becomes
\begin{equation}
\delta S=
\alpha
\int
\left(\theta f+\dot f\right)
d\lambda\,d^2A,
\end{equation}
where $\alpha = \mathrm{const}$ is the entropy density in the Einstein limit, 
$\theta$ is the
expansion of the null congruence generating the local horizon, $\lambda$ is an
affine parameter along the horizon generators, and the dot denotes
differentiation with respect to $\lambda$.

The analysis of Ref.~\cite{Eling:2006aw} showed that the standard Clausius
relation is no longer sufficient once curvature-dependent entropy corrections
are included. Indeed, the additional contribution proportional to $\dot f$
prevents the horizon from remaining in local thermodynamic equilibrium, making
it necessary to replace the equilibrium Clausius relation by the entropy
balance law
\begin{equation}
dS=\frac{\delta Q}{T}+d_iS,
\end{equation}
where $d_iS$ denotes the internal entropy production associated with
irreversible gravitational processes. This additional contribution restores the
compatibility of the thermodynamic construction with the conservation of the
matter energy-momentum tensor.

Within this generalized framework, the resulting gravitational equation of
state takes the form
\begin{equation}
fR_{ab}-f_{;ab}
+\left(\Box f-\frac{\mathcal{L}}{2}\right)g_{ab}
=
\frac{2\pi}{\hbar\alpha}T_{ab},
\end{equation}
where $f_{;ab}\equiv\nabla_a\nabla_bf$ is the covariant Hessian of $f$,
$\Box f\equiv g^{ab}\nabla_a\nabla_bf$ is the corresponding
d'Alembertian operator, and $\mathcal{L}(R)$ satisfies
$f=\frac{d\mathcal{L}}{dR}$.
Remarkably, this equation coincides with the field equations obtained by
varying the corresponding $f(R)$ gravitational action, thereby establishing a
direct connection between curvature-dependent horizon entropy and the dynamics
of higher-curvature gravity theories.

The analysis of Ref. \cite{Eling:2006aw} therefore demonstrates that, when the
horizon entropy acquires an explicit curvature dependence, the thermodynamic
description of spacetime generally departs from local equilibrium. Within this
framework, irreversible entropy production is not introduced phenomenologically
but emerges naturally as a consequence of the underlying gravitational
dynamics. The resulting non-equilibrium formulation consistently reproduces the
field equations of curvature-based modified gravity theories, thereby
establishing a direct connection between horizon thermodynamics and extended
gravitational dynamics. These ideas were subsequently generalized to 
cosmological apparent horizons,
leading to non-equilibrium thermodynamic derivations of the modified
Friedmann equations and their cosmological applications
\cite{Akbar:2006mq}, which will be discussed in detail in a later section.

More generally, the non-equilibrium thermodynamic paradigm has been further 
developed in several directions, both from a conceptual and
a phenomenological perspective. In particular, the physical origin of the
entropy-production term has been further clarified by relating it to
gravitational dissipative processes~\cite{Chirco:2009dc}. Alternative
equilibrium formulations of modified gravity have also been proposed through
appropriate redefinitions of the effective gravitational energy-momentum
tensor and the associated horizon entropy~\cite{Guedens:2011dy,Bamba:2009id}. 
Moreover, the formalism has
been extended beyond metric $f(R)$ gravity to broader classes of
higher-curvature and scalar-tensor theories, considerably enlarging the scope
of spacetime thermodynamics in modified 
gravity \cite{Cai:2006rs,Dey:2016zka,Bamba:2011jq,Bamba:2016aoo,
Bhattacharjee:2025xeb}.

\subsection{Gravitational entropy and the cosmological arrow of time}
While gravitational entropy is commonly associated with black-hole and
cosmological horizons, an equally important and longstanding question concerns
the existence of an intrinsic entropy of the gravitational field itself. This
problem is particularly relevant in cosmology, where the Universe evolved from
an initially smooth and nearly homogeneous state into the highly structured
configuration observed today. Unlike ordinary thermodynamic systems, gravity
drives the growth of inhomogeneities through gravitational instability,
suggesting that this process should be accompanied by an increase of
gravitational entropy. Despite decades of research, however, no unique or
universally accepted definition of gravitational entropy is currently
available \cite{Bolejko:2017yqk}.

A major conceptual breakthrough was provided by Penrose's Weyl Curvature
Hypothesis \cite{Penrose:1979}, according to which the remarkably low entropy
of the early Universe is related to the highly ordered state of the free
gravitational field near the initial singularity. Since the Weyl tensor,
\begin{equation}
C_{abcd}
=
R_{abcd}
-\frac{2}{D-2}
\left(
g_{a[c}R_{d]b}
-
g_{b[c}R_{d]a}
\right)
+\frac{2R}{(D-1)(D-2)}
g_{a[c}g_{d]b},
\end{equation}
encodes the tidal (free) gravitational field, Penrose argued that the
development of tidal inhomogeneities during cosmic evolution should be closely
related to the growth of gravitational entropy and hence to the emergence of
the cosmological arrow of time.

Motivated by this idea, several proposals have been developed to quantify
gravitational entropy. Early approaches explored scalar invariants constructed
from the Weyl tensor \cite{Goode:1985ab}, while later investigations focused on
the Bel-Robinson tensor, which is widely regarded as describing the
super-energy of the free gravitational field
\cite{Robinson:1997}.  Among the existing proposals, the covariant
construction introduced by Clifton, Ellis and Tavakol (CET)
\cite{Clifton:2013dha} represents one of the most extensively developed
thermodynamic frameworks. In this approach, gravitational entropy is defined
through the Gibbs-like relation
\begin{equation}
T_{\rm grav}\, dS_{\rm grav}
=
dU_{\rm grav}
+
p_{\rm grav}\, dV,
\end{equation}
where $T_{\rm grav}$, $S_{\rm grav}$, $U_{\rm grav}$, $p_{\rm grav}$ and $V$
denote the effective gravitational temperature, entropy, internal energy,
pressure and local volume element, respectively. The resulting formalism is 
consistent with the
Bekenstein-Hawking entropy for the Schwarzschild spacetime and has been
successfully applied to gravitational collapse, anisotropic cosmologies and
inhomogeneous Lema\^itre-Tolman-Bondi models
\cite{Clifton:2013dha,Sussman:2013xpa,Sussman:2015bea}.

Although the various proposals differ substantially in their mathematical
construction and range of applicability, they all share the common objective of
associating the irreversible growth of gravitational inhomogeneities with an
increase of entropy. In this sense, gravitational entropy complements the
thermodynamics of spacetime horizons by extending thermodynamic concepts to the
free gravitational field itself. At present, however, the identification of a
general and universally accepted measure of gravitational entropy remains an
open problem.

\subsection{Entropy production and decoherence of cosmological perturbations}

Another important role of entropy in cosmology concerns the quantum origin of 
primordial perturbations and their subsequent transition to the effectively
classical fluctuations observed in the cosmic microwave background (CMB) and
the large-scale structure of the Universe. Within the inflationary paradigm,
quantum vacuum fluctuations are continuously stretched beyond the Hubble
radius, where they evolve into highly squeezed quantum states and eventually
behave as classical stochastic perturbations
\cite{Mukhanov:1985rz,Sasaki:1986hm,Polarski:1995jg,Kiefer:1998qe,
Kiefer:2008ku}.

The evolution of scalar perturbations is governed by the Mukhanov-Sasaki
equation \cite{Mukhanov:1985rz,Sasaki:1986hm},
\begin{equation}
v_k''+\left(k^2-\frac{z''}{z}\right)v_k=0,
\end{equation}
where $v_k$ is the Fourier mode of the gauge-invariant Mukhanov-Sasaki
variable, $k$ is the comoving wavenumber, primes denote derivatives with
respect to conformal time $\eta$, and
$z=a\dot{\phi}/H$ in the case of single-field inflation, with $a$ the scale
factor, $\phi$ the inflaton field and $H$ the Hubble parameter. 

For super-Hubble modes ($|k\eta|\ll1$), the quantum state becomes highly
squeezed, with the squeezing parameter $r_k$ growing rapidly during the
inflationary expansion
\cite{Albrecht:1992kf,Polarski:1995jg,Kiefer:1998qe}. The corresponding
occupation number is $n_k=\sinh^2r_k$. As
inflation proceeds, one typically has
$r_k\gg1$,
which implies $n_k\gg1$ and gives rise to strong phase-space correlations,
allowing the perturbations to be accurately described by an effective
classical stochastic distribution, despite the global quantum state remaining
pure.

Entropy generation requires an additional coarse-graining prescription, which
may be implemented by tracing over inaccessible degrees of freedom or,
equivalently, by neglecting unobservable correlations
\cite{Kiefer:1998qe,Burgess:2006jn,Campo:2008ju,Campo:2008ij}. For Gaussian
squeezed states, the entropy associated with each perturbation mode is given by 
$S_k=(n_k+1)\ln(n_k+1)-n_k\ln n_k$,
where $S_k$ denotes the entropy of the $k$-th perturbation mode. In the limit
of strong squeezing, 
the above expression reduces to $
S_k\simeq2r_k$,
showing that the entropy increases linearly with the squeezing parameter.

An alternative formulation was developed by Campo and Parentani
\cite{Campo:2008ju,Campo:2008ij}, who constructed a coarse-grained entropy from
the hierarchy of correlation functions, thereby avoiding the explicit
introduction of an external environment. Although the detailed mechanism
underlying the quantum-to-classical transition remains an active area of
research, there is broad consensus that squeezing, decoherence and
coarse-graining together provide the essential ingredients for understanding
how primordial quantum fluctuations evolve into the effectively classical
perturbations that seed the CMB anisotropies and the large-scale structure of
the Universe (see also \cite{Martin:2015qta}).

\section{Spacetime thermodynamics conjecture}
\label{Spacetimethermodynamics}

The spacetime thermodynamics conjecture is based on the idea that gravitational 
dynamics may not be fundamental, but instead emerge as an 
effective, macroscopic description of underlying microscopic degrees of 
freedom. 
This perspective was motivated by the observation that spacetime horizons 
possess thermodynamic properties, such as temperature and entropy, 
closely 
analogous to those of ordinary thermodynamic systems. In this view, gravity 
appears not merely as a geometric interaction, but as a manifestation of 
thermodynamic relations applied to spacetime itself. Such an interpretation is 
strongly supported by the aforementioned connections between horizon mechanics, 
black-hole thermodynamics, and quantum field theory in curved spacetime.

A decisive step in this direction was taken by Jacobson, who demonstrated that 
the Einstein field equations can be derived from the Clausius 
relation~\eqref{Clre} applied locally to Rindler horizons, provided that entropy 
is 
proportional to horizon area. This result suggests that the field equations of 
General Relativity encode an equation of state for spacetime, rather than 
fundamental dynamical laws. Subsequent developments extended this thermodynamic 
interpretation to cosmological settings, where the apparent horizon of a 
FRW Universe plays the role of a causal boundary endowed 
with temperature and entropy. Within this framework, the Friedmann equations 
can 
be recovered by applying the first law of thermodynamics to the apparent 
horizon, assuming the standard Bekenstein-Hawking entropy and the associated 
horizon temperature.

In this section, we review the spacetime thermodynamics framework in its 
standard form. We begin by discussing the various notions of cosmological 
horizons and their physical significance, emphasizing the distinguished role of 
the apparent horizon in dynamical spacetimes. We then introduce the unified 
first law of thermodynamics for cosmological horizons and show how it encodes 
the flow of energy across the horizon. Finally, by combining these ingredients 
with the standard horizon entropy and temperature, we explicitly derive the 
Friedmann equations of General Relativity. This construction will serve as the 
theoretical basis for the next section, where we explore how modifying the 
entropy-area relation leads to generalized cosmological dynamics.

\subsection{Cosmological horizons}

In relativistic cosmology, horizons play a central role in determining which 
regions of spacetime can be causally connected and which physical information 
can be accessed by a given observer. Unlike black-hole horizons, cosmological 
horizons are observer-dependent and evolve dynamically with the expansion of 
the Universe. As a result, different notions of horizons arise, each capturing 
a distinct aspect of causal structure and information flow. Since spacetime 
thermodynamics is inherently linked to horizons, it is essential to clarify 
these concepts before formulating thermodynamic laws in a cosmological setting.

In order to formulate a thermodynamic description of cosmological dynamics, we 
consider a homogeneous and isotropic $(3+1)$-dimensional Universe described by 
the FRW metric
\begin{equation}
\label{2eq1}
 ds^2= -dt^2 + a^2(t) \left( \frac{dr^2}{1-kr^2} + r^2 d\Omega^2\right),
\end{equation}
where $a(t)$ is the scale factor and the spatial curvature constant 
$k=+1,0,-1$ corresponds to closed, flat, and open universes, respectively. The 
areal radius is given by $\tilde r=a(t)r$, and the two-dimensional metric on 
the 
$(t,r)$ subspace reads 
$h_{ab}=\mathrm{diag}(-1,a^2/(1-kr^2))$.
 In an FRW Universe, several horizons can be 
defined, 
depending on whether one focuses on past light cones, future causal boundaries, 
or locally defined trapping surfaces. The most commonly encountered 
cosmological 
horizons are briefly reviewed below.

\begin{enumerate}

\item{
\textit{Particle horizon.}  
The particle horizon characterizes the maximum comoving distance from which 
signals emitted since the beginning of the cosmic expansion could have reached 
a given observer by time $t$. It therefore defines the size of the observable 
Universe at that epoch. The particle horizon is given by
\begin{equation}
\label{parhor}
 \tilde r_P = a(t)\int^{t}_{0}\frac{dt'}{a(t')},
\end{equation}
where $a(t)$ is the scale factor. The existence of a particle horizon implies 
that regions of the Universe separated by distances larger than $\tilde r_P$ 
have never been in causal contact. This fact underlies the horizon problem of 
standard cosmology, which historically motivated the inflationary paradigm.
}

\item{
\textit{Cosmological event horizon (future event horizon).}  
The cosmological event horizon defines the boundary beyond which events will 
never be observable, even in the infinite future. It is therefore a genuinely 
global concept, depending on the entire future evolution of the scale factor. 
Its radius is given by
\begin{equation}
\label{fevhor}
 \tilde r_{FE} = a(t)\int^{\infty}_{t}\frac{dt'}{a(t')}.
\end{equation}
The event horizon exists only in accelerating universes for which the above 
integral converges, such as de Sitter or $\Lambda$-dominated cosmologies. In 
decelerating universes it does not exist. While the event horizon plays an 
important role in discussions of holographic dark energy and global entropy 
bounds, its teleological nature makes it less suitable for local or dynamical 
thermodynamic formulations.
}

\item{
\textit{Apparent horizon.}  
In a dynamical spacetime, the most physically relevant horizon for 
thermodynamic 
considerations is the apparent horizon. It is defined as a marginally trapped 
surface with vanishing expansion of outgoing null geodesics. In an FRW 
Universe, 
the apparent horizon is determined by the condition
\begin{equation}
h^{ab}\partial_a \tilde r \, \partial_b \tilde r = 0,
\end{equation}
where $\tilde r = a(t) r$ is the areal radius and 
$h_{ab}=\mathrm{diag}(-1,a^2/(1-kr^2))$ is the two-dimensional metric on the 
$(t,r)$ subspace. This yields the explicit expression
\begin{equation}
\label{aphor}
 \tilde r_A = \frac{1}{\sqrt{H^2 + k/a^2}},
\end{equation}
with $H\equiv \dot a/a$ the Hubble parameter.

The apparent horizon is a quasi-local concept and, unlike the event horizon, it 
always exists in FRW cosmologies, independently of the expansion history. 
Moreover, it can be shown that the apparent horizon is associated with a 
well-defined surface gravity and temperature, making it the natural candidate 
for formulating thermodynamic relations in 
cosmology~\cite{Hayward:1997jp,Cai:2005ra,Akbar:2006kj}. For this reason, it 
will play a 
central role in the spacetime thermodynamics framework discussed below.
}

\item{
\textit{Hubble horizon.}  
The Hubble horizon is defined as
\begin{equation}
\tilde r_H = \frac{1}{H},
\end{equation}
and is often interpreted as the scale beyond which comoving objects recede from 
an observer with superluminal recession velocities. Although frequently used as 
a rough estimate of causal scales, the Hubble horizon does not generally 
correspond to a true causal boundary, nor does it possess the geometric 
properties required for a consistent thermodynamic description in a dynamical 
spacetime.
}

\end{enumerate}

It is worth noting that in a spatially flat Universe ($k=0$), the apparent 
horizon coincides with the Hubble horizon, $\tilde r_A=\tilde r_H=1/H$. In 
special cases such as pure de Sitter spacetime, the future event horizon also 
coincides with these scales. However, in general cosmological evolutions these 
horizons differ, and their physical interpretation must be handled with care.

In the context of spacetime thermodynamics, the apparent horizon emerges as the 
most appropriate boundary on which to define entropy, temperature, and energy 
flux. Its local and dynamical nature allows the formulation of thermodynamic 
relations without reference to the global structure of spacetime, a feature 
that will be essential for deriving the Friedmann equations from the first law 
of thermodynamics in the following subsections.

\subsection{Unified first law of thermodynamics}

The key idea behind spacetime thermodynamics is that the evolution of a 
dynamical horizon can be described in terms of energy flow, work, and heat, in 
close analogy with ordinary thermodynamics. In a spherically symmetric 
spacetime, such as FRW, this correspondence can be made precise through the 
\emph{unified first law of thermodynamics}, originally formulated by 
Hayward~\cite{Hayward:1998ee} and later applied to cosmology 
in~\cite{Fischler:1998st,Bak:1999hd,Horava:2000tb,Cai:2005ra,Akbar:2006kj}.

A change of the apparent horizon induces both work and energy transfer. The 
\emph{work density} is defined as
\begin{equation}
\label{2eq5}
 W = -\frac{1}{2}T^{ab}h_{ab},
\end{equation}
where $T^{ab}$ denotes the projection of the $(3+1)$-dimensional 
energy-momentum 
tensor onto the two-dimensional $(t,r)$ subspace. Physically, $W$ represents 
the work done by the matter content of the Universe during an infinitesimal 
change of volume.
Additionally, the \emph{energy-supply vector}, which characterizes the energy 
flux crossing a 
surface of constant areal radius, is defined as
\begin{equation}
\label{2eq6}
 \Psi_a = T_a^{\ b}\partial_b \tilde r + W\,\partial_a \tilde r.
\end{equation}
This vector naturally decomposes the energy flow into contributions from pure 
energy transport and work, and it plays a central role in relating spacetime 
dynamics to thermodynamics.

Using the above definitions, one arrives at the relation
\begin{equation}
\label{2eq7}
\nabla E = A\Psi + W\nabla V,
\end{equation}
where $A=4\pi\tilde r^2$ and $V=\frac{4}{3}\pi\tilde r^3$ are the area and 
volume 
of a sphere with areal radius $\tilde r$. The quantity $E$ is the total energy 
contained within this sphere and is given by the Misner-Sharp energy
\begin{equation}
E = \frac{1}{2G}\tilde r
\left(1-h^{ab}\partial_a\tilde r\,\partial_b\tilde r\right).
\end{equation}
Equation~\eqref{2eq7} is known as the \emph{unified first law of 
thermodynamics}. 
It is a purely geometric identity, independent of the gravitational field 
equations, and expresses local energy conservation in a form directly 
analogous to the first law of thermodynamics.

To make contact with thermodynamics, one identifies the heat flow $\delta Q$ 
with the energy flux crossing the horizon. The energy-supply term 
in~\eqref{2eq7} can be rewritten as~\cite{Cai:2005ra}
\begin{equation}
\label{2eq9}
A\Psi=\frac{\kappa}{8\pi G}\nabla A
+\tilde r\nabla\!\left(\frac{E}{\tilde r}\right),
\end{equation}
where $\kappa$ is the surface gravity associated with the horizon, defined by
\begin{equation}
\label{2eq10}
\kappa = \frac{1}{2\sqrt{-h}}
\partial_a\!\left(\sqrt{-h}\,h^{ab}\partial_b\tilde r\right).
\end{equation}
On the apparent horizon, the second term in~\eqref{2eq9} vanishes identically, 
and the remaining term takes precisely the form $T dS$ once the horizon 
temperature $T=\kappa/(2\pi)$ and the entropy $S=A/(4G)$ are assigned. This 
identification provides a direct realization of the Clausius 
relation~\eqref{Clre} at the apparent horizon.

Finally, we consider the explicit form of the heat flow across the apparent 
horizon during an infinitesimal time interval $dt$. For a perfect fluid with 
energy-momentum tensor
\begin{equation}
T_{\mu\nu}=(\rho+p)U_\mu U_\nu + p g_{\mu\nu},
\end{equation}
the energy-supply vector becomes~\cite{Cai:2005ra}
\begin{equation}
\Psi_a=
\left(
-\frac{1}{2}(\rho+p)H\tilde r,
\frac{1}{2}(\rho+p)a
\right).
\end{equation}
As the Universe evolves, energy carried by the cosmic fluid crosses the 
apparent horizon. The total amount of energy crossing the horizon during the 
time interval $dt$ is therefore
\begin{equation}
\label{2eq12}
\delta Q = -dE
= A(\rho+p)H\tilde r_A\,dt,
\end{equation}
where $A=4\pi\tilde r_A^2$ is the area of the apparent horizon. This expression 
will serve as the starting point for deriving the Friedmann equations from the 
first law of thermodynamics in the following subsection.

\subsection{Friedmann equations in General Relativity}

A central element of the spacetime thermodynamics framework is the attribution 
of thermodynamic quantities, namely temperature and entropy, to the 
cosmological 
horizon. This attribution is not an ad hoc assumption, but follows directly 
from the interpretation of gravity as an emergent thermodynamic phenomenon, in 
which the first law of thermodynamics is applied to local causal horizons. In 
this perspective, originally developed in~\cite{Jacobson:1995ab,Bak:1999hd, 
Fischler:1998st,Horava:2000tb,Hayward:1997jp,Hayward:1998ee,
Padmanabhan:2003gd,
Padmanabhan:2009vy}, horizons act as causal boundaries rather than as physical 
walls separating a system from its environment. Consequently, the relevant 
thermodynamic quantities are associated with the horizon itself and with the 
energy flux crossing it.

A key observation is that the temperature associated with a horizon depends 
only on its surface gravity and is therefore largely insensitive to the 
underlying gravitational dynamics. For black holes, this leads to the universal 
relation $T=1/(2\pi r_h)$, independently of the specific form of the field 
equations~\cite{Gibbons:1977mu}. By analogy, the apparent horizon of an FRW 
Universe is assigned the temperature~\cite{Padmanabhan:2009vy}
\begin{equation}\label{thorizon}
T_h=\frac{1}{2\pi\tilde r_A},
\end{equation}
where $\tilde r_A$ is the radius of the apparent horizon. This identification 
is consistent with the local Rindler horizon picture, in which a comoving 
observer perceives a thermal spectrum due to the presence of a causal horizon.
It is important to stress that, although the apparent horizon evolves with 
cosmic time, the temperature~\eqref{thorizon} is defined quasi-locally. The 
first law is applied to infinitesimal processes, for which the horizon can be 
treated as approximately stationary. This is fully analogous to the treatment 
of slowly evolving black holes in black-hole thermodynamics.

In Einstein gravity, the entropy associated with a horizon is given by the 
Bekenstein-Hawking area law~\eqref{EEBekensteinHawking}, i.e.
\begin{equation}\label{sbh}
S_h=\frac{A}{4G},
\end{equation}
where $A=4\pi\tilde r_A^2$ is the area of the apparent horizon. As in the case 
of 
black holes, this entropy is independent of the details of the matter content 
and reflects the geometric nature of gravitational degrees of freedom. The 
validity of the area law in this context is supported by the close analogy 
between black-hole horizons and cosmological apparent horizons, as well as by 
the consistency of the resulting thermodynamic relations. 

A further assumption required to complete the thermodynamic description is the 
existence of local thermal equilibrium between the matter content of the 
Universe and the apparent horizon. After equilibrium is established, the 
temperature of the cosmic fluid equals the horizon temperature, namely $T=T_h$. 
This 
assumption is standard in spacetime thermodynamics and has been widely employed 
in the 
literature~\cite{Padmanabhan:2009vy,Frolov:2002va,Cai:2005ra,Akbar:2006kj,
Izquierdo:2005ku,Jamil:2010di}. Physically, it ensures that the entropy 
production is entirely associated with the horizon and not with irreversible 
processes in the bulk.

We now consider an expanding FRW Universe filled with a perfect fluid of energy 
density $\rho_m$ and pressure $p_m$. Applying the first law of thermodynamics 
to the apparent horizon,
\begin{equation}
-dE = T_h\,dS_h,
\end{equation}
and using the expression~(\ref{2eq12}) for the heat flow across the horizon 
derived in the 
previous subsection, together with~\eqref{thorizon} and~\eqref{sbh}, one 
obtains
\begin{equation}
\label{2eq14}
-4\pi G(\rho_m+p_m)=\dot H-\frac{k}{a^2},
\end{equation}
where we have used
\begin{equation}
\dot{\tilde r}_A=
-H\tilde r_A^3
\left(\dot H-\frac{k}{a^2}\right),
\qquad
dS=\frac{2\pi}{G}\tilde r_A\dot{\tilde r}_A\,dt.
\end{equation}

Equation~\eqref{2eq14} corresponds to the dynamical Friedmann equation. In 
order 
to recover the full set of cosmological equations, we further assume that the 
matter fluid satisfies the standard conservation equation,
\begin{equation}
\label{2eq16}
\dot\rho_m+3H(\rho_m+p_m)=0.
\end{equation}
Combining~\eqref{2eq14} and~\eqref{2eq16} and integrating with respect to time, 
we obtain
\begin{equation}
\label{2eq17}
H^2+\frac{k}{a^2}=\frac{8\pi G}{3}\rho_m+\frac{\Lambda}{3},
\end{equation}
where $\Lambda$ appears as an integration constant and can be interpreted as a 
cosmological constant.

Equations~\eqref{2eq14} and~\eqref{2eq17}, together with the conservation 
law~\eqref{2eq16}, constitute the complete set of Friedmann equations in 
Einstein 
gravity. Remarkably, these relations have been obtained without directly 
invoking the Einstein field equations, but solely from the application of the 
first law of thermodynamics to the apparent horizon. This result provides strong 
support for 
the spacetime thermodynamics conjecture and establishes a solid foundation for 
the extensions to modified entropy functionals that will be explored in the 
following section.

\subsection{Non-equilibrium horizon thermodynamics in modified gravity}

The equilibrium derivation presented in the previous subsection shows that, in
General Relativity, the Friedmann equations can be recast as a thermodynamic
identity at the apparent horizon. In modified gravity, however, the horizon
entropy generally acquires an explicit dependence on the gravitational
Lagrangian. Akbar and Cai examined this issue for metric $f(R)$ gravity and
showed that the thermodynamic interpretation of the cosmological field
equations requires, in general, an additional entropy-production 
term~\cite{Akbar:2006mq}. Their construction provides the cosmological 
counterpart
of the non-equilibrium spacetime thermodynamics developed for local Rindler
horizons by Eling, Guedens and Jacobson~\cite{Eling:2006aw}, and presented in 
subsection \ref{noneq} above.

For an $(n+1)$-dimensional FRW Universe filled with a perfect fluid of energy
density $\rho$ and pressure $P$, the modified Friedmann equations are written
as
\begin{align}
H^{2}+\frac{k}{a^{2}}
&=
\frac{16\pi G}{n(n-1)}
\left[
\frac{1}{F(R)}\rho
-\frac{1}{8\pi G F(R)}
\left(
\frac{1}{2}\bigl[f(R)-RF(R)\bigr]
+nH F_{,R}(R)\dot R
\right)
\right],
\label{fR_Friedmann_1}
\\[2mm]
\dot H-\frac{k}{a^{2}}
&=
-\frac{8\pi G}{(n-1)F(R)}
\left[
(\rho+p)
+\frac{1}{8\pi G}
\left(
d\!\left(F_{,R}(R)\dot R\right)
-Hf''(R)\dot R
\right)
\right],
\label{fR_Friedmann_2}
\end{align}
where $F(R)\equiv\frac{df(R)}{dR}$,
$F_{,R}(R)=dF/dR$, and an overdot denotes differentiation with respect to
cosmic time. The matter sector satisfies the usual continuity equation 
$\dot\rho+nH(\rho+p)=0$.

Following the black-hole result in $f(R)$ gravity, the entropy associated with
the apparent horizon is assumed to be given by the Wald expression 
$S=\frac{AF(R)}{4G}$. Making use of
this entropy together with the standard geometric and thermodynamic quantities
associated with the apparent horizon, the modified Friedmann equations can be 
rewritten in the form~\cite{Akbar:2006mq}
\begin{equation}
dE
=
T\,dS
+
W\,dV
-
T\frac{A}{4G}
\left[
H\tilde r_A^{\,2}
\left(
d\!\left(F_{,R}(R)\dot R\right)
-HF_{,R}(R)\dot R
\right)
+F_{,R}(R)\dot R
\right]dt\,,
\label{fR_noneq_identity}
\end{equation}
where $W=\frac{\rho-P}{2}$ is the usual work density. Definitely, in the 
Einstein limit, $f(R)=R$, one has $F(R)=1$ and
$F_{,R}(R)=0$. Consequently, the additional contribution in
Eq.~\eqref{fR_noneq_identity} identically vanishes, and the standard
equilibrium first law of horizon thermodynamics is recovered. The appearance
of the extra term is therefore a genuine consequence of the modified
gravitational dynamics rather than an independent thermodynamic assumption. 

Following the original proposal 
for local Rindler horizons~\cite{Eling:2006aw}, Akbar and Cai interpreted
this additional contribution as an internal entropy-production term generated
by the fact that the apparent-horizon thermodynamics of $f(R)$ gravity is
intrinsically out of equilibrium. The first law is therefore generalized to
\begin{equation}
dE=T\,dS+WdV+T\,d\bar S,
\label{fR_noneq_first_law}
\end{equation}
where the entropy production is given by
\begin{equation}
d\bar S
=
-\frac{A}{4G}
\left[
H\tilde r_A^2
\left(
d(F_{,R}\dot R)
-
HF_{,R}\dot R
\right)
+
F_{,R}\dot R
\right]dt.
\label{fR_entropy_production}
\end{equation}
The quantity $d\bar S$ represents the entropy generated internally by the
gravitational sector as the system evolves away from local thermodynamic
equilibrium. It is therefore not associated with the exchange of heat across
the apparent horizon, but rather with irreversible processes induced by the
additional curvature degrees of freedom of the underlying modified
gravitational theory.

For completeness, we note that the non-equilibrium thermodynamic description
outlined above has subsequently been extended in several directions. In
particular, it was shown that a broad class of modified gravity theories also
admits an equivalent equilibrium formulation upon a suitable redefinition of
the effective energy-momentum tensor~\cite{Bamba:2009id}. The same framework
has been applied to cosmological models exhibiting phantom-divide 
crossing~\cite{Bamba:2009ay}, extended to the Palatini formulation of $f(R)$ 
gravity~\cite{Bamba:2010kf}, and further generalized to other modified theories 
of gravity (see, e.g., the review~\cite{Bamba:2016aoo}).

\section{Applications of spacetime thermodynamics in cosmology}
\label{SpacetimethermodynamicsApplications}

As we showed in the previous section, the application of the first law of 
thermodynamics to the apparent horizon provides a powerful and internally 
consistent framework for deriving the Friedmann equations of General 
Relativity. The same spacetime thermodynamics construction has been 
successfully 
extended to a wide class of modified gravitational theories, where it correctly 
reproduces the corresponding cosmological equations once the appropriate 
horizon entropy is  
specified \cite{Cai:2006rs,Akbar:2006er,Paranjape:2006ca,Sheykhi:2007zp,
Jamil:2009eb,
Cai:2009ph,Saridakis:2009uu,Wang:2009zv,Jamil:2010di,Gim:2014nba,Fan:2014ala}.

A crucial point to emphasize is that this thermodynamic approach does not, by 
itself, generate new gravitational dynamics. In order to apply the 
gravity-thermodynamics conjecture to a given theory, one must already know the 
entropy functional associated with the horizon in that theory. In this sense, 
spacetime thermodynamics should be viewed as an effective macroscopic 
framework, 
which translates a prescribed entropy-area relation into modified cosmological 
equations, rather than as a fundamental principle that uniquely determines the 
underlying theory of gravity.

As we discussed in detail in Sec.~\ref{Entropydefinitions},  a variety of 
generalized entropy functionals 
have been proposed and extensively studied in the literature, both in black 
hole physics and in cosmology. 
These include generalizations of the standard 
Boltzmann-Gibbs entropy, such as Tsallis, R\'enyi, Sharma-Mittal, Kaniadakis, 
Barrow, Luciano-Saridakis and loop quantum gravity entropies, presented in 
subsections  \ref{EESectionEntanglementOrigins},  
\ref{EESectionQGEntropy} and \ref{entropymap}. Although originating from 
different 
physical considerations, ranging from non-extensive statistical mechanics and 
relativistic deformations to quantum-geometric effects, all these extended 
entropies share a set of basic consistency requirements. In particular, they 
reduce to the Bekenstein-Hawking entropy in an appropriate limit, are positive 
definite, are monotonically increasing functions of the standard area entropy, 
and satisfy a generalized version of the third law in the sense that they 
vanish when the Bekenstein-Hawking entropy vanishes~\cite{Nojiri:2022aof}.

In what follows, we apply the spacetime thermodynamics framework developed in 
Sec.~\ref{Spacetimethermodynamics} to these generalized entropy functionals. 
Keeping the geometric and 
thermodynamic setup fixed, we examine how each entropy proposal modifies the 
cosmological equations, and we discuss the resulting implications for the 
dynamics 
of the Universe. We begin with Tsallis entropy, the simplest and most widely 
studied non-extensive generalization, and then proceed to the other generalized 
entropy frameworks discussed in Sec.~\ref{Entropydefinitions}.

\subsection{Tsallis entropic cosmology}

Tsallis entropy~\cite{Tsallis:1987eu,Lyra:1997ggy,Wilk:1999dr,
Barboza:2014yfe,Saridakis:2018unr} generalizes standard thermodynamics to 
non-extensive systems with long-range interactions, while recovering 
Boltzmann-Gibbs statistics in the appropriate limit. It arises from the 
hypothesis of 
weak probabilistic correlations and their connection to ergodicity, in which 
the 
partition function diverges.  This non-additive generalized entropy, i.e. 
Tsallis 
entropy, has the form~(\ref{EETsallisBlackHoleDelta})~\cite{Tsallis:2012js}, 
which can be re-written as
\begin{equation} \label{tsallisentr}
S_{T}=\frac{\tilde \alpha}{4G} A^{\delta},
\end{equation}
where $\tilde \alpha$ is a positive constant and $\delta$ denotes the 
non-additivity parameter that quantifies the
non-extensivity. In the case where $\delta =1$ and $\tilde \alpha =1$, the 
standard Bekenstein-Hawking entropy is recovered.

In the non-extensive framework, using the first law of thermodynamics $-dE=TdS$ 
and the generalized Tsallis entropy~(\ref{tsallisentr}), leads 
to~\cite{Lymperis:2018iuz,Sheykhi:2018dpn}
\be \label{tsgfe1}
-\frac{(4\pi)^{2-\delta}G}{\tilde{\alpha}}(\rho_m+p_m)=\delta 
\frac{\dot{H}-\frac{k}{a^2}}{\left(H^2+\frac{k}{
a^2}\right)^{\delta -1}},
\ee
and
\be \label{tsgfe2}
\frac{2(4\pi)^{2-\delta}G}{3\tilde{\alpha}} \rho_m=\frac{ \delta 
}{2-\delta} \left(H^2+\frac{k}{a^2}\right) 
^{2-\delta}-\frac{\tilde{\Lambda}}{3\tilde{\alpha}},
\ee
which are the two modified Friedmann equations for the non-extensive scenario 
and where $\tilde{\Lambda}$ is an integration constant, which can be considered 
as the cosmological constant. As expected, for $\delta =1$ and $\tilde \alpha 
=1$, the above relations reduce to those of General Relativity.

From the cosmological equations~(\ref{tsgfe1}) and~(\ref{tsgfe2}), it can be 
seen that extra terms appear in the new Friedmann equations, which are 
quantified by the non-additive parameter $\delta$ and which can play the role 
of an effective dark-energy sector that could describe an interesting 
cosmological behaviour of the Universe. 

In the case of a flat Universe  ($k=0$)  filled with dust matter, 
namely $p_{m}=0$, analytical relations can be extracted for the dark-energy 
sector, i.e. for the dark-energy   density parameter, dark-energy 
equation-of-state parameter and the deceleration parameter. Furthermore, in the 
Tsallis non-extensive scenario, using the useful relation
\be \label{h2}
H=\frac{\sqrt{\Omega_{m0}} H_{0}}{\sqrt{a^3 (1-\Omega_{DE})}},
\ee
the analytical expressions of $\Omega_{DE}$ and $q$ are given 
by~\cite{Lymperis:2018iuz}
\begin{eqnarray} 
\label{omegaDEtsal}
\Omega_{DE}(z)=
1-H^{2}_{0}\Omega_{m0}(1+z)^3\left\{\frac{(2\!-\!\delta)}{\alpha 
\delta}\left[H^{2}_{0}\Omega_{m0}(1\!+\!z)^3+\frac{\Lambda}{3}\right]\right\}^{
\frac{1}{\delta -2}}\!
 \end{eqnarray}
and
\begin{equation}
 \label{qpar}
q(z)=-1+\frac{1}{2[1-\Omega_{DE}(z)]}\{3[1-\Omega_{DE}(z)]+(1+z)\Omega'_{DE}(z)
\}
\end{equation}
respectively.
\begin{figure}[t]
\centering
\includegraphics[width=7.2cm]{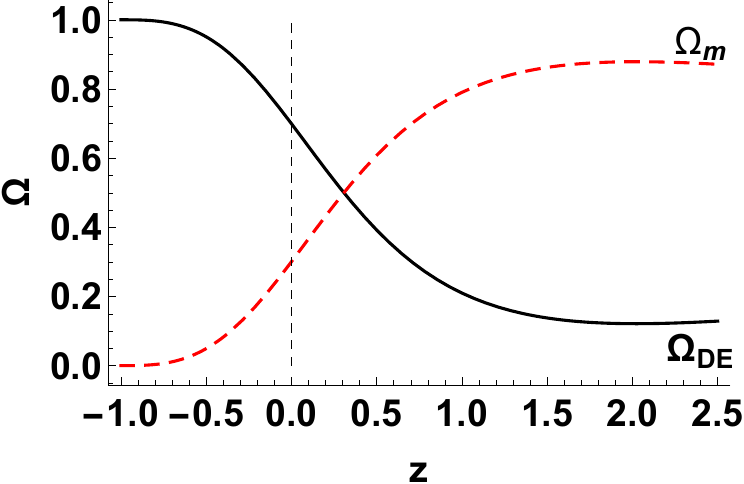}
\includegraphics[width=7.2cm]{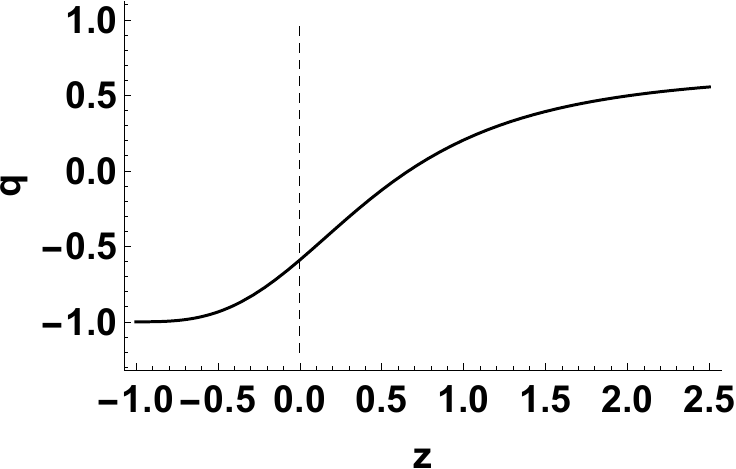}
\caption{{\it The evolution of the matter and dark-energy density parameters 
 (left panel), and of the deceleration parameter (right panel) as a 
function of the redshift $z$, for Tsallis entropic cosmology, with  $\alpha=1$ 
in units where $H_0=1$. We have set 
$\Omega_{m}(z=0)=\Omega_{m0}\approx0.3$ at the present time. The figures are 
from~\cite{Lymperis:2018iuz}. } }
\label{fig:fig1,3}
\end{figure} 
In Fig.~\ref{fig:fig1,3} we depict  the evolution of the matter and dark-energy 
  density parameters, as well as of the deceleration parameter. These 
graphs inform us that the Universe experiences the usual thermal history, with 
the sequence of the matter and dark-energy epochs, and with a transition from 
deceleration to acceleration  taking place at $z_{tr} \approx 0.45$, all in 
agreement with the observational cosmological behaviour.

The modified 
dark-energy equation-of-state parameter in the Tsallis scenario, using the 
relations~(\ref{h2}),(\ref{omegaDEtsal}), can be written in the 
form~\cite{Lymperis:2018iuz}
\be
\label{wDEfinal}
w_{DE}(z)=-1+\frac{\left\{3[1-\Omega_{DE}(z)]+(1+z)\Omega_{DE}'(z)\right\}
\left\{1-\alpha 
\delta 
\left[\frac{
H^{2}_{0}\Omega_{m0}(1+z)^3}{1-\Omega_{DE}(z)}\right]^{1-\delta}\right\}}{[
1-\Omega_{DE}
(z)]\left\{\frac{\Lambda 
[1-\Omega_{DE}(z)]}{H^{2}_{0}\Omega_{m0}(1+z)^3}+3\left\{1-\frac{\alpha 
\delta}{2-\delta}\left[\frac{H^{2}_{0}\Omega_{m0}(1+z)^3}{1-\Omega_{DE}(z)}
\right]^{ 
1-\delta}\right\} \right\}},
\ee
which shows an  interesting behaviour, deviating from the standard $\Lambda$CDM 
cosmology at intermediate times, and more specifically may experience the 
phantom-divide crossing and lie in the phantom regime. Nevertheless, at 
asymptotically large times, it will always stabilize at the cosmological 
constant value $-1$, namely, in a stable de Sitter phase.
\begin{figure}[t]
\centering
\includegraphics[width=7.7cm]{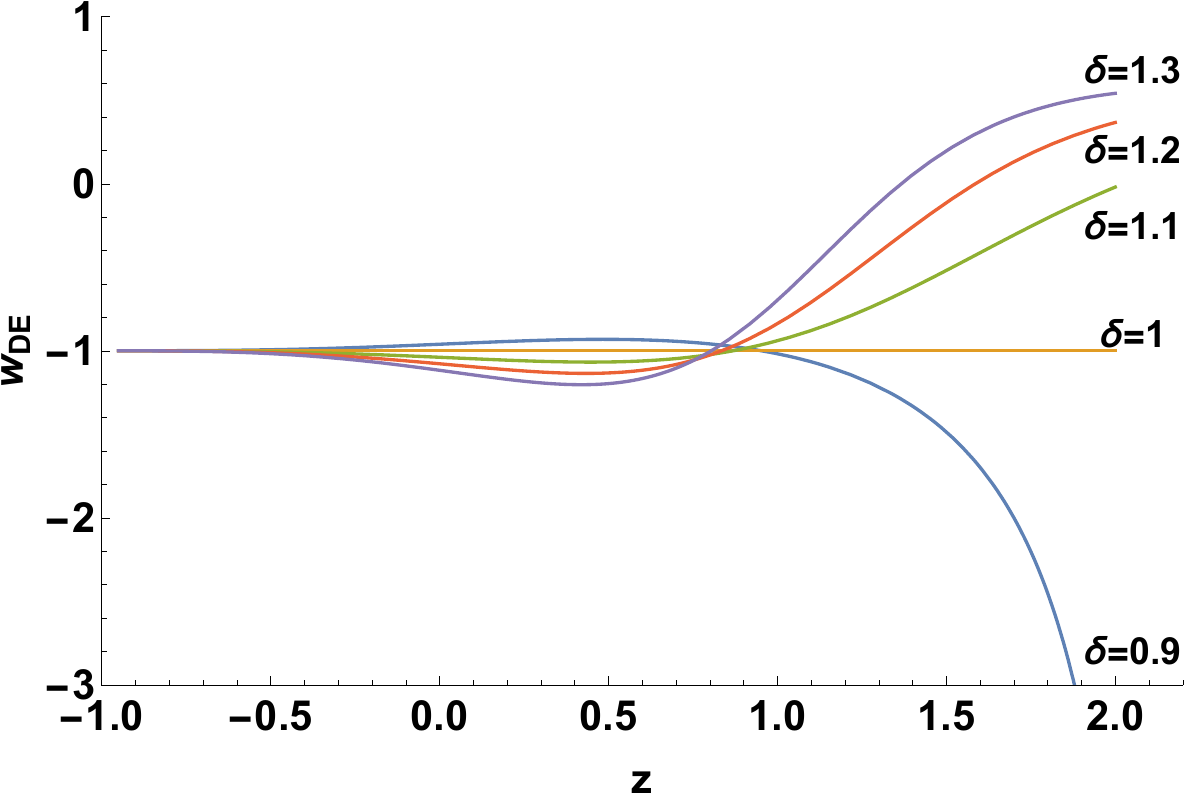}
\caption{\it{The evolution of the dark-energy equation-of-state parameter 
$w_{DE}$ as a 
function of the redshift $z$, for $\alpha=1$ in units where $H_0=1$, and various 
values 
of the nonextensive parameter $\delta$,  and for 
$\Omega_{m}(z=0)=\Omega_{m0}\approx0.3$ at present~\cite{Lymperis:2018iuz}.}}
\label{fig:fig2}
\end{figure}
Fig.~\ref{fig:fig2}  shows the effect of the Tsallis non-additive exponent 
$\delta$ on $w_{DE}$ for various values of the parameter. For $\delta >1$, the 
Universe experiences the phantom-divide crossing, and in the far future   
it approaches $w_{DE}=-1$ from below, reaching a de Sitter phase, with the 
equation-of-state parameter   acquiring larger values at earlier redshifts, 
while at intermediate times acquiring smaller values. For $\delta <1$, the 
dark-energy equation-of-state parameter experiences the opposite behaviour, i.e.
it initially lies in the phantom regime, crossing the phantom-divide, and   
is quintessence-like at present, approaching asymptotically to $-1$ from above 
in 
the far future. In general, the effect of the parameter $\delta$ which lies at 
the core of the modified scenario    results in  a dark-energy 
equation-of-state parameter, which can be quintessence-like, phantom-like, or 
experience the phantom-divide crossing during the 
evolution~\cite{Lymperis:2018iuz}. Moreover, in such a scenario the $H_0$ and 
$S_8$ tensions can be alleviated too \cite{Basilakos:2023kvk}.

\begin{figure}[t] 
\centering
\includegraphics[width=0.8\textwidth]{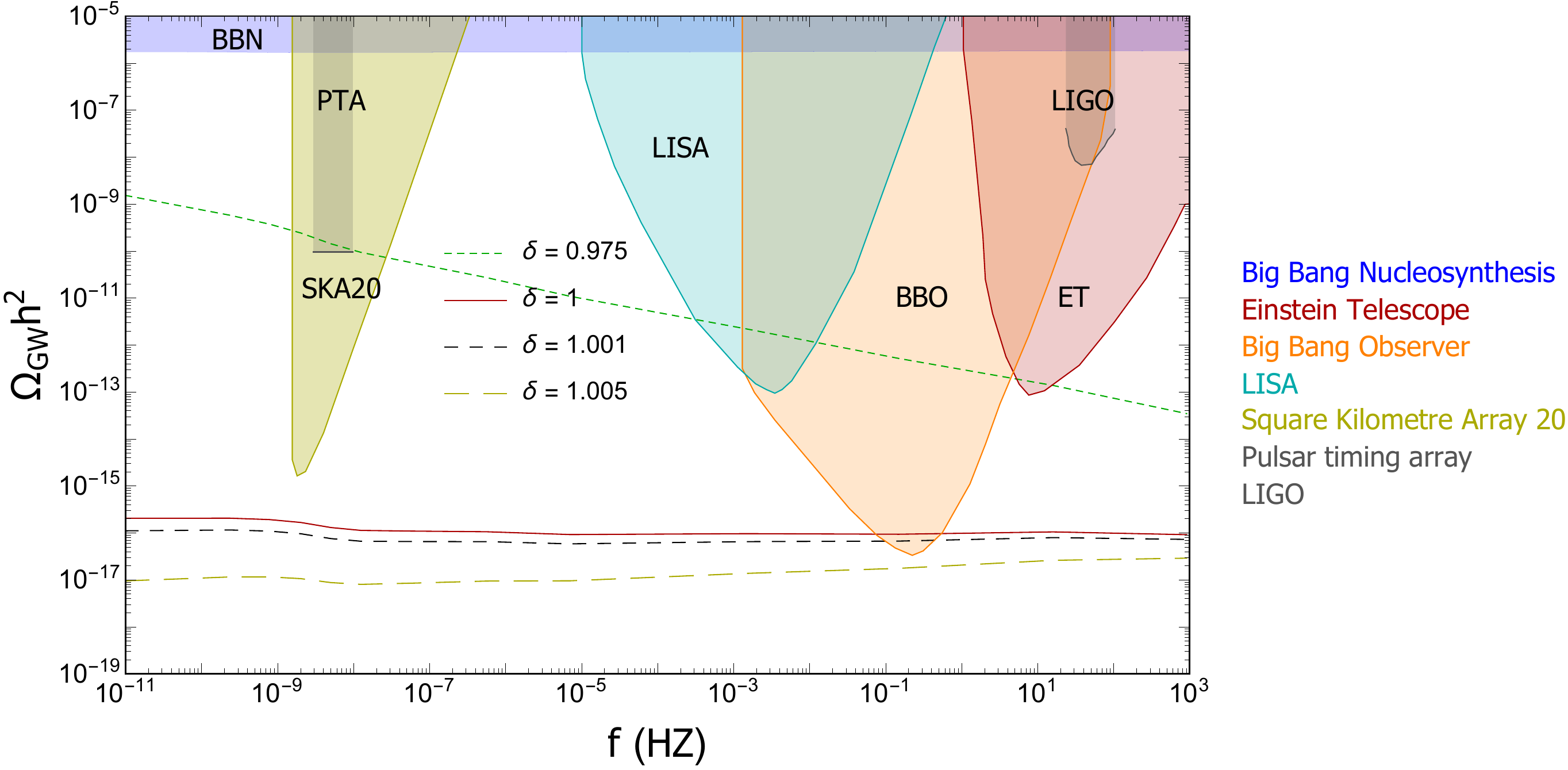}
\caption{{
\it{Energy density spectrum of PGWs as a function of frequency for 
different values of the Tsallis parameter $\delta$.
The colored regions show the projected sensitivities of the GW observatories 
listed on the right~\cite{Breitbach:2018ddu},
while the blue band denotes the BBN bound derived from constraints on the 
effective number of neutrinos~\cite{Boyle:2007zx,Stewart:2007fu}.
The gray regions indicate the parameter space excluded by current PTA and LIGO 
data~\cite{KAGRA:2021kbb}. The figure is adapted from~\cite{Jizba:2024klq}.}}}
\label{PGWTsallis}%
\end{figure}

We next examine possible imprints of Tsallis entropy in 
primordial gravitational waves. As it is known, in the case of standard 
cosmology the relic density 
evaluated at the present time   is
\begin{equation}
\Omega_{\rm GW}(\tau_0,k)h^2
\simeq
\frac{g_*(T_{hc})}{2}
\left(\frac{g_{*s}(T_0)}{g_{*s}(T_{hc})}\right)^{4/3}
\frac{P_T(k)\,\Omega_r (T_0) h^2}{24},
\end{equation}
where $g_*$ and $g_{*s}$ denote the effective numbers of relativistic degrees 
of 
freedom for energy and entropy, respectively, while $\Omega_r$ and $h$ 
represent 
the present-day radiation energy density parameter and the reduced Hubble 
constant.
Assuming a nearly scale-invariant primordial tensor spectrum, i.e.,
$P_T(k) = A_T\, (k/\tilde{k})^{n_T}$,
with pivot scale $\tilde{k} = 0.05\,\mathrm{Mpc}^{-1}$, spectral index $n_T 
\simeq 0$, and amplitude
$A_T = r A_s$, where $r \lesssim 0.034$ is the tensor-to-scalar ratio and
$A_s \simeq 2.1 \times 10^{-9}$ is the scalar amplitude~\cite{Tristram:2023haj},
standard gravity based on the Bekenstein-Hawking entropy predicts an almost 
flat
PGW spectrum over a wide range of frequencies, as can be seen in the solid red 
curve in Fig.~\ref{PGWTsallis}.
On the other hand,  in the case of modified Friedmann 
equations in Tsallis entropic cosmology, 
one finds 
\begin{equation}
 \Omega_{\mathrm{GW}}(\tau_0,k)
\ \simeq \ \Omega^{\mathrm{GR}}_{\mathrm{GW}}(\tau_0,k)\left( 
\frac{a_{\mathrm{hc}}}{a_{\mathrm{hc}}^{\mathrm{GR}}}\right)^4\left( 
\frac{H_{\mathrm{hc}}}{H_{\mathrm{hc}}^{\mathrm{GR}}}\right)^2   \,,
\label{eq:PGWBar} 
\end{equation} 
where the quantities labeled with  ``GR'' correspond to the standard 
expressions 
in General Relativity, whereas the quantities without this subscript represent 
their modified counterparts in Tsallis  cosmology.

The resulting curves are displayed in Fig.~\ref{PGWTsallis} for different 
values 
of the Tsallis parameter $\delta$. As we can see, for $\delta > 1$ 
(super-extensive regime),  one finds that the spectrum is progressively 
suppressed as $\delta$ increases. In particular, given the expected 
experimental 
sensitivity of the Big Bang Observer (BBO), a detection of PGWs would allow one 
to constrain deviations of $\delta$ from unity down to $\mathcal{O}(10^{-3})$ 
(dotted black curve).  Conversely, the absence of any detectable PGW signal 
would provide further evidence that General Relativity requires significant 
modifications for phenomenological viability. From Tsallis's perspective, such 
corrections can be effectively encoded in a modified entropy-area relation 
characterized by $\delta - 1 \ge 5\times10^{-3}$ (dot-dashed yellow curve).
On the other hand, for $\delta < 1$ (sub-extensive regime), the PGW spectrum is 
enhanced relative to the prediction of General Relativity (dashed green curve). 
If such a scenario was realized in Nature, PGW signatures could be detectable 
by next-generation experiments such as the Einstein Telescope (ET), LISA and 
SKA, in addition to BBO, even at frequencies below $10^{3}\,\mathrm{Hz}$. 
Remarkably, current Pulsar Timing Array (PTA) observations already allow one to 
exclude values such that $1-\delta\ge 0.025$ in this regime.
 
More recently, the cosmological implications of Tsallis entropy have been 
investigated in greater detail by considering not only the background 
expansion, 
but also the growth of matter perturbations and the abundance of collapsed 
structures~\cite{Mondal:2026xpt}. The resulting modified Friedmann dynamics 
preserves 
the standard transition from decelerated to accelerated expansion, occurring at 
$z_{\rm tr}\simeq0.64$, while deviations of the nonextensive parameter from its 
Bekenstein--Hawking value leave characteristic signatures in cosmographic and 
null-diagnostic quantities. At the perturbative level, the same modification 
affects the growth history of large-scale structures.

For further applications of non-additive Tsallis entropy, we mention that its 
implications in black-hole theory can be found 
in~\cite{Abreu:2020wbz,Lenzi:1998br,Liu:2022snq,Ahmed:2026zly,
Chunaksorn:2026lda,Ghaffari:2023vcw,
Sucu:2026nkw,Luciano:2023fyr} and in cosmology 
in~\cite{Lymperis:2018iuz,Barboza:2014yfe,
Abreu:2012msk,
Lima:2001vq,Abreu:2017hiy,Jawad:2019ouc,Kohyama:2006hu,Abreu:2018pua, 
Luciano:2021onl,
Abreu:2017fhw,Nojiri:2021czz,Wilk:2008gna,Luciano:2022ely, 
Moradpour:2020kss,Figliolia:2026sma,Ghoshal:2021ief,Sucu:2026wyq,
Alruwaili:2025rzu,Jizba:2023fkp,Luciano:2021mto,Shamari:2026dbj,
Luciano:2025hjn,Maqsood:2026krg}.

\subsection{R\'enyi  entropic cosmology}

The R\'enyi entropy~\cite{Renyi:1961EEE} is a one-parameter generalized 
entropy, which was proposed as an index of diversity and can be considered as a 
measure of the entanglement in information theory. Applying R\'enyi entropy to 
the black hole, one finds that its form can be written as 
in~(\ref{EERenyiTsallisRelation}), 
namely~\cite{Renyi:1961EEE,Czinner:2015eyk,Tannukij:2020njz,Promsiri:2020jga,
Samart:2020klx} 
\begin{equation} \label{renyientr}
S_{R}=\frac{1}{\alpha} \ln \left (1 +\alpha S_{BH} \right ),
\end{equation}
where in this case the parameter $\alpha$ quantifies the deviation from the 
standard statistical mechanics. In the case where $\alpha \rightarrow 0$, we 
have the usual Bekenstein-Hawking entropy.
Following the ``gravity-thermodynamics''  conjecture procedure, namely applying 
the first law of thermodynamics on apparent horizon but using the R\'enyi 
entropy instead of the usual Bekenstein-Hawking one, the two modified Friedmann 
equations can be extracted as \cite{Moradpour:2017ycq}
\begin{equation} \label{renyigfe1}
\frac{8\pi G}{3}\rho_{m}=H^{2}+\frac{k}{a^2}-\frac{\alpha \pi}{G} 
\ln{\left[\alpha \pi 
+G\left(H^{2}+\frac{k}{a^2}\right)\right]}-\frac{\Lambda}{3}
\end{equation}
and
\begin{equation} \label{renyigfe2}
-4\pi G (\rho_{m}+p_{m})=\frac{\dot{H}-\frac{k}{a^2}}{1+\frac{\alpha 
\pi}{G\left(H^{2}+\frac{k}{a^2}\right)}}.
\end{equation}
As expected, for $\alpha \rightarrow 0$ the standard Friedmann equations of 
General Relativity are recovered.

\begin{figure}[ht!]
\centering
\includegraphics[width=8.7cm]{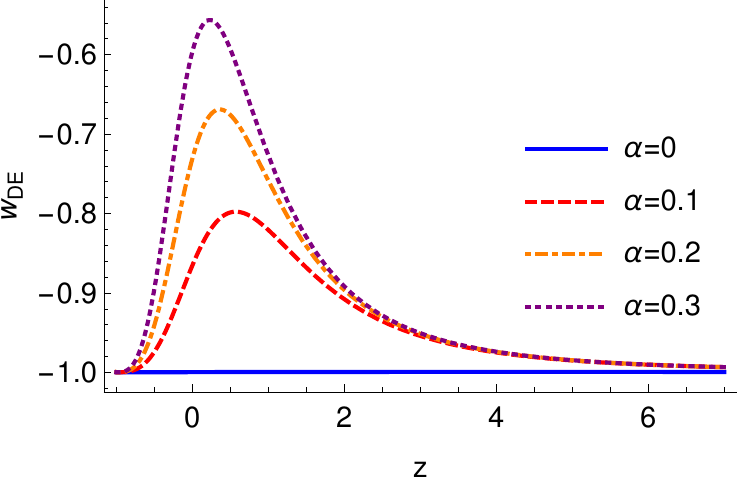}
\caption{\it{The evolution of the dark-energy equation-of-state parameter 
$w_{DE}$ as a 
function of the redshift $z$ in the modified scenario through R\'enyi entropy, 
in units where $H_0=1$, for various values 
of the nonextensive parameter $\alpha$, and for 
$\Omega_{m}(z=0)=\Omega_{m0}\approx0.3$ at present.  }}
\label{fig:figren}
\end{figure}

From the modified Friedmann equations~(\ref{renyigfe1}) and~(\ref{renyigfe2}), 
a 
dark sector arises which is quantified by the R\'enyi parameter $\alpha$ and 
can 
be described through analytical relations. For a flat geometry and dust matter, 
 
the dark-energy   density parameter can be proved to be of the form
\begin{equation} \label{renyiomegade}
\Omega_{DE}(z)=\frac{1-(1+\alpha \pi)\Omega_{m0}+\alpha \pi 
\Omega_{m0}(1+z)^{3}}{1+(1+\alpha \pi)\Omega_{m0}\left [(1+z)^{3}-1\right ]}.
\end{equation}
Using the solution for $\Omega_{DE}(z)$  one can use relation~(\ref{h2}), which 
holds in general, and extract the Hubble function $H(z)$, and then   the 
deceleration parameter using the general relation $q(z)\equiv 
-1-\frac{\dot H(z)}{H^2(z)}$. In the R\'enyi extended framework the sequence of 
the matter and dark-energy eras informs us that the Universe experiences the 
same thermal history, with the transition from deceleration to acceleration at 
$z_{tr} \approx 0.66$, similar to Fig.~\ref{fig:fig1,3} in the Tsallis 
modified scenario. The general form of the dark-energy equation-of-state 
parameter can be written as
\begin{equation} \label{renyiwde}
w_{DE}=-1-\frac{2\dot{H}\left (1-\frac{GH^{2}}{\alpha \pi +GH^{2}}\right 
)}{\Lambda+\frac{3\alpha \pi}{G}\ln(\alpha \pi +GH^{2})}, 
\end{equation} 
where its analytical relation can be provided using relations~(\ref{h2}) 
and~(\ref{renyiomegade}). Fig.~\ref{fig:figren} depicts the effect of the 
R\'enyi 
entropic parameter on the dark-energy equation-of-state parameter, for various 
values of the parameter $\alpha$. In the case where $\alpha = 0$ the 
$\Lambda$CDM cosmology is recovered. In the general case where $\alpha \neq 0$ 
and positive, and as the parameter $\alpha$ increases, $w_{DE}$ shows a 
dynamical behaviour, where it initially slightly lies in the quintessence 
regime, while at small redshifts, i.e. at intermediate times and at present, it 
deviates from the cosmological constant value. Finally, in the far future, 
namely as $z\rightarrow -1$, it always stabilizes at the cosmological constant 
value 
$-1$, namely in a dark-energy dominated, de Sitter phase, independently of the 
R\'enyi parameter $\alpha$. 

Black-hole theory implications of the R\'enyi entropy 
can be found 
in~\cite{Czinner:2015eyk,Biro:2013cra,Czinner:2017tjq,Promsiri:2020jga,
Mahapatra:2016iok,Barzi:2025etw, 
Kumar:2025hhy,Barzi:2025gwq,Sucu:2026wyq,Sucu:2026nti}, and 
cosmological implications 
in~\cite{Headrick:2010zt,Hung:2011nu,Chen:2013kpa,Chen:2013dxa,Akers:2018fow,
Chu:2016tps,Ghaffari:2019mrp,Nakaguchi:2016zqi,Anastasiou:2018mfk, 
Dong:2018lsk,Han:2017uco,Bao:2019aol,Nojiri:2021czz,Kruglov:2026imw, 
Mazumdar:2026iyt,Sheykhi:2025kfw,Kruglov:2026imw}.

\subsection{Sharma-Mittal entropic cosmology}

The Sharma-Mittal entropy constitutes a two-parameter generalization of both
Tsallis and R\'enyi entropies and is defined through
Eq.~(\ref{EESharmaMittalEntropyRenyiTsallisRelation}), 
where $R$ and $\delta$ characterize the departure from the standard
Bekenstein-Hawking entropy. As discussed in subsection~\ref{entropymap}, the 
limits $R\rightarrow\delta$ and $R\rightarrow0$
recover the Tsallis and R\'enyi entropies, respectively.

Applying the first law of thermodynamics to the apparent horizon of an FRW 
Universe, one
obtains the modified first Friedmann equation~\cite{Naeem:2023tcu}
\begin{equation}
\left(H^2+\frac{k}{a^2}\right)^{2-\delta}
+\frac{2\pi(R-\delta)}{G}
\left(\frac{4\pi}{A_0}\right)^{\delta-1}
\left(H^2+\frac{k}{a^2}\right)^{2(1-\delta)}
=\frac{8\pi G_{\rm eff}}{3}\left(\rho+\rho_\Lambda\right),
\label{SMfried1}
\end{equation}
where $\rho_\Lambda=\Lambda/(8\pi G_{\rm eff})$ and 
\begin{equation}
\frac{1}{G_{\rm eff}}
=\frac{\delta}{G(2-\delta)}
\left(\frac{4\pi}{A_0}\right)^{\delta-1}.
\label{SMGeff}
\end{equation}
The corresponding modified
acceleration equation is
\begin{align}
&(2-\delta)
\left(H^2+\frac{k}{a^2}\right)^{1-\delta}
\frac{\ddot a}{a}
+(2+\delta)
\left(H^2+\frac{k}{a^2}\right)^{2-\delta}+\frac{2\pi(1-\delta)(R-\delta)}{G}
\left(\frac{4\pi}{A_0}\right)^{\delta-1}
\left(H^2+\frac{k}{a^2}\right)^{1-2\delta}
\frac{\ddot a}{a}
\nonumber\\[2mm]
&+\frac{2\pi(2+\delta)(R-\delta)}{G}
\left(\frac{4\pi}{A_0}\right)^{\delta-1}
\left(H^2+\frac{k}{a^2}\right)^{2(1-\delta)}
=-8\pi G_{\rm eff}\left(p+p_\Lambda\right),
\label{SMfried2}
\end{align}
where $p_\Lambda=-\Lambda/(8\pi G_{\rm eff})$. For $R=\delta=1$, one has $G_{\rm 
eff}=G$, and the standard Friedmann
equations of General Relativity are recovered.

The additional terms controlled by the Sharma-Mittal parameters $R$ and
$\delta$ modify the cosmological dynamics and can be interpreted as an
effective contribution to the gravitational sector. In the presence of a
cosmological constant, the resulting framework naturally reproduces the
transition from an early matter-dominated, decelerating Universe to the
present accelerated expansion, while continuously recovering the standard
$\Lambda$CDM cosmology in the limit $R=\delta=1$. Moreover, the generalized
second law of thermodynamics is satisfied throughout the cosmological
evolution, and the same modified Friedmann equations can also be derived
within Padmanabhan's emergent-space paradigm, providing further support for
their thermodynamic interpretation~\cite{Naeem:2023tcu}. 

The above formulation is not the only realization of Sharma-Mittal entropic 
cosmology. 
Alternative approaches have also been proposed, most notably within Verlinde's 
entropic-force framework, where generalized Friedmann equations are derived by 
introducing an effective entropic pressure associated with a modified 
Sharma-Mittal entropy~\cite{Kolesnichenko:2022qsm,Gogoi:2026rit}. Although the 
underlying 
derivation differs from the horizon first-law approach adopted above, both 
frameworks share the common feature that the Sharma-Mittal deformation modifies 
the cosmological dynamics through additional entropy-induced terms, providing a 
thermodynamic mechanism capable of accounting for the late-time accelerated 
expansion without introducing an \emph{ad hoc} dark-energy component.

\subsection{Kaniadakis  entropic cosmology}

Kaniadakis entropy~\cite{Kaniadakis:2002zz,Kaniadakis:2005zk} is a one-parameter
generalization of the classical Boltzmann-Gibbs entropy, which arises from a 
coherent and self-consistent relativistic statistical theory, and is 
characterized by the single dimensionless parameter $K$, which quantifies the 
deviation from the case of standard statistical mechanics. The standard entropy 
is recovered in the limit where $K \rightarrow 0$, while the entropic parameter 
$K$ can vary in the range $-1 < K < 1$ 
~\cite{Abreu:2016avj,Abreu:2017fhw,Abreu:2017hiy,Abreu:2018mti,Yang:2020ria,
Sharma:2021zjx,
Abreu:2021avp,Drepanou:2021jiv,Lymperis:2021qty}. The generalized form of 
Kaniadakis entropy 
takes the form~(\ref{EEKaniadakisBekensteinHawkingEntropy}), 
namely~\cite{Moradpour:2020dfm}
\begin{equation} \label{kentropy}
S_{K}=\frac{1}{K} \sinh \left (KS_{BH} \right ),
\end{equation}
where $S_{BH}$ is the usual Bekenstein-Hawking entropy.

In this case the Friedmann equations of the resulting modified scenario have 
the 
form~\cite{Lymperis:2021qty}
\be \label{frwgfe1}
-4\pi G(\rho_{m}+p_{m})=\cosh{\left[K  
\frac{\pi}{G(H^2+\frac{k}{a^2})}\right]}\left (\dot{H}-\frac{k}{a^2} \right )
\ee
and
\begin{eqnarray} \label{frwgfe2}
\frac{8\pi G}{3}\rho_{m}= \cosh{\left[K  
\frac{\pi}{G(H^2+\frac{k}{a^2})}\right]}\left (H^{2}+\frac{k}{a^2} \right 
)-\frac{K\pi}{G} \text{shi}{\left[K  
\frac{\pi}{G(H^2+\frac{k}{a^2})}\right]}-\frac{\Lambda}{3},
\end{eqnarray}
where $\Lambda$ is the integration constant and the function $\text{shi}{(x)}$ 
is defined in general as 
$\text{shi}{(x)}=\int^{x}_{0}{\frac{\sinh(x')}{x'}dx'}$, which is an entire 
mathematical odd function of $x$ with no branch discontinuities.

As we can see, the modified Friedmann equations~(\ref{frwgfe1}) 
and~(\ref{frwgfe2}) contain extra terms compared to the standard cosmological 
equations of Einstein gravity, which are quantified by the single Kaniadakis
entropic parameter $K$, and effectively give rise to a dark energy sector. As 
expected for $K = 0$ the above modified equations reduce to the standard ones 
of 
General Relativity.

Furthermore, if the matter sector is dust ($w_{m}=0$) and in the flat case 
$k=0$, 
the dark-energy sector can be expressed by analytical solutions of the 
dark-energy density parameter $\Omega_{DE}$, the dark-energy equation-of-state 
parameter $w_{DE}$ and the deceleration parameter $q$. As a result of the 
dark-energy density parameter solution, which can be written in an analytical 
form as~\cite{Lymperis:2021qty}
\begin{eqnarray} 
 \label{omegaDEkan}
\Omega_{DE}(z)=
1+\frac{\epsilon_{1}}{2}\!\left 
[\frac{3}{\mathcal{A}^2}\mathcal{C}\!-\!\frac{6}{\mathcal{A}}\!-\!
\frac{5}{\mathcal{C}}
\right ]^{1/2}+\frac{\epsilon_{2}}{2}\!\left 
[\frac{12}{\mathcal{A}}\!-\!\frac{5}{\mathcal{C}}\!+\!\frac{3}{\mathcal{A}^2}
\mathcal{ C }\!-\!\frac{36\mathcal{B}}
{\mathcal{A}^{2}\left 
[\frac{3}{\mathcal{A}^2}\mathcal{C}\!-\!\frac{6}{\mathcal{A}}\!-\!
\frac{5}{\mathcal{C}}
\right ]^{1/2}}\right ]^{1/2}\!,
\end{eqnarray}
 with 
\begin{eqnarray}
\nonumber &&
\!\!\!\!\!\!\!\!\!\!\!\!\!\!\!\!\!\!
\mathcal{A}=\frac{K^2 \pi^2}{G^2 H^{4}_{0}\Omega^{2}_{m0}(1+z)^6}, \\ \nonumber
&& \!\!\!\!\!\!\!\!\!\!\!\!\!\!\!\!\!\!\!\!
\mathcal{B}= 
1+\frac{1-\Omega_{m0}}{\Omega_{m0}(1+z)^3}
-\frac{1}{2}\mathcal{A}\Omega_{m0}(1+z)^{3}(1+\frac{1}{9}\mathcal{A}\Omega^{2}_{
m0}(1+z)^{6}), \\ \nonumber
&& \!\!\!\!\!\!\!\!\!\!\!\!\!\!\!\!\!\!\!\!
\mathcal{C}=\!\left 
[9\mathcal{A}^{3}
\!+\!6\mathcal{B}^{2}\mathcal{A}^{2}\!+\!\mathcal{A}^{2} 
\sqrt{\frac{125 }{27}\mathcal{A
}^{2}\!+\!36
\left(\frac{3}{2}\mathcal{A}\!+\!\mathcal{B}^{2}
\right)^ { 2}}\right ]^{\frac{1}{3}}\!,
\end{eqnarray}
where $\epsilon_{1}, \epsilon_{2}=\pm1$, the Universe is undergoing the usual 
thermal history, with the sequence of matter and dark energy epochs, while it 
transits from deceleration to acceleration at a redshift $z_{tr} \approx 0.6$, 
in agreement with observations~\cite{Lymperis:2021qty}. The dark-energy 
equation-of-state parameter in the Kaniadakis scenario has the general 
form~\cite{Lymperis:2021qty}
\begin{equation}
\label{wDE}
w_{DE}\equiv\frac{p_{DE}}{\rho_{DE}}=-1-
 2\dot{H}\left [1-\cosh{\left(K  
\frac{\pi}{GH^2}\right)}\right ] 
\left\{\Lambda +3H^{2}\left [1-\cosh{\left(K  
\frac{\pi}{GH^2}\right)}\right 
] \right.
\left.
+\frac{3K\pi}{G} \text{shi}{\left(K  
\frac{\pi}{GH^2}\right)}\right\}^{-1}, 
\end{equation}
which can be calculated analytically by using the analytical expression of 
$\Omega_{DE}$~(\ref{omegaDEkan}), its differential relation and the 
relation~(\ref{h2}).
The evolution of the dark-energy equation-of-state parameter $w_{DE}$ shows  
that it always lies in the phantom regime, while at small redshifts  it 
deviates 
from the cosmological constant value. Additionally, at asymptotic late times it 
stabilizes in the cosmological constant value $-1$, i.e. the Universe always 
results in a dark-energy dominated, de Sitter phase~\cite{Lymperis:2021qty}.
\begin{figure}[t]
\centering
\includegraphics[width=8.7cm]{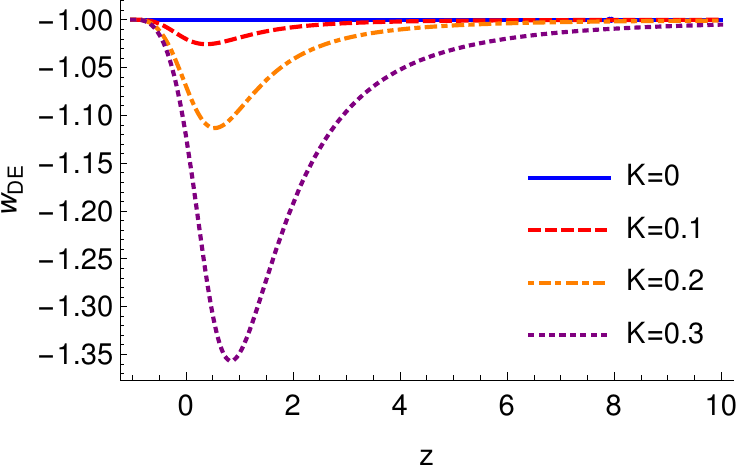}    \caption{{\it The  evolution 
of the effective dark energy equation-of-state 
parameter $w_{DE}$ in the case of Kaniadakis entropic cosmology, for different 
values of the Kaniadakis   parameter 
$K$, where $\Omega_{m0} \approx 0.3$ at the present time in units of 
$k_{_B}$. The figure is from~\cite{Lymperis:2021qty}.}}
\label{fig:multiwdeL}
\end{figure}
Fig.~\ref{fig:multiwdeL} shows the effect of the entropic parameter $K$ on the 
evolution of the equation-of-state parameter $w_{DE}$, for various values of 
$K$. As the parameter $K$ increases, the dark energy shows a dynamical 
behaviour, with $w_{DE}$ at larger redshifts lying slightly in the phantom 
regime, but at small redshifts and current time it deviates more significantly 
from $\Lambda$CDM cosmology. Finally, in the far future, it will always 
stabilize at the cosmological constant value $-1$, and the Universe always 
results in the de Sitter solution, independently of the Kaniadakis parameter 
$K$~\cite{Lymperis:2021qty}.

Applications of the extended Kaniadakis entropy  in black-hole solutions  can 
be found in~\cite{Sucu:2026wyq,Gogoi:2026nhu,Baruah:2024lcj,Alsaedi:2024ttj,
Kaczmarek:2024pbh,Sekhmani:2024kfj,Hazarika:2024lnx,Sadeghi:2024pme,
Luciano:2023bai} 
and 
in cosmology 
in~\cite{Abreu:2017hiy,Drepanou:2021jiv,Abreu:2018phv,Abreu:2017fhw,
Sharma:2021zjx,Abreu:2018mti,Hernandez-Almada:2021rjs,
Sadeghnezhad:2021ekw,Ghaffari:2021xja,
Benkrane:2026ssn,Liravi:2026xzm,Kolesnichenko:2025fuk,Zarandi:2025ytt,
Sheykhi:2025zre,Cruz:2025uuo,Salehi:2024kkj,Luciano:2024bco,Prasanthan:2024xsl,
Yarahmadi:2024lzd,Housset:2023jcm,Salehi:2023zqg,KordZangeneh:2023syq,
Lambiase:2023ryq,Blasone:2023yke,
Sheykhi:2023aqa,Luciano:2022eio,Luciano:2022knb,Dezhakam:2026gfd}.

\subsection{Barrow  entropic cosmology}

Barrow entropy, arises from quantum-gravitational effects that impose 
intricate, fractal structure on the surface of the black hole, and deviates 
from 
the standard cosmology through a single parameter which quantifies the 
deviation 
from the usual Bekenstein-Hawking entropy. Barrow entropy has the generalized 
form~(\ref{EEBarrowEntropyClean}), namely~\cite{Barrow:2020tzx}
\begin{equation}
S_{B}=\left (\frac{A}{A_{0}} \right )^{1+\Delta /2},
\end{equation}
where $A_{0}=4G$ is the Planck area and $\Delta$ is  the Barrow entropic 
parameter that quantifies the deviation from the Bekenstein-Hawking entropy.

In the case of  Barrow entropy the modified Friedmann equations can be proved 
to 
be~\cite{Saridakis:2020lrg}
\be \label{FRWgfe1}
-(4\pi)^{(1-\Delta/2)}A_{0}^{(1+\Delta/2)}(\rho_m+p_m)=2(2+\Delta) 
\frac{\dot{H}-\frac{k}{a^2}}{\left(H^2+\frac{k}{
a^2}\right)^{\Delta/2}}
\ee
and
\be \label{FRWgfe2}
\frac{ (4\pi)^{(1-\Delta/2)}A_{0}^{(1+\Delta/2)} }{6} \rho_m=\frac{2+\Delta 
}{2-\Delta} \left(H^2+\frac{k}{a^2}\right) 
^{1-\Delta/2}-\frac{{C}}{3} A_{0}^{(1+\Delta/2)},
\ee
with ${C}$   the integration constant that plays the role of the cosmological 
constant.  
Equations~(\ref{FRWgfe1}) and~(\ref{FRWgfe2}) contain new terms compared to the 
standard 
equations of General Relativity, and hence we   acquire a dark sector   that 
is quantified by the Barrow parameter $\Delta$. In the case where $\Delta = 0$, 
Eqs.~(\ref{FRWgfe1}),~(\ref{FRWgfe2}) reduce to the standard ones of 
$\Lambda$CDM paradigm.

  In the case of a flat Universe, namely $k=0$, and for dust matter, 
the resulting dark-energy sector can be expressed through analytical relations 
that present a dynamic behaviour. More specifically, the evolution of the 
dark-energy density parameter $\Omega_{DE}$    has the analytical 
form~\cite{Saridakis:2020lrg}
\begin{eqnarray} 
 \label{omegaDEbar}
\Omega_{DE}(z)=
1-H^{2}_{0}\Omega_{m0}(1+z)^3 
 \left\{\frac{
(2\!-\!\Delta)}{\beta 
(2+\Delta)}\left[H^{2}_{0}\Omega_{m0}(1\!+\!z)^3+\frac{\Lambda}{3}
\right]\right\}^{\frac{2}{\Delta -2}}\!.
 \end{eqnarray}
 This analytical solution  shows that the Universe experiences the usual 
thermal history, with the dark-energy epoch succeeding    the matter era, and 
additionally the transition from deceleration to acceleration is consistent 
with the observational behaviour. In this regime the dark energy 
equation-of-state parameter $w_{DE}$, using relations~(\ref{h2}) 
and~(\ref{omegaDEbar}), can be written as~\cite{Saridakis:2020lrg}
\be
\label{wDEfinalbis}
w_{DE}(z)=-1+\frac{\left\{3[1-\Omega_{DE}(z)]+(1+z)\Omega_{DE}'(z)\right\}
\left\{1-\beta 
(1+\Delta/2) 
\left[\frac{
H^{2}_{0}\Omega_{m0}(1+z)^3}{1-\Omega_{DE}(z)}\right]^{-\Delta/2}\right\}}{[
1-\Omega_{DE}
(z)]\left\{\frac{\Lambda 
[1-\Omega_{DE}(z)]}{H^{2}_{0}\Omega_{m0}(1+z)^3}+3\left\{1-\frac{\beta 
(2+\Delta)}{2-\Delta}\left[\frac{H^{2}_{0}\Omega_{m0}(1+z)^3}{1-\Omega_{DE}(z)}
\right]^{-\Delta/2}\right\} \right\}},
\ee
which can experience the phantom-divide crossing during the evolution.

\begin{figure}[t]
\centering
\includegraphics[width=8.7cm]{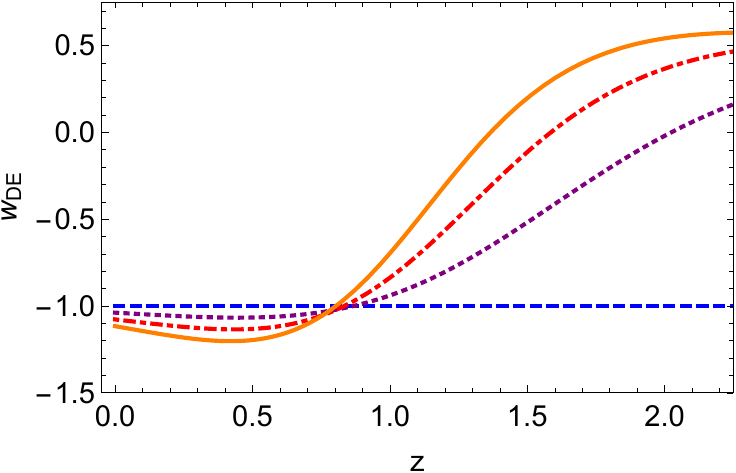}
\caption{{\it The evolution of  
$w_{DE}$ as a function of the redshift $z$ in the case of Barrow entropic 
cosmology,  for $A_0=1$, and for $\Delta=0$ (blue-dashed), $\Delta=0.2 $ 
(purple-dotted), $\Delta=0.4$ 
(red-dashed-dotted), and   $\Delta=0.6$ (orange-solid), with $ 
\Omega_{m0}\approx0.3$ at present time. The figure is 
from~\cite{Saridakis:2020lrg}.}}
\label{figwDEbarr}
\end{figure}

A more detailed examination of the effect of the Barrow exponent $\Delta$  
reveals the dynamical nature of the dark-energy. As     shown in 
Fig.~\ref{figwDEbarr}, which depicts $w_{DE}$ of the modified scenario for 
various 
values of parameter $\Delta$, as the Barrow entropic parameter $\Delta$ 
increases, and the quantum-gravitational deformation becomes more important, 
$w_{DE}$ at
larger redshifts acquires larger values, while at small redshifts and at 
the current 
time it acquires smaller ones. Hence, the Barrow parameter $\Delta$ plays a 
significant role in the dynamical nature of the dark energy. Furthermore, 
the dark energy sector can be  quintessence-like, phantom-like, or experience 
the phantom-divide crossing during the evolution~\cite{Saridakis:2020lrg}.

Further implications of Barrow entropy can be found 
in~\cite{Jawad:2022lww,Abreu:2020wbz,Abreu:2020rrh,Abreu:2020dyu,Sucu:2026hfc,
Zafar:2026ccw,NooriGashti:2026cwi,Yasir:2025viu,Sucu:2026tld,
Sucu:2026wyq,Abreu:2025efe,Sucu:2026ujm,Bora:2026uwn} within the framework of 
black-hole theory 
and in~\cite{Saridakis:2020cqq,
Barrow:2020kug,Sharma:2020ylh,Pradhan:2021cbj,
Sheykhi:2021fwh,Leon:2021wyx,Luciano:2023zrx,
Jusufi:2021fek,Luciano:2023roh,
Nojiri:2021jxf,Asghari:2021bqa,
Luciano:2022pzg,Altaibayeva:2026ycf,Sakalli:2025udo,Karabat:2025lqs,
Jayawiguna:2026ves} in cosmological contexts.

\subsection{String T-duality entropic cosmology}

String theory predicts the existence of a fundamental minimal length scale, 
which can be naturally implemented through T-duality symmetry. In this 
framework, quantum fluctuations of spacetime induce a non-vanishing zero-point 
length $l_0$, modifying the short-distance structure of gravity and 
regularizing 
the gravitational 
interaction~\cite{Padmanabhan:1996ap,Smailagic:2003hm,Spallucci:2005bm,
Fontanini:2005ik,Nicolini:2019irw}. The resulting deformation affects the 
entropy associated with horizons and leads to   modified cosmological dynamics. 
In particular, the horizon entropy receives corrections that originate from the 
modified Newtonian potential generated by T-duality 
effects~\cite{Jusufi:2022mir,Luciano:2024mcn}.

As discussed in subsection \ref{Correctedentropy1},  in this framework
the entropy associated with the apparent horizon acquires the differential form
(\ref{STDentropy}), namely  
\begin{equation}
dS_h=2\pi R\left(1+\frac{l_0^2}{R^2}\right)^{-3/2}dR,
\label{STDentropy22}
\end{equation}
where $R$ denotes the apparent horizon radius. In the limit $l_0\rightarrow0$ 
one recovers the standard area law.
Employing the gravity-thermodynamics conjecture and applying the first law of 
thermodynamics on the apparent horizon of a FRW Universe, it was shown that the 
above entropy deformation gives rise to modified Friedmann equations that 
encode 
the effects of the fundamental minimal length scale 
$l_0$~\cite{Luciano:2024mcn}. In particular, expanding perturbatively in the 
regime $l_0^2/R^2\ll1$, the modified cosmological dynamics can be expressed, at 
leading order, as
\begin{align}
\label{STDFR1}
    &H^2
=
\frac{8\pi}{3}\left(\rho+\rho_{DE}\right)\,,\\[2mm]
&\dot H=
-4\pi \left[\rho+p+\rho_{DE}+p_{DE}\right],
\label{STDFR2}
\end{align}
where we have defined the density and pressure of the effective dark-energy 
sector as
\begin{eqnarray}
\rho_{DE}&=&\frac{3}{8\pi}\left(\frac{\Lambda}{3}+\alpha H^4\right),
\\[2mm]
p_{DE}
&=&
-\frac{1}{8\pi}
\left[
\Lambda+\alpha H^2
\left(
4\dot H+3H^2
\right)
\right],
\end{eqnarray}
and the deformation parameter is defined by $\alpha\equiv\frac{3l_0^2}{4}$.

Compared to the standard Friedmann equation, Eq.~(\ref{STDFR1}) contains an 
additional quadratic contribution in the Hubble parameter, which represents the 
leading imprint of the minimal-length structure of spacetime induced by String 
T-duality. Such a correction becomes increasingly important at high energies and 
during the earliest stages of cosmic evolution, where terms scaling as higher 
powers of the Hubble rate can provide non-negligible contributions to the total 
energy budget. Nevertheless, owing to their rapid dilution as the Universe 
expands, contributions proportional to $H^4$ become quickly subdominant and 
cannot sustain the present accelerated expansion. More generally, even 
corrections scaling as $H^2$ decrease too rapidly with cosmic evolution to mimic 
an approximately constant dark-energy 
component~\cite{Maggiore:2010wr,Basilakos:2009wi}. This highlights the fact that 
minimal-length effects alone are primarily relevant in the high-energy regime 
and are not expected to account for the late-time acceleration of the Universe. 
Therefore, the standard $\Lambda$CDM cosmology is fully recovered in the limit 
$l_0\rightarrow0$, whereas non-vanishing values of $l_0$ induce departures from 
the conventional cosmological evolution and provide a phenomenological window 
into quantum-gravitational effects.

Hence, the effective equation-of-state parameter becomes
\begin{equation}
w_{DE} 
=
-1-\frac{4\alpha H^2\dot H}
{\Lambda}.
\label{omSTD}
\end{equation}
The phenomenological implications of String T-duality and the associated 
zero-point length have been extensively investigated in various gravitational 
settings. In particular, T-duality-inspired corrections have led to the 
construction of regular black-hole geometries and modified black-hole 
thermodynamics~\cite{Padmanabhan:1996ap}, while their cosmological consequences 
include the realization of singularity-free cosmological models, inflationary 
scenarios, and modifications of the late-time cosmic 
evolution~\cite{Padmanabhan:1996ap,Jusufi:2022mir,Luciano:2024mcn,
Sooraki:2025jvi,Luciano:2025dhb}. These developments  reveal the potential role 
of the minimal-length paradigm as a bridge between quantum gravity and 
observable gravitational phenomena.

\subsection{GUP entropic cosmology}

As discussed in subsection~\ref{GUPss}, the Generalized Uncertainty Principle
(GUP) induces the modified horizon-entropy relation~(\ref{GUP6}). Applying
this entropy to the apparent horizon of an FRW Universe within the
spacetime-thermodynamics framework leads to a cosmological scenario that is
formally equivalent to the T-duality construction discussed in the previous
subsection, upon an appropriate identification of the deformation parameter
$\beta$. Consequently, the modified Friedmann equations, the effective
dark-energy sector, and the associated equation-of-state parameter are directly
recovered from Eqs.~(\ref{STDFR1})-(\ref{omSTD}).

This correspondence shows that several cosmological effects usually associated
with String T-duality can be understood more generally as consequences of a
fundamental minimal-length scale, independently of its specific
quantum-gravity realization. At the same time, the formal equivalence between
the two approaches implies that they share similar phenomenological
limitations. In particular, the corrections generated by the minimum-length
scale become important primarily in the high-energy regime and rapidly decay
as the Universe expands. Hence, although they may significantly affect the
early cosmological evolution, they do not by themselves generate the nearly
constant dark-energy component required to account for the observed late-time
acceleration.

A richer phenomenology can arise by extending the minimal-length framework to
include also a maximum measurable length scale. As discussed in
subsection~\ref{GUPss}, this is realized within the Generalized and Extended
Uncertainty Principle (GEUP), defined by relation~(\ref{GEUPrelation}), with
the corresponding entropy-area relation given by~(\ref{GEUPentropy})
\cite{KMM,Bojowald:2011jd,Kouwn:2018rmp}. The presence of the additional
$\alpha$-dependent contribution leads to qualitatively different cosmological
behavior from that associated with the ordinary GUP parameter $\beta$.

In particular, when the GEUP entropy~(\ref{GEUPentropy}) is employed in the
thermodynamic derivation of the Friedmann equations, the $\alpha$-dependent
term generates a logarithmic correction to the effective dark-energy density.
This differs markedly from the $\beta$-dependent contribution, which produces
terms scaling as positive powers of the Hubble parameter and therefore
decreases rapidly during the cosmic expansion. Owing to its much slower
evolution, the logarithmic contribution can remain relevant at late times,
giving rise to an effective dark-energy component that varies only mildly
throughout the recent cosmological evolution. Thus, the simultaneous presence
of minimum- and maximum-length scales can lead to a substantially richer
cosmological phenomenology and may provide a more viable description of the
observed accelerated expansion than scenarios based exclusively on a minimum
length~\cite{Kouwn:2018rmp}.

\subsection{Luciano-Saridakis entropic cosmology}

Contrary to modified entropy  constructions that postulate modified area laws 
directly at the macroscopic level, Luciano-Saridakis entropy arises from a 
well-defined microscopic entropic functional and an 
associated generalized microstate counting~\cite{Luciano:2026ufu}. 
Hence, in the case of a gravitational system bounded 
by an area $A\sim L^2$, the resulting horizon entropy takes the 
holographic-like 
form~(\ref{LSentropy}), i.e.  
\begin{equation}
S_{\delta,\epsilon}
=\gamma_\delta A^\delta
+\gamma_\epsilon A^\epsilon,
\end{equation}
where $\gamma_\delta$ and $\gamma_\epsilon$ are positive constants with the 
appropriate dimensions, and $\delta$ and $\epsilon$  are the 
two   exponents.  
  
Application of the gravity-thermodynamics conjecture leads to the modified  
Friedmann equations
  \begin{equation}
\label{FMFE}
    H^2\,=\,\frac{8\pi G}{3}\left(\rho_m+\rho_{DE}\right),
\end{equation}
  \begin{equation}
\label{SMFE}
    \dot H\,=\,-4\pi G\left(\rho_m+p_m+\rho_{DE}+p_{DE}\right),
\end{equation}
where one introduces an effective dark 
energy sector with energy density and pressure~\cite{Luciano:2026ufu}
\begin{eqnarray}
 &&   \rho_{DE}\,=\,  \frac{3}{8\pi G} \left[H^2\left(1-\alpha_\delta 
H^{2\left(1-\delta\right)}-\alpha_\epsilon 
H^{2\left(1-\epsilon\right)}\right)\,+\,C
    \right]
    \label{EfDEdLS}\\
     && 
    p_{DE}=-\frac{1}{8\pi G}\Big\{
    \Lambda + 3H^2\left[1-\alpha_\delta H^{2(1-\delta)}-\alpha_\epsilon  
H^{2(1-\epsilon)}\right]  +\left.2\dot H \left[1-\alpha_\delta (2-\delta)  
H^{2(1-\delta)}-\alpha_\epsilon (2-\epsilon) H^{2(1-\epsilon)}
    \right]
    \right\},
    \label{EfDEpLS}\ \ \ \ \ \ 
\end{eqnarray}
 respectively, where   
$    \alpha_{\delta}\,\equiv\, \frac{\left(4\pi\right)^{\delta} 
 \gamma_\delta\hspace{0.2mm}\delta}{\left(2-\delta\right)\pi}
$ and  similarly  $\alpha_{\epsilon}\equiv\alpha_{\delta\rightarrow\epsilon}$.
 Note that 
 the Bekenstein-Hawking 
formula is recovered  for: i)
 $ \delta=1,    
\gamma_{\delta}=1/(4\ell_p^2),   \gamma_\epsilon=0 $, ii)
$\epsilon=1,   \gamma_{\delta}=0,  
 \gamma_\epsilon=1/(4\ell_p^2)$, or iii)
$\delta=\epsilon=1,   
\gamma_\delta=\gamma_\epsilon=1/(8\ell_p^2)$,
while   the generalized entropy 
$S_{\delta,\epsilon}$ recovers the $\delta$-Tsallis  and Barrow entropy  as a 
special case when $\delta = \epsilon$. 
 Additionally,   the effective dark-energy
equation
of state    is
\begin{equation}
\label{EoSLS}
   w_{DE}\equiv\frac{p_{DE}}{\rho_{DE}}\\[2mm]  =-1-\frac{2\dot 
H\left[1-\alpha_\delta (2-\delta) 
H^{2(1-\delta)}-\alpha_\epsilon (2-\epsilon)H^{2(1-\epsilon)}
    \right]}
    {\Lambda +3H^2\left(1-\alpha_\delta H^{2(1-\delta)}-\alpha_\epsilon 
H^{2(1-\epsilon)}\right) }\,,
\end{equation}
which reproduces the cosmological constant-like behavior $w_{\Lambda}=-1$  
in the aforementioned  limiting cases.
   
  \begin{figure}[t]
    \centering
\includegraphics[width=8.7cm]{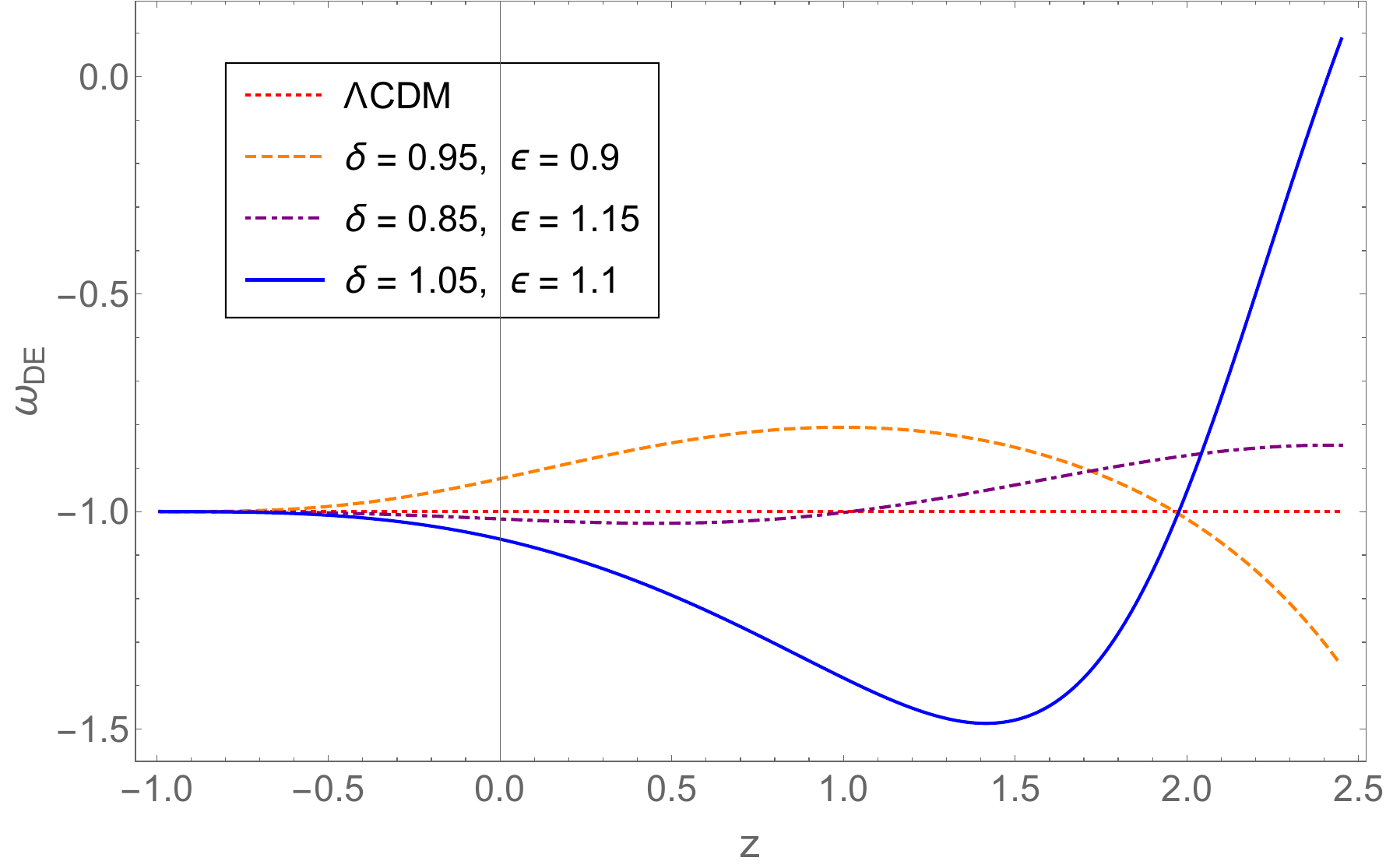}
\caption{\it{The  dark-energy equation-of-state parameter
$w_{DE}$ as a function of the redshift, for Luciano-Saridakis entropic 
cosmology, and 
for 
different combinations of $\delta$ and $\epsilon$. The figure is 
from~\cite{Luciano:2026ufu}.}}
\label{Fig2LS}
\end{figure}

\begin{figure}[t]
    \centering
    \includegraphics[width=0.6\textwidth]{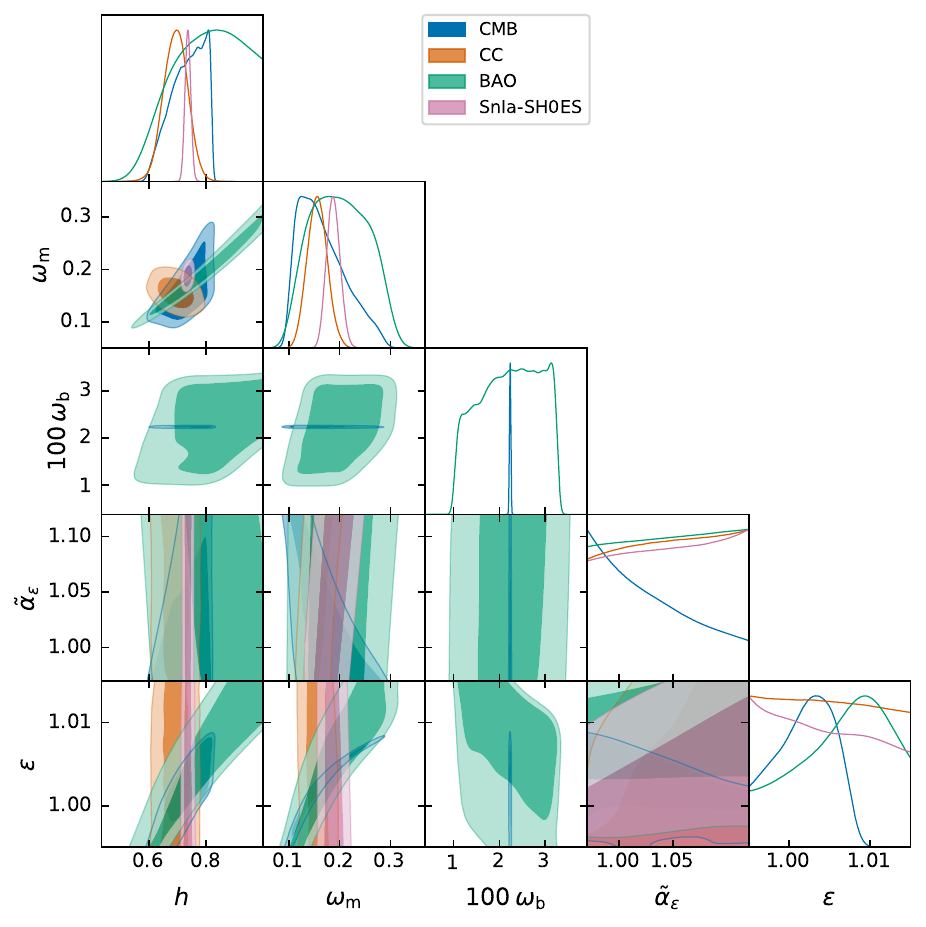}
    \caption{\it{Observational confrontation of Luciano-Saridakis entropic 
cosmology. The contour plots have been derived from the statistical analyses 
considering CMB (Planck  18 - Shift parameters), CC, SnIa + SH0ES (PPS), BAO 
(DESI DR2) datasets separately,   with 
$\alpha_\delta=0$,   where   $\tilde{\alpha}_\epsilon = 
\alpha_\epsilon H_0^{2(1-\epsilon)}$. The figure is 
from~\cite{Leizerovich:2026pfy}.
}}
\label{fig:results_separate_contours_lsec}
\end{figure}

The evolution of the effective dark-energy equation-of-state parameter is shown 
in Fig.~\ref{Fig2LS}. As can be seen, different combinations of the entropic 
exponents lead to qualitatively distinct cosmological behaviors. In particular, 
for $\delta,\epsilon>1$ the effective dark energy evolves from a 
quintessence-like regime to a phantom one, whereas the opposite transition may 
occur for $\delta,\epsilon<1$. Intermediate choices of the parameters can lead 
to an evolution very close to that of a cosmological constant. Nevertheless, in 
all cases the asymptotic future evolution approaches a de Sitter phase with 
$w_{DE}\rightarrow -1$. We mention that  the evolution of the matter and 
dark-energy density parameters 
follows the expected thermal history of the Universe, namely a transition from 
matter domination to late-time dark-energy domination, similarly to the 
standard cosmological scenario.
Finally, in Fig.~\ref{fig:results_separate_contours_lsec} we present the 
observational confrontation of the model with data from  Cosmic
Chronometers, Pantheon$^+$ Type Ia supernovae calibrated with SH0ES, BAO
measurements from DESI DR2 and compressed Planck 2018 CMB information.
 In summary, the Luciano-Saridakis generalized entropy provides a richer 
entropic cosmology, extending the 
phenomenology of Tsallis-, Barrow-, and Rényi-inspired cosmological models 
through the introduction of two independent holographic scaling exponents.

\subsection{Loop Quantum Gravity  entropic cosmology}

As discussed in subsection \ref{EESectionEntanglementOrigins}, the 
incorporation 
of non-extensive statistics into
the microscopic counting of LQG horizon states leads to the generalized
horizon entropy (\ref{SPCG22}), with the dependence on the Barbero-Immirzi
parameter encoded through~(\ref{LQGLambda}). We now investigate the
cosmological dynamics that arise when this entropy is applied to the apparent
horizon within the spacetime-thermodynamics framework.

 For a generic entropy functional $S(A)$, the Clausius relation
$-dE=T_h\,dS$, with $T_h=1/(2\pi\tilde r_A)$, leads to the generalized
dynamical Friedmann   
equation
\begin{equation}\label{eq:LQG_second_Friedmann_general}
\dot H-\frac{k}{a^2}
=
-\frac{\pi}{S'(A)}(\rho+p),
\qquad
A=4\pi\tilde r_A^2,
\qquad
\tilde r_A^{-2}=H^2+\frac{k}{a^2},
\end{equation}
and inserting~(\ref{SPCG22}) we   find
\begin{equation}\label{eq:LQG_second_Friedmann}
\dot H-\frac{k}{a^2}
=
-\frac{4\pi L_p^2}{\Lambda(\gamma)}
\exp\!\left[-(1-q)\Lambda(\gamma)\frac{\pi\tilde r_A^2}{L_p^2}\right]
(\rho+p).
\end{equation}

In order to cast the first Friedmann equation in a familiar form, we introduce 
an effective energy density $\rho_{\rm eff}$ through the standard geometric 
relation
\begin{equation}\label{eq:LQG_rhoeff_def}
H^2+\frac{k}{a^2}=\frac{8\pi L_p^2}{3}\,\rho_{\rm eff}.
\end{equation}
Combining Eq.~\eqref{eq:LQG_second_Friedmann} with the matter conservation law
$
\dot\rho+3H(\rho+p)=0,
$
one obtains an implicit modified first Friedmann equation that relates $\rho$ 
to $x\equiv H^2+k/a^2$ (equivalently to $\tilde r_A$). Writing the result in a 
compact form, one finds
\begin{equation}\label{eq:LQG_first_Friedmann_implicit}
\rho
=
\frac{3\,\Lambda(\gamma)}{8\pi L_p^2}
\left[
x\,\exp\!\left(\frac{C}{x}\right)
-
C\,{\rm Ei}\!\left(\frac{C}{x}\right)
\right]
+\rho_{\Lambda},
\qquad
x\equiv H^2+\frac{k}{a^2},
\qquad
C\equiv (1-q)\Lambda(\gamma)\frac{\pi}{L_p^2},
\end{equation}
where ${\rm Ei}(z)$ is the exponential integral function and $\rho_{\Lambda}$ 
is an integration constant (which can be interpreted as a cosmological-constant 
contribution). Equations~\eqref{eq:LQG_rhoeff_def} 
and~\eqref{eq:LQG_first_Friedmann_implicit} together determine $\rho_{\rm eff}$ 
as 
a function of the physical $\rho$ (and of the parameters $q,\gamma$), while 
Eq.~\eqref{eq:LQG_second_Friedmann} provides the corresponding dynamical 
equation.

Finally, for completeness, the effective pressure emerging in this framework 
can be written (after rearrangement) as~\cite{Liu:2021dvj}
\begin{equation}\label{eq:LQGpressure}
\begin{aligned}
p_{eff}= &- \frac{\rho}{3} +  \frac{\rho L^2_p}{1-q} \frac{1}{2 \pi 
\tilde{r}^2_A } 
\left[1- \exp \left( -(1-q)\Lambda (\gamma) \frac{\pi \tilde{r}^2_A }{ L^2_p}  
\right) \right] \\
& - \rho \left[1- \Lambda (\gamma) \exp \left( -(1-q)\Lambda (\gamma) 
\frac{\pi \tilde{r}^2_A }{ L^2_p}  \right) \right]
+ \frac{\rho}{2 \pi (1-q)} \left[1- \exp \left( -(1-q)\Lambda (\gamma) 
\frac{\pi \tilde{r}^2_A }{ L^2_p}  \right) \right].
\end{aligned}
\end{equation}

\subsection{Modified cosmology from non-extensive entropy with 
varying exponent}

This case constitutes the extension of the non-extensivity context, where the 
exponent of the non-extensive thermodynamics has a running behavior, depending  
on the energy scale. The conjecture of a running entropic exponent arises from 
quantum field theoretical considerations, when renormalization group is 
applied. 
While entropy corresponds to physical degrees of freedom, the degrees of 
freedom 
will depend on the scale, due to the renormalization of a quantum theory and 
quantum gravity. Hence, in general the exponent $\delta$ of Tsallis entropy can 
have a running behavior, and the entropy takes the form~\cite{Nojiri:2019skr}
\begin{equation}
\label{varTsal}
S = \frac{A_0}{4 G} \left(\frac{A}{A_0} \right)^{\delta (x)},
\end{equation}
where $x=\frac{H_1^2}{H^2}$, and where the energy scale is quantified by the 
value of the Hubble parameter $H$. More specifically, $H_1$ is a parameter with 
units of $H$ that sets the reference scale, and $A_0$ is a constant with 
dimensions through 
$A_0 = \frac{4\pi}{H_1^2}$. In the case where 
$\delta (x)=1$ the standard Bekenstein-Hawking entropy is recovered.

Following the aforementioned procedure of the  gravity-thermodynamics 
conjecture, in the case of a running exponent $\delta$, the modified Friedmann 
equations can be written as~\cite{Nojiri:2019skr}
\begin{equation}
\label{mgfevar2}
\frac{8\pi G}{3} \rho_{m} = \left. - H_1^2 \left\{ x^{\delta(x) - 2} + 2 \int^x 
dx x^{\delta(x) -3} \right\} 
\right|_{x=\frac{H_1^2}{H^2}} - \frac{\Lambda}{3} \, 
\end{equation}
and
\begin{equation}
\label{mgfevar1}
- 4\pi G \left( \rho_{m} + p_{m} \right) = \left\{ \delta 
+ \left[ \frac{H_1^2}{H^2} \ln \left( \frac{H_1^2}{H^2} \right) \right]
\delta' 
\right\} \left( \frac{H_1^2}{H^2} \right)^{\delta -1} \dot H  \, ,
\end{equation}
where $\delta'(x)\equiv\partial \delta(x)/\partial x$ and where prime denotes 
the 
derivative of a function with respect to $x$.

\begin{figure}[t]
\centering
\includegraphics[scale = 0.7] {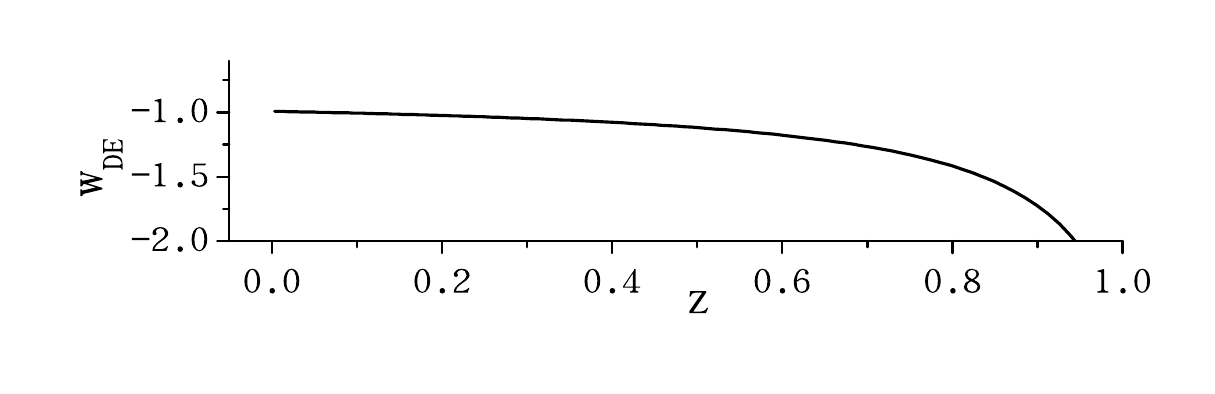} 
\caption{{\it The evolution of the   dark-energy equation-of-state 
parameter 
$w_\mathrm{DE}$  as a function of the redshift $z$, for the modified scenario 
through non-extensive thermodynamics with varying  exponent $\delta (x)$, for 
the parameter choices $\Lambda=0$, $b_1=0$, $n=4$, $H_1=1.1$ and $b_2=1$, 
in units where $8\pi G=1$. The figure is from~\cite{Nojiri:2019skr}.}}
\label{fig:wdevarexp}
\end{figure}

Due to the resulting new extra terms in the Friedmann equations of the extended 
scenario, an effective dark-sector arises, which is quantified by the varying 
entropic exponent $\delta (x)$. In this framework, analytical solutions of the 
observable quantities can be provided, which constitute the pillar for the 
examination of the cosmic evolution. Hence,   analytical expressions of the 
dark energy density parameter $\Omega_{DE}$ and of the equation-of-state 
parameter $w_{DE}$ can be provided as~\cite{Nojiri:2019skr}
\begin{equation}\label{varomegade}
\Omega_{DE}=\frac{\Lambda}{3H^2}-c\left( \frac{3 - n}{n-1} \right) \left( 
\frac{H_1^2}{H^2} \right)^{2-n}+1 
\end{equation}
and 
\begin{equation} \label{varwde}
w_{DE}=-1-\frac{2\dot{H}-2c\left(3 - n\right) \left( \frac{H_1^2}{H^2} 
\right)^{2-n} \dot{H}}{\Lambda-3c\left( \frac{3 - n}{n-1} \right) \left( 
\frac{H_1^2}{H^2} \right)^{2-n} H^2+3H^2},
\end{equation}
respectively,   with  $\Lambda$ the  cosmological constant  and where  $H$ 
is given by~(\ref{h2}). Note that 
in order to extract  relations~(\ref{varomegade}),~(\ref{varwde}) one has to 
consider  a specific ansatz 
for $\delta(x)$, that quantifies the aforementioned  
physical behavior of the extended entropy, namely~\cite{Nojiri:2019skr}
\begin{equation}
\label{Tslls17Bdelta}
\delta(x) = \frac{ \ln \left[c \left( x^{3-n} + \alpha(x) b_2 x^{2-n} + b_1 
b_2^2 
x^{1-n}\right)
\right] }{\ln x}\,,
\end{equation}
where 
\begin{equation}
\label{Tslls17Balpha}
\alpha (x)\equiv \frac{n (3 - n)}{(1-n)^2} + \frac{n^2}{(1+n)(2-n)} b_1 b_2^2 
x^{-n-1}\,, 
\end{equation}
with $n$, $b_1$, $b_2$ the model parameters, and 
$ c \equiv \left\{ \frac{3 - n}{(1-n)^2} + \frac{b_1}{1+n} 
\right\}^{-1}b_2^{n-2} $. 
From~(\ref{varomegade}) we deduce that  the dark-energy density parameter 
$\Omega_{DE}$ presents the usual thermal history of the Universe, 
with the sequence of matter and dark energy epochs (similar to that of  
Fig.~\ref{fig:fig1,3}  above).  Additionally,  from the analytical solution of 
the 
deceleration parameter through the relation $q=-1-\frac{\dot H}{H^2}$, we see  
that the transition from deceleration to acceleration happens at $z\approx 
0.6$, 
in 
agreement with observations. The cosmological 
behavior of the equation-of-state parameter is depicted in 
Fig.~\ref{fig:wdevarexp},  which is around $-1$  at present, as required by 
observations, while in the past it always lies in the phantom regime. We stress 
that in this example  the above behavior is obtained for $\Lambda=0$, i.e. 
without considering an explicit cosmological constant, namely it arises 
purely from the 
extra terms of the modified 
cosmology~\cite{Nojiri:2019skr,Nojiri:2019itp}. 

\subsection{Modified cosmology from Barrow entropy with varying anomalous 
dimension}
\label{BarRunCos}

Recently, a formally analogous extension has been investigated within the 
framewo
rk of Barrow entropy by promoting the Barrow anomalous dimension from a con
stant parameter to a running quantity depending on the cosmologica
l energy scale~\cite{Luciano:2026nhr}. As in the vary ing-Tsallis case, the 
energy scale is quantified through the dimensionless variable
$x=H_1^2/H^2$, where $H_1$ sets the reference scale. Owing to the correspondence 
between the $\delta$-Tsallis and Barrow entropy functionals, the resulting 
modified cosmological equations have the same formal structure as 
Eqs.~(\ref{mgfevar2}) and (\ref{mgfevar1}), upon the replacement
$\delta\rightarrow1+\Delta/2$.

As a specific ansatz for the running anomalous dimension, the parametrization
\begin{equation}
\label{runningBarrow}
\Delta(x)=\Delta_0+\frac{\Delta_1}{\ln x},
\end{equation}
was proposed, where $\Delta_0$ and $\Delta_1$ are free parameters to be 
constrained observationally. This choice is motivated by several complementary 
arguments. If the Barrow anomalous dimension encodes quantum-gravitational 
effects, as originally suggested in the black-hole 
context~\cite{Barrow:2020tzx}, it is natural to expect it to acquire a scale 
dependence, similarly to running couplings in quantum field theory. 
Furthermore, 
logarithmic corrections to the Bekenstein-Hawking entropy are ubiquitous in 
quantum-gravity approaches, while logarithmic running naturally arises in 
renormalization-group analyses, semiclassical gravity and effective field 
theory 
through one-loop quantum corrections. The parametrization~(\ref{runningBarrow}) 
therefore provides the minimal extension of the constant Barrow model while 
correctly recovering it in the limit $\Delta_1\rightarrow0$.

Within this framework, the modified Hubble parameter 
reads~\cite{Luciano:2026nhr}
\begin{equation}
\label{eq:genH}
\frac{H(z)}{H_0}=
\left[
\left(\frac{2-\Delta_0}{2+\Delta_0}\right)
\left(\frac{H_0}{H_1}\right)^{\Delta_0}
e^{-\Delta_1/2}
\Omega(z)
\right]^{\frac{1}{2-\Delta_0}},
\end{equation}
where
\begin{equation}
x=
\left[
\left(\frac{2-\Delta_0}{2+\Delta_0}\right)
\left(\frac{H_0}{H_1}\right)^2
e^{-\Delta_1/2}
\Omega(z)
\right]^{\frac{2}{\Delta_0-2}},
\end{equation}
and $\Omega(z)$ is the total normalized energy density, including matter, 
radiation and the cosmological constant.

\begin{figure}
\centering
\includegraphics[width=0.5\hsize,clip]{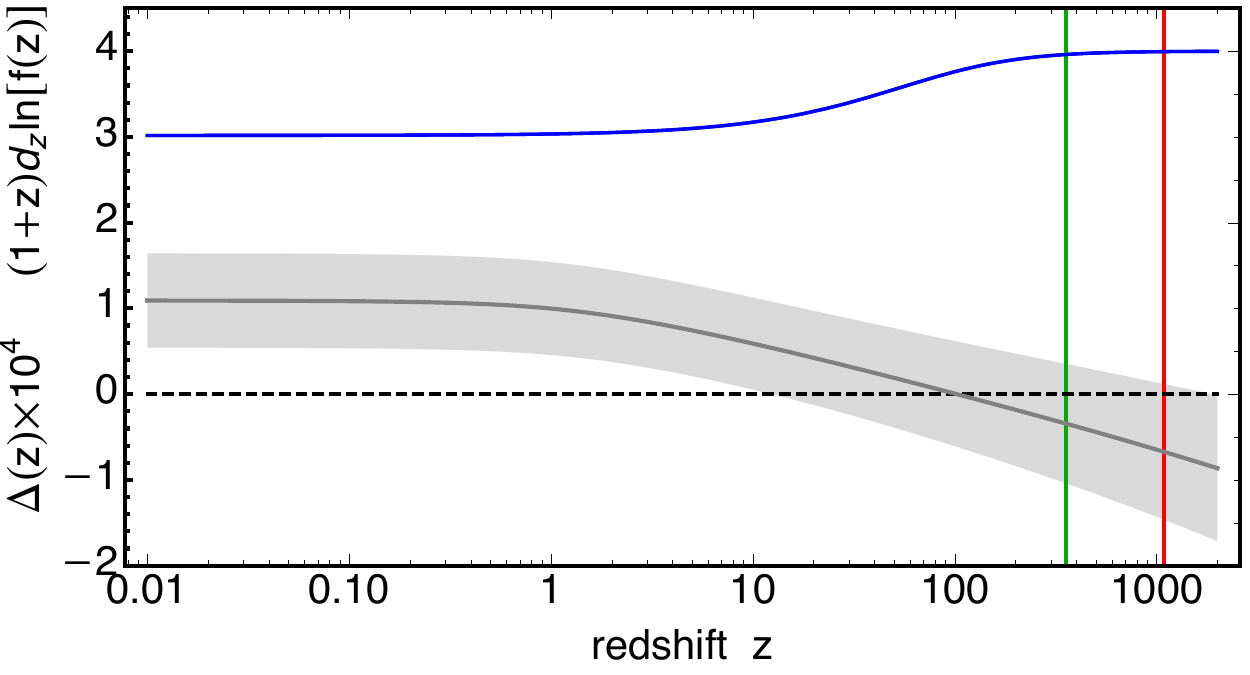}
\caption{\emph{Evolution of $\Delta(z)$, scaled by a factor of $10^4$, 
reconstructed from the analysis of Ref.~\cite{Luciano:2026nhr} (gray curve with 
the corresponding uncertainty bands). The dashed horizontal line denotes the 
standard case $\Delta=0$. For comparison, the neutrino matter-radiation 
transition function $(1+z),d_z\ln[f(z)]$ is also shown (blue curve). The 
vertical red and green lines indicate the photon decoupling ($z_\star$) and 
neutrino transition ($z_\nu$) redshifts, respectively. Figure from 
Ref.~\cite{Luciano:2026nhr}.}}
\label{changesign}
\end{figure}

The observational analysis based on OHD, Pantheon+, DESI DR2 BAO and CMB shift 
parameters shows that the running anomalous dimension remains extremely small 
throughout the cosmic evolution, while displaying a nontrivial scale dependence. 
In particular, as illustrated in Fig.~\ref{changesign}, $\Delta(z)$ undergoes a 
sign change during the early Universe, evolving from negative values at high 
redshift to positive values at late times. The transition takes place at $z_{\rm 
tr}\sim10^2$, close to the epochs of photon decoupling and neutrino decoupling. 
Interestingly, a qualitatively similar behavior has also been reported in the 
context of generalized holographic dark energy based on Barrow entropy (see 
Sec.~\ref{Varyingentexp}).

From a physical perspective, the sign reversal may signal the transition between 
two distinct microscopic regimes. At early times, when ultraviolet 
quantum-gravity effects dominate, a negative anomalous dimension corresponds to 
a reduction of the effective horizon degrees of freedom, possibly associated 
with dimensional reduction or other microscopic quantum effects. As the Universe 
expands and infrared physics becomes dominant, $\Delta$ becomes positive, 
indicating an enhancement of the effective gravitational degrees of freedom 
associated with the increasingly complex horizon geometry.

\begin{figure}[t]
{\includegraphics[width=0.62\hsize,clip]{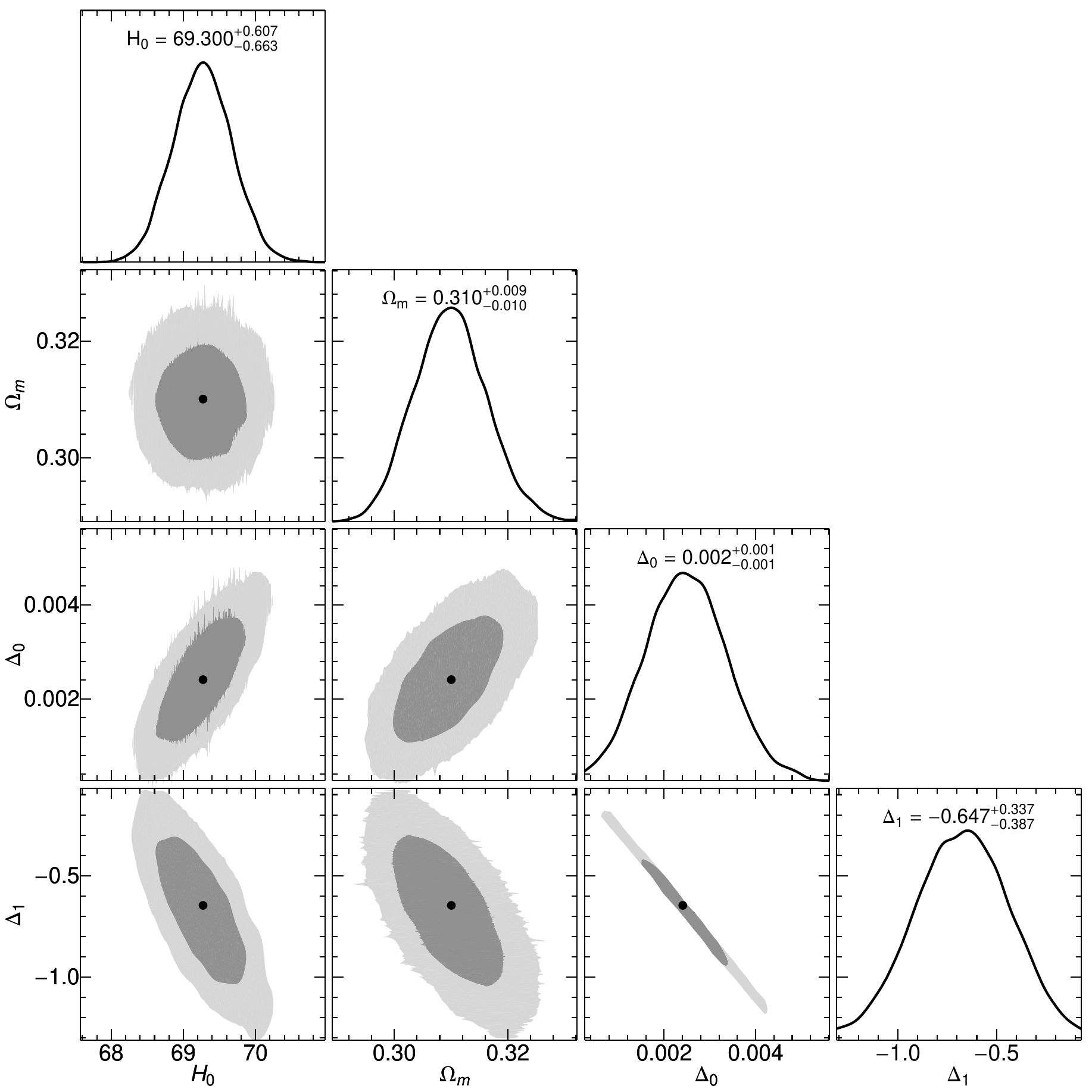}\hfill
\includegraphics[width=0.32\hsize,clip]{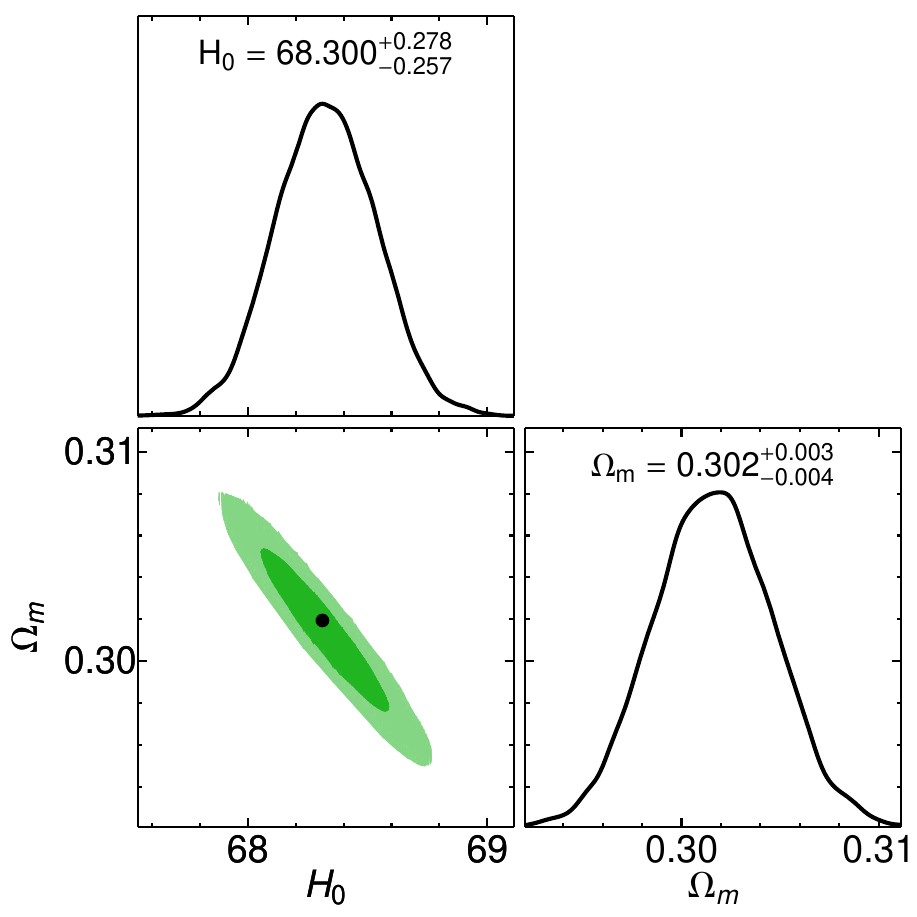}}
\caption{\emph{Marginalized MCMC posterior distributions for the running-Barrow 
model (left) and the $\Lambda$CDM model (right). The darker and lighter shaded 
regions correspond to the $1\sigma$ and $2\sigma$ confidence intervals, 
respectively, while the black dots denote the best-fit values. The figure is  
from Ref.~\cite{Luciano:2026nhr}.}}
\label{ConPlotBarRunning}
\end{figure}

Another interesting outcome concerns the Hubble tension. Although Bayesian model 
comparison still mildly favors the $\Lambda$CDM paradigm, the running-Barrow 
model predicts
$H_0=69.27^{+0.62}_{-0.65}$ km s$^{-1}$ Mpc$^{-1}$ (see 
Fig.~\ref{ConPlotBarRunning}), an intermediate value between the Planck 
determination,
$H_0^{\rm P}=67.36\pm0.54$ km s$^{-1}$ Mpc$^{-1}$~\cite{Planck:2018vyg}, and the 
SH0ES measurement,
$H_0^{\rm R}=73.04\pm1.04$ km s$^{-1}$ Mpc$^{-1}$~\cite{Riess:2021jrx}, 
suggesting a possible partial alleviation of the Hubble tension.

In summary, these results indicate that allowing the Barrow anomalous dimension 
to evolve with the cosmological scale leads to a richer cosmological 
phenomenology while remaining compatible with current observations. We finally 
note that the idea of scale-dependent Barrow-like entropy deformations has 
recently attracted growing attention in different gravitational contexts. In 
particular, running entropic parameters have been investigated both in 
cosmology 
and in black-hole thermodynamics, further supporting the relevance of this 
framework for probing possible quantum-gravitational 
effects \cite{DiGennaro:2022ykp,Basilakos:2023seo,Abreu:2024uvp,
Yerokhin:2026xwn}.

\subsection{Modified cosmology from  generalized 
mass-to-horizon entropy}

In this subsection we will present a modified cosmological scenario that is 
obtained  by adopting the gravity-thermodynamics correspondence, however  
employing the generalized mass-to-horizon entropy relation~(\ref{snentropy}), 
namely 
\begin{equation}
S_{n}=\gamma \frac{2n}{n+1} r_a^{\,n-1} S_{BH},
\label{snentropyBB}
\end{equation}
with
 $n$  a non-negative constant, and $\gamma$   a parameter with dimensions 
$[L]^{1-n}$ (in natural units $\hbar = k_{B} = c = 1$).  Assuming that the 
horizon temperature retains its standard Hawking form and inserting everything 
into the first law of thermodynamics gives~\cite{Basilakos:2025wwu}
\begin{equation}
\label{fried1}
-4\pi G (\rho_m + p_m)
= \gamma n \left(H^2 + \frac{k}{a^2}\right)^{\frac{1-n}{2}}
\left(\dot H - \frac{k}{a^2}\right),
\end{equation}
which corresponds to the modified second Friedmann equation. Integrating this 
equation and using the matter conservation law
$\dot\rho_m + 3H(\rho_m + p_m)=0$,
we finally arrive at
\begin{equation}
\label{fried2}
\frac{8\pi G}{3}\rho_m
= \frac{2\gamma n}{3-n}
\left(H^2 + \frac{k}{a^2}\right)^{\frac{3-n}{2}}
- \frac{\Lambda}{3},
\end{equation}
where $\Lambda$ arises as an integration constant. Equation~(\ref{fried2}) thus 
plays the role of the first Friedmann equation for the present scenario.

As expected, the application of spacetime thermodynamics with a generalized 
mass-to-horizon entropy leads to modified Friedmann equations and, 
consequently, 
to a modified cosmological dynamics. Restricting ourselves to a spatially flat 
Universe ($k=0$), these equations can be cast into the standard form
\begin{eqnarray}
\label{FR1}
H^2 &=& \frac{8\pi G}{3}\left(\rho_m + \rho_{DE}\right),\\
\label{FR2}
\dot H &=& -4\pi G \left(\rho_m + p_m + \rho_{DE} + p_{DE}\right),
\end{eqnarray}
by introducing an effective dark-energy density and pressure, respectively, as
\begin{eqnarray}
\label{rhode}
\rho_{DE} &=& \frac{3}{8\pi G}
\left[\frac{\Lambda}{3}
+ H^2
- \frac{2\gamma n}{3-n} H^{3-n}\right],\\
\label{pde}
p_{DE} &=& -\frac{1}{8\pi G}
\left[\Lambda + (2\dot H + 3H^2)
- 2\gamma n H^{1-n}
\left(\dot H + \frac{3}{3-n}H^2\right)\right].
\end{eqnarray}
Moreover, the corresponding equation-of-state parameter of the effective 
dark-energy sector is therefore given by 
\begin{equation}
\label{wde}
w_{DE} \equiv \frac{p_{DE}}{\rho_{DE}}
= -1 - \frac{2\dot H\left(1-\gamma n H^{1-n}\right)}
{\Lambda + 3H^2 - \frac{6\gamma n}{3-n}H^{3-n}}.
\end{equation}
Note that in the particular case $n=\gamma=1$, where the standard 
entropy is recovered, the above equations reduce to those of the $\Lambda$CDM 
paradigm.

Introducing the density parameters  and using the redshift as the independent 
variable, one obtains
 \be \label{omegademass}
\Omega_{DE}(z)=1-\left \{\frac{3-n}{2\gamma n}\left  
(\sqrt{\Omega_{m0}}H_{0}\right )^{n-1}(1+z)^{\frac{3(n-1)}{2}}\left 
[1+\frac{\Lambda}{3\Omega_{m0}H^{2}_{0}(1+z)^3}\right ]\right 
\}^{\frac{2}{n-3}},
\ee
which provides the analytical solution for the effective dark energy density 
parameter. Moreover, applying it at 
current  time gives the useful relation
\be \label{lambdacons}
\Lambda =\frac{6\gamma n}{3-n}H^{3-n}_{0}-3\Omega_{m0}H^{2}_{0}.
\ee
Differentiating~(\ref{omegademass}) we acquire
\be \label{omegaprime}
\Omega^{'}_{DE}(z)=3\left(\mathcal{A} 
\mathcal{B}\right)^{\frac{2}{n-3}}(1+z)^{2}\left[-1-\frac{6H^{2}_{0}\Omega_{m0}
(1+z)^{3}}{\mathcal{B}(n-3)}\right],
\ee
where
\begin{eqnarray}\nonumber
\!\!\!\!\!\!\!\!\!\!\!
\mathcal{A}
&=&-\frac{(n-3)\left(H_{0}\sqrt{\Omega_{m0}}\right)^{n-3}}{6\gamma n}, 
\\ 
\nonumber
\!\!\!\!\!\!\!\!\!\!\!\!\!\!\!\!\!\!\!\!
\mathcal{B}&=&\Lambda +3H^{2}_{0}\Omega_{m0}(1+z)^3.
\end{eqnarray}
Lastly,  the equation-of-state parameter is given as~\cite{Basilakos:2025wwu}
\be \label{wdez}
w_{DE}(z)=-1-\frac{{18}\left(\mathcal{A} 
\mathcal{B}\right)^{-\frac{2}{n-3}}\left(1-\gamma 
n\mathcal{C}^{1-n}\right)H^{4}_{0}\Omega^{2}_{m0}(1+z)^{3}}{\mathcal{B}
(n-3)\left[\Lambda -\frac{6\gamma n\mathcal{C}^{3-n}}{3-n}+3\left(\mathcal{A} 
\mathcal{B}\right)^{-\frac{2}{n-3}}H^{2}_{0}\Omega_{m0}\right]},
\ee
where
\begin{equation}
\mathcal{C}=\sqrt{\frac{H^{2}_{0}\Omega_{m0}}{\left(\mathcal{A} 
\mathcal{B}\right)^{\frac{2}{n-3}}}}.
\end{equation} 

In summary, we were able to extract analytical expressions for the important 
observable quantities. Hence, one can use them in order to  investigate in more 
detail the cosmological behavior in the scenario at hand. As 
can be explicitly demonstrated, the present framework successfully reproduces 
the standard succession of matter- and dark-energy dominated eras, while 
asymptotically evolving toward a de Sitter phase at late times 
\cite{Basilakos:2025wwu,Mondal:2026mgh,Shameeem:2026cwn,Sheykhi:2025ydd,
Denkiewicz:2025txx}. The transition from decelerated to 
accelerated 
expansion takes place at $z\approx0.6$, in agreement with current observational 
constraints. Additionally,  the effective dark-energy equation-of-state 
parameter remains close to -1  at the present epoch, displays a non-trivial 
dynamical evolution in the past, including a crossing of the phantom divide, 
and gradually converges to the cosmological-constant value in the asymptotic 
future. Finally, the scenario at hand is consistent with baryogenesis 
constraints~\cite{Luciano:2025fqg} and can successfully describe the  growth of 
 structures too~\cite{Luciano:2025tio}. For similar considerations,  
see~\cite{Anand:2025cer,Jusufi:2025rlr,Anand:2025rjg,Khodahami:2026qdt,
Luciano:2025ovj,Sheykhi:2026dxy}. 

Interestingly, a more detailed background analysis of this framework has 
recently shown that the standard thermal history of the Universe imposes 
remarkably stringent constraints on the generalized mass-to-horizon entropy 
parameters~\cite{Prasanthan:2026boc}. In particular, requiring consistency with 
Big Bang nucleosynthesis, matter-radiation equality, and CMB bounds on the 
early 
dark-energy fraction confines the viable parameter space to a narrow 
neighborhood of the Bekenstein-Hawking limit. Nevertheless, within this allowed 
region the model successfully reproduces the standard radiation-, matter-, and 
dark-energy-dominated sequence, while naturally approaching a $\Lambda$CDM-like 
cosmological evolution at late times.

More recently, the same framework has also been investigated in the primordial 
inflationary regime, providing a complementary probe of the generalized 
mass-to-horizon relation \cite{Sheykhi:2026dxy}. While the MHR modification does 
not alleviate the observational tension of representative power-law inflationary 
potentials, a markedly different behavior emerges for Starobinsky inflation. In 
this case, primordial observables are sensitive to departures from the 
Bekenstein--Hawking limit, with the scalar power-spectrum normalization 
providing, under a fixed-normalization prescription and for $N=60$, the 
stringent constraint $0.960\lesssim n\lesssim1.040$. 

\subsection{Modified cosmology in Cotton gravity}
\label{Cottongravitysub}

Cotton gravity is a recent extension of General Relativity in which the
gravitational dynamics is governed by the Cotton tensor, a geometrical tensor
constructed from derivatives of the Ricci tensor and encoding the conformal
properties of spacetime, rather than by the Einstein tensor~\cite{Harada}.
A particularly useful formulation is provided by the Codazzi parametrization,
which recasts the theory in terms of a divergence-free Codazzi tensor acting as
an additional geometric source in the gravitational field
equations~\cite{Mantica:2023stl}.

In an FRW Universe, the Codazzi tensor assumes the perfect-fluid form
\begin{equation}
C_{\mu\nu}
=
\mathcal{A}(t)u_\mu u_\nu
+
\mathcal{B}(t)g_{\mu\nu}
+
\frac{\Lambda}{3}g_{\mu\nu},
\end{equation}
where $u_\mu$ is the comoving four-velocity of the cosmic fluid. The functions
$\mathcal{A}(t)$ and $\mathcal{B}(t)$ completely characterize the homogeneous
and isotropic Codazzi sector and act as additional geometric contributions to
the effective energy density and pressure. They satisfy the Codazzi condition
$\dot{\mathcal{B}}=-H\mathcal{A}$, where $H$ is the Hubble rate.

Within this framework, it has recently been shown that the modified
gravitational dynamics admits a thermodynamic interpretation. In particular,
by applying the first law of thermodynamics at the apparent horizon of an FRW
Universe, the cosmological equations of Cotton gravity can be recast in a
Friedmann-like form, provided that the standard Bekenstein-Hawking entropy is
replaced by a generalized entropy functional incorporating the effects of the
Codazzi sector~\cite{Ghaffari:2026tik}. Specifically, the horizon entropy can
be written as
\begin{equation}
S=S_{BH}+\Delta S,
\end{equation}
where $S_{BH}$ is the standard Bekenstein-Hawking entropy and
\begin{equation}
\label{Delta S FRW}
\Delta S
=
\frac{\pi}{G}\int \tilde{r}_A^{4}\dot{\mathcal{B}}\,dt
=
-\frac{\pi}{G}\int \tilde{r}_A^{4}H\mathcal{A}(t)\,dt .
\end{equation}

Hence, the entropy correction is controlled by the Codazzi function
$\mathcal{A}(t)$, weighted by the evolution of the apparent horizon and the
Hubble rate. Its sign therefore depends on the behavior of the Codazzi sector
throughout the cosmological evolution. In an expanding Universe, with $H>0$,
if $\mathcal{A}(t)$ remains positive throughout the evolution, the correction
is negative and the horizon entropy is reduced relative to the
Bekenstein-Hawking value. Such behavior may signal a nonstandard thermodynamic
regime and has been discussed in connection with exotic components such as
phantom-like fluids~\cite{Gonzalez-Diaz:2004clm}. Conversely, if
$\mathcal{A}(t)$ remains negative, the entropy correction is positive and the
horizon entropy is enhanced. The standard limit is recovered for
$\mathcal{A}(t)=0$, for which the correction vanishes identically. Thus, the
modified horizon entropy provides a direct thermodynamic probe of the Codazzi
sector and encodes information about the additional geometric structure of
Cotton gravity.

Beyond cosmology, the thermodynamic formulation of Cotton gravity has also
been extended to static and spherically symmetric spacetimes, where the
Codazzi sector induces non-trivial corrections to the horizon entropy and
establishes a connection between the additional geometric structure, matter
content, and black-hole thermodynamics~\cite{Ghaffari:2026tik}. These results
further motivate the investigation of Cotton gravity in black holes and other
compact objects, and more generally at the interface between modified gravity
and horizon thermodynamics
~\cite{Sussman:2023wiw,Sussman:2024mzs,Clement:2024xmr,
Capozziello:2025kws,Khodadi:2025gsm,Mantica:2024mun,Ghaffari:2025qmv}.

\section{Holographic dark energy}
\label{Holographicdarkenergy}

The holographic dark energy paradigm constitutes a direct and 
ambitious attempt  to connect the observed late-time acceleration of the 
Universe with fundamental principles expected to govern a consistent theory of 
quantum gravity. Rather than introducing a new fundamental field or modifying 
the gravitational dynamics by hand, holographic dark energy builds upon the 
holographic principle, a concept that emerged from black-hole thermodynamics 
and 
quantum gravity, and elevates it to a guiding principle for cosmology. In this 
sense, holographic dark energy represents a conceptually economical framework, 
where the dark energy density is determined by global properties of spacetime 
and horizon physics, instead of local degrees of freedom.

Over the last two decades, holographic dark energy has evolved into a rich and 
well-developed theoretical framework, attracting   attention due to its 
ability to address deep conceptual issues, such as the origin and magnitude of 
dark energy, while remaining phenomenologically viable. Starting from its basic 
formulation, where the infrared cutoff is identified with a cosmological 
horizon, the model has been extensively explored at both the background and 
perturbative levels, confronted with observational data, and extended in 
multiple directions. In this section, we provide a systematic and 
self-contained presentation of holographic dark energy, beginning with its 
conceptual and theoretical foundations, followed by its basic formulation, 
observational status, and general cosmological consequences. This discussion 
will set the stage for the subsequent section, where various extensions and 
generalizations of the holographic dark energy framework are examined in detail.

\subsection{Conceptual and theoretical motivation}
\label{Intro}

The holographic principle constitutes one of the most interesting ideas that 
have emerged from the study of quantum gravity, and it is widely regarded as a 
key guiding principle toward a consistent microscopic description of spacetime. 
Its 
origin can be traced back to the seminal developments in black-hole 
thermodynamics, where it was realized that the entropy of a black hole scales 
with the area of its event horizon rather than with the enclosed 
volume~\cite{Bekenstein:1973ur,Hawking:1975vcx}. This   departure from 
the 
extensive behavior familiar from local quantum field theory suggests that the 
fundamental degrees of freedom of a gravitational system are encoded on its 
boundary.

As we discussed in Sec.~\ref{Entropydefinitions}, building upon these 
insights, ’t~Hooft proposed that the information content of 
a 
black hole, and more generally of a quantum gravitational system, could be 
fully described by degrees of freedom living on a lower-dimensional 
boundary~\cite{tHooft:1993dmi}. Susskind subsequently elevated this observation 
to a general principle, emphasizing its relevance within string 
theory~\cite{Susskind:1994vu}. A concrete and mathematically precise 
realization 
of this idea was later achieved through the AdS/CFT correspondence, where a 
gravitational theory in a higher-dimensional spacetime is shown to be exactly 
dual to a conformal field theory defined on its 
boundary~\cite{Maldacena:1997re}. 
Taken together, these developments established the holographic principle as a 
general statement about the nature of spacetime, namely that  the physics 
within a given 
region can, in principle, be fully encoded by degrees of freedom residing on 
its boundary.

Over the past decades, the holographic principle has found applications across 
a remarkably broad range of physical systems, including strongly coupled gauge 
theories, nuclear and high-energy physics, condensed matter systems, and 
quantum information theory~\cite{Liu:2006he,Hartnoll:2009sz,Takayanagi:2012kg}. 
Its potential relevance to cosmology, however, carries a special significance, 
since cosmology probes spacetime at the largest possible scales, where quantum 
gravitational effects, although subtle, may leave observable imprints. This 
connection   raises the question of whether the holographic principle 
can provide insight into some of the known puzzles of modern 
cosmology.

The main puzzle is the observed accelerated expansion of the Universe and the 
nature of dark energy, since as we mentioned above, high-precision 
observations of distant type-Ia supernovae, later corroborated 
by cosmic microwave background and large-scale structure data, revealed that 
the cosmic expansion is currently 
accelerating~\cite{SupernovaSearchTeam:1998fmf,
SupernovaCosmologyProject:1998vns,
Peebles:2002gy}. Additionally, one faces the coincidence problem, namely the 
fact that  
the 
energy densities of matter and dark energy are of the same order precisely at 
the present cosmic 
epoch~\cite{Peebles:2002gy,Padmanabhan:2002ji,Copeland:2006wr,Frieman:2008sn,
Caldwell:2009ix,Silvestri:2009hh,Li:2011sd,Bamba:2012cp,Padmanabhan:2002si,
Padmanabhan:2004qc}. These issues strongly suggest that dark energy may be 
intimately connected to physics beyond classical General Relativity, and 
possibly to quantum gravity itself.

In this context, the holographic principle offers a compelling conceptual 
framework. If the number of fundamental degrees of freedom within a given 
region of spacetime is bounded by its boundary area, then the vacuum energy 
contained in that region cannot be arbitrarily large. In other words, the 
holographic principle implies an intrinsic relation between ultraviolet and 
infrared cutoffs, which in turn constrains the total vacuum energy that can be 
stored within a cosmological volume. Hence, this observation provides a 
natural connection 
between quantum gravity considerations and the dark energy problem.

The first explicit application of the holographic principle to cosmology and 
dark energy was proposed by Li in 2004~\cite{Li:2004rb}. The resulting 
framework, 
known 
as holographic dark energy (HDE), posits that the dark energy density is 
determined by an infrared length scale $L$ associated with the Universe, in 
combination with the reduced Planck mass $M_p=1/\sqrt{8\pi G}$. By identifying 
$L$ with the future event horizon, the model yields a dynamically evolving dark 
energy component that is consistent with the holographic bound and capable of 
driving the observed cosmic acceleration. Remarkably, HDE may naturally 
alleviate
some aspects of the fine-tuning and coincidence problems, while remaining 
closely tied to fundamental principles of quantum gravity \cite{Addazi:2021xuf}.

Owing to its solid theoretical motivation and its phenomenological viability, 
holographic dark energy has attracted considerable attention and has been 
explored in a wide range of cosmological and astrophysical contexts. These 
include studies of spatial curvature, cosmic perturbations, neutrino physics, 
black-hole thermodynamics, and extensions to modified gravity 
theories~\cite{Zhang:2015rha,Myung:2007tx,Nojiri:2022ljp,Setare:2008mjw}. In 
what 
follows, 
we review the holographic dark energy paradigm, emphasizing its theoretical 
foundations, its cosmological implications, and its various extensions.

\subsection{Holographic dark energy: the basic model} 
\label{HDE}

Among the many proposals put forward to explain the origin of dark energy, 
holographic dark energy (HDE) occupies a distinguished position, since it is 
directly based on the holographic principle and therefore on fundamental 
expectations from quantum gravity~\cite{Li:2004rb}. As emphasized in the 
previous subsection, the holographic principle suggests that the number of 
independent degrees of freedom describing a gravitational system is bounded by 
the area of its boundary rather than by its volume. Consequently, any effective 
description of vacuum energy in cosmology should respect this intrinsic 
holographic bound.

To formulate this idea quantitatively, let us assume that the Universe is 
characterized by a cosmological length scale $L$, which plays the role of an 
infrared cutoff. According to the holographic principle, all physical 
quantities, including the dark energy density, should be expressible in terms 
of 
boundary quantities. In the absence of additional dimensionful parameters, the 
only relevant scales are the reduced Planck mass $M_p=1/\sqrt{8\pi G}$, which 
describes the strength of gravity, and the length scale $L$, which captures the 
size of the accessible region of the Universe.

From dimensional analysis, the most general form of the dark energy density,
consistent with these assumptions, can be written as~\cite{Li:2004rb}
\begin{equation}
\label{rhodecomplete}
\rho_{de}=C_1 M_p^4+C_2 M_p^2 L^{-2}+C_3L^{-4}+\dots\, .
\end{equation}
 Note that the parameters $C_i$ could, in 
principle, exhibit a non-trivial dynamics. However, for phenomenological 
purposes, one can  set $C_j$, 
$j=2,3,\dots,$ to constants, by reabsorbing their time dependence into a 
suitable 
redefinition of $L$. 
We mention here that  the 120-order-of-magnitude discrepancy with 
the observational value of vacuum 
energy~\cite{Peebles:2002gy}, and theoretical inconsistencies with the 
holographic principle in the 
local quantum field theory~\cite{Cohen:1998zx} 
suggest that the $C_1$ term should not contribute to $\rho_{de}$. 
In particular, keeping the holographic principle and the
Bekenstein bound in mind, we deduce that, for an effective quantum field theory 
in a box of
size $L$ with UV cutoff $\Lambda$, the total
entropy should not exceed the Bekenstein limit, i.e. $L^3\Lambda^3\lesssim 
S_{BH}=\pi L^2M_p^2$, where $S_{BH}$ denotes the entropy
of a black hole with radius $L$. This sets a relation between the UV cutoff 
$\Lambda$ and the length $L$, which acts as an IR cutoff. 

Nevertheless, based on the validity of effective quantum field theory, 
an even stronger constraint on the IR cutoff was proposed 
in \cite{Cohen:1998zx}, requiring 
the total energy in a region of size $L$ to be less than the mass of a black 
hole of the same size. Since the maximum energy density in the effective theory 
is $\Lambda^4$, the UV-IR relationship turns into $L^3\Lambda^4\lesssim L 
M_p^2$, which gives the upper bound $\rho_{de}\sim \Lambda^4\lesssim 
M_p^2L^{-2}$ on the   zero-point energy density. Therefore, by comparison 
with 
Eq.~\eqref{rhodecomplete}, it follows that the $C_1$ term is expected not to 
contribute to $\rho_{de}$.
Lastly, 
the 
third and any other term in the above expansion turn out to be 
negligible compared to the second one. 

In summary, considering all these aspects together, 
Eq.~\eqref{rhodecomplete} can be finally simplified to~\cite{Li:2004rb}
\be
\label{rhosimp}
\rho_{de}=3C^2M_p^2L^{-2}\,,
\ee
for a suitable redefinition of $C_2$ in terms of the new constant $C$. We 
notice 
that this constant cannot be inferred from the theoretical framework
of the HDE model, but can only be constrained by fitting the observational data 
(see Sec.~\ref{Obs} for more discussion).

The next ingredient in the construction of HDE is the choice of the IR cutoff. 
The original proposals of the Hubble scale $L=1/H$ and the particle horizon 
were 
proved unsuitable to describe the correct EoS of DE~\cite{Hsu:2004ri} and the 
accelerating expansion of the Universe, respectively. On the other hand, a 
competitive DE model could be obtained by setting $L$ equal to the \emph{future 
event horizon}~(\ref{fevhor}),  namely
\be
\label{feh}
L=a\int_t^{\infty}\frac{dt'}{a}=a\int_a^{\infty}\frac{da'}{Ha'^2}\,,
\ee
which represents the spacetime boundary beyond which future events become 
causally disconnected from a fixed observer. 

\begin{figure}[t]
\begin{center}
\includegraphics[width=8.7cm]{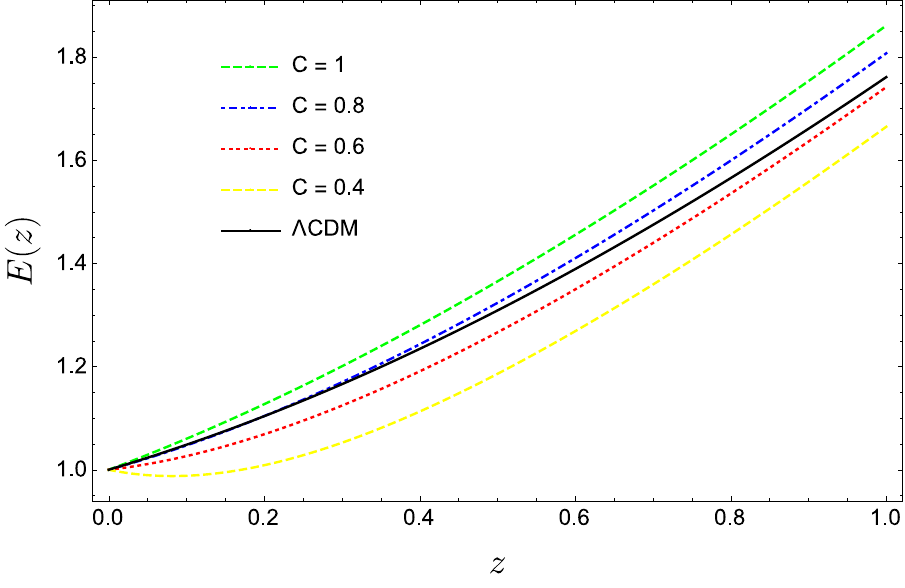}
\caption{{\it{The Hubble function evolution  versus $z$, in the case of the 
basic model of holographic dark energy, for different values of $C$. We set 
$\Omega_{m0}\simeq0.31$ and $\Omega_{de0}\simeq0.69$ according to the latest 
Planck data~\cite{Planck:2018vyg}. The $\Lambda$CDM plot (black solid curve) is 
shown for comparison. The figure is adapted from~\cite{Wang:2016och}.}}}
\label{Hzbasic}
\end{center}
\end{figure}

In the above setting, by assuming a flat Universe dominated by 
(non-interacting) 
HDE and dust matter, one can write the first Friedmann equation  as 
\begin{eqnarray} 
3M_p^2H^2=\rho_m+\rho_{de},
\label{Fr1eqHDE}
\end{eqnarray}
and thus 
the redshift
evolution of the Hubble parameter of the HDE scenario can be reconstructed   as 
\be
E(z)=\sqrt{\frac{\Omega_{m0}\left(1+z\right)^3}{1-\Omega_{de}(z)}}\,,
\ee
where $\Omega_{de}=\rho_{de}/\rho_c=\left[C/(LH)\right]^2$
and $\rho_c=3M_p^2H^2$ is the critical density. Derivation with respect to 
$\log 
a$ (denoted by a prime) yields the differential equation 
\be
\label{diffOm}
\Omega_{de}'=2\Omega_{de}\left(-\frac{H'}{H}-1+\frac{\sqrt{\Omega_{de}}}{C}
\right),
\ee
where we have used the definition~\eqref{feh}. On the other hand, 
from~\eqref{Fr1eqHDE}, we acquire
\be
 \frac{H'}{H}=\frac{
3-\Omega_{de}}{2}-\frac{\Omega_{de}^{3/2}}{C}\,,
\ee
which can be replaced into~\eqref{diffOm} to finally give
\be
\label{diffOm2basic}
\frac{d\Omega_{de}}{dz}=-\frac{\Omega_{de}\left(1-\Omega_{de}\right)}{1+z}
\left(1+\frac{2\sqrt{\Omega_{de}}}{C}\right).
\ee
Since, by definition, the fractional energy density $0<\Omega_{de}<1$, one can 
immediately infer that $\Omega_{de}$ always increases as $z\rightarrow-1$. 
Thus, the expansion driven by HDE will never reverse, leaving the Universe safe 
from any future re-collapse or Big Crunch. 
The redshift evolution of $H(z)$ of the HDE model is plotted in 
Fig.~\ref{Hzbasic} 
for different values of $C$.

Let us now turn our attention to the equation-of-state parameter of HDE. 
Towards this end, we consider 
the energy conservation equations for both dust matter and dark energy, which 
are
\be
\label{ConsEq}
\rho'_m+3\rho_m=0\,,\,\qquad\,\rho'_{de}+3\left(1+w_{de}\right)\rho_{de}=0\,.
\ee
From the latter relation, by deriving Eq.~\eqref{rhosimp} with respect to $\log 
a$, we easily obtain
\be
\label{wdeMine}
w_{de}=-\frac{1}{3}-\frac{2}{3}\frac{\sqrt\Omega_{de}}{C}\,.
\ee
This expression enables us to distinguish two different regimes.
In the early Universe, HDE is sub-dominant and we have 
$\Omega_{de}\ll1\Longrightarrow w_{de}\simeq-1/3$, which  implies 
$\Omega_{de}\sim 
a^{-2}$. On the other hand, 
as the Universe evolves  HDE becomes increasingly important and ends up 
dominating at late times, which yields $\Omega_{de}\simeq1\Longrightarrow 
w_{de}\simeq-1/3-2/(3C)$. Therefore, HDE with the cutoff~\eqref{feh} drives 
an accelerated expansion of the Universe provided that $C>0$.

Holographic dark energy at late times can further exhibit three
  cases, 
depending on the specific value of $C$.  i) When  $C=1$ we 
obtain $ w_{de}=-1$, thus recovering the cosmological 
constant  behavior. On the other hand, when 
  $C<1$  we find $w_{de}<-1$, and thus in this scenario  we obtain a phantom 
Universe, and a possible Big Rip  
fate~\cite{Caldwell:1999ew,Caldwell:2003vq}.
Finally, for $C>1$ we have $ w_{de}>-1$,  thus  HDE   
behaves as   
quintessence~\cite{Zlatev:1998tr}, and the Universe undergoes an 
eternal cosmic expansion.

Before moving on to the study of supplemental  arguments for HDE, it is 
interesting to discuss how this paradigm deals with the cosmological 
coincidence 
and fine-tuning problems. Concerning the former, we start by noticing that a 
common
reformulation is to explain why the ratio between the dark energy and radiation 
densities 
turns out to be so small at the onset of the radiation dominated 
era~\cite{Steinhardt:1999nw}. Following~\cite{Li:2004rb}, we assume that the 
energy budget during inflation consists only of  HDE and 
inflation energy, with the second component being nearly constant and 
completely 
decaying into radiation at the end of inflationary epoch. For 
$E_{inf}\simeq10^{14}\,\mathrm{GeV}$, straightforward calculations give 
$\rho_{de}/\rho_{r}\simeq10^{-52}$ as long as inflation lasts for the minimal 
number  of e-folds $N\simeq60$~\cite{Li:2004rb}. Therefore, HDE and inflation 
provide an elegant way out of the coincidence problem for a suitable value of 
$N$.
On the other hand, while the holographic assumption potentially implies a 
reduced need for fine-tuning, a definitive answer to this problem remains a 
subject of theoretical exploration.  
Tentative solutions are addressed in~\cite{Cohen:1998zx} by scaling dark energy 
through 
cosmological scale or in~\cite{Zhang:2009un} by associating dark energy with 
the 
spacetime scalar curvature rather than Planck or other high energy physical 
scales. However, since such models do not set the IR cutoff to the future event 
horizon, they suffer from some of the other conceptual shortcomings discussed 
above Eq.~\eqref{feh}.

We close this subsection by   briefly presenting  alternative dark energy 
models 
that are still compatible with HDE. The fact that these explorations stem from 
independent 
conceptual principles further bolsters the role of holography in understanding 
the nature of DE.

\paragraph{Vacuum entanglement energy.}
In the realm of quantum information theory, a model  which identifies dark 
energy as the \emph{vacuum entanglement 
energy} of the Universe  was proposed 
in~\cite{Lee:2007zq}. The basic ingredient is the entanglement entropy of the 
quantum field vacuum with a horizon, which is generally expressed in the form 
$S_{ent}=\beta R_h^2/l^2$. Here, $\beta\sim\mathcal{O}(1)$ is a numerical 
constant, $R_h$ the future event horizon with Gibbons-Hawking temperature 
$T_{ent}=1/(2\pi R_h)$, and $l$ the UV cut-off of quantum gravity. By defining 
the vacuum entanglement energy as the ``thermal energy'' related to the 
entanglement entropy, i.e. $d E_{ent}=T_{ent} d S_{ent}$, after integration one 
arrives at the energy density $\rho_{ent}=3d^2 M_p^2/R_h^2$, where 
$d=\sqrt{\beta N_{dof}/}(2\pi l M_p)$~\cite{Lee:2007zq}. Here, $N_{dof}$ is the 
number of spin degrees of freedom of quantum fields in $R_{h}$. Thus, 
such a model is consistent with the HDE form of Eq. \eqref{rhosimp}. 
Moreover, for $l\sim1/M_p$,  $N_{dof}\sim\mathcal{O}(10^2)$ as in the Standard 
Model and $\beta\simeq 0.3$ according to lattice simulation, we infer 
$d\sim\mathcal{O}(1)$, which is the naturally expected estimate in HDE. 

\paragraph{Holographic gas.}
From the condensed matter perspective, non-perturbative systems are often 
described by weakly interacting quasi-particle excitations. The idea of 
extending this paradigm to gravitational systems, and in particular to dark 
energy (where 
the Newton's constant appears in the denominator of the entropy formula 
$S=A/(4G)$) was first addressed in~\cite{Li:2008qh}. The study revolves around 
the investigation of the phenomenological features of a gas of 
\emph{holographic 
particles}, which are conjectured to represent the quasi-particle excitations 
of 
a strongly correlated gravitational system. Under this assumption, 
the degeneracy of a holographic particle with fixed momentum $k$ is expressed 
as 
$w=w_0 k^a V^b M_p^{3b-a}$, where $V$ is the volume of the system, while 
$w_0,a$ 
and $b$ are dimensionless constants. By leveraging the holographic idea, one 
can 
take the scaling $S\propto V^{2/3}$ and $T\propto V^{-1/3}$ for the entropy and 
temperature, respectively,  which yield the proportionality relation $a=3b-2$. 
In turn, the energy density of the system reads $\rho=E/V=\left(a+3\right) S 
T/\left[\left(a+4\right)V\right]$~\cite{Li:2008qh}.   

For the case of a cosmological holographic gas, the obtained density can be 
further manipulated by using the Gibbons-Hawking expressions of temperature and 
entropy, namely $T=1/(2\pi R)$ and $S=8\pi^2M_p^2R^2$, with $R$ being the 
Universe's radius. Therefore, we are led to the HDE-like form 
$\rho_{hg}=3c^2M_p^2R^{-2}$, 
where $c^2=\left(a+3\right)/\left(a+4\right)$. Moreover, statistical arguments 
on the convergence of the partition function require $a>-3$ and 
thus $ 
0<c^2<1$, which results in a phantom-like behavior of the holographic gas at 
late times.

\paragraph{Casimir energy.}

The Casimir effect is perhaps the most striking manifestation of zero-point 
energy~\cite{Casimir:1947kzi}. In its simplest form, it predicts the emergence 
of an attractive force between two parallel plates in vacuum. The suggestive 
possibility that the Casimir energy of the electromagnetic field in static de 
Sitter 
space could be the source of dark energy was considered in~\cite{Li:2009pm}. 
Specifically, it was shown that for a de Sitter Universe of radius $L$, the 
Casimir energy density scales as $\rho_{Cas}\sim M_p^2L^{-2}$, 
which in fact exhibits a HDE form. 
Finally, the cosmological 
uses 
of Casimir energy were reviewed in~\cite{Elizalde:2003bz}.

\paragraph{Holographic dark energy from action principle.}

In~\cite{Li:2012xf} the HDE model in the FRW geometry was derived for the first 
time by applying the variational principle to the action
\be
S=\frac{1}{16\pi G}\int dt \left[\sqrt{-g}\left(R-\frac{2 
c}{a^2(t)L^2(t)}\right)-\lambda(t)\left(\dot L(t)+\frac{N(t)}{a(t)}\right)
\right]+S_M\,,
\ee
where $N(t)$ is the redshift factor, $R$ the Ricci scalar, $\sqrt{-g}=Na^3$, 
$L$ 
a local cutoff and $S_M$ the action of all matter fields. Then, variation with 
respect to $N, a, L$ and the Lagrange multiplier $\lambda$ allows one to infer 
the 
corresponding equations of motion and constraints on $L$ and $\lambda$ 
(see~\cite{Li:2012xf} for computational details). Two results do stand out. On 
the  
one hand, it is shown that $aL$ corresponds exactly to the future event 
horizon. 
In other words, the usage of the future event horizon as an IR cut-off is not 
an 
input, but naturally follows from the equations of motion of such a variational 
derivation. On the other hand, the solution of these equations at sufficiently 
late times gives the energy density of dark energy $\rho_{de}=\left(8\pi 
G\right)^{-1}\left(\frac{C}{a^2L^2}+\frac{\lambda}{2a^4}\right)$. The first 
term 
is in the HDE-like form, while the $\lambda(0)$ term behaves the same way as 
radiation and can thus be  regarded as a ``dark radiation'' 
component~\cite{Hamann:2010bk}. See also~\cite{Dehyadegari:2026drv} for a 
recent consideration.

 \subsection{Observational status and phenomenology}
\label{Obs}

As discussed below Eq.~\eqref{wdeMine}, the cosmological behavior of 
holographic 
dark 
energy is largely controlled by the dimensionless parameter $C$, which 
determines the late-time equation of state and, consequently, the ultimate fate 
of the Universe. From a theoretical point of view, the holographic framework 
provides only qualitative guidance on the expected magnitude of $C$, typically 
suggesting that it should be of order unity. However, the precise value of this 
parameter, and in particular whether $C$ is smaller, equal, or larger than 
unity, can only be established through confrontation with observations.

Over the past twenty years, substantial progress has been achieved in 
constraining the basic holographic dark energy model using a variety of 
cosmological probes. Early analyses already indicated that HDE can fit the 
background expansion history of the Universe comparably well to the standard 
$\Lambda$CDM scenario, while exhibiting a richer dynamical structure. 
Subsequent 
studies incorporating high-precision cosmic microwave background data, baryon 
acoustic oscillations, type-Ia supernovae, and large-scale structure 
observations have significantly tightened the allowed parameter space.

In particular, combined analyses involving Planck temperature and polarization 
data, often supplemented by weak-lensing information, tend to favor values of 
$C$ smaller than unity at the $1\sigma$ confidence level. This regime 
corresponds 
to a phantom-like equation of state for holographic dark energy at late times, 
implying that the effective equation-of-state parameter crosses below 
$w_{de}=-1$ and that the cosmic expansion may eventually culminate in a Big Rip 
singularity. The inclusion of lensing data generally strengthens this trend, 
reflecting the sensitivity of structure formation to the detailed dynamics of 
dark energy.

\begin{figure}[!]
\centering
\includegraphics[width=9cm]{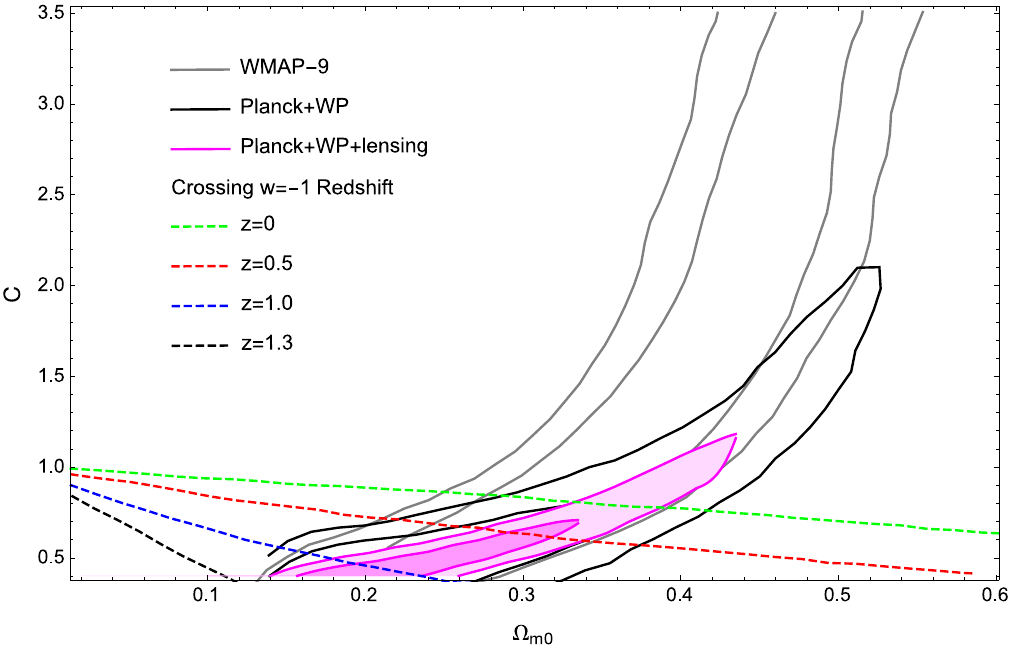} 
\caption{{\it{Marginalized $1\sigma$ and $2\sigma$ confidence-level contours in 
the $\Omega_m$-$C$ 
plane, for the basic model of holographic dark energy, 
namely~(\ref{diffOm2basic}). The figure is from~\cite{Li:2013dha}.}}}
\label{Constobs}
\end{figure}

At the same time, the present-day matter density parameter is consistently 
constrained to lie in the range $\Omega_{m0}\simeq0.26$-$0.28$ at the 
$1\sigma$ confidence level, in close agreement with independent determinations 
within $\Lambda$CDM. In Fig.~\ref{Constobs} we present  representative 
marginalized 
$1\sigma$ and $2\sigma$ confidence contours in the $\Omega_m$-$C$ plane, 
which reveal both the degeneracies and the preferred parameter region for the 
basic HDE model.

More refined observational investigations have explored potential systematic 
effects and model dependencies in greater detail. For instance, the analysis 
of~\cite{Wang:2013zca} revisited constraints on holographic dark energy using 
the 
SNLS3 supernova dataset, allowing for a redshift-dependent evolution of the 
supernova color-luminosity parameter $\beta$ and employing different 
light-curve fitting techniques. The results show that assuming a constant 
$\beta$ yields best-fit values $\Omega_{m0}\approx0.274$ and $C\approx0.687$, 
whereas permitting a linear redshift dependence $\beta(z)$ shifts the best-fit 
parameters to $\Omega_{m0}\approx0.288$ and $C\approx0.768$. This sensitivity 
underscores the importance of carefully accounting for astrophysical and 
observational systematics when constraining dynamical dark energy models such 
as HDE.

From a phenomenological perspective, the observationally preferred values of 
$C$ suggest that holographic dark energy behaves dynamically and departs from a 
pure cosmological constant, while remaining fully compatible with current data 
within statistical uncertainties. Nevertheless, it should be emphasized that 
the basic HDE model also faces challenges. In particular, the tendency toward a 
phantom regime raises questions regarding theoretical consistency and the 
treatment of future singularities, while mild tensions with certain datasets 
have motivated the exploration of extended scenarios. 
These considerations have led to a broad class of generalizations, including 
interacting holographic dark energy, extensions within modified gravity, and 
alternative choices of infrared cutoff. Such developments aim to preserve the 
holographic motivation while improving phenomenological flexibility and 
theoretical robustness. These extended frameworks will be discussed in detail 
in the next section.

\subsection{Holographic dark energy in non-flat Universe}

Although current observations mildly favor a nearly spatially flat Universe, 
the 
possibility of a small but nonvanishing curvature has played an important role 
in the historical development of cosmological models. 
In this respect, examining holographic dark energy in non-flat 
FRW geometries is both interesting and necessary.

A systematic extension of HDE to a closed Universe was first presented 
in~\cite{Huang:2004ai}, generalizing the original proposal of~\cite{Li:2004rb}. 
For positive spatial curvature ($k>0$), the relation between the comoving 
radial 
coordinate $r(t)$ and the future event horizon $R_h$ follows from
\begin{equation}
\int_0^{r(t)}\frac{dr}{\sqrt{1-kr^2}}
=\int_t^{\infty}\frac{dt}{a(t)}
=\frac{R_h}{a(t)}\,,
\end{equation}
which leads to
\begin{equation}
\label{rt}
r(t)=\frac{1}{\sqrt{k}}
\sin\!\left(\frac{\sqrt{k}R_h}{a}\right).
\end{equation}
Accordingly, the infrared cutoff is naturally chosen as $L=a(t)r(t)$.
Despite the geometric modifications, the relation
\begin{equation}
HL=\frac{C}{\sqrt{\Omega_{de}}}
\end{equation}
remains valid. 
Using this result, one finds
\begin{equation}
\dot L = \frac{C}{\sqrt{\Omega_{de}}}
-\cos\!\left(\frac{\sqrt{k}R_h}{a}\right),
\end{equation}
which leads to the equation of state
\begin{equation}
w=-\frac{1}{3}\left[
1+\frac{2}{C}\sqrt{\Omega_{de}}
\cos\!\left(\frac{\sqrt{k}R_h}{a}\right)
\right].
\end{equation}

Assuming the closure condition
$\Omega_k+\Omega_m+\Omega_r+\Omega_{de}=1$, 
the evolution of the holographic dark energy density is governed by
\begin{equation}
\frac{\Omega'_{de}}{\Omega_{de}^2}
=(1-\Omega_{de})
\left[
\frac{2}{C}\frac{1}{\sqrt{\Omega_{de}}}
\cos\!\left(\frac{\sqrt{k}R_h}{a}\right)
+\frac{\Omega_{de0}}{\Omega_{de0}+a\Omega_{k0}}
\frac{1}{\Omega_{de}}
\right].
\end{equation}
Although spatial curvature introduces additional mathematical complexity, the 
qualitative behavior of the model closely resembles that of the flat case. 
In particular, for $C=1$ the dark energy density asymptotically approaches a 
constant, while for $C>1$ it is gradually diluted by cosmic expansion. 
The formalism can be extended to open geometries by replacing 
$\sin(x)$ with $\sinh(x)$ and $\sqrt{k}$ with $\sqrt{|k|}$~\cite{Gong:2004cb}.

An alternative treatment of curvature effects was proposed 
in~\cite{Zhang:2014ija}, 
where the future event horizon itself is used as the infrared cutoff, yielding 
the same equation of state as in the flat case. 
Importantly, this approach also allowed for improved observational constraints 
using combined SN, BAO, CMB, and $H_0$ datasets, resulting in values of $C$ and 
$\Omega_{k0}$ consistent with Planck observations~\cite{Planck:2018vyg}.

\subsection{Physical and cosmological consequences of holographic dark energy}
\label{Implic}

Having established the basic formulation of holographic dark energy, we now 
turn 
to a discussion of its broader cosmological implications. 
Beyond providing a dynamical alternative to the cosmological constant, HDE 
exhibits a number of distinctive theoretical and phenomenological features that 
set it apart from conventional dark energy models. 
These features concern the role of spatial curvature, the interplay with 
particle physics sectors such as neutrinos, and the behavior of cosmological 
perturbations. 
Together, they demonstrate that HDE is not just a phenomenological ansatz, 
but a framework with rich internal structure and non-trivial observational 
consequences.

\paragraph{Neutrino Physics.}

Neutrinos constitute a fundamental component of the cosmic energy budget and 
play a crucial role in structure formation and cosmic evolution. 
Despite significant progress, both the absolute neutrino mass scale and the 
mass hierarchy remain unresolved.

The impact of neutrino physics within the HDE framework was first explored 
in~\cite{Li:2012spm}. 
It was shown that HDE leads to tighter upper bounds on the sum of neutrino 
masses 
compared to the standard $\Lambda$CDM model. 
Moreover, the data exhibited a mild preference for the inverted hierarchy. 
Subsequent analyses~\cite{Wang:2016tsz} refined these results, finding that 
while 
the minimum $\chi^2$ slightly favors the normal hierarchy, the statistical 
significance remains insufficient for a definitive conclusion. 
These studies highlight the sensitivity of HDE to particle physics parameters 
and reveal its potential as a framework linking cosmology and neutrino 
phenomenology.

\paragraph{Density fluctuations.}
\label{Densfluc}

Unlike a pure cosmological constant, holographic dark energy is dynamical, and 
its perturbations can affect the evolution of gravitational potentials and 
large-scale structure. 
Early stability analyses based on the sound speed formalism~\cite{Myung:2007pn} 
provided useful insight, but such methods are not fully adequate for HDE, whose 
energy density is intrinsically nonlocal due to its dependence on the future 
event horizon.

\begin{figure}[t]
\includegraphics[width=8.7cm]{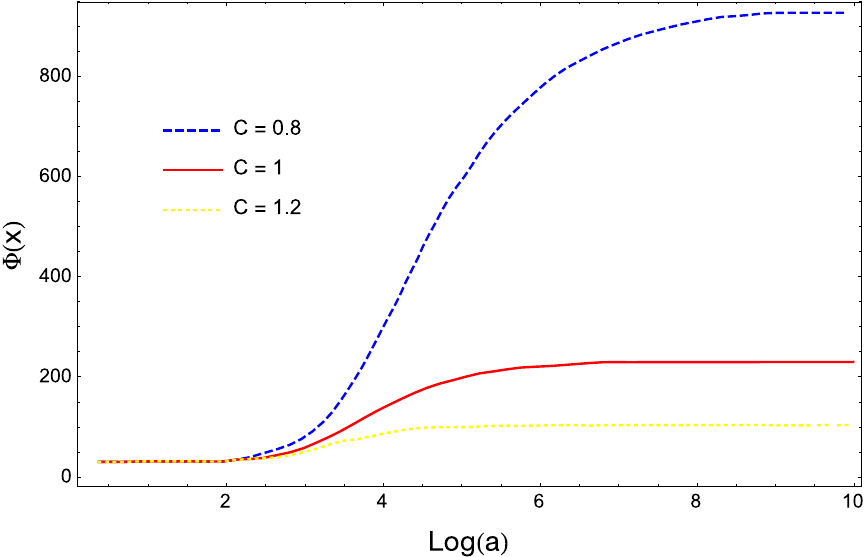}
\includegraphics[width=8.7cm]{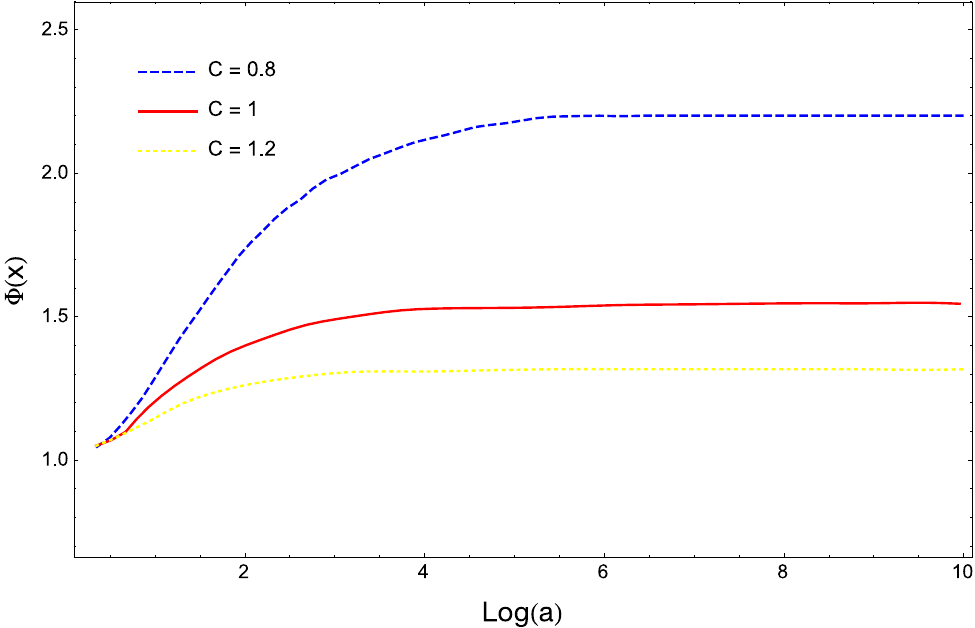}
\caption{{\it{Gravitational potential $\Phi$ as a function of $\log(a)$ in the 
$kr_{h0}\ll1$ limit. The left panel refers to the matter dominated era 
($\Omega_{de}|_{x=0}=0.01$), while the right panel to the dark energy dominated 
era 
($\Omega_{de}|_{x=0}=0.71)$. In all cases, we observe that perturbations are 
bounded from above, which means that HDE 
fluctuations are stable at late times.
}}}
\label{PertubHDEbasic}
\end{figure}

A more refined treatment was developed in~\cite{Li:2008zq}, where scalar 
perturbations were attributed to fluctuations of the event horizon itself. 
Working in the Newtonian gauge,
\begin{equation}
ds^2=-\left(1+2\Phi\right)dt^2
+a^2(t)\left(1-2\Phi\right)d\mathbf{x}^2,
\end{equation}
the perturbed future event horizon can be consistently defined and varied.
At leading order, the resulting fluctuation of the holographic dark energy 
density is
\begin{equation}
\delta\rho_{de}
=-2\rho_{de}\frac{\delta R_h}{R_h}.
\end{equation}
Substituting this relation into Einstein’s equations yields an 
integro-differential equation for the gravitational potential. 
Analytical solutions can be obtained in both the sub-Hubble and super-Hubble 
limits. 
In the former case, perturbations decay with time, while in the latter they 
approach a constant value. 
Although a temporary growth may occur near horizon crossing, it remains 
bounded, 
ensuring the overall stability of the model. In Fig.~\ref{PertubHDEbasic} we 
depict the evolution of the gravitational potential, showing 
that   perturbations are bounded from above, which means that HDE 
fluctuations are stable at late times.

These results demonstrate that holographic dark energy is free from classical 
instabilities at the perturbative level. 
A fully quantum treatment of initial fluctuations would require a consistent 
quantum gravity framework and remains an open direction for future research.

\paragraph{Varying gravitational constant.}

A natural question that arises within the holographic dark energy framework is 
whether the presence of a dynamical infrared cutoff, intimately linked to 
horizon 
physics, may also signal a departure from the standard assumption of a constant 
gravitational coupling. 
Indeed, since HDE effectively probes gravity at cosmological distances, while 
being sensitive to ultraviolet physics through the holographic bound, it is 
reasonable to expect that remnants of quantum gravitational effects could 
manifest themselves as a slow evolution of Newton’s constant.

Motivated by early ideas originating from Kaluza-Klein theories and 
scalar-tensor 
extensions of gravity, holographic dark energy models with a time-dependent 
gravitational coupling were developed 
in~\cite{Jamil:2009sq,Lu:2009iv,Alavirad:2013nfy}. 
In this setting, the basic HDE evolution equation is modified to
\begin{equation}
\Omega'_{de}
=
\Omega_{de}\left(1-\Omega_{de}\right)
\left[
1+\frac{2\sqrt{\Omega_{de}}}{C}
\right]
-
\Omega_{de}\left(1-\Omega_{de}\right)\frac{G'}{G}\,,
\end{equation}
where primes denote derivatives with respect to $\ln a$, and the quantity
$\Delta_G\equiv G'/G$ parametrizes the running of the gravitational coupling.

For simple parametrizations, such as $\Delta_G=\mathrm{const}$, the equation 
can 
be solved analytically, allowing for direct confrontation with observations. 
Using combined SN, BAO, CMB and Hubble data, it was found that in a flat 
Universe
\begin{equation}
\Delta G=-0.0016\pm0.0049\qquad (1\sigma),
\end{equation}
corresponding to $C=0.80^{+0.16}_{-0.13}$~\cite{Lu:2009iv}. 
For a non-flat background, the constraints become
\begin{equation}
\Delta G=-0.0025^{+0.0080}_{-0.0050}\qquad (1\sigma),
\end{equation}
with $C=0.80^{+0.19}_{-0.14}$. 

These results indicate that a mild variation of $G$ is compatible with current 
data, while the holographic parameter $C$ remains close to the values preferred 
in the minimal HDE scenario.

\paragraph{Inflationary model.}

Holographic dark energy  can exhibit a non-trivial interplay with the 
inflationary epoch, too. 
One particularly appealing idea is that a sufficiently long period of inflation 
may alleviate the cosmological coincidence problem by naturally diluting 
the initial HDE density~\cite{Wang:2016och}. 
This possibility was already hinted at in the original proposal 
of~\cite{Li:2004rb}, and later explored in detail in~\cite{Chen:2006qy} and 
especially in~\cite{Nojiri:2019kkp}.

Assuming that sub-Hubble HDE fluctuations are decaying modes (as we did in 
the density fluctuations discussion above), one may consistently neglect 
intrinsic HDE perturbations 
during inflation. 
Within a minimal single-field slow-roll setup, the background equations take 
the 
form
\begin{eqnarray}
3M_p^2H^2
&=&
\frac{\dot\varphi^2}{2}+V(\varphi)+3C^2M_p^2R_h^{-2}\,,\\[2mm]
0
&=&
\ddot\varphi+3H\dot\varphi+V_\varphi\,,
\end{eqnarray}
while the HDE density parameter obeys
\begin{equation}
\Omega'_{de}
=
-2\Omega_{de}\left(1-\Omega_{de}\right)
\left(1-\frac{\sqrt{\Omega_{de}}}{C}\right).
\end{equation}
Then, the implicit solution is given by
\begin{equation}
\label{Omegade}
-2\ln a + \mathrm{const.}
=
\frac{1}{C^2-1}\Big[
-C^2\ln(1-\Omega_{de})
+C^2\ln\Omega_{de}
+C\ln\!\left(\frac{1+\sqrt{\Omega_{de}}}{1-\sqrt{\Omega_{de}}}\right) \\
+2\ln(C-\sqrt{\Omega_{de}})
-\ln\Omega_{de}
\Big].
\end{equation}
At early times, where $\Omega_{de}\ll1$, this yields the scaling
$\Omega_{de}\sim a^{-2}$.
Numerical solutions show that for $C<1$ the evolution encounters a divergence 
at 
$\Omega_{de}=C^2$, associated with a phantom-like regime and a future big rip, 
whereas for $C\le1$ a prolonged HDE-dominated phase emerges.

\begin{figure}[t]
\centering
\includegraphics[scale=.45]{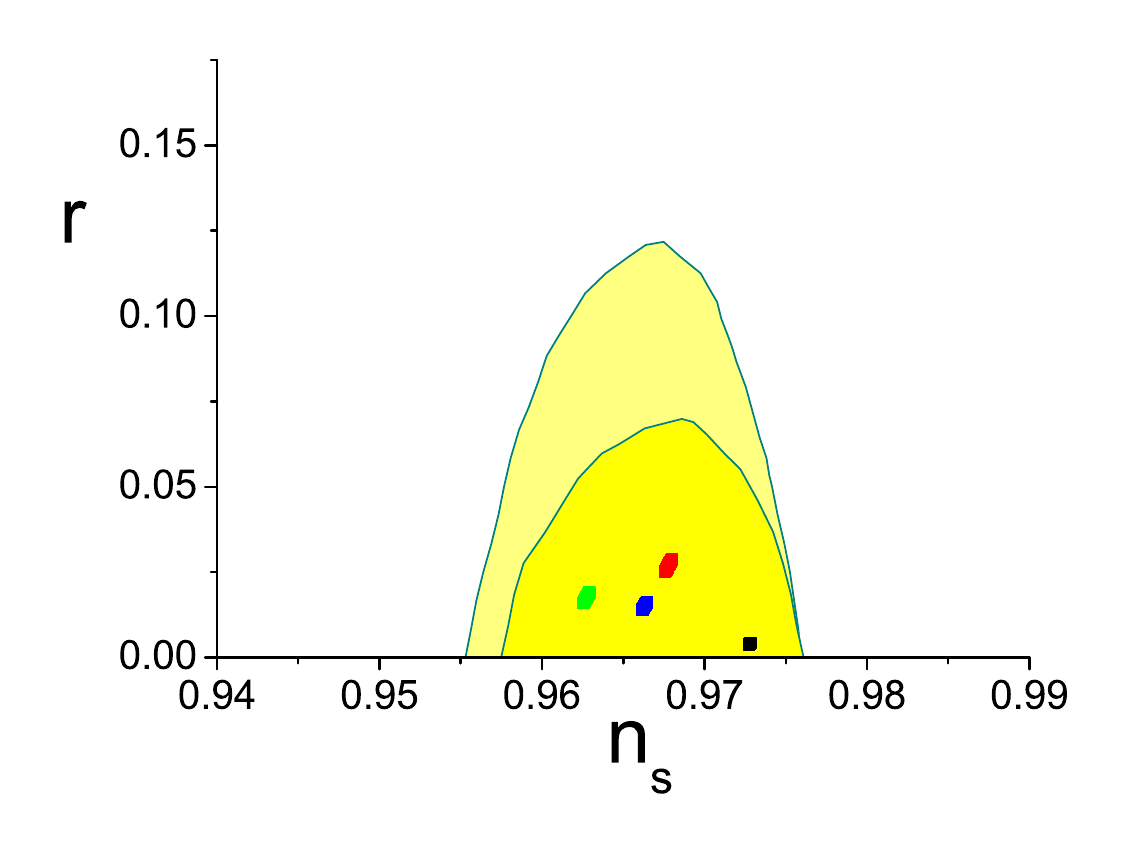}
\caption{{\it{ The  predictions of holographic 
inflation with the  cutoff $L \equiv \sqrt{  R_h^2 + 
1/\Lambda_\mathrm{UV}^2}$,  on the
$n_{\mathrm{s}}-r$ plane,  for 
$C=1.007$,$\Lambda_\mathrm{UV}=20$ (black points),
$C=1.009$,$\Lambda_\mathrm{UV}=19$ (red points),
$C=1.009$,$\Lambda_\mathrm{UV}=18$ (blue points), and
$C=1.01$,$\Lambda_\mathrm{UV}=18$ (green points), 
in Planck units, and for e-folding number varying between $N=50$ and 
$N=60$. Additionally, we depict the 
1$\sigma$ (yellow) and 2$\sigma$ (light yellow) contours for 
Planck 2018 results (Planck $+TT+lowP$)~\cite{Planck:2018jri}. The figure is 
from~\cite{Nojiri:2019kkp}.
}}}
\label{rnsnNojiri}
\end{figure}

Additionally, perturbations can be treated consistently by considering the 
gravitational 
potential $\Phi$, whose evolution equation reads
\begin{equation}
\ddot\Phi
+\left(H-\frac{2\ddot\varphi}{\dot\varphi}\right)\dot\Phi
+\left(4\dot H-H\frac{2\ddot\varphi}{\dot\varphi}
+\frac{\dot\varphi^2}{M_p^2}\right)\Phi
-\frac{\nabla^2}{a^2}\Phi
=0.
\end{equation}
This equation can be cast into a Bessel-type form and solved 
analytically~\cite{Chen:2006qy}.
Since the comoving curvature perturbation is no longer conserved, a nearly 
conserved quantity is introduced, namely
\begin{equation}
\mathcal{R}
=
\frac{2M_p^2H^2}{\dot\varphi^2}
\left(\frac{\dot\Phi}{H}+\Phi\right)
\exp\!\left[
2C^2\!\int_t^{t_{LS}}\!\frac{dt}{R_h^2H}
\left(1-\frac{1}{R_hH}\right)
\right],
\end{equation}
with power spectrum
\begin{equation}
\mathcal{P}_\mathcal{R}
=
\frac{H^4}{4\pi^2\dot\varphi^2}
\exp\!\left[
4C^2\!\int_t^{t_{LS}}\!\frac{dt}{R_h^2H}
\left(1-\frac{1}{R_hH}\right)
\right].
\end{equation}
Accordingly, the spectral index receives the correction
\begin{equation}
\label{correction}
\delta n_s
=
-\frac{10C^2}{R_h^2H^2}
\left(1-\frac{1}{R_hH}\right),
\end{equation}
which, although not favored by current CMB data, is not decisively excluded. 
Finally, as it was found in~\cite{Nojiri:2019kkp}, and presented in 
Fig.~\ref{rnsnNojiri}, the resulting spectral index and tensor-to-scalar ratio 
of 
holographic inflation, in the case where the cutoff is $
L \equiv \sqrt{  R_h^2 + \frac{1}{\Lambda_\mathrm{UV}^2}}$,
are in agreement with  observations.

\paragraph{Black-hole physics.}

The holographic origin of HDE naturally raises speculations about its possible 
relation to black-hole physics. 
A tentative connection was proposed in~\cite{Myung:2007tx}, based on the idea 
that 
the HDE density may effectively arise from an ensemble of black holes.
Further insight was provided in~\cite{Balazs:2006sd}, where black holes were 
argued to admit a dual description in terms of a weakly interacting quantum 
gas. 
Within this picture, the effective equation-of-state parameter becomes
\begin{equation}
\omega_{QG}=-\frac{1}{2},
\end{equation}
which interpolates between pressureless black holes ($\omega_{BH}=0$) and the 
radiation-like behavior expected from AdS/CFT ($\omega_{CFT}=1/3$). 
Although still too large to describe the observed dark energy, this result 
offers an intriguing hint that black-hole microphysics may play a role in the 
holographic description of cosmic acceleration.

\paragraph{Big rip singularity.}
\label{BRS}

As we mentioned above, current observational data favor values $C\simeq0.8$ in 
the minimal HDE model, 
implying a phantom-like equation of state and a future big rip singularity. 
In such a scenario, the expansion rate diverges in finite time and all bound 
structures are eventually destroyed by tidal forces.

The fate of the Universe in this regime was first examined 
in~\cite{Elizalde:2005ju}. 
Since the Big Rip corresponds to a finite cosmic lifetime $t_s$, one may 
naturally identify $t_s$ as an effective infrared cutoff. 
As an illustrative example, consider
\begin{equation}
\frac{L_\Lambda}{C}
=
\frac{2t_s\left(\frac{L_p+L_f}{\pi t_s}\right)^2}
{\left[1+\left(\frac{L_p+L_f}{\pi t_s}\right)^2\right]^2},
\end{equation}
which admits the solution
\begin{equation}
H=\frac{1}{2}\left(\frac{1}{t}+\frac{1}{t_s-t}\right),
\qquad
a(t)=\sqrt{\frac{t}{t_s-t}}.
\end{equation}
Furthermore, quantum gravitational effects can be incorporated via the 
conformal anomaly, 
leading to the regularized equation
\begin{equation}
\label{QGeff}
\frac{3H^2(1-C^2)}{8\pi G}
=
-6b'H^4,
\end{equation}
where $b'$ depends on the particle content of the effective theory. 
For ordinary matter we have $b'<0$, yielding the de Sitter attractor
\begin{equation}
H^2=\frac{1-C^2}{-16\pi G b'}.
\end{equation}
Additionally, alternative resolutions of the Big Rip have been explored in 
brane-world 
scenarios~\cite{Zhang:2009xj}, where the effective holographic parameter becomes
\begin{equation}
\label{Crmeff}
C_{\rm eff}
=
C\sqrt{1+3C^2M_p^2R_h^{-2}\rho_c^{-1}}.
\end{equation}
In this case, the future event horizon never shrinks to zero, and the Universe 
approaches a final de Sitter phase. Lastly, we mention that other mechanisms 
for avoiding the Big Rip involve interactions between dark 
energy and dark matter, leading to the interacting holographic dark energy 
scenario. This scenario, as well as other extended models, will be discussed in 
the next section.

\section{Extended Models of Holographic Dark Energy}
\label{ExtendedHolographicdarkenergy}

Although the original formulation of holographic dark energy provides a
remarkably simple and physically motivated framework for explaining late-time
cosmic acceleration, it is by no means unique. Both theoretical considerations
and phenomenological challenges have motivated a broad class of extensions of
the basic HDE scenario, and a substantial amount of research has been 
dedicated to them~\cite{Gong:2004fq,Wang:2005jx, Kim:2005at, 
Wang:2005ph,Hu:2006ar, 
Setare:2006wh, Sadjadi:2006qb, Zimdahl:2007zz,Sadjadi:2007ts, Setare:2007at, 
Setare:2007we,Wu:2007fs, Zhang:2007uh,Kim:2007dp, Feng:2007wn, Setare:2007fb, 
 Karwan:2008ig, MohseniSadjadi:2008na,
Cruz:2008er, Jamil:2009zzd, Wang:2009zzd, Wu:2009zzl,Setare:2009ti, 
Setare:2009mc, Sheykhi:2009dz, Sheykhi:2009zv, Karami:2009ux,Setare:2010aa, 
Wu:2010zzg, Jamil:2010xq, Rozas-Fernandez:2010qan, Jamil:2010sk, Yu:2010kh, 
Debnath:2010za, Duran:2010hi, Duran:2010ky, Khodam-Mohammadi:2011oji, 
Bolotin:2013jpa,Karami:2011js, Adabi:2011eh, Chimento:2011dw, 
Farajollahi:2011pb,Mazumder:2011kj, Biswas:2011ki, Liu:2011gx, 
Chimento:2011pk,Fu:2011ab,Sharif:2012zza, Ghose:2012ulq, Chimento:2012hn, 
Forte:2012ww, Chimento:2012zz, Sharif:2012ua, Pan:2012ki, Pasqua:2012apa, 
Debnath:2013kua,Chattopadhyay:2013oaa, Fayaz:2013pja, Chimento:2013se, 
Taji:2013dfa,Sadeghi:2013vfa, Arevalo:2013tta, Pankunni:2013jhv, 
Arevalo:2014zoa,Kiran:2014qra, Li:2014eba,Som:2014hja, Kiran:2014nta, 
Sarkar:2015uzp, Adhav:2015fca, Darabi:2015zwa, Ranjit:2015pwa, Kiran:2015llc, 
Lepe:2015qhq,Ramesh:2016qim,Raju:2016rso,
Hossienkhani:2017uku,  Karami:2010aq,
 Setare:2010zy, Karami:2010bys,
 Chen:2011rz,Zhai:2011pp,
Reddy:2016qnl, 
Darabi:2016igc, Herrera:2016uci, Feng:2016djj, Jawad:2016uty, Reddy:2016hmd, 
Felegary:2016znh,Li:2017usw, Hossienkhani:2018pip, 
Srivastava:2018zaj, Chirde:2018jhn, George:2018myt, Feng:2018yew, 
Sadri:2018rcp,Sadri:2018lzz, Belkacemi:2018dgy, AbdollahiZadeh:2019cqi, 
Mishra:2019uqm, Sadri:2019qxt, Sharma:2019bgp, Nayak:2019njd,Aditya:2019bbk, 
Sinha:2019axe, Zhang:2019zxv, Sharma:2020glf, Mamon:2020wnh, Cid:2020kpp, 
Saha:2020vxn, Bhattacharjee:2020rqk,Mamon:2020spa, Eser:2020dcv, Shekh:2021mjw, 
Dubey:2021lmm, Kim:2021rwd, Saleem:2021iju, Bolanos:2021wmn, Saha:2022oph, 
Escamilla-Rivera:2022baz, Koussour:2022sdy,Astashenok:2022lsf, Luciano:2022viz, 
Landim:2022jgr, Shekh:2022ykf, Luciano:2022hhy, Li:2023ubd, Remya:2023vhh,
Hatkar:2023hjm,Sharma:2023toq, Cid:2023yvw, Sultana:2024che, Mandal:2024euw, 
Chunlen:2024unl, Yarahmadi:2024afr, Khan:2024kzz, Adhikary:2024sax, 
Abdelrashied:2025hsi, Yun:2025cgi, Guin:2025xki, Shen:2025cjm, 
Pooya:2025wyd,
Ito:2004qi, Ke:2004nw, 
Enqvist:2004ny, Huang:2004mx,Myung:2004ch, Shen:2004ck, 
Kao:2005xp, Gao:2005ldz,Guberina:2005fb, Zhang:2005yz,Pavon:2005yx, 
Gong:2005ya,  Zhang:2005hs, Nojiri:2005pu, Kim:2005gk, 
Guberina:2005mp,Chang:2005ph,
Guberina:2006fy, Li:2006ci, BeltranAlmeida:2006is, Yi:2006bw, 
Shalyt-Margolin:2006lav, Setare:2006vz, Setare:2006sv,
  Zhang:2006qu, Simpson:2006wd, Pavon:2006qm, Setare:2006yj, 
Guberina:2006qh, Setare:2006xu,Zhang:2007sh,
Sun:2007rh, El-Nabulsi:2007lla,
Banerjee:2007zd, Ng:2007bp,Setare:2007hq, 
Zhang:2007zga,Saridakis:2007cy, Saridakis:2007ns,Saridakis:2007wx,
Wei:2007ig, 
Zhang:2007an, Feng:2007zzc, Ma:2007pd, Horvat:2007ic, 
Gao:2007ep, Wu:2007tn,Sen:2008tr, 
Setare:2008bb, Medved:2008vh, Myung:2008pi, 
Nayak:2008pn, Xu:2008sn, Kim:2008pi, Horvat:2008gk, 
Gong:2008br,   Bisabr:2008gu, Wu:2008gg, Paul:2008rm, Setare:2008pc, 
Setare:2008hm, Granda:2008tm,Granda:2008ic, Ma:2009uw, 
Chattopadhyay:2009tg, 
  Wei:2009kp, Wei:2009au,   Jamil:2009ia, Feng:2009hr, 
Li:2009bn, Granda:2009dia, Xu:2009ssa, MohseniSadjadi:2009va, 
Gong:2009dc, Xu:2009ys,  Zhang:2009xj,
Mazumder:2009zza,   Granda:2009zx, Granda:2009di, 
Li:2009zs, Rozas-Fernandez:2009jvg, Karami:2009je,   Liu:2009ha, 
Karami:2009we, Micheletti:2009jy, Nozari:2009zk, 
Setare:2010wt,  Lepe:2010vh,   Xu:2010zzf, Karami:2010pxa, 
 Horvat:2010wg, Lee:2010ew,
 Wang:2010kwa, Sadjadi:2010az, Cui:2010dr, 
Dutta:2010mp, Setare:2010md, Saaidi:2010jq,  Malekjani:2010jb, 
Paul:2010ff, Biswas:2010yf,Jamil:2010ed,   Sun:2010cs, 
Radicella:2010vf, Liu:2010am, Micheletti:2010cm, 
Cardenas:2010wx, Saaidi:2010fq, Setare:2010dr, Malekjani:2010nk, 
Ebrahimi:2010xz,   Sun:2010zr, Sheykhi:2011yjb, Sheykhi:2011egx, 
  Granda:2011zm, Yang:2011zza, 
Saadat:2011zza, Saadat:2011zzb, Sharif:2011ig, Saadat:2011zz,
Bandyopadhyay:2011za, Lu:2011zzh, Paul:2011oba,  delCampo:2011jp, 
 Gough:2011eq, 
Bouhmadi-Lopez:2011qvd,
 Khatua:2011ec, 
Sheykhi:2011cn,  Khodam-Mohammadi:2011vja, 
 Cruz:2011wx, Aviles:2011sfa, Mazumder:2011qd, Houndjo:2011fb, 
HamaniDaouda:2011uag, Mathew:2011aa,  Zhitnitsky:2011aa, Amani:2011tv, 
Belkacemi:2011zk, Karami:2012ra, Saadat:2012zzb, Khatua:2012zz, 
Saadat:2012zzd, Farajollahi:2012uj, Huang:2012nz, Huang:2012gd, Huang:2012xma, 
Zhang:2012qra, Sharif:2012dr,  Gough:2012nbu, Zhang:2012uu, 
Ling:2012gu, Huang:2012xm, Xu:2012aw,
Duran:2012yr,Kim:2012ik, 
Malekjani:2012gh, Malekjani:2012bw, 
Khodam-Mohammadi:2012zuk, Li:2012fj, Chattopadhyay:2012eu, Jawad:2012xy, 
 Duran:2013dr, Saadat:2013hva, Yu:2013enz, 
Sarkar:2013dxa, Pasqua:2013sfa, Farajollahi:2013qea, Borah:2013mna, 
  Chattopadhyay:2013mta, Samanta:2013vna, Zhang:2013qgz, 
Saaidi:2013yfa, Sharif:2013cga, Pasqua:2013iga, Chattopadhyay:2013mwa, 
Sharif:2013qil,   Bouhmadi-Lopez:2013mji,
Malekjani:2013xsa, Xu:2013mic, Zhang:2013mca, Chimento:2013qja, Pasqua:2013lha, 
Huang:2013una, Chattopadhyay:2013vya, Farooq:2013ava, 
  Majumder:2013fza,
Shan:2013vfa, Debnath:2013woa, Naderi:2013jwa, Cadoni:2013gza, 
Mathew:2013fka, Viaggiu:2013qfa, Huang:2013xca, Belkacemi:2014vnw, 
Huang:2014ama, Belkacemi:2014aqa, Sharif:2014ylv, Sarkar:2014tra, 
Borah:2014gca, 
Fayaz:2014swa, Chattopadhyay:2014oba, Sharif:2014qra, Salti:2014aha, 
Sarkar:2014ysa, Pasqua:2014oya, Adhav:2014lma, Jawad:2014qma, Fayaz:2014bja, 
Samanta:2014ida, Das:2014qsa, Farajollahi:2014hzp, Aghamohammadi:2014cza, 
Sarkar:2014dda, Chattopadhyay:2014yda, Praseetha:2014deq, Cui:2014sma, 
Sarkar:2014ywa, Ghosh:2014mva, Debnath:2014mea, Li:2014pua, Borah:2014xea, 
Majeed:2014gfa, Ghaffari:2014pxa, Chattopadhyay:2014jua, Sarkar:2014sba, 
Chaubey:2014gja,   Zeng:2014xza,  Borah:2014asa, 
Borah:2014vea, Zhang:2014sqa, Amani:2014nea, Pan:2014afa, 
  AzizurRahman:2014pla, Naderi:2014awa,  Roy:2014wqa, 
Li:2014zxa,  Oliveros:2014kla, Debnath:2014doa, Pasqua:2014owa, 
Sharif:2015fya, Zubair:2015opa, Rao:2015efk, Salako:2015xra, 
Ghaffari:2015foa,
Hu:2015qoa,Fayaz:2015yka, Horvat:2015cwa, Zeng:2015cha, 
Jawad:2015xoa, Albarran:2015tga, Sheykhi:2015nba, Maity:2015mys,  
Singh:2015wya, Banerjee:2015kva, Mahata:2015nga,  
DavoodSadatian:2015zab, Landim:2015hqa,  
Praseetha:2015xuj, Umadevi:2015kba, JamilAmir:2015vcv, Chattopadhyay:2015jvk, 
Cui:2015oda, Das:2015jya, Pasqua:2015ywa, Mehrabi:2015kta, Li:2015bza, 
Pasqua:2015bpm,Pasqua:2015bfz, Praseetha:2015bjf, 
George:2015lok,Chattopadhyay:2016enn, Rao:2016hge,  
Das:2016nlj,Nastase:2016sji, Wang:2016pwd,   Singh:2016kyv, 
Chattopadhyay:2016tdy,
Chattopadhyay:2016mnl, Forte:2016ben, Pasqua:2016wrm, Hossienkhani:2016sou, 
Khurshudyan:2016uql, Rani:2016wpa,   Pavao:2016rhe,
Rudra:2016unu,  He:2016rvp, Mukherjee:2016lor, Saha:2016bjs, 
  Singh:2016xmc, BhaskaraRao:2016acr, Rao:2016dsa, Reddy:2016ywh, 
Reddy:2016rgi,   Jawad:2016tne, Chattopadhyay:2016hzt, 
Darabi:2016mjg, Zadeh:2016vgc, Reddy:2016xrl, Santhi:2016ipz, Komatsu:2016vof, 
Khurshudyan:2016gmb,Sheykhi:2016emj, 
Sadri:2016hlc, Samanta:2017oyk, VijayaSanthi:2017zna, 
Mete:2017qvn,Srivastava:2020riu,
Reddy:2017kgy, Katore:2017exd, Wu:2017ukk, Fayaz:2017vyp, Rao:2017lnt, 
Santhi:2017xoi, Rao:2017emm, Dymnikova:2017dba, Nojiri:2017opc, Zhao:2017urm, 
Pourhassan:2017cba, AlMamon:2017tbm, Sheykhi:2017cid, Srivastava:2017dvt, 
Luongo:2017yta, Bouhmadi-Lopez:2017kvc, Saridakis:2017rdo, Mukherjee:2017oom, 
Zhao:2017zcj, Godonou:2017ugt, BhaskarRao:2017gli,   Sadri:2017vwl, 
Ali:2017epx, 
Hassan:2018uoy, Rani:2018huv, 
Cruz:2018xzn,   Chakrabarti:2018gou, Makarenko:2018blx, 
Moradpour:2018ivi, Singh:2018cip, Tavayef:2018xwx,
Akhlaghi:2018knk, Singh:2018uyg,   Forte:2018pff, 
Zadeh:2018osy, Srivastava:2018gye, Sharif:2018qxb,
Zadeh:2018poj, Sharif:2018yps, 
VijayaSanthi:2018ojp,   Ghaffari:2018wks, Aditya:2018lkw, 
DasuNaidu:2018tvz, Majumdar:2018ups, Salehi:2018czt, Rao:2018xvg, 
Aditya:2018zvx,  Malekjani:2018qcz,  
Ghaffari:2018rzs, Jawad:2018juh, Lee:2018lgs, Ishikawa:2018hqh,
  Cruz:2018lcx, Younas:2018kmy, Aly:2019otq, Vinutha:2019pul, 
Sireesha:2019qpx, Mahanta:2019tpo, Kadam:2019wpa,Ahmed:2019qsi,
Khadekar:2019byt,AbdollahiZadeh:2019lsx,Sharma:2021dqj,
 Sharif:2019seo,Ghaffari:2019scd,Sharif:2019guu,
Naidu:2019nhh,Jawad:2019doj,DAgostino:2019wko,AbdElrashied:2019nwh,
Pawar:2019ftm,Sharma:2019mtn,Katore:2019kyl,Srivastava:2019oif,Singh:2019qgz,
Li:2019bqb,  Nojiri:2019yzg,
Sadri:2019yqs, Dubey:2019kzh,Rodriguez-Benites:2019lgy, 
 Shaikh:2019lle,Zhang:2019ple,
Varshney:2019fzj,Dixit:2019nfl,Aly:2019wtr,Huang:2019hex, 
Dubey:2019wyz,Ghaffari:2019qcv,Korunur:2019rhg,Granda:2019agf,
Chakrabarti:2019zey,Maity:2019qbv,Astashenok:2019rwt, 
 Iqbal:2019ooy,Shaikh:2019ppk,Singh:2019xdk, 
Raut:2020aps,Bharali:2020fxd,Waheed:2020cxw,Hatkar:2020xsw, 
ChandraDubey:2020tng,Ghaffari:2020nnk,Golanbari:2020coz,Aly:2020uli, 
Chakraborty:2020jsq,Dubey:2020kwx,Dai:2020rfo,Dubey:2020ckn, 
Kritpetch:2020vea,Pradhan:2020cnt,Saridakis:2020zol,
Kumar:2020vnk,Anagnostopoulos:2020ctz, 
DivyaPrasanthi:2020bec,Ebrahimi:2020dkh,  
Sharma:2020lmm,Maity:2020cde,
Srivastava:2020sbp,AlMamon:2020usb,Chattopadhyay:2020mqj,Nojiri:2020wmh, 
Jawad:2020jtq,Kim:2020cbm, Ens:2020bxh,
Chunlen:2020nru,Srivastava:2020hng,Dabrowski:2020atl,
Sharma:2020rvx,
 Shaikh:2020ndr,
Sharma:2020mmn,Srivastava:2020cyk,Maity:2020wqo,Yadav:2020wsd,Das:2020rmg, 
Dubey:2020vho,daSilva:2020bdc,Ali:2020gfx,Bhattacharjee:2020ixg, 
VijayaSanthi:2020feh,Singh:2020xbg,Prasanthi:2021ihf, 
Chakraborty:2021ljt,Saha:2021ngs,Mohammadi:2021wde,Gusu:2021zsh,
Zubair:2021yrq,Colgain:2021beg,Bhardwaj:2021chg,Shekh:2021bgh,
Chakraborty:2021uzp,Jawad:2021xsr,Adhikary:2021xym,Shaikh:2021iaq,
Pradhan:2021crw,Nojiri:2021iko,Liu:2021heo,
Das:2021jvp,Sarkar:2021izd,Upadhyay:2021atf,Pourojaghi:2021den,
Bargach:2021foj,Rani:2021hvh,Huang:2021zgj,Qiu:2021cww,Ram:2021kjg,
Sardar:2021eaj,Kumar:2021xgi,Shekh:2021ule,Kaur:2021dix,Kim:2021bbj,
Zubair:2021gve,Ganz:2021dmx,Bhardwaj:2021nzd,Varshney:2021xvg,
Dheepika:2021fqv,Pankaj:2021nkg,Pandey:2021fvr,Ghosh:2021zpb,Saridakis:2021qzv,
Hernandez-Almada:2021aiw,Pradhan:2021tij,VijayaSanthi:2021wyx,Varshney:2021rbq,
Delgado:2021lzi,Lou:2021gwk,Bharali:2021pmm,Saha:2021mlz,Sobhanbabu:2021vzw,
Dixit:2021phd,
Telali:2021jju,Nandhida:2021vxl,Pradhan:2021bse,Maity:2022gdy,
Mahanta:2022xog,Saha:2022vcb,Korunur:2022ifb,Bharali:2022kbl,Zhang:2022wco,
Zhao:2022bxw,Nakarachinda:2022mlz,Upadhyay:2022jwa,Paul:2022doh,Shaikh:2022ynt,
Zarandi:2022and,Koussour:2022nsu,
Pinki:2022aht,Kumar:2022acs,Saleem:2022eti,Sadeghi:2022fow,
Remya:2022frs,Mohammadi:2022vru,Koussour:2022rsv,Srivastava:2022nex,
Oliveros:2022biu,VijayaSanthi:2022hef,Pradhan:2022jjz,Biswas:2022udk,
Jawad:2022qab, 
Bhardwaj:2022uhf, Fazlollahi:2022kbv,
Pandey:2022rtu,Rani:2022upi,Sharma:2022poz,Rezaei:2022bkb,Boulkaboul:2022mfb,
Manoharan:2022qll,Santhi:2022pyp,Astashenok:2022pni,
Koussour:2022xiy,
Ghaffari:2022skp,Das:2022igz,Ali:2022twr,Chanda:2022tpk,Koussour:2022eec, 
AlMamon:2022mlu,Kumar:2022pfn,Zubair:2022vnd,Jimenez-Aguilar:2022hty, 
Kaur:2022lgk,Luciano:2022ffn,Cardona:2022pwm,Alam:2022oae, 
Gupta:2022vye,Sheykhi:2022fus,Santhi:2022bqt,Dubey:2022hkh,Sadeghi:2022mrm, 
P:2022amn,VijayaSanthi:2022xjb,Garg:2022sbc,Singh:2022ubm, 
FeiziMangoudehi:2022rwj,Sharma:2022dzc,Bogdanova:2022dtd,AmetMemet:2022hjd, 
Ram:2022isr,Mandal:2022ocg,Luciano:2023wtx,Kaur:2023jhx, 
Boulkaboul:2023yks,Pawar:2023por,Bousder:2023cnf,A:2023qvh,Sania:2023fjx, 
Feng:2023cbl,Salehi:2023byk,Nojiri:2023nop,VijayaSanthi:2023ipb, 
Astashenok:2023jfp, Shaikh:2023awd,
Sobhanbabu:2023erj,AlMamon:2023zek,Jawad:2023aog,Kim:2023oxg,Mule:2023cfa, 
Fazlollahi:2023kkj,Shaikh:2023gvc,Sharma:2023til,Shekh:2023wka,Mahanta:2023qvp, 
Wankhade:2023ufc,Bharali:2023wha,Lymperis:2023prf,Sharma:2023asq,  
Wang:2023gov,Ghaderi:2023cih,Garattini:2023wgk,Narasimharao:2023qou,
Shekh:2023mwx,Manoharan:2023yus,Waheed:2023flg,Korunur:2023qsu,Korunur:2023qry, 
Cruz:2023xjp,Basilakos:2023seo,Nakarachinda:2023jko,
Pankaj:2023bkj,A:2024hez,A:2024wrd,Huang:2024xqk,Abaca:2024khy,
Chokyi:2024xff,Xu:2019hhs,Blekman:2024tci,
Trivedi:2024rhp,VijayaSanthi:2024qgy,Tita:2024jzw, 
Mahanta:2024hzi,Aditya:2024lia,Rao:2024yrp,
Solanke:2024fqf,He:2024bll,Bharali:2024tqs,Fazlollahi:2024lrt,Trivedi:2024dju,
Sharma:2024ywq, 
Trivedi:2024jcy,Mukhopadhyay:2024vcq,Karte:2024wej,Altaibayeva:2024cyc, 
Saha:2024ugn,Manoharan:2024thb,Sobhanbabu:2024zmv,Prasanthi:2024cly, 
Dhore:2024gfv,Ali:2024xfk,VanRaamsdonk:2024sdp,Ghosh:2024jvs,
Astashenok:2024tdg,
Khadekar:2024vkl,Fang:2024yni,Goyal:2024jco,Tang:2024gtq,Aktas:2024rgc, 
Hatkar:2024xlr,Sultana:2024urf,Trivedi:2024inb,Yun:2024xzk, 
Enkhili:2024mlx,Devi:2024gcr,Motaghi:2024rag,Myrzakulov:2024qtd,Sharif:2024fli, 
 Dhore:2024xeo,Murali:2024vos,Mahanta:2024qml,Sobhanbabu:2024bnt, 
Samaddar:2024web,Blitz:2024nil,Samaddar:2024uwx,Astashenok:2024jje, 
Chokyi:2024nis,Yarahmadi:2024oqv,Mahmoudifard:2024gmn,Maity:2024tkq,
Kaur:2024hqe,Li:2024qus,Baziar:2024qbt,Brevik:2024ozg, 
Bakry:2024xjx,Myrzakulov:2024jvg,Mahanta:2024iel, 
Santos:2024cvx,
Mahanta:2024xyj,Han:2024sxm,Ens:2024zzs,Li:2024bwr,Dubey:2024utn,Gupta:2024qyn, 
Ualikhanova:2024xxe,Huang:2025tqr,Myrzakulov:2025tcd,Kaur:2025vbm, 
Manoharan:2025nju,Bidlan:2025pzi,Singha:2025trb, 
Kapil:2025mpt, Saleem:2025fwd, Cimdiker:2025vfn, 
Varshney:2025tkb,Paul:2025vem,Luciano:2025fox,Solanke:2025jsh,Malakar:2025qos, 
Hounmenou:2025rra,Yarahmadi:2025ema,Luciano:2025elo,Chinnappalanaidu:2025hzq, 
Raut:2025piz,Pradhan:2025rir,Aditya:2025psr,Rathore:2025ptu,Zhang:2025oki, 
 Tita:2025qaa,Thakran:2025xwo,Mehta:2025laj,Das:2025nds, 
Zapata:2025ngr,Chaudhary:2025khr,Paul:2025non,Astashenok:2025ktx,
Prasanthi:2025xif,Plaza:2025nip,Shaikh:2025oyp, 
Shaikh:2025mlw,Pasqua:2025clh,Das:2025ccm, Pasqua:2025oxv, 
Luciano:2025ykr,Altaibayeva:2025rhf,Raut:2025ezy, Bekova:2025qco, 
Sheykhi:2025gkx,Chakraborty:2025jfi,Yarahmadi:2025fml,Kapil:2025pyt,
Santos:2025fdp, Singh:2026qfj, Aditya:2025ehz,Murali:2025qqo,
Prasanthan:2024fwg, Sultana:2024fvb,Gonzalez-Espinoza:2026iyh, 
Khan:2026xqf,Prasanthi:2026byc,Prasanthi:2026xgv,
Tamri:2026pxo,Huang:2026zuv,
Cruz:2025ebg,Shaikh:2025vce,Li:2025vqt,Yarahmadi:2025luc,Kotal:2025fqh, 
Garcia-Bellido:2025hji,Li:2025ivi,Sharif:2025aul,Mazumdar:2025jfv,
RodriguezMeza:2025kiq,Dubey:2025hpd,Luciano:2026vhm,Das:2026wry,Ghosh:2026rll,
Abdullghani:2026iju,Siquieri:2026dby, 
Sadeghnezhad:2025zxa,Mukherjee:2026jmz,Sultana:2026kup,Satyanarayana:2026ead,
Bolotin:2026hot,Aditya:2026jmu,Mahanta:2026fmq,
Mahanta:2026xkt,Maity:2026uok,
Prasanthi:2026dhd,Ajmal:2026fhf,Ibrar:2025hah,Dubey:2025nuf,Khapekar:2026zkc,
Maity:2025cxd,Feng:2008rs,Feng:2008kz,Xu:2008rp,Feng:2008hk,
Xu:2009xi,
Feng:2009ag,Feng:2009ai,Feng:2009jr,Suwa:2009gm,Kim:2010pdn,
Xu:2010gg,Zhang:2010im,Yang:2011us,Zhang:2011zze,
Chattopadhyay:2011zz,Chattopadhyay:2011mpa,Bhattacharya:2011xa,
Chattopadhyay:2011er,Karwan:2011sh,Wang:2011km,Broda:2011np,
Pasqua:2011gh,Mathew:2012md,Saridakis:2025ltr,
 Pasqua:2013ewa,
delCampo:2013hka,Silva:2013yaa,Myung:2013cqa,
Aguilera:2013tmp, 
Chattopadhyay:2014ixa,Aly:2015zya,Yu:2015sla,
Salti:2017ywf,Izaurieta:2026edz,
Singh:2018yau,Jesus:2019nwi,George:2019vko,Saleem:2020fjf,
Anchordoqui:2020sqo,Hossienkhani:2021emv,Kumar:2021avx,
Rudra:2022qbv,Sultana:2022pmj,Pasqua:2023rdp,Shekh:2023baf,
Alvarenga:2024yqa,Bubuianu:2024zsm,Satyanarayana:2024gxa,
Scomparin:2025jub, Errehymy:2026nna,Karimi:2026mzi,Yun:2026kra,
Wu:2025vfs,Chokyi:2026ryf,Molavi:2019mlh,Maurya:2026dky,Yun:2026gua,
Sanyal:2025udg,Nisar:2026seu,Aditya:2026utq,Sardar:2024kcw,Chaudhary:2026uzs,
NooriGashti:2024dvq,NooriGashti:2024tog,Yarahmadi:2025ujq,Zafar:2026rnu,
Bidlan:2026slr,Huang:2026szy,Rizwan:2026gxh,
Alshammari:2026wnf,Singh:2026ebh,Nashed:2026rah,Nashed:2026vwv,Naik:2026lfb,
Bidlan:2026xaa,Meetei:2026eja,Meetei:2026uhe,Gore:2026tbh,Huang:2026khc}. 
These 
extended models aim 
either to improve the
theoretical consistency of the holographic construction, to alleviate known
cosmological tensions, or to enlarge the phenomenological flexibility of the
model so as to better accommodate observational 
data. 

From a theoretical perspective, the standard HDE model relies on a number of
assumptions that are not uniquely fixed by first principles. These include the
choice of the infrared cutoff scale, the assumption of minimal coupling between
dark energy and the rest of the cosmic fluid,  the restriction to Einstein
gravity as the underlying gravitational framework, and the consideration of 
standard  Bekenstein-Hawking entropy relation. Relaxing any of these
assumptions naturally leads to generalized holographic scenarios with distinct
dynamical properties. At the same time, from an observational viewpoint,
extended models have been explored in order to address issues such as the
coincidence problem, the possible evolution of the dark energy equation-of-state
parameter, and the compatibility of holographic dark energy with precision
cosmological data.

The extensions of holographic dark energy considered in the literature can be
broadly classified into several categories. A first class involves interactions
between holographic dark energy and other cosmic components, most notably dark
matter. Such interactions are well motivated both phenomenologically and from
effective field theory considerations, and can significantly modify the
background and perturbation dynamics. A second class is based on alternative
choices of the infrared cutoff, leading to holographic-inspired dark energy
models that differ from the original proposal while preserving its underlying
holographic motivation. A third class embeds holographic dark energy within
extended theories of gravity, where modifications of the gravitational sector
alter the cosmological evolution associated with the holographic energy density.
Finally, reconstruction approaches attempt to map holographic dark energy into
effective scalar-field or modified-gravity descriptions, providing a useful
bridge between phenomenological models and more fundamental theories.

In the present section we review these extended holographic dark energy models,
emphasizing their theoretical foundations, cosmological dynamics, and
observational implications. For clarity and coherence, we organize the
discussion according to the physical mechanism underlying each extension, rather
than following a purely chronological presentation. Extensions based on
generalized entropy expressions, which constitute a particularly rich and
rapidly developing subfield, will be treated separately in a dedicated section.

\subsection{Interacting holographic dark energy}
\label{Interact}

Interacting holographic dark energy (IHDE) models constitute one of the most 
natural and extensively studied extensions of the original HDE framework. 
The possibility that dark energy and dark matter exchange energy during cosmic 
evolution has been explored for a long time, motivated primarily by the 
desire to alleviate the fine-tuning and coincidence problems of standard 
cosmology~\cite{Wetterich:1987fm,Wetterich:1994bg}. 
Within the holographic context, such interactions acquire additional 
theoretical interest, since holographic dark energy is not associated with a 
fundamental local field, but rather with a global vacuum energy determined by 
horizon physics. This nonlocal nature renders the standard quantum field theory 
arguments against dark-sector couplings inapplicable, opening the way to a 
consistent interacting scenario. Interacting  holographic dark energy leads to 
rich  cosmological behavior, and has been studied in detail ~\cite{Wang:2005jx, 
Kim:2005at, Wang:2005ph,Hu:2006ar, Setare:2006wh, 
Sadjadi:2006qb, Zimdahl:2007zz, Sadjadi:2007ts, Setare:2007at, Setare:2007we, 
Wu:2007fs, Zhang:2007uh,Kim:2007dp, Feng:2007wn, Setare:2007fb, 
Saridakis:2007wx, Karwan:2008ig,   MohseniSadjadi:2008na, 
Cruz:2008er, Jamil:2009zzd, Wang:2009zzd, Wu:2009zzl, Setare:2009ti, 
Setare:2009mc, Sheykhi:2009dz, Sheykhi:2009zv, Karami:2009ux,Setare:2010aa, 
Wu:2010zzg, Jamil:2010xq, Rozas-Fernandez:2010qan, Jamil:2010sk, Yu:2010kh, 
Debnath:2010za, Duran:2010hi, Duran:2010ky, Khodam-Mohammadi:2011oji, 
Bolotin:2013jpa,
Karami:2011js, Adabi:2011eh, Chimento:2011dw, Farajollahi:2011pb,
Mazumder:2011kj, Biswas:2011ki, Liu:2011gx, Chimento:2011pk,Fu:2011ab, 
Sharif:2012zza, Ghose:2012ulq, Chimento:2012hn, Forte:2012ww, Chimento:2012zz, 
Sharif:2012ua, Pan:2012ki, Pasqua:2012apa, Debnath:2013kua,
Chattopadhyay:2013oaa, Fayaz:2013pja, Chimento:2013se, Taji:2013dfa, 
Sadeghi:2013vfa, Arevalo:2013tta, Pankunni:2013jhv, Arevalo:2014zoa, 
Kiran:2014qra, Li:2014eba,Som:2014hja, Kiran:2014nta, Sarkar:2015uzp, 
Adhav:2015fca, Darabi:2015zwa, Ranjit:2015pwa, Kiran:2015llc, Lepe:2015qhq, 
Ramesh:2016qim,Raju:2016rso, Hossienkhani:2017uku, Reddy:2016qnl, 
Darabi:2016igc, Herrera:2016uci, Feng:2016djj, Jawad:2016uty, Reddy:2016hmd, 
Felegary:2016znh,Li:2017usw, Hossienkhani:2018pip, 
Srivastava:2018zaj, Chirde:2018jhn, George:2018myt, Feng:2018yew, 
Sadri:2018rcp,Sadri:2018lzz, Belkacemi:2018dgy, AbdollahiZadeh:2019cqi, 
Mishra:2019uqm, Sadri:2019qxt, Sharma:2019bgp, Nayak:2019njd,Aditya:2019bbk, 
Sinha:2019axe, Zhang:2019zxv, Sharma:2020glf, Mamon:2020wnh, Cid:2020kpp, 
Saha:2020vxn, Bhattacharjee:2020rqk,Mamon:2020spa, Eser:2020dcv, Shekh:2021mjw, 
Dubey:2021lmm, Kim:2021rwd, Saleem:2021iju, Bolanos:2021wmn, Saha:2022oph, 
Escamilla-Rivera:2022baz, Koussour:2022sdy,Astashenok:2022lsf, Luciano:2022viz, 
Landim:2022jgr, Shekh:2022ykf, Luciano:2022hhy, Li:2023ubd, Remya:2023vhh, 
Hatkar:2023hjm,Sharma:2023toq, Cid:2023yvw, Sultana:2024che, Mandal:2024euw, 
Chunlen:2024unl, Yarahmadi:2024afr, Khan:2024kzz, Adhikary:2024sax, 
Abdelrashied:2025hsi, Yun:2025cgi, Guin:2025xki, Shen:2025cjm, 
Pooya:2025wyd,Shen:2026ynj,Huang:2026sip,Amiriborkhani:2026bfi}.

\subsubsection{Theoretical background}

Interacting dark energy models were initially proposed as phenomenological 
extensions of $\Lambda$CDM, but they soon attracted attention as effective 
descriptions of modified gravity theories and unified dark sector models. 
A comprehensive overview of interacting dark energy can be found 
in~\cite{Wang:2016lxa}. 
From a quantum field theoretic perspective, it has been argued that long-range 
forces and radiative stability severely restrict direct couplings between dark 
matter and a quintessence-like scalar 
field~\cite{DAmico:2016jbm,Marsh:2016ynw}. 
However, these arguments do not apply to holographic dark energy, whose vacuum 
origin and horizon dependence place it outside the standard local field theory 
framework.

The interacting holographic dark energy scenario has been investigated in a 
variety of settings, while a thermodynamic interpretation in non-flat 
spacetime was developed in~\cite{Setare:2008bb}. Below we summarize the main 
results relevant 
for cosmological evolution.

\paragraph{IHDE in a non-flat Universe.}

In a FRW Universe with spatial curvature, the Friedmann 
equation reads
\begin{equation}
3M_p^2H^2=\rho_{dm}+\rho_b+\rho_r+\rho_k+\rho_{de}\,.
\end{equation}
Allowing for energy exchange between dark matter and holographic dark energy, 
the conservation equations take the form
\begin{equation}
\label{conseq}
\dot\rho_{dm}+3H\rho_{dm}=Q\,,\qquad 
\dot\rho_{de}+3H(1+w)\rho_{de}=-Q\,,
\end{equation}
where $Q$ denotes the interaction term. The total energy density of the dark 
sector remains conserved.
Since the microscopic origin of the interaction is unknown, $Q$ is typically 
parameterized phenomenologically as
\begin{equation}
Q=H\left(\Gamma_1\rho_{dm}+\Gamma_2\rho_{de}\right),
\end{equation}
with $\Gamma_i$ dimensionless constants constrained by observations. 
Alternative 
forms can be found in~\cite{Wang:2016och}.

Following~\cite{Zhang:2012uu}, the dark energy pressure can be expressed as
\begin{equation}
p_{de}=-\left(\frac{2}{3}\frac{\dot H}{H^2}+1\right)\rho_c
-\frac{1}{3}\rho_r+\frac{1}{3}\rho_k\,.
\end{equation}
Combining this with the conservation equations yields the coupled evolution 
equation
\begin{equation}
2(\Omega_{de}-1)\frac{\dot H}{H}+\dot\Omega_{de}
+H(3\Omega_{de}-3+\Omega_k-\Omega_r)
=-H\Omega_I\,,
\qquad 
\Omega_I\equiv\frac{Q}{H\rho_c}.
\end{equation}
This must be supplemented by
\begin{equation}
\frac{\dot\Omega_{de}}{2\Omega_{de}}+H+\frac{\dot H}{H}
=\sqrt{\frac{\Omega_{de}H^2}{C^2}-\frac{k}{a^2}},
\end{equation}
which follows from the holographic relation with infrared cutoff $L=a r(t)$.
Together, these equations lead to the master system governing IHDE evolution:
\begin{eqnarray}
\label{dedz}
\frac{1}{E(z)}\frac{dE(z)}{dz}&=&-\frac{\Omega_{de}}{1+z}
\left(\sqrt{\frac{\Omega_{de}}{C^2}+\Omega_k}
+\frac{\Omega_k-\Omega_r-3+\Omega_I}{2\Omega_{de}}+\frac{1}{2}\right),\\[2mm]
\label{domdz}
\frac{d\Omega_{de}}{dz}&=&-\frac{2\Omega_{de}(1-\Omega_{de})}{1+z}
\left(\sqrt{\frac{\Omega_{de}}{C^2}+\Omega_k}
-\frac{\Omega_k-\Omega_r+\Omega_I}{2(1-\Omega_{de})}+\frac{1}{2}\right).
\end{eqnarray}

\paragraph{Equation of state and phantom behavior.}

\begin{figure}[t]
\includegraphics[width=8.7cm]{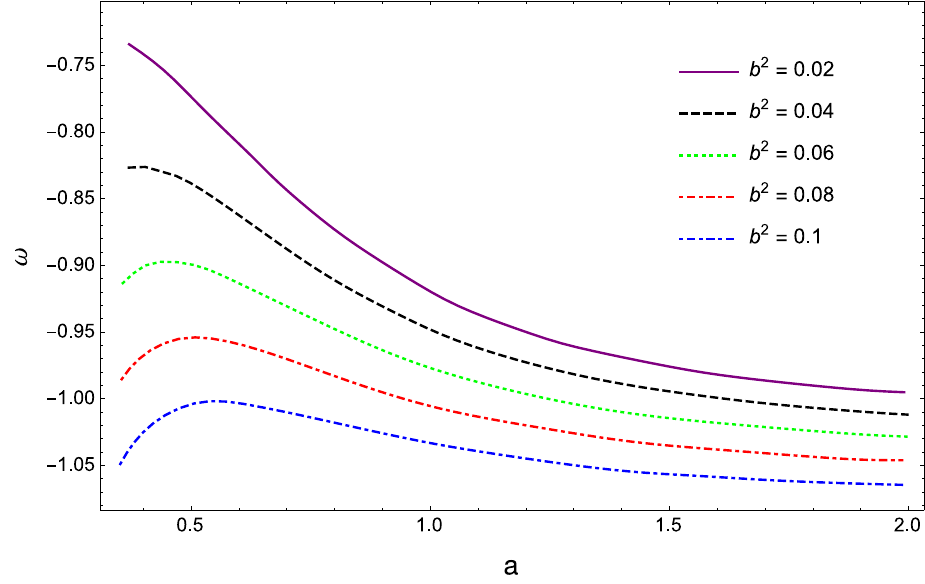}
\caption{{\it{Evolution of   the equation-of-state parameter of interacting 
HDE versus the scale factor $a$, for 
$C=1$, and for the interaction form $Q=3b^2H\rho_c$, with 
  different choices of 
$b^2$. The figure is from~\cite{Wang:2005jx}.}}}
\label{wevolIHDE}
\end{figure}

To illustrate the effect of interaction, consider a flat Universe filled with 
dust matter and HDE. Introducing the ratio
$r\equiv\rho_m/\rho_{de}$,
the conservation equations yield
\begin{equation}
\label{revol}
\dot r=3Hrw+\frac{(1+r)Q}{\rho_{de}}\,.
\end{equation}
Using
\begin{equation}
r=\frac{1-\Omega_{de}}{\Omega_{de}},\qquad 
\dot r=-\frac{\dot\Omega_{de}}{\Omega_{de}^2},
\end{equation}
one obtains
\begin{equation}
\label{wIHDE}
w=-\frac{\Omega'_{de}}{3\Omega_{de}(1-\Omega_{de})}
-\frac{Q}{3H(1-\Omega_{de})\rho_{de}}\,.
\end{equation}
For the commonly adopted interaction $Q=3b^2H\rho_c$~\cite{Wang:2005jx}, this 
reduces to
\begin{equation}
w=-\frac{1}{3}-\frac{2\sqrt{\Omega_{de}}}{3C}-\frac{b^2}{\Omega_{de}}\,.
\end{equation}
This expression shows that the interaction can drive the equation-of-state 
parameter across the phantom divide, even when the non-interacting model 
remains non-phantom. Finally, in Fig.~\ref{wevolIHDE} we present the evolution 
of the equation-of-state parameter of interacting HDE, for various values of 
the interaction parameter $b$.

\paragraph{Coincidence problem.}

In the absence of interaction and for constant $w$, one finds 
$r=r_0 a^{3w}$, implying that $r\sim\mathcal{O}(1)$ only near the present 
epoch, 
which constitutes the coincidence problem. Interacting models offer a possible 
resolution by allowing $r$ to approach a stable constant value at late times.
For instance, choosing $Q=\Gamma\rho_{de}$~\cite{Pavon:2005yx}, one finds
\begin{equation}
\dot r=3Hr\left(w+\frac{1+r}{r}\frac{\Gamma}{3H}\right).
\end{equation}
Although an exactly constant $r$ cannot be achieved in HDE due to the 
future-event-horizon cutoff, it was shown in~\cite{Hu:2006ar} that a stable 
late-time solution with $r=\mathcal{O}(1)$ can still be obtained, significantly 
alleviating the coincidence problem.

\paragraph{Thermodynamic consistency.}

Finally, the thermodynamic viability of IHDE has been investigated within the 
framework of the generalized second law of thermodynamics. Defining 
the effective 
equations of state
\begin{equation}
\omega_{de}^{eff}=w+\frac{\Gamma}{3H},\qquad 
\omega_m^{eff}=-\frac{1}{r}\frac{\Gamma}{3H},
\end{equation}
the continuity equations take their standard form. The entropy of matter, dark 
energy, and the horizon can then be evaluated using the horizon temperature 
$T=1/(2\pi L)$ and the Bekenstein-Hawking area law. Lastly, as shown 
in~\cite{Setare:2007at}, the total entropy 
$S_m+S_{de}+S_L$ is a non-decreasing function of time for suitable parameter 
choices, indicating that interacting holographic dark energy can be consistent 
with the generalized second law of thermodynamics.

\subsubsection{Observational constraints}

We now turn to the observational status of interacting holographic dark energy. 
Since IHDE introduces additional degrees of freedom beyond the minimal HDE 
model, mainly through the interaction strength, it is essential to 
confront these scenarios with cosmological data. 
Below we summarize the main results, referring the reader to the comprehensive 
analysis of~\cite{Wang:2016och} and references therein for further details.

\paragraph{Interaction strength.}

The primary quantity constrained by observations in IHDE models is the coupling 
strength between dark matter and holographic dark energy. 
Early analyses addressed this issue by adopting specific phenomenological forms 
of the interaction term and confronting them with combined datasets.

In~\cite{Wu:2007fs}, the interaction $Q=9b^2 M_p^2H^2$ was considered, and the 
model was tested against the Gold04 and ESSENCE type-Ia supernova samples, the 
baryon acoustic oscillation parameter $A$ from SDSS, and the CMB shift 
parameter 
$R$ from WMAP3. The results indicated that the non-interacting case $b^2=0$ is 
favored at the $1\sigma$ confidence level, casting doubt on the necessity of 
introducing a dark-sector interaction.
Similar conclusions were reached in~\cite{Feng:2007wn}, where the interaction 
$Q=3b^2H(\rho_m+\rho_{de})$ was examined using a combination of SNIa data, CMB 
shift parameter, BAO measurements, $H(z)$ observations, and lookback time data. 
Once again, the best-fit results showed a clear preference for negligible or 
vanishing interaction strength.

The robustness of these findings was further tested by allowing for nonzero 
spatial curvature. In~\cite{Li:2009zs}, several interaction forms were analyzed 
using the Constitution SNIa compilation, WMAP5 CMB data, and SDSS BAO 
measurements. The inclusion of curvature did not alter the conclusion that 
interacting scenarios are generally disfavored. These results were subsequently 
confirmed by independent numerical investigations~\cite{Ma:2009uw,Feng:2016djj}.

\paragraph{Statefinder diagnostic.}

Beyond direct parameter estimation, geometrical diagnostics provide a useful 
tool for discriminating between dark energy models. 
The statefinder diagnostic, introduced in~\cite{Sahni:2002fz}, employs the 
dimensionless pair $\{r,s\}$ defined as
\begin{equation}
r=\frac{\dddot a}{aH^3}\,,\qquad 
s=\frac{r-1}{3\left(q-\frac{1}{2}\right)},
\end{equation}
where $a$ is the scale factor and $q$ the deceleration parameter. 
Different cosmological models trace distinct trajectories in the $(r,s)$ plane, 
while the spatially flat $\Lambda$CDM model corresponds to the fixed point 
$\{r,s\}=\{1,0\}$.

\begin{figure}[t]
\begin{center}
\includegraphics[width=8.7cm]{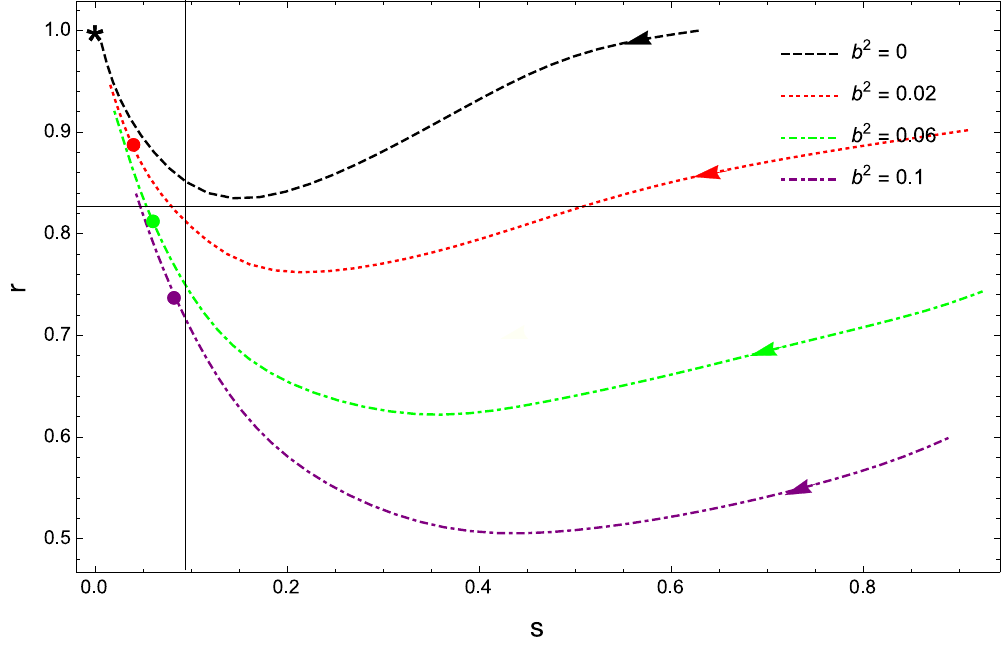}
\caption{{\it{The statefinder diagrams for the interacting HDE model   with  
interaction form $Q=3b^2H\rho_c$, with fixed 
$\Omega_{de0}=0.73$, $C=1$ and different interaction strengths. The evolution 
direction of the statefinder curves is indicated by the arrows. The solid lines 
mark the point $(r,s)=(0.82,0.09)$, which gives the present values of 
statefinder parameters corresponding to $H(z)$ and Pantheon data 
sets~\cite{Solanki:2023yoa}. The figure is from~\cite{Zhang:2007uh}.}}}
\label{EvTr}
\end{center}
\end{figure}

The statefinder diagnostic has been extensively used to distinguish dark energy 
models~\cite{Alam:2003sc,Zhang:2005rj,Wang:2008fx}. 
In the context of interacting holographic dark energy, this method was applied 
in~\cite{Zhang:2007uh}. For the interaction 
$Q=3b^2H(\rho_m+\rho_{de})$, the statefinder parameters read
\begin{equation}
r=1-\frac{3}{2}\Omega_{de}w'
+3\Omega_{de}w\left(1-\frac{\sqrt{\Omega_{de}}}{C}\right),\qquad 
s=1+w-\frac{w'}{3w}+\frac{b^2}{\Omega_{de}},
\end{equation}
with
\begin{equation}
w'=
(1-\Omega_{de})
\left(b^2-\frac{\Omega_{de}^{3/2}}{3C}\right)
\left[
\frac{1}{\Omega_{de}}
-\frac{3b^2}{\Omega_{de}(1-\Omega_{de})}
+2\frac{C}{\sqrt{\Omega_{de}}}
\right].
\end{equation}

The resulting trajectories $r(s)$ for $C=1$ and various values of $b^2$ are 
shown in Fig.~\ref{EvTr}. The $\Lambda$CDM fixed point is marked by a star, 
while 
the present-day values are indicated by dots. 
In the absence of interaction, the HDE trajectory asymptotically approaches the 
$\Lambda$CDM point. In contrast, once interaction is introduced, the 
trajectories 
deviate from this fixed point, with stronger couplings leading to larger 
departures. This behavior provides further support to the conclusion that a 
significant interaction between dark matter and holographic dark energy is not 
favored by current observations.

\paragraph{Cosmic age crisis.}

Another observational consistency test concerns the so-called cosmic age 
problem, which arises when the inferred age of the Universe at a given redshift 
is shorter than the age of its oldest observed objects. 
The cosmic age at redshift $z$ is given by
\begin{equation}
t(z)=\int_{z}^{\infty}\frac{dz'}{(1+z')H(z')},
\qquad 
T_{cos}(z)\equiv H_0 t(z)
=\int_{z}^{\infty}\frac{dz'}{(1+z')E(z')},
\end{equation}
and consistency requires $T_{cos}(z)\ge T_{obj}(z)\equiv H_0 t_{obj}(z)$. 
Equivalently, the ratio
\begin{equation}
\tau(z)\equiv\frac{T_{cos}(z)}{T_{obj}(z)}
\end{equation}
must satisfy $\tau(z)\ge1$ at all redshifts.

A well-known example is the old quasar APM 08279+5255 at $z=3.91$, whose age is 
estimated to be $t_{obj}(3.91)\gtrsim2.0\,\mathrm{Gyr}$~\cite{Hasinger:2002wg}. 
Remarkably, many standard cosmological models, including $\Lambda$CDM, predict 
$\tau(3.91)<1$ for this object~\cite{Friaca:2005ba}, leading to the cosmic age 
crisis.

The original HDE model was first examined in this context in~\cite{Wei:2007ig}, 
where it was shown that accommodating APM 08279+5255 requires adopting a 
significantly lower Hubble constant. 
The issue was revisited in the interacting HDE framework in~\cite{Cui:2010dr}. 
Allowing for suitable interaction terms and coupling strengths, the ratio 
$\tau(3.91)$ can exceed unity, thereby resolving the age problem for this 
object. This result suggests that dark-sector interactions, although not 
favored 
by current data, may offer a potential mechanism for alleviating certain 
cosmological tensions.

\subsection{Holographic dark energy with different IR cutoffs}
\label{Other}

In the original holographic dark energy model~\cite{Li:2004rb}, the infrared 
cutoff is identified with the future event horizon of the Universe. Although 
this choice is phenomenologically successful and well motivated within the 
holographic framework, it is not dictated by a unique or fundamental principle. 
As a result, considerable effort has been devoted to exploring alternative dark 
energy models inspired by holography, where different cosmological length or 
time scales are adopted as the infrared cutoff. Such constructions preserve the 
spirit of holography, namely the connection between ultraviolet and infrared 
physics, while leading to qualitatively different cosmological dynamics.

In what follows, we review some representative dark energy models arising from 
alternative choices of the infrared cutoff and discuss their main properties in 
comparison with the standard HDE scenario.

\subsubsection{Agegraphic dark energy}

Agegraphic dark energy (ADE) constitutes one of the earliest alternatives to 
holographic dark energy based on a different infrared scale. The central idea 
is to associate the dark energy density with the age of the Universe, motivated 
by the Karolyhazy uncertainty relation and quantum fluctuations of spacetime 
~\cite{Cai:2007us,Wei:2007ty,Wei:2007xu}. In this approach, the characteristic 
time scale of the FRW Universe plays the role of the 
infrared cutoff.

The original version of ADE, which employed the cosmic time as the relevant 
scale~\cite{Cai:2007us}, was shown to be unable to account for the observed 
late-time acceleration of the Universe. This shortcoming motivated an improved 
formulation, where the conformal time is adopted as the infrared 
cutoff~\cite{Wei:2007ty}. Within this framework, the dark energy density is 
given by
\begin{equation}
\label{ADE}
\rho_{de}=\frac{3n^2M_p^2}{\eta^2}\,,
\end{equation}
where
\begin{equation}
\eta=\int_0^t\frac{dt'}{a(t')}
=\int_0^a\frac{da'}{H a'^2}
\end{equation}
denotes the conformal time. The dimensionless parameter $n$ encapsulates 
uncertainties related to spacetime curvature effects, the number of quantum 
fields, and possible deviations from the idealized assumptions of the model. 
Since the present conformal time satisfies $\eta_0\sim H_0^{-1}$, the magnitude 
of $\rho_{de}$ today is naturally of the observed order, provided that $n$ is 
of order unity.
Moreover, from Eq.~\eqref{ADE}, the fractional dark energy density and its 
evolution 
equation follow as
\begin{eqnarray}
\Omega_{de}&=&\frac{n^2}{H^2\eta^2}\,,\\[2mm]
\Omega'_{de}&=&\Omega_{de}\left(1-\Omega_{de}\right)
\left(3-\frac{2}{n}\frac{\sqrt{\Omega_{de}}}{a}\right),
\end{eqnarray}
where a prime denotes differentiation with respect to $\ln a$. Additionally, 
the  
corresponding 
equation-of-state parameter reads
\begin{equation}
\label{EoSADE}
w=-1+\frac{2}{3n}\frac{\sqrt{\Omega_{de}}}{a}\,.
\end{equation}

An important distinction between ADE and holographic dark energy is that the 
former involves only a single free parameter,  namely $n$. Indeed, during the 
matter-dominated era one has $\eta\propto\sqrt{a}$, which implies 
$\rho_{de}\propto a^{-1}$. Substituting this scaling into the continuity 
equation yields $w=-2/3$, which, when inserted into Eq.~\eqref{EoSADE}, leads 
to 
the consistency relation
\begin{equation}
\Omega_{de}=\frac{n^2a^2}{4}\,.
\end{equation}
This behavior indicates that the fractional dark energy density grows 
monotonically with the scale factor, thereby offering a natural explanation of 
why dark energy becomes dominant only at late times. In this sense, ADE 
ameliorates the coincidence problem without invoking interactions between dark 
sectors.

\begin{figure}[t]
\begin{center}
\includegraphics[width=7.7cm]{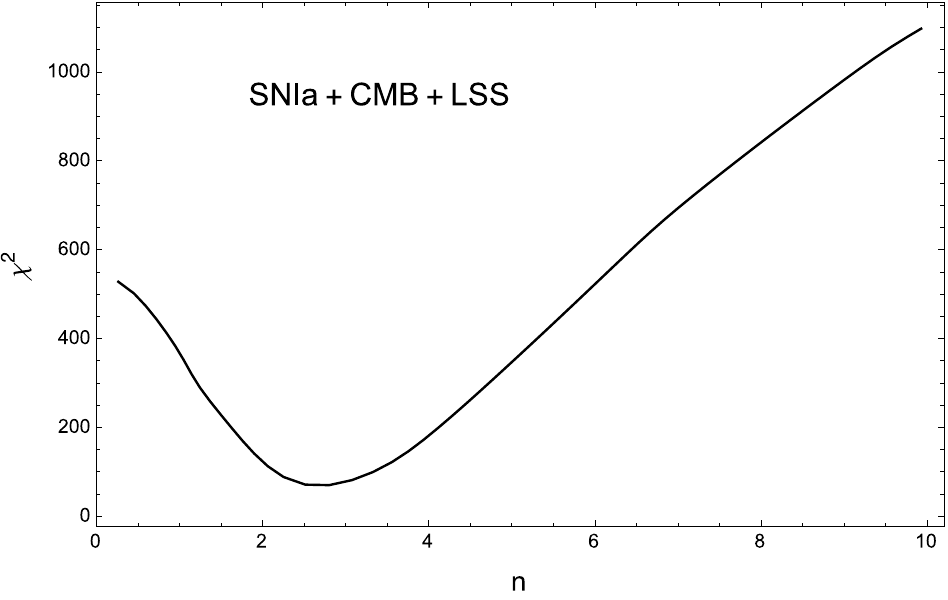}
\hspace{4mm}\includegraphics[width=7.5cm]{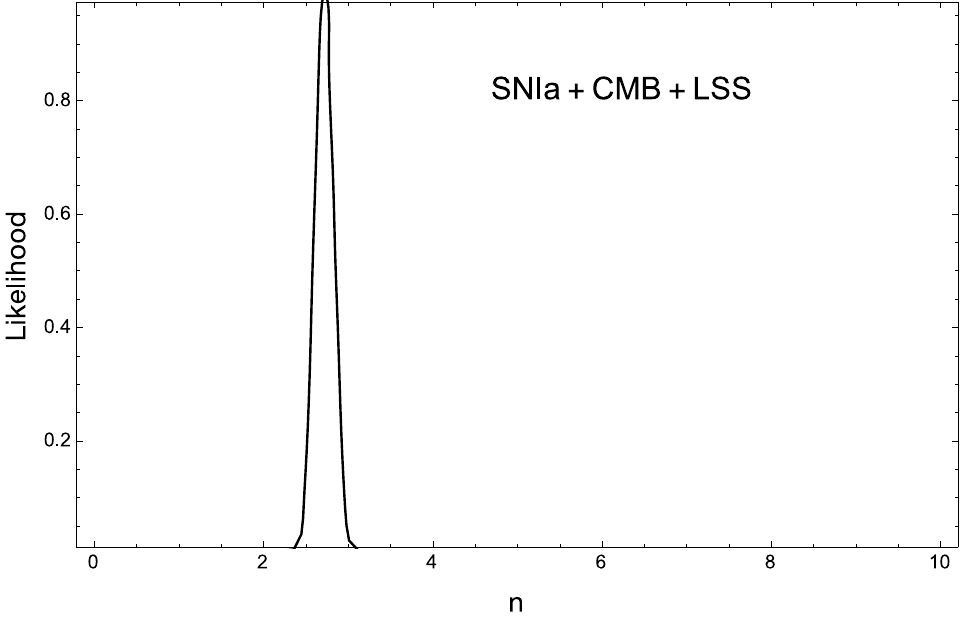}
\caption{{\it{The $\chi^2$ (left plot) and the corresponding likelihood (right 
panel) 
as a function of $n$, for the agegraphic dark energy model, using SNIa, CMB and 
BAO measurements. The figures are from~\cite{Wei:2007xu}.}}}
\label{ADEplot}
\end{center}
\end{figure}

The observational viability of ADE was examined in~\cite{Wei:2007xu}. In Fig. 
~\ref{ADEplot} we depict the  $\chi^2$ 
 and the corresponding likelihood of the observational confrontation. In 
particular, using 
type-Ia supernova data, the parameter $n$ was constrained to 
$n=2.954^{+0.264}_{-0.245}$ at the $1\sigma$ confidence level. Including CMB 
and 
large-scale structure information further tightened the bounds to 
$n=2.716^{+0.111}_{-0.109}$. 

Despite its appealing conceptual features and 
rich phenomenology~\cite{ Sharma:2020ylh, Hossienkhani:2017uku,
Wei:2007ut, Wei:2007zs, Wu:2007wu, 
Wei:2007ty, 
Zhang:2007ps, Wei:2007xu, Neupane:2007ra, Kim:2007iv, Zhang:2008mb,Lee:2008zzw, 
Kim:2008hz, Wu:2008jt, Setare:2009jf, Cui:2009ns, Sheykhi:2009gfu, 
Sheykhi:2009rk, Sheykhi:2009yva, Sheykhi:2009kj,Sheykhi:2009sz, Karami:2009ns,
Liu:2009xb, Setare:2009zzb, Zhang:2009qa, Sheykhi:2009ui, Karami:2009wd, 
Sheykhi:2009em, Setare:2010anr, Sheykhi:2010jn,Farooq:2010qr, Wu:2010zzd, 
Karami:2010tr, Karami:2010aa, Jamil:2010kg, Jamil:2010vr, Malekjani:2010uv, 
Karami:2010aq, Jamil:2010wa,Zhang:2010im, Karami:2010qe, Sun:2010hv, 
Setare:2010zy, Karami:2010bys, Setare:2010zzb, Lemets:2010qz, Li:2010ak, 
Karami:2011vw,Sun:2011vg, Sheykhi:2011zz, Chen:2011rz, Khatua:2011sh, 
Karami:2011bm, Zhai:2011pp, Farajollahi:2012mm, Saaidi:2012qp, Li:2012xm, 
Karami:2012ra,Farajollahi:2012uh, Farajollahi:2012zz, Liu:2012kha, 
Zhang:2012pr, 
Xu:2013mfa, Xu:2013vba, Payandeh:2013foa, Pasqua:2013ega, Jawad:2013uil, 
Farajollahi:2013isk,Zhang:2013lea, Majumder:2013fza, Pasqua:2013gto, 
Debnath:2013woa, Wu:2014osz, Fayaz:2014bja, Aly:2014ssa, Xu:2014gda, 
Aly:2015qra, Jawad:2015vra, Maity:2015mys,Jawad:2015wea, Xu:2015ata, 
Fayaz:2016jql, Hossienkhani:2016pzw, Saha:2016bjs, 
Setare:2016mfv, Malekjani:2016edh, 
Kumar:2016los,Rezaei:2017hon,Sharma:2020lmm,
AbdollahiZadeh:2018heh,Ravanpak:2018sfk,Sharma:2021dqj,
AbdollahiZadeh:2018ubg, Ravanpak:2018gpi, Ravanpak:2019zdg, Saba:2019tna, 
Maity:2019qbv, Xu:2019hhs,Srivastava:2020riu, Pourbagher:2020zkm, 
Maity:2020wqo, 
Sharma:2020mzl, Srivastava:2021apm, 
Huang:2021rpf,
Sardar:2021eaj, Saha:2022vcb, Fazlollahi:2022dwz,FeiziMangoudehi:2022yvu, 
Kumar:2022dwm, Pankaj:2022lnk, Sobhanbabu:2022woo, Huang:2022hiv, Xu:2022uwv, 
Kumar:2022bfw, Pinki:2023plx, Kumar:2023njq, Sheykhi:2023woy,Xu:2023kbx, 
Hernandez-Marquez:2023erq, Sharma:2024nko, Chakraborty:2024ilj, Ajmal:2024hqs, 
Singha:2025trb, Naeem:2025ybg, Sharif:2025oxp, Kotal:2025bof, Sharif:2025qwy, 
Mazumdar:2025vdk, Kotal:2025hus, Sadatian:2025glu, Huang:2025hdg, 
Maity:2025cxd,Maity:2026uok,Maity:2026mak}, 
agegraphic dark energy  is 
generally found to be disfavored relative to both $\Lambda$CDM and holographic 
dark energy when confronted with the full set of cosmological observations. 
A detailed comparison with other dark energy models will be presented in 
Sec.~\ref{Comp}.

\subsubsection{Ricci  holographic dark energy}

From a geometric and covariant perspective, a particularly natural choice for 
the infrared cutoff is provided by the Ricci scalar curvature of spacetime. 
Since the Ricci scalar encodes local information about the expansion rate and 
its time variation, it offers an appealing alternative to horizon-based 
cutoffs, while remaining fully compatible with the principles of General 
Relativity~\cite{Gao:2007ep,Zhang:2009un,delCampo:2013hka}. This idea leads to 
the 
so-called Ricci dark energy (RDE) model.

In a FRW Universe, the Ricci scalar is given by
\begin{equation}
R=-6\left(\dot H+2H^2+\frac{k}{a^2}\right),
\end{equation}
where  $k$ is the 
spatial curvature constant. By identifying the infrared cutoff with the Ricci 
scalar, the dark energy density is constructed as
\begin{equation}
\rho_{de}=-\frac{\alpha R}{16\pi}
=\frac{3\alpha}{8\pi}\left(\dot H+2H^2+\frac{k}{a^2}\right),
\end{equation}
where the dimensionless parameter $\alpha$ governs the strength of the Ricci 
contribution and ultimately determines the cosmic evolution.

Within this framework, the Friedmann equation can be recast into a differential 
equation for the Hubble rate. Introducing the variable $x\equiv\ln a$, one 
obtains
\begin{equation}
H^2=\frac{8\pi G}{3}\rho_{m0}e^{-3x}
+\left(\alpha-1\right)k e^{-2x}
+\alpha\left(\frac{1}{2}\frac{dH^2}{dx}+2H^2\right).
\end{equation}
Defining the dimensionless expansion rate $E(a)\equiv H(a)/H_0$, this 
equation 
admits the solution
\begin{equation}
\label{ERde}
E^2(a)=\Omega_{m0}a^{-3}
+\Omega_{k0}a^{-2}
+\frac{\alpha}{2-\alpha}\Omega_{m0}a^{-3}
+f_0 a^{-\left(4-\frac{2}{\alpha}\right)},
\end{equation}
where the integration constant $f_0$ is fixed by the normalization condition 
$E_0=1$, yielding
\begin{equation}
f_0=1-\Omega_{k0}-\frac{2}{2-\alpha}\Omega_{m0}.
\end{equation}
Furthermore, from Eq.~\eqref{ERde}, the fractional energy density of Ricci dark 
energy can 
be 
read directly as
\begin{equation}
\Omega_{de}
=\frac{\alpha}{2-\alpha}\Omega_{m0}a^{-3}
+f_0 a^{-\left(4-\frac{2}{\alpha}\right)}.
\end{equation}
This expression reveals that RDE naturally consists of two components, namely  
one that 
scales like pressureless matter and another with a distinct scaling behavior 
controlled by the parameter $\alpha$.

\begin{figure}[t]
\begin{center}
\includegraphics[width=8.7cm]{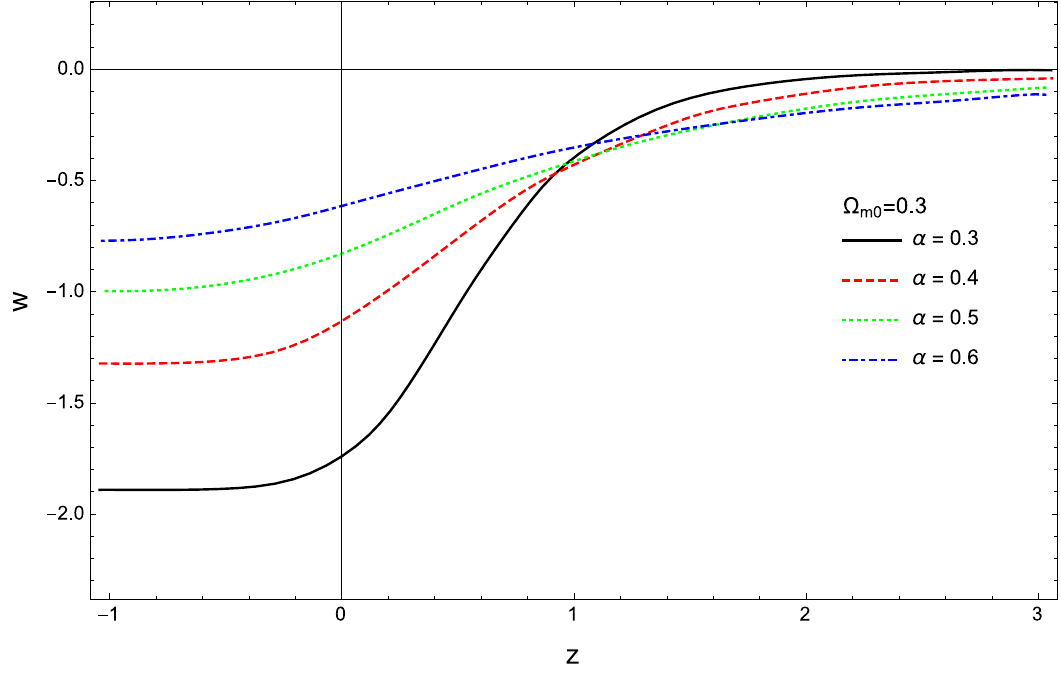}
\caption{{\it{The evolution of the   equation-of-state parameter    
of Ricci holographic dark energy,  for various values of the parameter $\alpha$.
For $\alpha\le1/2$ it lies in the  quintessence regime, while for 
$\alpha<1/2$  it experiences the phantom-divide crossing. The figure is 
from~\cite{Zhang:2009un}.}}}
\label{RDEfig}
\end{center}
\end{figure}

The cosmological implications of RDE crucially depend on the value of 
$\alpha$. For $1/2\le\alpha\le1$, the effective equation-of-state parameter 
lies in the range $-1\le w\le-1/3$, corresponding to a quintessence-like 
behavior capable of driving late-time acceleration. In the special case 
$\alpha=1/2$, the model effectively mimics a cosmological constant 
supplemented 
by an additional matter-like component, leading the Universe asymptotically to 
a 
de Sitter phase. 
For $\alpha<1/2$, the Ricci dark energy exhibits a quintom-like evolution 
\cite{Cai:2009zp}, 
transitioning from a quintessence regime at early times to a phantom regime at 
late times. This dynamical crossing of the phantom divide is a distinctive 
feature of RDE and has attracted considerable interest in the literature. In 
Fig. \ref{RDEfig} we present the evolution of the equation-of-state parameter 
of  RDE, for various values of the parameter $\alpha$, where the above behavior 
can be clearly seen.

Finally, observational constraints place the parameter $\alpha$ close to the 
quintom-favored regime. In particular, type-Ia supernova data yield 
$\alpha=0.394^{+0.152}_{-0.106}$ at the $1\sigma$ confidence level, while a 
combined analysis of SNIa, CMB, and BAO data tightens the bound to 
$\alpha=0.359^{+0.024}_{-0.025}$~\cite{Zhang:2009un}. These results indicate 
that Ricci dark energy provides a viable and competitive alternative to both 
$\Lambda$CDM and standard holographic dark energy, and that is why it has been 
studied in 
detail~\cite{Feng:2008rs,Feng:2008kz,Xu:2008rp,Feng:2008hk,Zhang:2009un,
Xu:2009xi,
Feng:2009hr,Feng:2009ag,Feng:2009ai,Feng:2009jr,Suwa:2009gm,Kim:2010pdn,
Lepe:2010vh,Xu:2010gg,Zhang:2010im,Yang:2011us,Zhang:2011zze,
Chattopadhyay:2011zz,Chattopadhyay:2011mpa,Chimento:2011dw,Bhattacharya:2011xa,
Chattopadhyay:2011er,Chimento:2011pk,Karwan:2011sh,Wang:2011km,Broda:2011np,
Fu:2011ab,Belkacemi:2011zk,Pasqua:2011gh,Chimento:2012zz,Mathew:2012md,
Izaurieta:2026edz,Saridakis:2025ltr,
Saadat:2013hva,Bouhmadi-Lopez:2013mji,Chimento:2013se,Pasqua:2013ewa,
Chimento:2013qja,delCampo:2013hka,Silva:2013yaa,Myung:2013cqa,Arevalo:2013tta,
Aguilera:2013tmp,Pankunni:2013jhv,Mathew:2013fka,Belkacemi:2014aqa,
Chattopadhyay:2014ixa,Salti:2014aha,Li:2014eba,Aly:2015zya,Yu:2015sla,
Albarran:2015tga,Forte:2016ben,Herrera:2016uci,Rao:2017lnt,Salti:2017ywf,
Bouhmadi-Lopez:2017kvc,George:2018myt,Aditya:2018lkw,DasuNaidu:2018tvz,
Singh:2018yau,Jesus:2019nwi,Sadri:2019yqs,George:2019vko,Saleem:2020fjf,
Cid:2020kpp,Anchordoqui:2020sqo,Hossienkhani:2021emv,Kumar:2021avx,
Rudra:2022qbv,Sultana:2022pmj,Pasqua:2023rdp,Cid:2023yvw,Shekh:2023baf,
Alvarenga:2024yqa,Bubuianu:2024zsm,Satyanarayana:2024gxa,
Scomparin:2025jub, 
Sanyal:2025udg,Wu:2025vfs,Chokyi:2026ryf,Mishra:2026vbo,Chokyi:2026eqo}.

\subsubsection{Ricci-Gauss-Bonnet holographic dark energy}

In the standard formulation of Ricci dark energy of the previous paragraph, the 
infrared   cutoff is constructed by employing the Ricci scalar.   However, 
it is well understood that when curvature invariants are introduced within a 
given modification, theoretical consistency requires that all invariants of the 
same dimensional order should be taken into account. In this respect, within 
FRW 
geometry the Gauss-Bonnet invariant,
$
G=R^2-4R_{\mu\nu}R^{\mu\nu}+R_{\mu\nu\rho\sigma}R^{\mu\nu\rho\sigma}$,
is of the same order as $R^2$. Consequently, in a holographic dark energy 
framework where the Ricci scalar is employed to define the IR cutoff, a 
contribution proportional to $\sqrt{|G|}$ should also be included.

Motivated by these considerations, in~\cite{Saridakis:2017rdo} the author 
proposed a holographic dark energy scenario in which the inverse squared IR 
cutoff is defined as
\begin{equation}\label{GBHDEL}
\frac{1}{L^2}=-\alpha R+\beta \sqrt{|G|},
\end{equation}
where $\alpha$ and $\beta$ are free model parameters. Evidently, setting 
$\beta=0$ one recovers the usual Ricci dark energy model, while the choice 
$\alpha=0$ yields a purely Gauss-Bonnet holographic dark energy contribution.

Substituting~(\ref{GBHDEL}) into the standard holographic dark energy 
expression $\rho_{DE}=\frac{3c}{\kappa^2 L^2}$, one obtains the energy density 
of the Ricci-Gauss-Bonnet holographic dark energy  as~\cite{Saridakis:2017rdo}
\begin{equation}\label{rhoHDE}
 \rho_{DE}=\frac{3}{\kappa^2}\left(-\alpha R+\beta \sqrt{|G|}\right),
\end{equation}
where, for simplicity, the constant $c$ has been absorbed into the definitions 
of $\alpha$ and $\beta$.
Since  in a spatially flat FRW spacetime, the Ricci scalar and the 
Gauss-Bonnet invariant take the form
$ R=-6\left(2H^2+\dot{H}\right)$ and $ G=24H^2\left(H^2+\dot{H}\right),$ 
respectively,     the 
Ricci-Gauss-Bonnet holographic dark energy density can be written as
\begin{equation}\label{rhoHDE2}
 \rho_{DE}=\frac{3}{\kappa^2}\left[6\alpha\left(2H^2+\dot{H}\right)
 +2\sqrt{3}\,\beta\,H\,\sqrt{|H^2+\dot{H}|}\right].
\end{equation}
  Restricting to the standard dust matter 
case, and using the  density parameters, one     
obtains~\cite{Saridakis:2017rdo}
\begin{equation}
\Omega_{DE}
=3\alpha\left[1+\frac{\Omega_{DE}'}{1-\Omega_{DE}}\right]
+2\sqrt{3}\,\beta
\sqrt{\left|\frac{\Omega_{DE}'}{1-\Omega_{DE}}-1\right|},
\label{OmDEdifeq}
\end{equation} 
with primes
denoting derivatives with respect to  $x=\ln a=-\ln(1+z)$,
which constitutes the differential equation governing the evolution of 
Ricci-Gauss-Bonnet holographic dark energy in a spatially flat Universe filled 
with dust matter.
This equation admits an implicit analytical solution, which can be written as
\begin{eqnarray}
 &&  \epsilon_{\pm}\frac{\gamma_+}{\delta\zeta_+}\,
 \text{arctanh}\!\left(
 -\frac{\sqrt{6\alpha^2+\beta^2-\alpha\,\Omega_{DE}}}{\zeta_+}
 \right) 
 -\epsilon_{\pm}\frac{\gamma_-}{\delta\zeta_-}\,
 \arctan\!\left(
 -\frac{\sqrt{6\alpha^2+\beta^2-\alpha\,\Omega_{DE}}}{\zeta_-}
 \right)\nonumber\\
 && 
 -24\epsilon_{\pm}\frac{\beta}{\delta}
 \sqrt{6\alpha^2+\beta^2-\alpha}\,
 \text{arctanh}\!\left(
 \frac{\sqrt{6\alpha^2+\beta^2-\alpha\,\Omega_{DE}}}
 {\sqrt{6\alpha^2+\beta^2-\alpha}}
 \right) 
 -\frac{\sqrt{3}(3\alpha^2+2\beta^2-\alpha)}{\delta}
 \ln\!\left[
 \frac{9\alpha^2-12\beta^2-6\alpha\,\Omega_{DE}+\Omega_{DE}^2}
 {(\Omega_{DE}-1)^2}
 \right]\nonumber\\
 && 
 -\frac{2(9\alpha-1)\beta}{\delta}
 \text{arctanh}\!\left(
 \frac{\Omega_{DE}-3\alpha}{2\sqrt{3}\beta}
 \right)
 =2\ln a+x_0,
 \label{gensol}
\end{eqnarray}
where $\epsilon_{\pm}=\pm1$ labels the two solution branches, and with
\begin{eqnarray}
 && \gamma_{\mp}=2\Big[\mp9\alpha^3
 +\beta^2(\pm1+2\sqrt{3}\beta)
 \mp\alpha\beta(\pm2\sqrt{3}+15\beta) 
 +3\alpha^2(4\sqrt{3}\beta\pm1)\Big],\nonumber\\
 && \delta=\frac{1-6\alpha+9\alpha^2-12\beta^2}{\sqrt{3}},\nonumber\\
 && \zeta_{\mp}=\sqrt{\mp3\alpha^2+2\sqrt{3}\alpha\beta\mp\beta^2}.
\end{eqnarray}
The integration constant $x_0$ is fixed by evaluating~(\ref{gensol}) at the 
present epoch, namely at $a=a_0=1$ and $\Omega_{DE}=\Omega_{DE0}$.
Finally, the  dark-energy equation-of-state 
parameter $w_{DE}$ is given by~\cite{Saridakis:2017rdo} 
\begin{eqnarray}\label{wDEgenfin}
&&
\!\!\!\!\!\!\!\!\!\!\!\!\!\!\!\!\!\!\!\!\!\!\!\!\!\!\!\!\!\!\!\! 
w_{DE}
=-1+\Omega_{DE}^{-1}\Bigg\{
\alpha\left[
3+\frac{2\Omega_{DE}'}{1-\Omega_{DE}}
-\frac{2(\Omega_{DE}')^2}{(1-\Omega_{DE})^2}
-\frac{\Omega_{DE}''}{1-\Omega_{DE}}
\right]-\frac{\beta}{\sqrt{3}(1-\Omega_{DE})^2}
\frac{1}{\sqrt{\left|\frac{\Omega_{DE}'}{1-\Omega_{DE}}-1\right|}}
\nonumber\\
&&\qquad\qquad \ \ \ 
 \times
\Big[
2\big(\Omega_{DE}'-3(1-\Omega_{DE})\big)
(\Omega_{DE}'+\Omega_{DE}-1) 
+\Omega_{DE}''(1-\Omega_{DE})+(\Omega_{DE}')^2
\Big]
\Bigg\},
\end{eqnarray}
which is known as long as   $\Omega_{DE}$   is   
known from 
the implicit solution~(\ref{gensol}).

In summary, relation~(\ref{gensol}) provides the two analytical branches 
describing the evolution of Ricci-Gauss-Bonnet holographic dark energy as a 
function of $\ln a$.  
As can be inferred, the model is capable of reproducing the standard thermal 
history of the Universe, including the transition from decelerated to 
accelerated expansion at $z\simeq0.45$, in agreement with observational 
expectations. Moreover, at late times the Universe asymptotically evolves 
toward a fully dark-energy-dominated state. Finally, one can show that, 
depending on the region of parameter space, the dark energy equation-of-state 
parameter may cross the phantom divide either before or after the present epoch,
remain entirely within the quintessence regime, or asymptotically approach 
$w_{DE}=-1$, corresponding to a de Sitter phase~\cite{Saridakis:2017rdo,
Iqbal:2018maa,Ahmed:2019qsi,Lohakare:2022mkd,
Pradhan:2024fjl,Altaibayeva:2025rhf,Dubey:2025hpd,Chattopadhyay:2026dly,
Molavi:2019mlh,Rudra:2022qbv,Sanyal:2025udg,Joshi:2026lih,
Chattopadhyay:2026dly,Joshi:2026ref}.

\subsubsection{Holographic dark energy involving the  Hubble horizon as the 
characteristic length scale}

As discussed in Sec.~\ref{HDE}, the Hubble radius $H^{-1}$ by itself is not a 
suitable infrared cutoff for the original holographic dark energy construction, 
since it fails to generate an accelerating Universe when combined with the 
standard UV-IR relation. Nevertheless, this conclusion does not exclude more 
general frameworks in which the Hubble scale enters the definition of the 
effective cutoff together with additional dynamical quantities. In this sense, 
generalized HDE models involving the Hubble parameter cannot be ruled out 
\emph{a priori}.

A particularly simple and well-motivated extension was proposed by 
Granda and Oliveros~\cite{Granda:2008dk}, based on the assumption that the dark 
energy density depends on both the Hubble parameter and its time derivative, 
namely
\begin{equation}
\label{GO}
\rho_{de}=3M_p^2\left(\alpha H^2+\beta \dot H\right),
\end{equation}
where $\alpha$ and $\beta$ are dimensionless constants. This prescription 
is 
known as the Granda-Oliveros (GO) cutoff. From a physical perspective, this 
form 
effectively incorporates local information about the cosmic expansion rate and 
its evolution, while remaining consistent with dimensional analysis.
Upon adopting~\eqref{GO}, the Friedmann equation becomes
\begin{equation}
H^2
=\frac{1}{3M_p^2}\left(\rho_{m0}e^{-3x}+\rho_{r0}e^{-4x}\right)
+\alpha H^2
+\frac{\beta}{2}\frac{dH^2}{dx},
\end{equation}
where $x\equiv\ln a$. It is straightforward to verify that this structure is 
closely related to that of Ricci dark energy. Indeed, in a spatially flat FRW 
background, the Ricci scalar is proportional to a linear combination of $H^2$ 
and $\dot H$, rendering the GO model a natural generalization of the Ricci 
cutoff. The extension of this construction to non-flat geometries is more 
involved and has been addressed in~\cite{Karami:2009je}, where the evolution of 
the deceleration parameter, equation-of-state parameter, and dark energy 
density was analyzed in detail, while other relevant studies have been 
performed in \cite{Chander:2026zxg}.

Interestingly, viable cosmological models employing the Hubble horizon as IR 
cutoff have also been developed within more elaborate theoretical frameworks. 
In~\cite{Gong:2009dc}, a consistent HDE model with Hubble cutoff was 
constructed 
inspired by the Dvali-Gabadadze-Porrati (DGP) braneworld 
scenario~\cite{Dvali:2000hr}. 
In that case, the UV cutoff is not imposed in four dimensions but instead 
arises 
from the higher-dimensional black-hole formation bound. Alternative approaches 
include models with a time-dependent holographic parameter 
$C$~\cite{Xu:2009ys}, as well as extensions of standard HDE in scalar-tensor 
theories such as Brans-Dicke gravity with a potential 
term~\cite{Liu:2009ha}, or in General Relativity supplemented by suitable 
interactions between dark energy and dark 
matter~\cite{Jamil:2010ed,Khodam-Mohammadi:2012zuk,Duran:2012yr,
Ghaffari:2014pxa,Pasqua:2016wrm,Koussour:2022nsu,Kaur:2023jhx,Shekh:2023wka,
Manoharan:2024thb,Dubey:2024utn,Gupta:2024qyn,Mehta:2025laj}.

\subsubsection{Alternative infrared length scales}

Beyond horizon-based and Hubble-related cutoffs, several alternative infrared 
length scales have been proposed within the holographic framework. In 
Ref.~\cite{Guberina:2005mp}, a generalized HDE model was formulated by allowing 
for an energy-dependent Newton’s constant. In this scenario, the dark energy 
density takes the form
$\rho_{de}(\mu)\sim \mu^2 G_N^{-1}(\mu)$, where $\mu$ denotes the IR energy 
scale. The evolution of $G_N$ is determined under the assumption that the 
total 
stress-energy tensor $G_N T_{\mathrm{total}}^{\mu\nu}$ and the matter tensor 
$T_{\mathrm{matt}}^{\mu\nu}$ are conserved separately, leading to non-trivial 
cosmological dynamics.

Another possibility is to identify the apparent horizon as the infrared 
cutoff~\cite{Sheykhi:2009zv}. While this choice is appealing from a 
thermodynamic standpoint, the resulting model does not naturally produce 
late-time acceleration unless an interaction between dark energy and dark matter 
is 
introduced. As an alternative, one may consider a conformal-age-like length 
scale
\begin{equation}
L=\frac{1}{a^4(t)}\int_0^t dt' \, a^3(t'),
\end{equation}
as proposed in~\cite{Huang:2012nz}. Phenomenological analyses indicate that 
this 
model is consistent with Big Bang nucleosynthesis constraints on early dark 
energy and successfully predicts the transition from decelerated to accelerated 
expansion at late times~\cite{Huang:2012gd}.
A closely related construction employs the total comoving horizon of the 
Universe,
$\eta=\int_0^t \frac{dt'}{a(t')}$, as the infrared scale. This scenario was 
investigated in~\cite{Huang:2012xma}, where it was shown that the resulting 
$\eta$HDE model provides a plausible resolution to both the fine-tuning and 
coincidence problems.

More generally, one may envisage additional holographic dark energy models 
based 
on other cosmological quantities acting as infrared cutoffs, such as scales 
associated with the growth of large-scale structure or curvature-related 
invariants of FRW geometry. Although such constructions remain largely 
unexplored, they offer promising directions for future research within the 
holographic paradigm.

\subsection{Holographic dark energy in extended gravity} 
\label{Extended}

The holographic dark energy   paradigm, by construction, establishes a 
non-trivial relation between ultraviolet and infrared scales through the 
holographic principle. Since such a UV-IR correspondence is absent in standard 
Einstein gravity at the level of the action, it is natural to expect that HDE 
may find a more fundamental realization within gravitational theories beyond 
General Relativity. In this sense, HDE does not merely represent an effective 
dark energy component, but rather points toward possible infrared modifications 
of gravity that encode holographic information.

Over the last decades, a substantial body of work has explored the interplay 
between HDE and extended gravitational frameworks, motivated by the idea that 
modifications of gravity may offer a more natural explanation of cosmic 
acceleration without introducing a cosmological constant. In many cases, HDE 
acts as a guiding principle for constructing viable infrared modifications of 
gravity, while in others it provides a phenomenological probe of the underlying 
theory. Following the review in~\cite{Wang:2016och}, we summarize below the 
main 
classes of extended gravity theories in which HDE has been investigated, and we 
highlight their key cosmological implications.

\paragraph{Brans-Dicke and scalar-tensor theories}

Brans-Dicke theory extends General Relativity by introducing a scalar field 
that mediates the gravitational interaction and renders Newton's constant 
dynamical~\cite{Brans:1961sx}. This framework naturally accommodates evolving 
cosmological backgrounds and provides a fertile ground for implementing 
holographic ideas, since both the gravitational coupling and the dark energy 
sector acquire a dynamical character.

Extended HDE in Brans-Dicke theory was studied in 
Refs.~\cite{Gong:2004fq,Banerjee:2007zd,Setare:2006yj,Xu:2008sn}. By assuming a 
power-law ansatz for the Brans-Dicke scalar field,
$\phi/\phi_0 \sim a^{\zeta}$, and solving the modified Friedmann equations, one 
obtains generalized expressions for the equation-of-state parameter of HDE and 
the deceleration parameter. Observational constraints on the variation of 
Newton’s constant restrict the exponent to $\zeta<0.14$, which still includes 
Einstein gravity as the limiting case $\zeta\rightarrow0$.

More general scalar-tensor extensions, such as $f(R,\phi)$ gravity, were 
examined in~\cite{Bisabr:2008gu}. It was shown that for a wide class of scalar 
potentials the effective equation of state of HDE remains negative, allowing 
for a consistent transition from decelerated to accelerated expansion. These 
models illustrate how holographic dark energy can coexist with scalar-mediated 
gravitational interactions.

\paragraph{Braneworld models}

HDE has also been extensively studied within braneworld scenarios, where 
gravity 
propagates in higher dimensions while matter fields are confined to a 
four-dimensional brane. In the Randall-Sundrum framework, it was shown 
in~\cite{Zhang:2009xj} that the presence of extra dimensions may soften the 
future singularity associated with HDE models with $C<1$ (see also the 
discussion below Eq.~\eqref{Crmeff}).

Another prominent realization is the Dvali-Gabadadze-Porrati (DGP) 
braneworld~\cite{Dvali:2000hr}, in which gravity leaks into an extra dimension 
at 
large scales. Two main approaches have been adopted in the literature. In the 
first, HDE is treated as a four-dimensional energy component obeying the 
standard holographic relation, and its evolution is studied for different 
choices of the infrared cutoff~\cite{Wu:2007tp}. In the second approach, the 
holographic relation itself is generalized to five dimensions, leading to a 
modified expression for the dark energy density. Such higher-dimensional HDE 
models were explored 
in
\cite{Saridakis:2007cy,Saridakis:2007ns,Saridakis:2007wx,Farajollahi:2014hzp}, 
where it was shown that the conventional four-dimensional behavior is recovered 
in the low-energy limit, while significant deviations may arise when the bulk 
geometry is finite.
Although the DGP model alone faces observational challenges, its combination 
with HDE improves the agreement with cosmological 
data~\cite{Farajollahi:2014hzp}, 
suggesting that holography may play a stabilizing role in braneworld cosmology.

\paragraph{Loop quantum gravity}

The implementation of HDE in loop quantum gravity and in braneworld models with 
timelike extra dimensions was investigated in~\cite{Zhang:2007an}. Remarkably, 
the resulting Friedmann equation closely resembles that obtained in models with 
spacelike extra dimensions, with the crucial difference being a negative sign 
in 
front of the quadratic term in the energy density.

Despite this seemingly minor modification, the cosmological consequences are 
profound. In particular, the Hubble parameter may vanish at a finite value of 
the dark energy density, opening the possibility of a cyclic Universe that 
undergoes successive phases of expansion and contraction. This behavior 
illustrates how quantum gravitational corrections may dramatically alter the 
global dynamics of HDE cosmology.

\paragraph{Ho\v{r}ava-Lifshitz gravity}

Ho\v{r}ava-Lifshitz gravity was proposed as a candidate ultraviolet completion 
of gravity by introducing anisotropic scaling between space and 
time~\cite{Horava:2009uw}. By breaking Lorentz invariance at high energies, the 
theory achieves power-counting renormalizability while recovering General 
Relativity in the infrared.

The cosmological implications of Ho\v{r}ava-Lifshitz gravity with an effective 
dark energy sector were first studied in~\cite{Saridakis:2009bv}, and 
subsequent 
analyses incorporated HDE within this framework~\cite{Setare:2010wt}. The 
resulting evolution of the dark energy density exhibits a non-trivial 
dependence on the anisotropic scaling parameter, leading to a richer 
phenomenology compared to the standard HDE model. In particular, the interplay 
between holographic dynamics and Lorentz-violating effects modifies the 
late-time 
cosmic evolution in an observationally testable way.

\paragraph{Induced gravity}

Induced (or emergent) gravity is based on the idea that Einstein gravity arises 
dynamically through spontaneous symmetry breaking~\cite{Zee:1978wi}. In this 
framework, the gravitational action contains a scalar potential that drives the 
theory toward General Relativity at low energies.

The impact of induced gravity on HDE was examined in~\cite{Sun:2007rh} by 
considering various choices for the infrared cutoff, including the Hubble 
radius, particle horizon and event horizon. It was found that only the future 
event horizon leads to a viable cosmological evolution consistent with 
late-time 
acceleration, reinforcing the special role played by this cutoff in holographic 
dark energy models.

\paragraph{Minimal supergravity}

Supergravity provides a unified framework combining supersymmetry with General 
Relativity. The minimal four-dimensional supergravity theory was constructed 
in~\cite{Freedman:1976xh}, and its connection to HDE was explored 
in~\cite{Landim:2015hqa, Hatkar:2023hjm}. 
In this setting, a model of HDE interacting with dark matter was formulated in 
terms of a single chiral superfield.

Since extended supergravity theories arise as low-energy limits of string 
theory, the incorporation of HDE within minimal supergravity suggests that the 
holographic dark energy paradigm may have deeper roots in fundamental quantum 
gravity. This connection further supports the idea that holography, dark energy 
and high-energy physics are intrinsically linked.

\subsection{Reconstruction of effective theories from holographic dark energy}

Despite their conceptual differences, scalar-field models of dark energy and 
modified theories of gravity can be reconstructed so as to reproduce the 
cosmological dynamics dictated by holographic dark energy. In this sense, HDE 
can be regarded as an effective description that encodes the infrared behavior 
of a more fundamental theory. Additionally, reconstruction techniques provide 
a useful connection 
between phenomenological HDE models and underlying field-theoretic or 
gravitational frameworks.

In what follows, we briefly review representative reconstruction scenarios in 
which the background evolution of HDE is used as an input in order to determine 
the form of scalar-field potentials or modified gravitational actions. We focus 
on the methodology, key results, and physical interpretation, rather than on 
model-specific technicalities.

\subsubsection{Scalar-field dark energy}

Reconstruction of scalar-field dark energy models proceeds by introducing a 
suitable scalar potential such that the resulting cosmological evolution 
matches that of holographic dark 
energy~\cite{Guberina:2005fb,Kim:2005gk,Zhang:2006qu,
Setare:2007eq,Setare:2007jw,
Zhang:2007es,Setare:2008pc,Karami:2009we,Nojiri:2005pu}. 
While the case $C>1$ allows for a straightforward reconstruction using 
canonical scalar fields~\cite{Zhang:2006av}, it is disfavored by observations. 
The phenomenologically relevant regime is instead $C<1$, which corresponds to 
a phantom-like equation of state.

In this case, HDE can be reconstructed by means of a phantom scalar field, 
characterized by a negative kinetic term 
$-\dot\phi^2$~\cite{Caldwell:1999ew,Setare:2007eq}. The corresponding energy 
density and 
pressure read~\cite{Wang:2016och}
\begin{equation}
\rho_\phi=-\frac{\dot\phi^2}{2}+V(\phi), \qquad 
p_\phi=-\frac{\dot\phi^2}{2}-V(\phi),
\end{equation}
where $V(\phi)$ denotes the scalar potential, and the equation-of-state 
parameter 
takes the form
\begin{equation}
w_\phi=\frac{-\dot\phi^2-2V(\phi)}{-\dot\phi^2+2V(\phi)}.
\end{equation}
For a positive potential, one immediately obtains $w_\phi<-1$, in agreement 
with the HDE behavior for $C<1$.

By identifying the scalar-field energy density with that of HDE, one finds 
after some algebra~\cite{Wang:2016och}
\begin{eqnarray}
\label{Phantom}
&&V(\phi)=\frac{M_p^2}{2}\left[2\dot H+3H^2\left(1+\Omega_{\phi}\right)
+\frac{k}{a^2}\right], \\
&&
\dot\phi^2=M_p^2\left[2\dot H+3H^2\left(1-\Omega_{\phi}\right)
+\frac{k}{a^2}\right].
\end{eqnarray}
In order to express the Hubble parameter and its derivative as functions of the 
scalar field, a convenient ansatz is
\begin{equation}
\phi=t, \qquad H=f(t),
\end{equation}
together with the slow-roll equation governing the scalar-field dynamics,
\begin{equation}
-3H\dot\phi+V'(\phi)=0.
\end{equation}
This yields
\begin{equation}
f(\phi)=\frac{V'(\phi)}{3},
\end{equation}
where the prime denotes differentiation with respect to $\phi$. Substituting 
into Eq.~\eqref{Phantom}, one obtains
\begin{eqnarray}
\label{phantom2}
&&V(\phi)=\frac{M_p^2}{2}\left[2f'(\phi)+3f^2(\phi)\left(1+\Omega_{\phi}\right)
+\frac{k}{a^2}\right], \\
&&
1=\dot\phi^2=M_p^2\left[2f'(\phi)+3f^2(\phi)\left(1-\Omega_{\phi}\right)
+\frac{k}{a^2}\right].
\end{eqnarray}
Hence, matching the scalar-field quantities to those of HDE finally 
gives~\cite{Wang:2016och}
\begin{eqnarray}
V(\phi)&=&\frac{1}{2}\left(1-w_{de}\right)\rho_{de}
=\frac{3H^2\Omega_{de}}{16\pi G}
\left[\frac{4}{3}
+\frac{2\sqrt{\Omega_{de}-C^2\Omega_k}}{3C}
+\frac{b^2(1+\Omega_k)}{\Omega_{de}}\right],\\[2mm]
1=\dot\phi^2&=&-\left(1+w_{de}\right)\rho_{de}
=\frac{H^2\Omega_{de}}{4\pi G}
\left[-1+\frac{\sqrt{\Omega_{de}-C^2\Omega_k}}{C}
+\frac{3b^2(1+\Omega_k)}{2\Omega_{de}}\right].
\end{eqnarray}
Finally, combining with Eq.~\eqref{phantom2}, one arrives at the reconstructed 
potential
\begin{equation}
V(\phi)=3M_p^2f^2(\phi)
\left[1+\frac{2}{\left(6M_p^2f^2(\phi)-1\right)
\pm\sqrt{\left(1-6M_p^2f^2(\phi)\right)^2
+24M_p^2f^2(\phi)}}\right].
\end{equation}

It is worth noting that the instabilities typically associated with phantom 
fields may be avoided by considering generalized scalar theories, such as 
Galileon models~\cite{Deffayet:2010qz,Kobayashi:2010cm} or, more generally, 
Horndeski theory~\cite{Horndeski:1974wa}. Reconstruction of HDE within these 
frameworks remains largely unexplored and constitutes an interesting direction 
for future research.

\subsubsection{Modified gravity}

Reconstruction techniques can also be applied to modified theories of gravity, 
where the goal is to determine a gravitational action whose cosmological 
solutions reproduce the HDE background 
evolution~\cite{Wu:2007tn,Setare:2008hm,HamaniDaouda:2011uag,Houndjo:2011fb,
Karami:2011np,Chattopadhyay:2012eu}. As a representative example, we 
follow~\cite{Wu:2007tn} and consider $f(R)$ gravity, although the same strategy 
can be 
generalized to other extensions of General Relativity.

The field equations of $f(R)$ gravity can be written in the form (setting 
$M_p=1$ for simplicity)
\begin{equation}
\label{Gmunu}
G_{\mu\nu}=T_{\mu\nu}^{(\mathrm{curv})}+T_{\mu\nu}^{(\mathrm{m})},
\end{equation}
where the effective curvature energy-momentum tensor and the matter tensor are 
given by
\begin{eqnarray}
T_{\mu\nu}^{(\mathrm{curv})}&=&\frac{1}{f'(R)}
\left\{\frac{g_{\mu\nu}}{2}\left[f(R)-Rf'(R)\right]
+f'(R)^{;\alpha\beta}\right\}
\left(g_{\mu\alpha}g_{\nu\beta}-g_{\mu\nu}g_{\alpha\beta}\right),\\[2mm]
T_{\mu\nu}^{(\mathrm{m})}&=&\frac{\tilde T_{\mu\nu}^{(\mathrm{m})}}{f'(R)},
\end{eqnarray}
with $\tilde T_{\mu\nu}^{(\mathrm{m})}$ the standard matter tensor of General 
Relativity.
The modified Friedmann equations then read
\begin{eqnarray}
H^2+\frac{k}{a^2}&=&\frac{1}{3}
\left[\rho_{\mathrm{curv}}+\frac{\rho_{\mathrm{m}}}{f'(R)}\right],\\[2mm]
2\frac{\ddot 
a}{a}+H^2+\frac{k}{a^2}&=&-\left(p_{\mathrm{curv}}+p_{\mathrm{m}}\right),
\end{eqnarray}
while the total energy-momentum conservation equation remains unchanged. After 
some manipulation, one obtains
\begin{equation}
\dot H=-\frac{1}{2f'(R)}
\left\{
3H_0^2\Omega_m(1+z)^3
+\ddot R f''(R)
+\dot R\left[\dot R f'''(R)-Hf''(R)\right]
\right\}.
\end{equation}
This expression can be recast as a third-order differential equation for $f(R)$,
\begin{equation}
\mathcal{C}_3(z)\frac{d^3f}{dz^3}
+\mathcal{C}_2(z)\frac{d^2f}{dz^2}
+\mathcal{C}_1(z)\frac{df}{dz}
=-3H_0^2\Omega_{m0}(1+z)^3,
\end{equation}
where the coefficients $\mathcal{C}_n$ depend on $\Omega_{de}$ and its 
derivatives. Solving this equation numerically allows one to reconstruct the 
functional form of $f(R)$ that reproduces the HDE cosmological evolution 
(see~\cite{Wu:2007tn} for details).

\subsection{Observational comparison and model selection}
\label{Comp}

Given the large variety of dark energy models proposed in the literature, a 
crucial question concerns their relative performance when confronted with 
observational data. Since different models are characterized by different 
numbers of free parameters, a direct comparison based solely on the minimum 
$\chi^2_{\mathrm{min}}$ is generally inadequate. Models with a larger parameter 
space may achieve a better fit simply due to increased flexibility, without 
necessarily providing a more accurate description of the underlying physics.

To address this issue, several information-theoretic criteria have been 
developed that balance goodness of fit against model complexity. Among the most 
widely used are the Akaike Information Criterion (AIC)~\cite{Akaike:1974vps} 
and 
the Bayesian 
Information Criterion (BIC)~\cite{Schwarz:1978tpv}, defined respectively as
\begin{equation}
\mathrm{AIC}=-2\log \mathcal{L}_{\mathrm{max}}+2k\,, \qquad
\mathrm{BIC}=-2\log \mathcal{L}_{\mathrm{max}}+k\log N\,,
\end{equation}
where $\mathcal{L}_{\mathrm{max}}$ is the maximum likelihood, $k$ denotes the 
number of model parameters, and $N$ is the number of data points. Smaller 
values 
of AIC or BIC indicate a statistically preferred model. In the special case of 
Gaussian errors, one has 
$\chi^2_{\mathrm{min}}=-2\log\mathcal{L}_{\mathrm{max}}$, 
making explicit the role of the parameter penalty. As a result, models with 
additional parameters that do not significantly improve the fit are naturally 
disfavored~\cite{Liddle:2004nh}.
In practice, it is convenient to adopt $\Lambda$CDM as a reference model and 
define the relative quantities
\begin{equation}
\Delta \mathrm{AIC}\equiv \mathrm{AIC}_{\mathrm{model}}-\mathrm{AIC}_{\Lambda 
\mathrm{CDM}}, \qquad
\Delta \mathrm{BIC}\equiv \mathrm{BIC}_{\mathrm{model}}-\mathrm{BIC}_{\Lambda 
\mathrm{CDM}}.
\end{equation}

A complementary approach is provided by Bayesian Evidence (BE), which is 
defined 
as~\cite{Mukherjee:2005wg}
\begin{equation}
\mathrm{BE}=\int 
\mathcal{L}(\mathbf{d}|\theta,M)\, p(\theta|M)\, d\theta\,,
\end{equation}
where $\mathcal{L}(\mathbf{d}|\theta,M)$ is the likelihood of the data 
$\mathbf{d}$ given model $M$ and parameters $\theta$, while $p(\theta|M)$ 
denotes the prior distribution. Since BE corresponds to the likelihood averaged 
over the parameter space, it naturally incorporates both the quality of the fit 
and the size of the parameter space. In contrast to AIC and BIC, larger values 
of BE indicate a preferred model. As before, it is customary to consider the 
difference
\begin{equation}
\Delta \log\mathrm{BE}\equiv 
\log\mathrm{BE}_{\mathrm{model}}-\log\mathrm{BE}_{\Lambda \mathrm{CDM}}.
\end{equation}

\paragraph{Holographic dark energy models.}

A representative comparison among dark energy models inspired by the 
holographic framework was performed in~\cite{Li:2009bn}, focusing on the 
original 
holographic dark energy (HDE), agegraphic dark energy (ADE), and Ricci dark 
energy (RDE). The resulting parameter constraints and Bayesian evidence values 
are summarized in Table~\ref{Tab}. While the original HDE exhibits a reasonable 
fit to the data, ADE and RDE are significantly disfavored, particularly when 
Bayesian evidence is taken into account. 
\begin{table}[ht!]
\centering
\begin{tabular}{||c c c||} 
 \hline
 HDE & ADE & RDE \\ [0.5ex]
 \hline\hline
 $\Omega_{m0}=0.277^{+0.022}_{-0.021}$ & $n=2.807^{+0.087}_{-0.086}$ & 
 $\Omega_{m0}=0.324^{+0.024}_{-0.022}$ \\ [1ex]
 $C=0.818^{+0.113}_{-0.097}$ &  & $\alpha=0.371^{+0.023}_{-0.023}$ \\ [1ex]
 $\chi^2_{\mathrm{min}}=465.912$ & $\chi^2_{\mathrm{min}}=481.694$ & 
 $\chi^2_{\mathrm{min}}=483.130$ \\ [1ex]
 $\Delta\log\mathrm{BE}=-0.86$ & $\Delta\log\mathrm{BE}=-5.17$ & 
 $\Delta\log\mathrm{BE}=-8.14$ \\ [0.8ex]
 \hline
\end{tabular}
\caption{Parameter constraints and Bayesian evidence for selected holographic 
dark energy models~\cite{Li:2009bn}.}
\label{Tab}
\end{table}

\begin{figure}[t]
\begin{center}
\includegraphics[width=8.8cm]{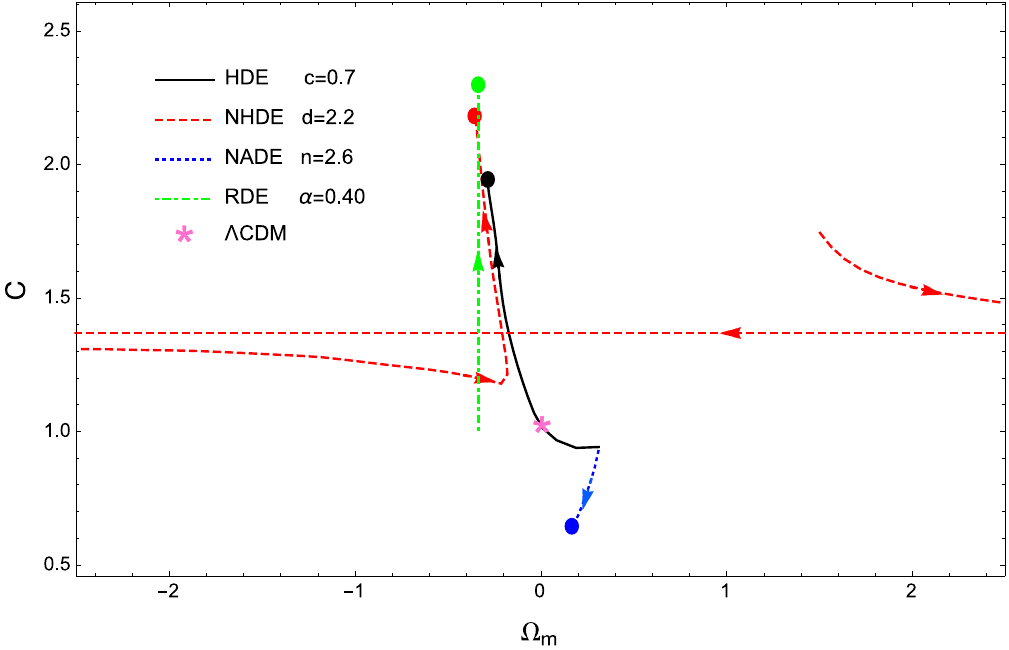}
\caption{{\it{Statefinder diagnostic of   four holographic dark energy models 
(basic holographic dark
energy (HDE), Ricci holographic dark
energy (RDE), 
new holographic dark
energy (NHDE) model, and  new agegraphic dark energy
(NADE))
in 
the $r$-$s$ 
plane, for $\Omega_{m0}=0.27$. Dots and arrows indicate the present value 
$\{r_0,s_0\}$ and the evolution direction for each model, respectively. The 
$\Lambda$CDM model value $(1,0)$ is denoted by a star and displayed for 
comparison. The figure is from~\cite{Cui:2014sma}.}}}
\label{CompAnalys}
\end{center}
\end{figure}

Further comparative studies using statefinder diagnostics were carried out 
in~\cite{Cui:2014sma}, where the evolutionary trajectories in the $r(r,s)$ 
plane 
were shown to efficiently discriminate between HDE, new holographic dark energy 
(NHDE), ADE, and RDE, particularly at low redshifts. Subsequent improvements 
based on the statefinder hierarchy and the fractional growth parameter were 
presented in~\cite{Zhang:2014sqa}, allowing degeneracies among holographic 
models to be broken even at higher redshifts. In Fig. \ref{CompAnalys} we 
depict the statefinder diagnostic of four holographic dark energy models, and 
as we can see, one can indeed distinguish between the various scenarios.

\paragraph{Holographic dark energy versus alternative dark energy scenarios.}

Beyond comparisons within the holographic class, several works have confronted 
holographic dark energy models with alternative cosmic 
scenarios~\cite{Li:2010xjz,Wei:2010wu}. A comprehensive analysis presented 
in~\cite{Xu:2016grp} 
considered ten different dark energy models using current observational data. 
While $\Lambda$CDM remains the statistically preferred description, models such 
as the generalized Chaplygin gas (GCG)~\cite{Bento:2002ps}, $w$CDM, and 
$\alpha$DE~\cite{Dvali:2003rk} exhibit comparable performance. HDE, the new 
generalized Chaplygin gas (NGCG)~\cite{Zhang:2004gc}, and the CPL 
parametrization~\cite{Chevallier:2000qy,Linder:2002et} provide acceptable fits 
but are not 
favored, whereas ADE, DGP, and RDE seem to be disfavored by current 
data.

\section{Holographic dark energy with modified entropies}
\label{HolographicdarkenergyModified}

The holographic dark energy scenario provides a particularly natural setting 
in 
which generalized entropy concepts can be implemented at the cosmological 
level. 
Since the holographic principle itself is intrinsically entropic in origin, any 
modification of horizon entropy immediately translates into a corresponding 
deformation of the holographic energy density. In this approach, entropy does 
not alter the gravitational field equations, as in spacetime thermodynamics, 
but 
instead reshapes the holographic bound that constrains the vacuum energy sector.

As we discussed in Sec.~\ref{Entropydefinitions}, motivated by developments 
in non-extensive statistics, quantum gravity, and 
generalized information measures, a variety of entropy functionals extending 
the 
Bekenstein-Hawking area law have been proposed and systematically explored 
within holographic dark energy models. These constructions lead to a rich 
phenomenology, allowing for departures from the standard $L^{-2}$ scaling of 
the 
dark energy density, while preserving the conceptual foundations of holography. 
In the following subsections we review the most extensively studied 
realizations 
of holographic dark energy based on modified entropies, focusing on their 
defining relations and cosmological implications.

\subsection{Tsallis holographic dark energy}

Tsallis holographic dark energy (Tsallis HDE)~\cite{Saridakis:2018unr} 
constitutes one of the 
earliest 
and
most theoretically motivated extensions of the original holographic dark energy
scenario. Its construction is based on the observation that gravitational
systems are intrinsically non-extensive, due to the long-range nature of the
gravitational interaction, and therefore the standard Boltzmann-Gibbs entropy
may not provide an adequate thermodynamic description at cosmological scales.
Within this perspective, Tsallis non-extensive entropy offers a natural
generalization of horizon thermodynamics, introducing a new dimensionless
parameter $\delta$ that quantifies the degree of non-extensivity.

In the Tsallis framework, the entropy associated with a horizon of area $A$
is given by~(\ref{EETsallisBlackHoleDelta}), namely~\cite{Tsallis:2012js} 
\begin{equation}
S_T = \gamma\, A^\delta ,
\end{equation}
where $\gamma$ is a constant and $\delta=1$ restores the standard
Bekenstein-Hawking entropy. Applying holographic arguments to this generalized
entropy leads to a modified holographic dark energy density of the form
\begin{equation}
\rho_{DE} = B L^{2\delta-4},
\end{equation}
where $L$ denotes the infrared cutoff of the theory and $B$ is a constant with
appropriate dimensions. As in the standard holographic dark energy model, the
choice of the IR cutoff is crucial. In order to preserve consistency with the
original holographic construction and to recover standard HDE as a limiting
case, the infrared cutoff is taken to be the future event horizon
\begin{equation}
R_h = a \int_t^{\infty} \frac{dt'}{a(t')} = a \int_a^{\infty} \frac{da'}{H 
a'^2}.
\end{equation}
With this choice, Tsallis HDE reduces smoothly to standard holographic dark
energy for $\delta=1$, while $\delta=2$ yields an effective cosmological
constant. We mention here that  in~\cite{Tavayef:2018xwx} Tsallis HDE was 
formulated using the Hubble horizon as 
an IR cutoff, however this approach  has the serious 
disadvantage that it 
does not have standard holographic dark energy as a 
sub-case in the limit where Tsallis entropy becomes standard entropy. That is 
why in this subsection we focus on Tsallis HDE using the future event horizon 
as the IR cutoff~\cite{Saridakis:2018unr}.

We consider a spatially flat FRW Universe filled with
pressureless matter and Tsallis holographic dark energy. The Friedmann equations
are
\begin{align}
3 M_p^2 H^2 &= \rho_m + \rho_{DE}, \\
-2 M_p^2 \dot H &= \rho_m + \rho_{DE} + p_{DE},
\end{align}
with $\rho_m$ the matter energy density and $p_{DE}$ the Tsallis HDE pressure.
Introducing the density parameters
\begin{equation}
\Omega_m = \frac{\rho_m}{3M_p^2H^2}, \qquad
\Omega_{DE} = \frac{\rho_{DE}}{3M_p^2H^2},
\end{equation}
and assuming dust matter, one obtains a first-order differential equation that
governs the evolution of $\Omega_{DE}$ as a function of $x=\ln a$. For
$\delta\neq1$ this equation exhibits explicit $x$-dependence and therefore does
not admit an analytic solution, requiring numerical treatment. Importantly, for
$\delta=1$ the equation reduces exactly to that of standard holographic dark
energy.

Combining the above relations, we finally obtain the evolution equation for the
Tsallis holographic dark energy density parameter~\cite{Saridakis:2018unr}
\begin{eqnarray}\label{OdediffeqTsallis}
\frac{\Omega_{DE}'}{\Omega_{DE}(1-\Omega_{DE})}
=2\delta-1+
Q
(1-\Omega_{DE})^{\frac{1-\delta}{2(2-\delta)}} 
(\Omega_{DE})^{\frac{1}{2(2-\delta)}}
e^{\frac{3(1-\delta)}{2(2-\delta)}x},
\end{eqnarray}
where the constant $Q$ is defined as
\begin{equation}\label{Qdef}
Q\equiv 2(2-\delta)\left(\frac{B}{3M_p^2}\right)^{\frac{1}{2(\delta-2)}}
\left(H_0\sqrt{\Omega_{m0}}\right)^{\frac{1-\delta}{\delta-2}} .
\end{equation}
Equation~(\ref{OdediffeqTsallis}) fully determines the cosmological evolution 
of 
Tsallis
holographic dark energy in a spatially flat Universe filled with dust matter, as
a function of the variable $x=\ln a$. It is worth stressing that the structure 
of
this equation crucially depends on the value of the non-extensive parameter
$\delta$.

In the special case $\delta=1$, the explicit dependence on $x$ disappears and
Eq.~(\ref{OdediffeqTsallis}) reduces exactly to the evolution equation of 
standard
holographic dark energy~\cite{Li:2004rb}, namely 
\begin{equation}
\Omega_{DE}'\big|_{\delta=1}
=
\Omega_{DE}(1-\Omega_{DE})
\left(1+2\sqrt{\frac{3M_p^2\Omega_{DE}}{B}}\right),
\end{equation}
where complete equivalence is achieved upon identifying
$B=3c^2M_p^2$. In this case, the evolution equation admits an implicit analytic
solution~\cite{Li:2004rb}, allowing for a fully analytical treatment of the
cosmological dynamics.

By contrast, when $\delta\neq1$, Eq.~(\ref{OdediffeqTsallis}) exhibits an 
explicit
$x$-dependence, rendering an analytic solution unattainable. Consequently, the
cosmological evolution of Tsallis holographic dark energy must be investigated
numerically. This feature reflects the genuinely generalized nature of the
Tsallis entropy framework and leads to a richer phenomenology compared to the
standard holographic dark energy scenario.

The equation-of-state parameter of Tsallis holographic dark energy,
$w_{DE}=p_{DE}/\rho_{DE}$, follows from the conservation equation of the dark
energy sector and can be expressed as~\cite{Saridakis:2018unr} 
\begin{equation}
w_{DE}=\frac{1-2\delta}{3}
-\frac{Q}{3}
(\Omega_{DE})^{\frac{1}{2(2-\delta)}} 
(1-\Omega_{DE})^{\frac{\delta-1}{2(\delta-2)}}
e^{\frac{3(1-\delta)}{2(\delta-2)}x}.
\end{equation} 
This expression reveals the richer phenomenology of Tsallis 
HDE, i.e.
depending on the value of $\delta$, the dark energy component can behave as
quintessence-like, phantom-like, or exhibit the phantom-divide crossing.

The cosmological evolution implied by Tsallis HDE reproduces the standard 
thermal
history of the Universe, namely a matter-dominated era followed by a late-time
dark-energy-dominated phase. The transition from deceleration to acceleration
occurs at redshift $z \sim 0.5$, in agreement with observations. The 
deceleration
parameter
\begin{equation}
q = -1 - \frac{\dot H}{H^2}
= \frac{1}{2} + \frac{3}{2}\left(w_{DE}\Omega_{DE}\right)
\end{equation}
confirms the emergence of late-time acceleration without requiring additional
interactions or modifications of gravity.

\begin{figure}[t]
\includegraphics[scale=0.42]{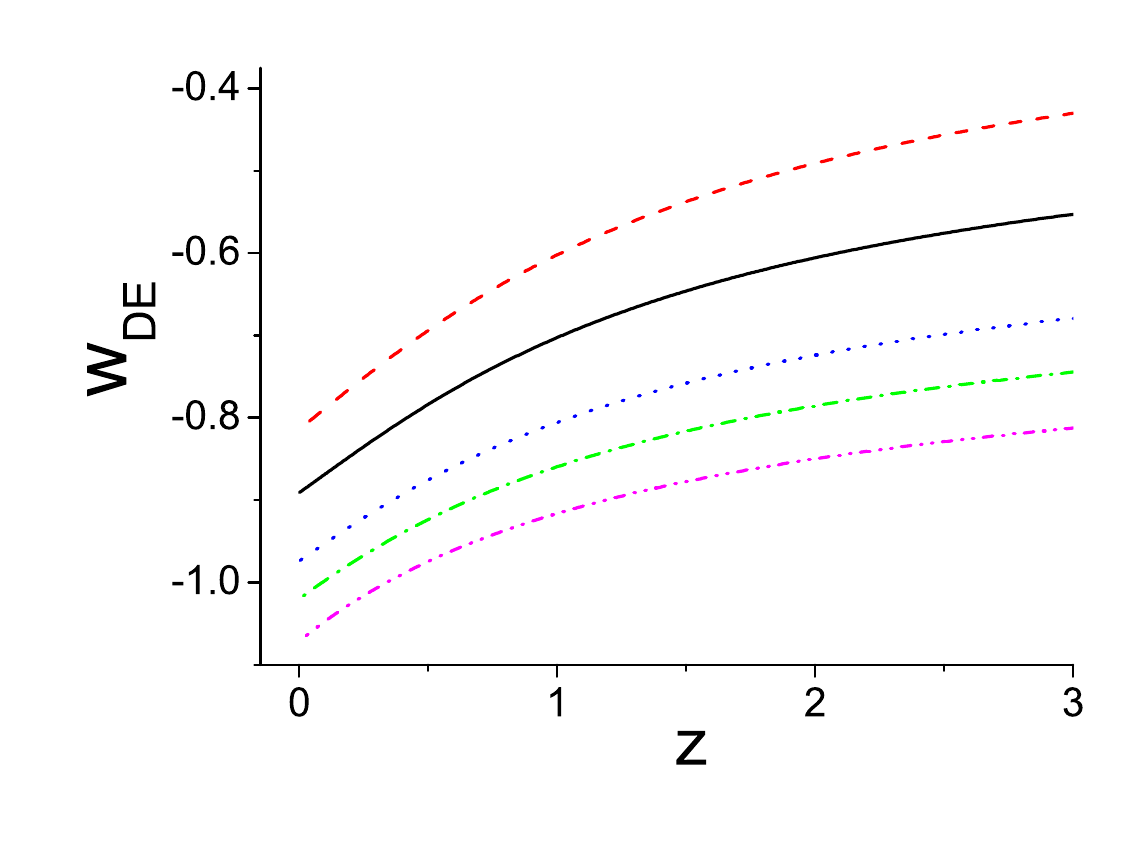}
\caption{
{\it The evolution of the equation-of-state parameter $w_{DE}$ of Tsallis
holographic dark energy as a function of the redshift $z$, for $B=3$ and
$\delta=0.8$ (red-dashed), $\delta=1$ (black-solid), $\delta=1.2$
(blue-dotted), $\delta=1.3$ (green-dashed-dotted), and $\delta=1.4$
(magenta-dashed-dot-dotted). In all cases
$\Omega_{DE}(z=0)\approx0.7$ is imposed. The figure is 
from~\cite{Saridakis:2018unr}. }}
\label{wzplot}
\end{figure}

In Fig.~\ref{wzplot} we present the evolution of the equation-of-state parameter
$w_{DE}(z)$ for several values of the Tsallis exponent $\delta$. Increasing
$\delta$ drives $w_{DE}$ toward more negative values, allowing the model to
enter the phantom regime while remaining consistent with observational bounds.
Notably, standard holographic dark energy is exactly recovered for $\delta=1$.

\begin{figure*}[t]
\includegraphics[width=0.58\textwidth]{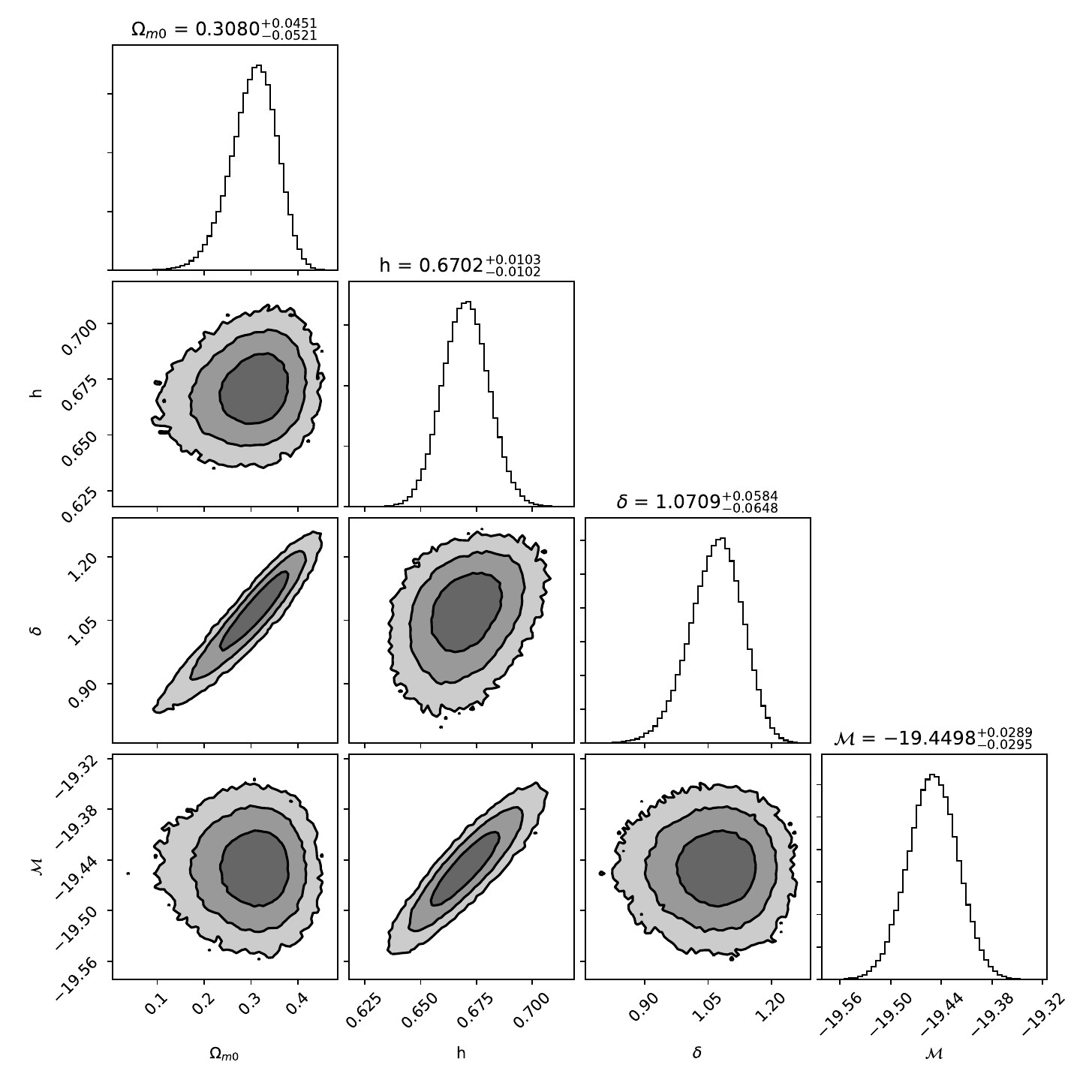}
\caption{
{\it The $1\sigma$, $2\sigma$, and $3\sigma$ confidence contours for Tsallis
holographic dark energy obtained from SNIa and $H(z)$ data, together with the
corresponding one-dimensional marginalized posterior distributions. The figure 
is from~\cite{Saridakis:2018unr}.}}
\label{SNdata}
\end{figure*}

Beyond background evolution, Tsallis HDE has been confronted with observational
data using Type Ia supernovae and cosmic chronometer $H(z)$ measurements. The
results demonstrate very good agreement with observations, yielding matter
density and Hubble parameter values consistent with Planck constraints. The
corresponding marginalized confidence regions are shown in
Fig.~\ref{SNdata}. As we observe, the non-extensive parameter $\delta$ 
exhibits a mild
preference for values close to, but not exactly equal to, unity, while the
standard holographic limit $\delta=1$ remains well within the allowed
confidence intervals (see also \cite{Cruz:2026zzp} on theoretical constraints 
arising from consistency with the generalized second law).

In summary, Tsallis holographic dark energy provides a theoretically 
well-motivated
and phenomenologically rich extension of the standard holographic dark energy
scenario. It preserves holographic consistency, it naturally incorporates
non-extensive thermodynamics, and it offers enhanced flexibility in describing 
the
late-time acceleration of the Universe without sacrificing observational
viability~\cite{Saridakis:2018unr,Tavayef:2018xwx,Zadeh:2018poj,
Ghaffari:2018wks,
Ghaffari:2018rzs,Jawad:2018juh,Aly:2019otq,AbdollahiZadeh:2019lsx,
Sharif:2019seo,Jawad:2019doj,Sharma:2019mtn,Sadri:2019qxt,Aly:2019wtr,
Huang:2019hex,Dubey:2019kzh,Dixit:2019nfl,
Dubey:2019wyz,Ghaffari:2019qcv,Korunur:2019rhg,Maity:2019qbv,Aditya:2019bbk, 
Zhang:2019zxv,Iqbal:2019ooy,Waheed:2020cxw,ChandraDubey:2020tng,Aly:2020uli, 
Mamon:2020wnh,Saha:2020vxn,Sharma:2020lmm,AlMamon:2020usb,Jawad:2020jtq, 
Ens:2020bxh,Yadav:2020wsd,daSilva:2020bdc,Bhattacharjee:2020ixg, 
VijayaSanthi:2020feh,Mohammadi:2021wde,Zubair:2021yrq,Shekh:2021bgh,
Jawad:2021xsr,Pradhan:2021crw,Liu:2021heo,Zubair:2021gve,Varshney:2021xvg,
Dheepika:2021fqv,Pankaj:2021nkg,Pandey:2021fvr,Varshney:2021rbq,
Sobhanbabu:2021vzw,Koussour:2022sdy,Saleem:2022eti,Sadeghi:2022fow,
Astashenok:2022lsf,Pandey:2022rtu,Das:2022igz,Ali:2022twr,Zubair:2022vnd,
Sadeghi:2022mrm,FeiziMangoudehi:2022rwj,AmetMemet:2022hjd,Astashenok:2023jfp,
Shaikh:2023gvc,Mahanta:2023qvp,Sharma:2023asq,Sharma:2023toq,
Chokyi:2024xff,Srivastava:2020sbp,
Trivedi:2024rhp,Sharma:2024ywq,Dhore:2024gfv,Astashenok:2024tdg,Aktas:2024rgc, 
Sultana:2024che,Sharif:2024fli,Astashenok:2024jje,Baziar:2024qbt,
Mahanta:2024iel,Ualikhanova:2024xxe,Luciano:2025elo,Pradhan:2025rir,
Astashenok:2025ktx, 
Yarahmadi:2025fml,Yarahmadi:2025luc,Das:2026wry,Ghosh:2026rll,
Abdullghani:2026iju,Siquieri:2026dby,Luciano:2026vhm,Neelima:2026mmb,
Sadeghnezhad:2025zxa,Mukherjee:2026jmz,Sultana:2026kup,Satyanarayana:2026ead,
Bolotin:2026hot,Zafar:2026rnu,Borah:2026gyw,Chandra:2026hsh,Astashenok:2026ayf,
Addo:2026icc}.

\subsection{R\'{e}nyi holographic dark energy}

Let us now turn to a particularly interesting realization of holographic dark 
energy that is based on R\'{e}nyi entropy. As discussed earlier, generalized 
entropies arise naturally in systems characterized by long-range interactions 
and correlations, with gravity being a prime example. In this context, 
R\'{e}nyi 
entropy occupies a distinguished position, since it can be viewed as an 
additive 
entropy constructed from the non-additive Tsallis form, while still encoding 
non-trivial deviations from the standard Boltzmann-Gibbs framework.

More specifically, R\'{e}nyi entropy can be expressed as a logarithmic function 
of the Tsallis entropy, and when the latter is identified with the 
Bekenstein-Hawking area law, one arrives at a modified horizon entropy of the 
form~(\ref{EERenyiTsallisRelation})~\cite{Renyi:1961EEE,Czinner:2015eyk,
Tannukij:2020njz,Promsiri:2020jga,
Samart:2020klx}, which can be written as
\begin{equation}
S_R=\frac{1}{\delta}\ln\!\left(1+\frac{\delta}{4}A\right),
\end{equation}
where $\delta$ parametrizes the deviation from the standard extensive case. 
This 
construction preserves additivity at the level of the total entropy, while 
simultaneously introducing corrections that become relevant whenever the 
horizon 
area is sufficiently small.

Implementing the holographic principle with this entropy input leads to a 
modified dark-energy density, which in a flat FRW 
Universe takes the characteristic form~\cite{Komatsu:2016vof,Moradpour:2018ivi}
\begin{equation}
\rho_{\rm DE}=\frac{3C^2 H^2}{8\pi\left(1+\frac{\delta\pi}{H^2}\right)},
\end{equation}
with $C^2$ a dimensionless constant. 
At this point, the physical content of the 
construction becomes transparent. In the limit $\delta\rightarrow0$, the 
correction term disappears and one recovers the standard holographic dark 
energy 
density $\rho_{\rm DE}\propto H^2$. Hence, R\'{e}nyi holographic dark energy 
  represents a smooth and well-controlled deformation of the original 
holographic scenario. Finally, the Friedmann equation 
becomes~\cite{Komatsu:2016vof,Moradpour:2018ivi}
\begin{eqnarray} 
\frac{H^2(z)}{H_0^2}=\Omega_m(1+z)^3+\frac{(1+\frac{\delta\pi}{H_0^2}
)(1-\Omega_m) } { 1+\frac { \delta\pi}{H^2}}\frac{H^2(z)}{H_0^2}.
\end{eqnarray}

An important qualitative feature of  R\'{e}nyi holographic dark energy  is that 
the modification effectively 
introduces a dynamical suppression factor, governed by the ratio 
$\delta\pi/H^2$. At early times, when $H$ is large, the correction is 
negligible 
and the model behaves similarly to ordinary holographic dark energy. At late 
times, however, as the Hubble rate decreases, the denominator deviates from 
unity and the dark-energy density acquires a non-trivial evolution, which can 
naturally drive the Universe toward accelerated expansion.

Assuming standard energy-momentum conservation for the dark-energy sector, the 
corresponding pressure can be derived consistently, allowing one to determine 
the effective equation-of-state parameter. One then finds that  R\'{e}nyi 
holographic dark energy  can 
interpolate between a matter-like behavior at high redshifts and an asymptotic 
de Sitter phase in the far future, with the possibility of crossing the phantom 
divide depending on the value of the parameter $\delta$. In this sense, the 
R\'{e}nyi framework provides additional freedom compared to ordinary 
holographic 
dark energy, without the need to introduce extra degrees of freedom or explicit 
interactions.

From a phenomenological perspective,  R\'{e}nyi holographic dark energy  
exhibits several appealing properties. 
The transition redshift from deceleration to acceleration can be brought into 
agreement with observationally inferred values, and, notably, the model has 
been 
shown to possess an improved stability behavior compared to both ordinary 
holographic dark energy and other entropy-corrected scenarios. In particular, 
the squared sound speed can remain positive over extended cosmological epochs, 
indicating the absence of classical instabilities even during matter-dominated 
phases.

In summary, R\'{e}nyi holographic dark energy constitutes a theoretically 
well-motivated and phenomenologically viable extension of the holographic 
paradigm. By modifying only the entropy input while leaving the holographic 
logic intact, it provides a minimal yet powerful framework capable of enriching 
the cosmological dynamics and offering a smoother and more stable evolution 
than 
its standard 
counterpart~\cite{Komatsu:2016vof,Moradpour:2018ivi,Jawad:2018juh,Maity:2019qbv,
Iqbal:2019ooy,Sharma:2020glf,Dubey:2020ckn,Bhattacharjee:2020rqk,Sharma:2020rvx,
Ali:2020gfx,Prasanthi:2021ihf,Saha:2021ngs,Dubey:2021lmm,Shekh:2021bgh,
Sardar:2021eaj,Bhardwaj:2021nzd,Lou:2021gwk,Saha:2021mlz,Bharali:2022kbl,
Zhang:2022wco,Saha:2022oph,VijayaSanthi:2022hef,Shekh:2022ykf,Manoharan:2022qll,
Alam:2022oae,Shaikh:2023gvc,Wankhade:2023ufc,Nakarachinda:2023jko,
VijayaSanthi:2024qgy,Bharali:2024tqs,Abdelrashied:2025hsi,Das:2025nds,
Prasanthi:2025xif, 
Khan:2026xqf,Prasanthi:2026byc,Abdullghani:2026iju,Prasanthi:2026xgv,
Tamri:2026pxo,Huang:2026zuv,Abdelrashied:2026lco,JnanaPrasuna:2026ndi}.

\subsection{Sharma-Mittal holographic dark energy}

We now turn to the construction of holographic dark energy based on the
Sharma-Mittal entropy, which constitutes a two-parameter generalization
encompassing both the Tsallis and R\'enyi entropic formalisms as limiting cases.
This framework allows for a controlled interpolation between different
non-extensive entropy prescriptions, and thus provides a natural extension of
entropy-based holographic dark energy models.

The Sharma-Mittal entropy associated with a horizon of area $A$ is given by 
Eq.~(\ref{EESharmaMittalEntropyRenyiTsallisRelation}), namely
\begin{equation}
S_{\rm 
SM}=\frac{1}{R}\left[\left(1+\frac{\delta}{4}A\right)^{\frac{R}{\delta}}-1\right
],
\label{SM_entropy}
\end{equation}
where $\delta$ and $R$ are free parameters characterizing the degree of
non-extensivity. As we discussed in subsection~\ref{entropymap}, the limits 
$R\rightarrow\delta$ and
$R\rightarrow0$ recover the Tsallis and R\'enyi entropies respectively.

Invoking the holographic principle, which relates the ultraviolet and infrared
cutoffs through the entropy bound, the corresponding dark energy density takes
the form
$
\rho_{D}=\frac{3c^{2}}{8\pi L^{4}}\,S_{\rm SM},
$
where $c^{2}$ is a dimensionless parameter encoding uncertainties of the
effective theory. Employing the Hubble horizon as the infrared cutoff, one
obtains the explicit Sharma-Mittal holographic dark energy density
\begin{equation}
\rho_{D}=\frac{3c^{2}H^{4}}{8\pi R}
\left[\left(1+\frac{\pi\delta}{H^{2}}\right)^{\frac{R}{\delta}}-1\right].
\label{SMHDE_density}
\end{equation}
Hence, by considering a spatially flat FRW Universe 
filled with
pressureless matter and Sharma-Mittal holographic dark energy, the first
Friedmann equation reads
$
H^{2}=\frac{8\pi G}{3}\left(\rho_{m}+\rho_{D}\right)
$, where $\rho_{D}$ will be given by~(\ref{SMHDE_density}).

Differentiating Eq.~(\ref{SMHDE_density}) with respect to cosmic time and
combining with the Friedmann equation, one finally finds the  evolution 
equation for the Sharma-Mittal holographic dark energy, 
namely~\cite{Jawad:2018juh,Upadhyay:2021atf}
\begin{align}
\Omega_{D}'=&-
\Bigg[
3(\Omega_{D}-1)
-\frac{c^{2}\pi 
H^{2}}{\left(1+\frac{\pi\delta}{H^{2}}\right)^{\frac{R}{\delta}}}
+\pi\delta\Omega_{D}+H^{2}\Omega_{D}
\Bigg] \times
\Bigg[
2\pi\delta\Omega_{D}
-\frac{\pi 
c^{2}H^{2}}{\left(1+\frac{\pi\delta}{H^{2}}\right)^{\frac{R}{\delta}}}
-\pi\delta+2H^{2}\Omega_{D}-H^{2}
\Bigg]^{-1},
\label{SMHDE_Omega}
\end{align}
which fully determines the cosmological dynamics of the
Sharma-Mittal holographic dark energy scenario. Additionally,  the resulting 
equation-of-state parameter    acquires the explicit form
\begin{equation}
w_{D}=
\Omega_{D}^{-1}\left[
1-(\Omega_{D}-1)\,
\frac{\pi\delta+H^{2}}
{\pi c^{2}H^{2}\left(1+\frac{\pi\delta}{H^{2}}\right)^{\frac{R}{\delta}}
-(2\Omega_{D}-1)(\pi\delta+H^{2})}
\right],
\label{SMHDE_EoS}
\end{equation}
revealing directly how the non-extensive parameters $R$ and $\delta$ deform the
effective dark energy behavior. The Sharma-Mittal holographic dark energy leads 
to 
interesting cosmological phenomenology and has been studied in detail 
\cite{Jawad:2018juh,Maity:2019qbv,Iqbal:2019ooy,Dubey:2020vho,Shekh:2021bgh,
Upadhyay:2021atf,Sardar:2021eaj,Upadhyay:2022jwa,Shaikh:2023xpc,
Korunur:2023qry,Aditya:2024lia,NooriGashti:2024dvq,Sardar:2024kcw,
NooriGashti:2024tog,Sharif:2025aul,Mukherjee:2026jmz,Abdullghani:2026iju,
Nisar:2026seu,Aditya:2026utq,Rani:2025gvb,Nisar:2026gnr,Kadali:2026lcy}. 

Besides its cosmological applications, the Sharma-Mittal entropy has recently 
attracted renewed interest in gravitational physics, providing a unified 
framework to investigate black-hole thermodynamics, information-theoretic 
aspects of gravity, and infrared modifications of the gravitational 
interaction~\cite{Rani:2024qju,Dashtianeh:2024wgt, Rani:2025mip}. In 
particular, 
it has been shown to satisfy the Bekenstein bound under suitable conditions 
and, 
within Verlinde's entropic gravity, to naturally give rise to MOND-like 
behavior 
for a specific relation between the deformation 
parameters~\cite{Benkrane:2026oos}.

\subsection{Kaniadakis holographic dark energy}

Kaniadakis holographic dark energy~\cite{Drepanou:2021jiv} constitutes a 
natural 
extension of the
standard holographic dark energy scenario, obtained by replacing the
Bekenstein-Hawking entropy with the generalized entropy emerging from
Kaniadakis relativistic statistics.  
Kaniadakis entropy is a one-parameter deformation of Boltzmann-Gibbs entropy,
motivated by relativistic considerations, and is characterized by a
dimensionless parameter $K$ quantifying deviations from standard 
statistics~\cite{Kaniadakis:2002zz,Kaniadakis:2005zk}. When applied to 
black-hole
thermodynamics, it leads to the entropy 
expression~(\ref{EEKaniadakisBekensteinHawkingEntropy}), 
namely~\cite{Moradpour:2020dfm}
\begin{equation}
S_K=\frac{1}{K}\sinh(K S_{BH}),
\end{equation}
which reduces smoothly to the Bekenstein-Hawking entropy in the limit
$K\rightarrow0$. Since any realistic modification must remain close to the
standard area law, one expects $|K|\ll1$, and hence an expansion yields
\begin{equation}
S_K=S_{BH}+\frac{K^2}{6}S_{BH}^3+\mathcal{O}(K^4),
\end{equation}
where the second term represents the leading Kaniadakis correction.

Inserting this entropy into the holographic bound and saturating it leads to
the energy density of Kaniadakis holographic dark energy~\cite{Drepanou:2021jiv}
\begin{equation}
\label{KHDEreview}
\rho_{DE}=3c^2 M_p^2 L^{-2}+K^2 M_p^6 L^{2},
\end{equation}
with $c$ a constant,
which explicitly contains the standard holographic contribution plus a new
term induced by the generalized entropy. 
The limit $K\rightarrow0$ restores
exactly the usual 
holographic dark energy, ensuring the internal consistency
of the construction.
Adopting a spatially flat FRW Universe filled with
pressureless matter and Kaniadakis holographic dark energy, the Friedmann
equations read
\begin{eqnarray}
\label{Fr1Kaniad}
3M_p^2 H^2 &=& \rho_m+\rho_{DE},\\
-2M_p^2\dot H &=& \rho_m+p_m+\rho_{DE}+p_{DE},
\label{Fr2Kaniad}
\end{eqnarray}
supplemented by the standard matter conservation equation
$\dot\rho_m+3H\rho_m=0$. Introducing the density parameters
$\Omega_m=\rho_m/(3M_p^2H^2)$ and $\Omega_{DE}=\rho_{DE}/(3M_p^2H^2)$, one finds
$\Omega_m+\Omega_{DE}=1$.

Using the defining relation of the infrared cutoff and working with
$x\equiv\ln a$ as the independent variable, one obtains a first-order
differential equation governing the evolution of the dark energy density
parameter, namely
\begin{equation}
\Omega_{DE}'=\Omega_{DE}(1-\Omega_{DE})
\left\{
3-\frac{2(\mathcal{A}-2K^2M_p^4\mathcal{B})}{\mathcal{A}}
\left[
1-\sqrt{3}\left(\frac{\Omega_{DE}}{\mathcal{A}\mathcal{B}}\right)^{1/2}
\right]
\right\},
\end{equation}
where  
\begin{eqnarray}\nonumber
\!\!\!\!\!\!\!\!\!\!\!
\mathcal{A}
&=&
\frac{3e^{-3x}H^{2}_{0}\Omega_{m0}\Omega_{DE}}{1-\Omega_{DE}}, 
\\ 
\nonumber
\!\!\!\!\!\!\!\!\!\!\!\!\!\!\!\!\!\!\!\!
\mathcal{B}&=&\frac{\mathcal{A}-\sqrt{\mathcal{A}^{2}-12c^{2}K^{2}M^{4}_{p}}}{
2K^{2}M^{4}_{p}}.
\end{eqnarray}
  This
equation fully determines the background evolution of Kaniadakis holographic
dark energy. Importantly, in the limit $K\rightarrow0$ it reduces exactly to
the corresponding evolution equation of standard holographic dark 
energy~\cite{Li:2004rb}, which admits an analytic implicit solution. Note that 
for 
$K\neq0$ the
equation must be solved numerically.

The equation-of-state parameter of Kaniadakis holographic dark energy follows
from the conservation of the dark energy sector,
$
\dot\rho_{DE}+3H\rho_{DE}(1+w_{DE})=0,
$
leading to    
\begin{equation}\label{wDEKaniad}
w_{DE}=-1-2\left (\frac{\Omega_{DE}}{3\mathcal{A}^{3}} \right 
)^{\frac{1}{2}}\left 
(\frac{-3c^{2}+K^{2}M^{4}_{p}\mathcal{B}^{2}}{\mathcal{B}^{\frac{3}{2}}}\right 
)\left [-1+\frac{\sqrt{3}}{3}\left (\frac{\mathcal{A}\mathcal{B}}{\Omega_{DE}} 
\right )^{\frac{1}{2}}\right ],
\end{equation}
and thus it is straightforwardly obtained once $\Omega_{DE}(x)$ is known.
As expected,
the standard holographic result
$w_{DE}=-\frac{1}{3}-\frac{2}{3}\frac{\sqrt{\Omega_{DE}}}{c}$
is recovered when $K\rightarrow0$. For non-vanishing $K$, however, the
equation-of-state parameter can exhibit a much richer behavior, allowing for
quintessence-like evolution, phantom behavior, or phantom-divide crossing,
depending on the values of $c$ and $K$.

\begin{figure}[t]
    \centering
    \includegraphics[width=0.5\textwidth]{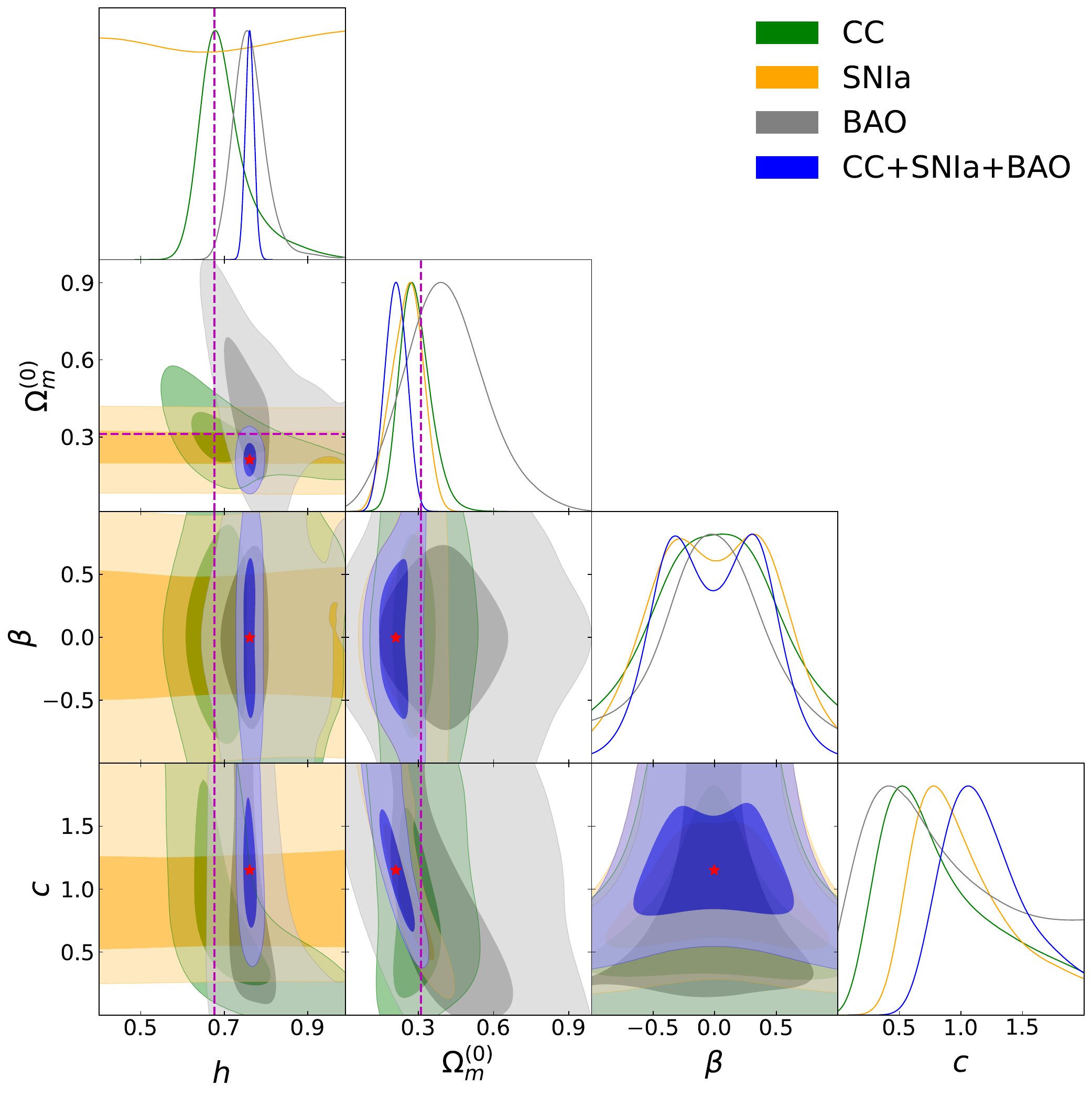}
    \caption{{\it{Observational constraints on  Kaniadakis-holographic dark 
energy, alongside the corresponding 1D posterior distribution of the free 
parameters, where  $\beta=\frac{K M_p^2}{H_0^2}$ and 
$h=H_0/(100\text{Km/s/Mpc})$. The stars 
correspond to the mean values  and the dashed lines 
denote the best-fit values for $\Lambda$CDM cosmology~\citep{Planck:2018vyg}.  
The figure is from~\cite{Hernandez-Almada:2021aiw}. }}}
 \label{figKHDE}
\end{figure}

 Numerical
studies show that Kaniadakis holographic dark energy can reproduce the standard
thermal history of the Universe, including the transition from deceleration to
acceleration at redshifts compatible with observations, while asymptotically
driving the Universe toward a dark-energy-dominated phase. Additionally, 
it is in agreement with cosmic chronometers, SN Ia, and  
BAO data, as can be seen in Fig.~\ref{figKHDE}.

In summary, Kaniadakis holographic dark energy provides a theoretically 
consistent
and phenomenologically rich extension of the standard holographic dark energy
scenario, rooted in relativistic generalized statistics and admitting the
latter as a well-defined limiting case \cite{Drepanou:2021jiv,
Hernandez-Almada:2021aiw,Jawad:2021xsr,Sharma:2021zjx,
Ghaffari:2021xja,Korunur:2022ifb,Sadeghi:2022fow,Rani:2022upi,Dubey:2022hkh,
P:2022amn,Singh:2022ubm,Sania:2023fjx,Jawad:2023aog,Chokyi:2024xff,Rao:2024yrp,
Prasanthi:2024cly,Fang:2024yni,Murali:2024vos,Luciano:2025ykr,Kapil:2025pyt,
Li:2025ivi,Singh:2026qfj, Aditya:2025ehz,Murali:2025qqo,
Prasanthan:2024fwg, 
Sultana:2024fvb,Gonzalez-Espinoza:2026iyh,Bhupinder:2026rdt,Kalkan:2026waj,
Kapil:2026hcd,Aditya:2026elg,Malakar:2026gyg,Asvar:2026qgg}.

\subsection{Barrow holographic dark energy}

Barrow holographic dark energy (BHDE)~\cite{Saridakis:2020zol} is obtained by 
combining the usual
holographic reasoning with the Barrow entropy proposal, according to which
quantum-gravitational effects may induce a fractal deformation of the horizon
surface and hence a deformation of the entropy-area 
relation~\cite{Barrow:2020tzx}. In such a framework, the Bekenstein-Hawking 
entropy is
replaced by a generalized entropy that departs from strict area scaling, with
the deformation quantified by a new parameter $\Delta$ (with $\Delta=0$
recovering the standard case). Since holographic dark energy crucially relies
on the entropy associated with a horizon, this deformation immediately
propagates into the effective dark-energy density, offering a minimal but
non-trivial extension of the standard holographic dark energy scenario.

Starting from the holographic bound $\rho_{DE}L^4\leq S$ and employing Barrow
entropy~(\ref{EEBarrowEntropyClean}), namely~\cite{Barrow:2020tzx}
\begin{equation}
S_{B}=\left (\frac{A}{A_{0}} \right )^{1+\Delta /2},
\end{equation}
  one obtains the BHDE density~\cite{Saridakis:2020zol}
\begin{equation}
\label{rhoBHDE_gen}
\rho_{DE}=C\,L^{\Delta-2},
\end{equation}
where $C$ is a constant with dimensions $[L]^{-2-\Delta}$.
In the limit $\Delta=0$, Eq.~(\ref{rhoBHDE_gen}) reduces to standard
holographic dark energy $\rho_{DE}=3c^2M_p^2L^{-2}$ upon identifying
$C=3c^2M_p^2$.

\subsubsection{Constant Barrow exponent}

In what follows we consider a spatially flat FRW Universe filled with dust
matter ($p_m=0$) and BHDE. The Friedmann equations can be written in the 
standard form~(\ref{Fr1Kaniad}),~(\ref{Fr2Kaniad}), with $\rho_{DE}$ given 
by~(\ref{rhoBHDE_gen}), while  the matter conservation equation for the dust 
case gives $\rho_m=\rho_{m0}a^{-3}$. 
Employing BHDE,  and expressing the horizon
length in terms of $\Omega_{DE}$, one arrives at an evolution equation for
$\Omega_{DE}(x)$ with $x\equiv\ln a$. The resulting differential equation can be
cast into the form~\cite{Saridakis:2020zol}
\begin{eqnarray}
\label{Odediffeq_BHDE}
\frac{\Omega_{DE}'}{\Omega_{DE}(1-\Omega_{DE})}
&=&\Delta+1
+Q\,(1-\Omega_{DE})^{\frac{\Delta}{2(\Delta-2)}}
(\Omega_{DE})^{\frac{1}{2-\Delta}}
e^{\frac{3\Delta}{2(\Delta-2)}x},
\end{eqnarray}
where primes denote derivatives with respect to $x$ and the dimensionless
constant $Q$ is
\begin{equation}
\label{Qdef_BHDE}
Q\equiv (2-\Delta)\left(\frac{C}{3M_p^2}\right)^{\frac{1}{\Delta-2}}
\left(H_0\sqrt{\Omega_{m0}}\right)^{\frac{\Delta}{2-\Delta}}.
\end{equation}
Equation~(\ref{Odediffeq_BHDE}) determines the cosmological evolution of BHDE in
a flat Universe with dust matter. In the standard limit $\Delta=0$ the explicit
$x$-dependence disappears and one recovers the usual holographic dark energy
equation
\begin{equation}
\Omega_{DE}'\big|_{\Delta=0}=
\Omega_{DE}(1-\Omega_{DE})
\left(1+2\sqrt{\frac{3M_p^2\Omega_{DE}}{C}}\right),
\end{equation}
which admits an analytic implicit solution~\cite{Li:2004rb}. For $\Delta\neq0$,
Eq.~(\ref{Odediffeq_BHDE}) must be treated numerically.

Having determined $\Omega_{DE}(x)$, the dark-energy equation-of-state parameter
$w_{DE}\equiv p_{DE}/\rho_{DE}$ follows from the dark-energy conservation
equation
$\dot\rho_{DE}+3H\rho_{DE}(1+w_{DE})=0$, yielding
\begin{equation}
\label{wDE_BHDE}
w_{DE}=-\frac{1+\Delta}{3}
-\frac{Q}{3}\,
(\Omega_{DE})^{\frac{1}{2-\Delta}}
(1-\Omega_{DE})^{\frac{\Delta}{2(\Delta-2)}}
e^{\frac{3\Delta}{2(2-\Delta)}x}.
\end{equation}
Therefore, the expansion history is fully specified once
Eq.~(\ref{Odediffeq_BHDE}) is solved. In the standard limit $\Delta=0$,
Eq.~(\ref{wDE_BHDE}) reduces to the well-known expression of standard
holographic dark energy~\cite{Wang:2016och},
\begin{equation}
w_{DE}\big|_{\Delta=0}=
-\frac{1}{3}-\frac{2}{3}\sqrt{\frac{3M_p^2\Omega_{DE}}{C}}.
\end{equation}

\begin{figure}[t]
\centering\includegraphics[width=0.57\textwidth]{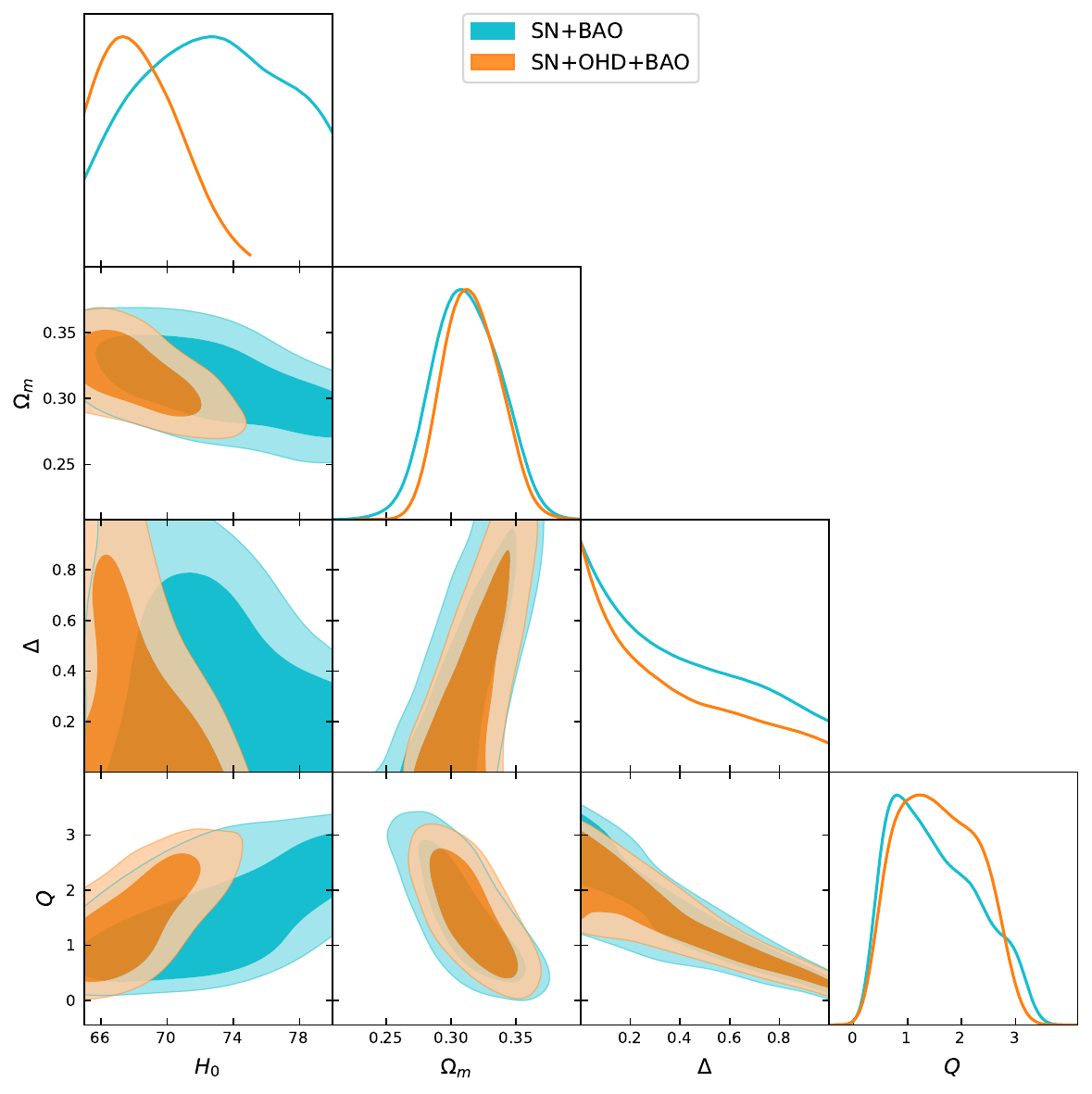}
\caption{\it{Observational constraints on  Barrow  
holographic dark energy, for the datasets     SN+BAO and  SN+OHD+BAO. The 
figure is from~\cite{Luciano:2025elo}. }}
\label{datafigBHDE}
\end{figure}

The BHDE scenario exhibits richer phenomenology than the standard holographic
model. The new parameter $\Delta$ controls the departure from the area-law
entropy and can significantly affect the evolution of $w_{DE}$: depending on
parameter values and initial conditions, BHDE can behave as quintessence-like
dark energy, can enter the phantom regime, or can allow for phantom-divide
crossing during the evolution. This flexibility is achieved while maintaining a
smooth and well-defined $\Delta\to0$ limit, which recovers standard holographic
dark energy. Additionally, BHDE is in agreement with  observational data from 
Supernova Type Ia (SNIa), Cosmic 
Chronometers (CC) and Baryon Acoustic oscillations (BAO), including the 
recently released DESI DR2 dataset, as can be seen in Fig.~\ref{datafigBHDE}.

More recently, the BHDE framework has been further tested by allowing the dark
matter sector to possess a small but non-vanishing pressure
\cite{Mamon:2026tnm}. Using DESI DR2 BAO measurements in combination with
Cosmic Chronometers and different SNIa compilations, it was found that the
pressureless dark-matter limit remains compatible with the data, while the
Barrow parameter $\Delta$ is only weakly constrained due to its degeneracy with
the dark-matter equation-of-state parameter. In the allowed region, the dark
energy equation of state remains above the phantom divide, and Bayesian model
comparison shows no significant statistical preference between BHDE and
$\Lambda$CDM.
Consequently, BHDE might provide a theoretically motivated and
phenomenologically versatile extension of holographic dark energy, suitable for
further confrontation with observations and for exploring the imprint of
quantum-gravitational horizon deformations at cosmological 
scales~\cite{Saridakis:2020zol,Anagnostopoulos:2020ctz,Mamon:2020spa,
Srivastava:2020cyk,Chakraborty:2021uzp,Adhikary:2021xym,Rani:2021hvh,
Huang:2021zgj,Nandhida:2021vxl,Nojiri:2021jxf,Maity:2022gdy,Zhao:2022bxw,
Paul:2022doh,Kumar:2022acs,Saleem:2022eti,Remya:2022frs,Koussour:2022rsv,
Srivastava:2022nex,Oliveros:2022biu,Pradhan:2022jjz,Luciano:2022viz,
Jawad:2022qab,Sharma:2022poz,Ghaffari:2022skp,Chanda:2022tpk,
Luciano:2022ffn, Luciano:2022hhy,Sheykhi:2022fus,Sharma:2022dzc,Luciano:2023wtx,
Boulkaboul:2023yks,A:2023qvh,Feng:2023cbl,Salehi:2023byk,AlMamon:2023zek,
Basilakos:2023seo,Pankaj:2023bkj,Huang:2024xqk,Mukhopadhyay:2024vcq,
Altaibayeva:2024cyc,Ghosh:2024jvs,Motaghi:2024rag,Yarahmadi:2024oqv,
Mahmoudifard:2024gmn,Yarahmadi:2024afr,Myrzakulov:2024jvg,Adhikary:2024sax,
Mahanta:2024xyj,Ens:2024zzs,Singha:2025trb,Luciano:2025elo,Guin:2025xki,
Bekova:2025qco,Yarahmadi:2025fml,Kotal:2025fqh,Luciano:2026vhm,Aditya:2026jmu,
Mahanta:2026fmq,Mahanta:2026xkt,
Prasanthi:2026dhd,Ajmal:2026fhf,Ibrar:2025hah,Dubey:2025nuf,Khapekar:2026zkc,
Maity:2025cxd,Yarahmadi:2025ujq,Bhardwaj:2021chg,Dixit:2021phd,Sarkar:2021izd,
Kumar:2026nod,Sudharani:2026zte,Mohanty:2026laq,Shahkarami:2026jub,
Kotal:2026lrd,Mahanta:2026gpl,Mamon:2026tnm,Tavakoli:2026hno,Aditya:2026fld,
Azarakhsh:2026wxj}.

\subsubsection{Varying Barrow exponent}
\label{Varyingentexp}

More generally, the BHDE framework has been extended by promoting the Barrow
exponent to a dynamical quantity. This possibility is physically motivated by
the expectation that quantum-gravitational deformations of the horizon should
be more relevant at early times and progressively fade during the cosmic
expansion, as also discussed in Sec. \ref{BarRunCos}.

The possibility of a running Barrow exponent was first advocated in
Ref.~\cite{DiGennaro:2022ykp}. The authors argued that, if the Barrow entropy
encodes quantum-gravitational corrections, the deformation parameter should
depend on the energy scale, approaching its maximal value in the ultraviolet
and vanishing at low energies. Otherwise, a constant nonzero Barrow exponent
would lead to sizeable deviations from the Bekenstein-Hawking area law even
for macroscopic black holes, where quantum-gravitational effects are expected
to be negligible. Within this framework, the running exponent naturally
induces an effective dynamical dark-energy component in the modified
Friedmann equations, which evolves from a negative value in the early universe
to a small positive value at late times, thus realizing a sign-switching dark
energy scenario with possible implications for the Hubble tension.

In Ref.~\cite{Basilakos:2023seo}, the Barrow exponent is assumed to depend
directly on the redshift, or equivalently on $x=\ln a$. The BHDE density 
preserves the form
\begin{equation}
\label{rhoBHDE_running}
\rho_{DE}=C R_h^{\Delta(x)-2},
\end{equation}
where $R_h$ is the future event horizon. However, the evolution equations now
acquire additional terms proportional to $\Delta'$, where primes denote 
derivatives with respect to $x$. In particular, the
dark-energy density parameter satisfies
\begin{equation}
\label{Odediffeq_BHDE_running}
\frac{\Omega_{DE}'}{\Omega_{DE}(1-\Omega_{DE})}
=
\sqrt{\Omega_{DE}}
\left(\frac{C}{3M_p^2}\right)^{-1/2}
\left[
\frac{P(1-\Omega_{DE})}{\Omega_{DE}}
\right]^{\frac{\Delta}{2(\Delta-2)}}
(2-\Delta)
+\Delta+1
+
\ln\left[
\frac{P(1-\Omega_{DE})}{\Omega_{DE}}
\right]^{\frac{\Delta'}{\Delta-2}},
\end{equation}
where
\begin{equation}
P(x)\equiv
\frac{C e^{3x}}{3M_p^2H_0^2\Omega_{m0}}.
\end{equation}
The corresponding equation-of-state parameter is
\begin{equation}
\label{wDE_BHDE_running}
w_{DE}=-\frac{\Delta+1}{3}
+\frac{(\Delta-2)\sqrt{\Omega_{DE}}}{3}
\left(\frac{C}{3M_p^2}\right)^{-1/2}
\left[
\frac{P(1-\Omega_{DE})}{\Omega_{DE}}
\right]^{\frac{\Delta}{2(\Delta-2)}}
-\frac{\Delta'}{3}
\ln\left[
\frac{P(1-\Omega_{DE})}{\Omega_{DE}}
\right]^{\frac{1}{2-\Delta}}.
\end{equation}
For $\Delta=\mathrm{const}$, these expressions consistently reduce to
Eqs.~(\ref{Odediffeq_BHDE}) and~(\ref{wDE_BHDE}). Several phenomenological
parametrizations were investigated, including linear, CPL-like, exponential,
trigonometric and hyperbolic-tangent forms. In particular, the latter allows
$\Delta$ to remain within its theoretical interval throughout the entire
cosmological evolution. The resulting scenarios reproduce the usual sequence
of matter and dark-energy domination, with the transition to acceleration
occurring around $z\simeq0.65$, while $w_{DE}$ can remain in the quintessence
regime or cross the phantom divide~\cite{Basilakos:2023seo,Basilakos:2026atu}.

A conceptually different formulation has recently been proposed in
Ref.~\cite{Yerokhin:2026xwn}. Instead of prescribing the running exponent
directly inside a global power-law entropy, the scale dependence is introduced
through the local logarithmic slope of a general horizon entropy,
\begin{equation}
S_h(L)=S_{BH}(L)F(L),\qquad
\chi(L)\equiv\frac{d\ln F}{d\ln L}\,.
\end{equation}
Accordingly, the local entropy-scaling dimension is
$d_S=2+\chi$, and the running Barrow exponent is defined through the local
relation
\begin{equation}
\frac{d\ln S_h}{d\ln L}
=2\delta+\delta\Delta_{\rm loc}(L),
\end{equation}
rather than by replacing $\Delta\rightarrow\Delta(L)$ in the global
power-law expression for the entropy. This construction preserves the
interpretation of $\Delta_{\rm loc}$ as the local scaling exponent and allows
one to reconstruct the complete entropy function. In particular, adopting the
interpolating profile
\begin{equation}
\Delta_{\rm loc}(L)=
\frac{\Delta_{\rm UV}}
{1+\left(L/L_t\right)^\kappa},
\end{equation}
which smoothly connects the ultraviolet and infrared regimes, one obtains
\begin{equation}
F(L)=F(L_\ast)
\left(\frac{L}{L_\ast}\right)^{\chi_{\rm UV}}
\left[
\frac{1+(L/L_t)^\kappa}
{1+(L_\ast/L_t)^\kappa}
\right]^{-\delta\Delta_{\rm UV}/\kappa},
\end{equation}
with $\chi_{\rm UV}=2\delta-2+\delta\Delta_{\rm UV}$.

At the cosmological level, the resulting system admits a unique
crossing of the phantom divide, provided that the entropy grows more slowly
than $L^4$. The crossing occurs at an extremum of the future event-horizon
radius and proceeds from the quintessence to the phantom regime. Moreover, the
standard matter- and radiation-dominated asymptotic behaviors are preserved,
whereas the admissible late-time trajectories may approach a de Sitter state
or evolve toward different future singularities, depending on the model
parameters. The thermodynamic analysis further indicates that, within the
adopted entropy decomposition, an additional dark-energy entropy contribution
is required to satisfy the generalized second law in the phantom regime.

\subsection{Luciano-Saridakis holographic dark energy}

This holographic dark energy model arises through application of  the usual 
holographic framework in the case of the two-parameter generalized  
Luciano-Saridakis entropic functional~(\ref{LSentropy}), 
namely~\cite{Luciano:2026ufu}
\begin{equation}
S_{\delta,\epsilon}
=\gamma_\delta A^\delta
+\gamma_\epsilon A^\epsilon.
\end{equation}
Such a two-exponent extended entropy arises   from a well-defined microscopic 
entropic 
functional and an 
associated generalized microstate counting  
through a controlled violation of the separability 
axiom~\cite{Luciano:2026ufu}.
The resulting holographic dark energy density reads~\cite{Luciano:2026eiy}
 \begin{eqnarray}
\label{GenHDELS}
&&\rho_{DE}=  
 \gamma_\delta 
L^{2(\delta-2)}
+\gamma_\epsilon 
L^{2(\epsilon-2)}.
\end{eqnarray} 
If one considers the infrared cutoff to be the Hubble horizon $H^{-1}$, 
then~(\ref{GenHDELS}) yields
 \begin{equation}
\label{GenHDEHubbleLS}
\rho_{DE}=  
 \gamma_\delta 
H^{2(2-\delta)}
+\gamma_\epsilon 
H^{2(2-\epsilon)}\,,
\end{equation}
and in this case the corresponding equation-of-state parameter  becomes 
 \begin{equation}\label{wderelationLS}
 w_{DE}=-1+\frac{2 \dot{H}[\gamma_\epsilon (\epsilon-2 ) H^{2\delta}+
 \gamma_\delta (\delta-2 ) H^{2\epsilon}]}
 {3H^2(\gamma_\epsilon  H^{2\delta} + \gamma_\delta H^{2\epsilon} )}.
\end{equation}
Note that in the special case where $\epsilon=2$, we obtain
$
 \rho_{DE}=  
 \gamma_\delta 
H^{2(2-\delta)}
+\gamma_\epsilon \,,
$
which corresponds to a correction to the $\Lambda$CDM scenario, while taking 
$\gamma_\delta=0$ 
recovers the $\Lambda$CDM scenario exactly. Similarly, in the special 
case $\delta=2$, we obtain
$\rho_{DE}=  
 \gamma_\epsilon
H^{2(2-\epsilon)}
+\gamma_\delta \,, 
$
which again corresponds to a correction to the $\Lambda$CDM scenario, 
recovering 
the $\Lambda$CDM 
paradigm for $\gamma_\epsilon=0$.

 The cosmological viability of this holographic scenario has recently been 
investigated through a comprehensive observational analysis using Cosmic 
Chronometers, the Pantheon++SH0ES Type Ia supernova compilation, DESI DR2 
baryon 
acoustic oscillations, and compressed Planck 2018 CMB shift parameters 
\cite{Leizerovich:2026qie}. 
The analysis shows that the model provides an excellent fit to current 
cosmological observations, with the preferred parameter region lying close to 
the $\Lambda$CDM limit, although moderate departures driven by the generalized 
entropic sector remain fully compatible with the data. Interestingly, the 
enlarged parameter space allows for regions in which the Pantheon++SH0ES and 
CMB 
datasets become simultaneously consistent, a feature that is not achieved 
within 
the standard $\Lambda$CDM framework. Furthermore, both the complete two-sector 
realization ($\gamma_\delta,\gamma_\epsilon\neq0$) and its reduced 
single-sector 
limit (obtained by setting either $\gamma_\delta=0$ or $\gamma_\epsilon=0$) 
provide statistically comparable descriptions of the observations, indicating 
that the generalized entropic framework offers a robust and flexible extension 
of holographic dark energy while naturally encompassing the standard 
cosmological scenario as a limiting case.

On the other hand, 
 if we consider the infrared cutoff to be the future event horizon 
  then~(\ref{GenHDELS}) yields~\cite{Luciano:2026eiy}  
  \begin{eqnarray}\label{Odediffeqfull}
&&
\!\!\!\!\!\!\!\!\!\!\!\!\!\!\!\! 
3M_p^2 h_0^3 \frac{\Omega_{DE}'}{ (\Omega_{DE}-1)}
= \Omega_{DE} \left\{3 (1-2 
\delta ) M_p^2 h_0 ^3+6 (\delta -2) M_p^2 h_0 ^2  
\sqrt{1-\Omega_{DE}} \left[\frac{3 M_p^2 h_0 ^2
   e^{3(1-\delta ) x}\Omega_{DE}}{\gamma_\delta (1-  
\Omega_{DE})}-\frac{\gamma_\epsilon}{\gamma_\delta} 
   e^{3(2-\delta ) x}\right]^{\frac{1}{4-2 \delta }}\right.\nonumber\\
&&
\ \ \ \ \ \ \ \ \ \ \ \ \ \ \ \ \ \ \ \ \ \ \  
\left.
+2 \gamma_\epsilon  (\delta
   -2) e^{7 x/2} \sqrt{1-\Omega_{DE}} \left[\frac{3 M_p^2 h_0 ^2 e^{x-2 
\delta  x} \Omega_{DE}}{\gamma_\delta (1-  
\Omega_{DE})}-\frac{\gamma_\epsilon}{\gamma_\delta}   e^{-2
   (\delta -2) x}\right]^{\frac{1}{4-2 \delta }}-2\gamma_\epsilon    (\delta 
-2) 
h_0  e^{3 x}\right\}\nonumber\\ 
&& \ \ \ \ \ \ \ \ \ \ \ \ \ \ \ \ \ \  \
-2 \gamma_\epsilon   (\delta -2) e^{3 x}
   \left\{e^{x/2} \sqrt{1-\Omega_{DE}} \left[\frac{3 M_p^2 h_0 ^2 e^{x-2 
\delta  x} \Omega_{DE}}{\gamma_\delta (1-  
\Omega_{DE})}-\frac{\gamma_\epsilon}{\gamma_\delta}   e^{-2
   (\delta -2) x}\right]^{\frac{1}{4-2 \delta }}-h_0 \right\},
\end{eqnarray}
 with $h_0=H_0\sqrt{\Omega_{m0}}$, which is  
the differential equation that determines the 
evolution of  the generalized holographic dark energy density parameter
in a flat Universe 
and for dust matter, as a function of  $x\equiv \ln a$.  For 
$\gamma_\epsilon=0$, this equation recovers those of  Tsallis 
  and Barrow 
  holographic dark energy given in~(\ref{OdediffeqTsallis}) 
and~(\ref{Odediffeq_BHDE}) above, while in the case $\gamma_\epsilon=0$ and 
$\delta=1$, 
the model coincides with standard holographic dark energy 
expression~(\ref{diffOm2basic}).
 Finally,  the 
holographic dark energy equation-of-state parameter  is written 
as~\cite{Luciano:2026eiy}   
\begin{eqnarray}\label{wDEter}
 &&
 \!\!\!\! \!\!\!\! \!\!\!\! \!\!\!\! \!\!\!\! \!\!\!\! 
 w_{DE}= -1+2(2-\delta)\gamma_\delta \left[\frac{3 M_p^2 h_0^2\Omega_{DE} 
e^{-3x}-\gamma_\epsilon (1-\Omega_{DE})}{\gamma_\delta(1-\Omega_{DE})}
\right]^{\frac{2\delta-5}{2(\delta-2)}}\frac{(1-\Omega_{DE})e^{3x}}{9M_p^2 
h_0^2\Omega_{DE}}\nonumber\\
 && \ \ \ \ \ \ \ \ \ \ \ \ \ \ \ \ \ \ 
 \left[\left(\frac{3 M_p^2 h_0^2\Omega_{DE} 
e^{-3x}-\gamma_\epsilon 
(1-\Omega_{DE})}{\gamma_\delta(1-\Omega_{DE})}\right)^{ {\frac{1}
{2(\delta-2)}}}
-\frac{(1-\Omega_{DE})^{ {1/2}}e^{3x/2}}{h_
0}
\right],
\end{eqnarray} 
which once again recovers the result of Tsallis and Barrow holographic dark 
energy for 
  $\gamma_\epsilon=0$, and that of the  basic holographic dark 
energy for  
  $\gamma_\epsilon=0$ and $\delta=1$.

    \begin{figure}[t]
  \hspace{-1.cm}
\includegraphics[scale=0.36]{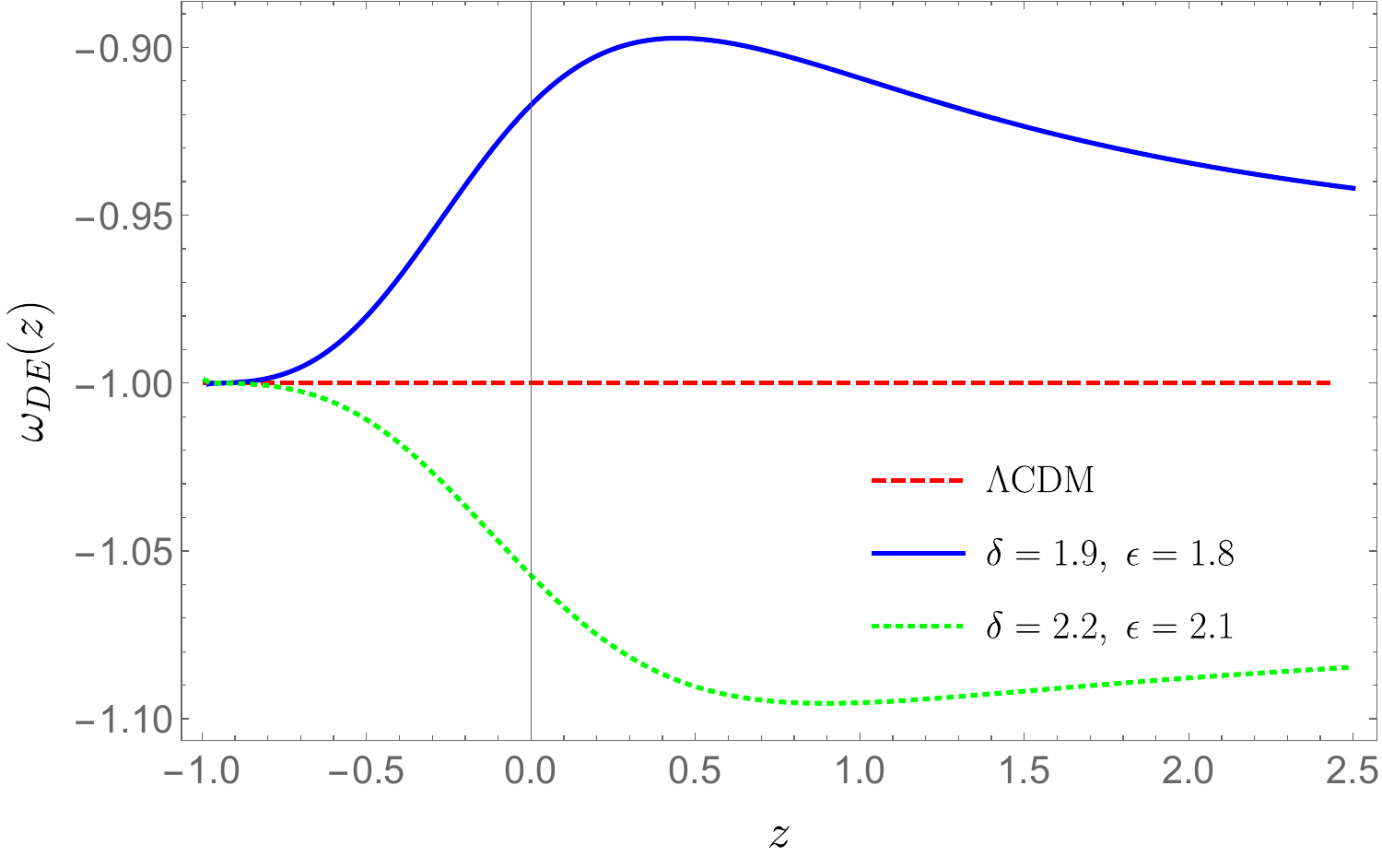}
\caption{
{\it{Evolution of the Luciano-Saridakis holographic dark energy 
equation-of-state parameter 
$w_{DE}$, considering the Hubble horizon as the 
IR cutoff, for $\delta=1.9$, $\epsilon=1.8$ (solid blue curve) and 
$\delta=2.2$, 
$\epsilon=2.1$ (dotted green curve), in Planck units. Additionally, the red 
dashed 
curve marks  the $\Lambda$CDM prediction. The figure is 
from~\cite{Luciano:2026eiy}.
}} }
\label{wdehubLS}
\end{figure}
 In Fig.~\ref{wdehubLS} we depict the evolution of 
$w_{DE}(z)$ for various choices of the exponents $\delta$ and $\epsilon$. As 
one 
can see,  depending on these parameters, the model can lie in 
 the quintessence or   phantom  regime.

\subsection{Viaggiu holographic dark energy}

An interesting extension of the holographic dark energy paradigm arises from 
the application of the Viaggiu entropy~\cite{Viaggiu:2014woa,Viaggiu:2015cra}, 
which incorporates the
effects of cosmological expansion into the entropy-area relation. As we 
mentioned in subsection~\ref{entropymap}, contrary to
the standard Bekenstein-Hawking entropy, which is derived in asymptotically
flat spacetimes, Viaggiu argued that in an expanding Friedmann Universe the
formation of trapped surfaces requires additional energy due to the Hubble
flow. By combining the corresponding trapped-surface theorems with the
Bekenstein entropy bound, the entropy associated with a spherical region of
volume $V$ and area $A$ acquires an extra contribution proportional to the
expansion rate, and takes the form~(\ref{Viaggiu1}), namely 
\begin{equation}
S_V=\frac{A}{4G}+\frac{3VH}{2G},
\end{equation}
where $H$ is the Hubble parameter. As we mentioned, the second term reflects 
the dynamical
degrees of freedom associated with the cosmic expansion and vanishes in the
static limit.

Employing this generalized entropy in the holographic construction
  one obtains the Viaggiu holographic dark energy (VHDE)
scenario~\cite{Saha:2019qnx,Saha:2026lnb,Halder:2026wvg}. In particular, when 
the future event horizon 
$R_E=a\int_t^\infty dt/a$ is chosen as the infrared cutoff, the resulting dark
energy density becomes
\begin{equation}
\rho_d=\frac{\delta^2}{8R_E^2}(1+2HR_E),
\end{equation}
where $\delta$ is a model parameter. Introducing 
the dark-energy
density parameter $\Omega_d$, its evolution is governed by
\begin{eqnarray}
\label{omddreview}
\frac{\Omega_{d}^{'}}{\Omega_{d}(1-\Omega_{d})}=
\frac{\frac{6\Omega_d}{\pi \delta^2}
+\left(1+\sqrt{1+\frac{3\Omega_d}{\pi \delta^2}}\right)}
{(2-\Omega_d)\left(1+\sqrt{1+\frac{3\Omega_d}{\pi \delta^2}}\right)
-\frac{\frac{3\Omega_d}{\pi \delta^2}(1-\Omega_d)}
{\sqrt{1+\frac{3\Omega_d}{\pi \delta^2}}}}.
\end{eqnarray}
Additionally, the corresponding dark-energy equation of state follows as
\begin{eqnarray}
\label{wdreview}
w_d=
-\frac{\Omega_d'}{3\Omega_d(1-\Omega_d)} 
=
-\frac{1}{3}
\left[
\frac{\frac{6\Omega_d}{\pi \delta^2}
+\left(1+\sqrt{1+\frac{3\Omega_d}{\pi \delta^2}}\right)}
{(2-\Omega_d)\left(1+\sqrt{1+\frac{3\Omega_d}{\pi \delta^2}}\right)
-\frac{\frac{3\Omega_d}{\pi \delta^2}(1-\Omega_d)}
{\sqrt{1+\frac{3\Omega_d}{\pi \delta^2}}}}
\right],
\end{eqnarray}
while the deceleration parameter is given by
\begin{equation}
q=\frac{1}{2}\left(1+3w_d\Omega_d\right).
\end{equation}

The cosmological evolution predicted by the model exhibits the standard
sequence of radiation-, matter-, and dark-energy-dominated epochs, with the
density parameters evolving in a manner qualitatively similar to conventional
holographic dark energy scenarios. However, the additional entropy contribution
associated with the Hubble expansion modifies the detailed behavior of the
dark-energy sector and the equation-of-state parameter. Depending on the value
of $\delta$, the model can exhibit quintessence-like or near-phantom behavior
at late times, while naturally leading to an accelerating Universe. Recent
analyses employing contemporary cosmological datasets, including DESI DR2,
indicate that the VHDE framework remains compatible with observational
constraints and provides a viable example of how generalized horizon entropy
can directly affect holographic dark-energy 
phenomenology~\cite{Saha:2019qnx,Saha:2026lnb,Halder:2026wvg,
Jhunjhunwala:2026kmg,Saha:2026fie}.

Specifically, the statefinder hierarchy and fractional growth diagnostics show
that VHDE can be clearly distinguished from $\Lambda$CDM, with composite null
diagnostics efficiently breaking the degeneracy among different values of
$\delta$~\cite{Saha:2026fie}. Moreover, the model remains in the freezing regime
without crossing the phantom divide, although the negative squared sound speed
points to a possible classical instability under perturbations.

\subsection{Power-law holographic dark energy}

Power-law holographic dark energy arises from employing a power-law corrected
black-hole entropy of the form (\ref{powerlawentropy}), motivated by quantum 
entanglement 
considerations~\cite{Telali:2021jju}.
Specifically, as we discussed in subsection~\ref{Correctedentropy1}, when 
excited states are mixed with the ground state in the
entanglement entropy, the resulting entropy deviates from the standard
Bekenstein-Hawking area law by a power-law correction~\cite{Das:2007mj}.
This modification provides a theoretically well-motivated framework for
constructing an extended holographic dark energy scenario.

Starting from the holographic bound $\rho_{DE}L^4\leq S$ and using the power-law
corrected entropy~(\ref{powerlawentropy}), namely
\begin{equation}
S=S_{BH}+c\,S_{BH}^{-\gamma},
\end{equation}
one obtains, upon saturation of the bound, the power-law holographic dark energy
density
\begin{equation}
\label{rhoPLHDE}
\rho_{DE}=c_1\pi M_p^2 L^{-2}
+c_2(\pi M_p^2)^{-\gamma}L^{-2\gamma-4},
\end{equation}
where $c_1$ and $c_2$ are dimensionless parameters, while $\gamma$ quantifies 
the
strength of the power-law correction. The standard holographic dark energy model
is recovered in the limit $c_2\rightarrow0$, ensuring consistency with the
Bekenstein-Hawking entropy.

We consider a spatially flat FRW Universe filled with
pressureless matter and power-law holographic dark energy. The Friedmann
equations take the standard form~(\ref{Fr1Kaniad}),~(\ref{Fr2Kaniad}), with 
$\rho_{DE}$ given by~(\ref{rhoPLHDE}). Additionally, introducing the 
  density parameters,~(\ref{rhoPLHDE}) can be re-written as 
\begin{equation}
\Omega_{DE}H^2=C_1L^{-2}+C_2L^{-2\gamma-4},
\end{equation}
with $C_1\equiv c_1\pi/3$ and
$C_2\equiv c_2\pi^{-\gamma}M_p^{-2+2\gamma}/3$.
Differentiation with respect to $x=\ln a$, together with the background
equations, leads to a second-order differential equation governing the evolution
of $\Omega_{DE}(x)$. This equation reduces exactly to the corresponding equation
of standard holographic dark energy when $C_2=0$, while for $C_2\neq0$ it must 
be
solved numerically.
Moreover, the dark-energy equation-of-state parameter follows from the 
conservation
equation $
\dot\rho_{DE}+3H\rho_{DE}(1+w_{DE})=0,
$
leading to
\begin{equation}
\label{wPLHDE}
w_{DE}=-1+\frac{HQ-\tilde{\Omega}}{3H^2\Omega_{DE}},
\end{equation}
where the auxiliary quantities $Q$ and $\tilde{\Omega}$ are functions of
$\Omega_{DE}$ and its derivatives~\cite{Telali:2021jju}. In the limit 
$C_2\rightarrow0$,
Eq.~(\ref{wPLHDE}) reproduces the well-known result of standard holographic dark
energy,
\begin{equation}
w_{DE}=-\frac{1}{3}-\frac{2}{3}\frac{\sqrt{\Omega_{DE}}}{c},
\end{equation}
as expected.

Numerical investigations reveal that power-law holographic dark energy can
successfully describe the standard sequence of cosmological epochs, namely
matter domination followed by dark-energy domination and late-time accelerated
expansion. The model exhibits a significantly richer phenomenology compared to
standard holographic dark energy, since the additional parameter $\gamma$
allows for quintessence-like evolution, phantom behavior, or phantom-divide
crossing, depending on parameter choices.

\begin{figure}[t]
\centering
\includegraphics[width=8.7cm]{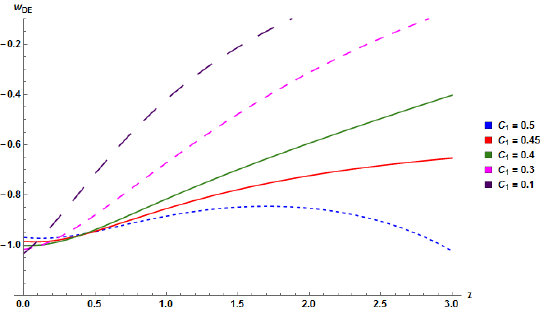}
\caption{\it{The   equation-of-state parameter $w_{DE}$ as a function 
of the redshift, for power-law holographic dark energy for fixed $\gamma=0.5$ 
and various values of
$C_1$, in units where $k_{_B}=c=\hbar=M_p=1$, imposing
$\Omega_{DE0}\approx0.7$. The figure is from~\cite{Telali:2021jju}.}}
\label{wDEparamC}
\end{figure}

The dependence of the dark-energy equation-of-state parameter on the model
parameters is illustrated in Fig.~\ref{wDEparamC}, where the redshift evolution
of $w_{DE}$ is shown for fixed $\gamma$ and various values of $C_1$. While the
present-day value of $w_{DE}$ remains close to $-1$ in agreement with
observations, its behavior at higher redshift is sensitive to the power-law
correction, highlighting the extended phenomenology of the model.

In summary, power-law holographic dark energy constitutes a theoretically 
motivated
 extension of the standard holographic scenario,
emerging naturally from quantum corrections to black-hole entropy and admitting
the latter as a well-defined limiting case, while it can lead to rich 
phenomenology~\cite{Telali:2021jju,Lymperis:2023prf,Ebrahimi:2010xz, 
Sheykhi:2011egx, Khodam-Mohammadi:2011vja, Jawad:2012xy, Chattopadhyay:2013mta, 
Borah:2014vea, Pasqua:2015bfz, Saha:2016bjs, 
Bhardwaj:2022uhf,Pasqua:2026uzf,Yadav:2026bds}.

\subsection{Logarithmic entropy-corrected holographic dark energy}

Let us now consider a particularly conservative deformation of holographic dark 
energy, namely the case where the underlying entropy-area relation is 
supplemented by the well-motivated logarithmic corrections that appear in 
various quantum-gravity settings.   If the horizon entropy 
receives subleading contributions, then the holographic bound itself is 
modified 
and the corresponding dark-energy density inherits additional 
terms~\cite{Wei:2009kp}.

In particular, adopting the corrected entropy-area relation of the 
class~(\ref{Logarithmic11}), namely 
\begin{equation}
S=\frac{A}{4G}+\tilde{\alpha}\ln\!\left(\frac{A}{4G}\right)
+\tilde{\beta}\,,
\end{equation}
where $\tilde{\alpha}$ and $\tilde{\beta}$ are dimensionless constants, and 
following the standard holographic reasoning  one obtains the energy density 
of the logarithmic entropy-corrected holographic dark energy (LECHDE) in the 
form~\cite{Wei:2009kp}
\begin{equation}
\rho_{\rm DE}=3n^2 m_p^2 L^{-2}
+\alpha\,L^{-4}\ln\!\left(m_p^2 L^2\right)+\beta\,L^{-4}\,,
\end{equation}
where the IR cutoff $L$ will be the future event horizon, 
  $n$ is the usual holographic parameter, and $\alpha,\beta$  are dimensionless 
constants that encode the strength of the correction terms.
Additionally, the Friedmann equation becomes
\begin{equation}
 H^\prime=\frac{-2}{3m_p^2 H^2}\left(H-\frac{1}{L}\right) \left(3m_p^2 
H^2-\frac{3}{2}n^2 m_p^2 L^{-2}- \frac{\alpha}{2}L^{-4}\right),\label{eq20}
\end{equation}
where we have used that 
$L^\prime=L-H^{-1}$ which holds when $L$ is the    future event horizon.

At this stage, it is important to pause and extract the physical content of the 
above expression. The first term is precisely the standard HDE contribution and 
will dominate whenever the infrared cutoff is sufficiently large. The remaining 
two terms scale as $L^{-4}$ (modulo the mild logarithmic running), and 
therefore 
they can be relevant only when $L$ is small. In other words, LECHDE is not 
designed to dramatically alter the late-time phenomenology of holographic dark 
energy, but rather  it provides a theoretically motivated completion that can 
become 
operative in the early Universe or in epochs where the relevant horizon scale 
is 
short. As the Universe expands and $L$ grows, the correction terms decay 
rapidly 
and the scenario smoothly reduces to ordinary HDE.

From a practical cosmological viewpoint, the presence of the $L^{-4}$ sector 
implies that the effective equation-of-state and the background evolution may 
receive non-trivial contributions in the ultraviolet regime, potentially 
supporting (for appropriate parameter choices) an early accelerated phase 
without the need to introduce an explicit inflaton degree of freedom. 
Nevertheless, once the system exits this regime and the hierarchy between the 
$L^{-2}$ and $L^{-4}$ terms is established, the evolution naturally returns to 
the standard holographic track, ensuring that the entropy correction acts as a 
genuine high-energy imprint rather than an artificial late-time distortion.

In summary, logarithmic entropy-corrected HDE constitutes a minimal and 
theoretically well-motivated extension of the holographic paradigm. It keeps 
the 
holographic logic intact, modifies only the entropy input in a controlled way, 
and introduces corrections that are significant precisely where one would 
expect 
quantum-gravity effects to be most 
visible~\cite{Wei:2009kp,Jamil:2010sk,Karami:2010aq,Setare:2010dr,
Ebrahimi:2010xz,
Khodam-Mohammadi:2011oji,Amani:2011tv,Farajollahi:2013qea,Pasqua:2013iga,
Sharif:2013qil,Sadeghi:2013vfa,Ghosh:2014mva,Borah:2014asa,Borah:2014vea,
Pasqua:2015bfz,Darabi:2016mjg,Saha:2022vcb}.

\subsection{Holographic dark energy from a generalized entropy}
\label{subsec:gen-entropy-hde}

In the present construction the starting point is a new four-parameter entropy
functional given in~(\ref{EEParametricEntropyModifiedStatisticsAll}), 
namely~\cite{Nojiri:2022dkr}
\begin{align}
S_\mathrm{g}\left[\alpha_+,\alpha_-,\beta,\gamma \right] = 
\frac{1}{\gamma}\left[\left(1 + \frac{\alpha_+}{\beta}~S\right)^{\beta} 
 - \left(1 + \frac{\alpha_-}{\beta}~S\right)^{-\beta}\right] \,,
\label{gen-entropy}
\end{align}
where $\alpha_+$, $\alpha_-$, $\beta$, and $\gamma$ are assumed positive, and
$S$ is the Bekenstein-Hawking entropy. The above expression has been 
constructed artificially, and for suitable
limits of the parameters, it reproduces  Tsallis, 
R\'enyi, Sharma-Mittal, Barrow and Kaniadakis  forms,
while still preserving basic thermodynamic requirements such as
$S_\mathrm{g}\rightarrow0$ for $S\rightarrow0$ and monotonicity in~$S$.

Applying the above generalized entropy in the holographic dark energy framework 
gives rise to the Friedmann equations
\begin{align}
H^2=&\, \frac{8\pi G}{3}\left(\rho + \rho_\mathrm{g}\right) + \frac{\Lambda}{3} 
\,,\nonumber\\
\dot{H}=&\,-4\pi G\left[\left(\rho + \rho_\mathrm{g}\right) + \left(p + 
p_\mathrm{g}\right)\right] ,
\end{align}
    where the generalized holographic dark energy density and pressure are 
given by~\cite{Nojiri:2022dkr}
\begin{align}
\rho_\mathrm{g} = \frac{3}{8\pi G}\left\{ H^2 - 
\frac{GH^4\beta}{\pi\gamma}\left[ \frac{1}{\left(2+\beta\right)}
\left(\frac{GH^2\beta}{\pi\alpha_-}\right)^{\beta}~
{}_2F_{1}\!\left(1+\beta, 2+\beta, 3+\beta, 
-\frac{GH^2\beta}{\pi\alpha_-}\right)
\right. \right. \nonumber\\ 
\left.\left. + 
\frac{1}{\left(2-\beta\right)}\left(\frac{GH^2\beta}{\pi\alpha_+}\right)^{-\beta
}
~
{}_2F_{1}\!\left(1-\beta, 2-\beta, 3-\beta, 
-\frac{GH^2\beta}{\pi\alpha_+}\right)
\right] \right\} \,,
\label{efective energy density}
\end{align}
\begin{align}
p_\mathrm{g} = \frac{\dot{H}}{4\pi 
G}\left\{\frac{1}{\gamma}\left[\alpha_{+}\left(1 + \frac{\pi 
\alpha_+}{GH^2\beta}\right)^{\beta - 1} 
+ \alpha_-\left(1 + \frac{\pi \alpha_-}{GH^2\beta}\right)^{-\beta-1}\right] - 
1\right\} - \rho_\mathrm{g} \,,
\label{effecyive pressure}
\end{align}
 where ${}_2F_1$ denotes the hypergeometric function and $\Lambda$ again appears
as an integration constant. Thus, the generalized entropy effectively generates 
an additional component
$\left(\rho_\mathrm{g},p_\mathrm{g}\right)$, which may be interpreted as a dark
sector induced by horizon thermodynamics.  As one can see,  the model 
retains enough structure to reproduce known entropic HDE scenarios as limits, 
while being flexible enough to capture new cosmological behaviors.

In the late-time Universe, where $GH^2\ll 1$ is extremely well 
satisfied, the above expressions are simplified to 
\begin{align}
\rho_\mathrm{g}=&\, \frac{3H^2}{8\pi G}\left[1 - 
\frac{\alpha_+}{\gamma_0(2-\beta)}\left(\frac{GH^2\beta}{\pi\alpha_+}\right)^{
1-\beta}\right] \,,
\nonumber\\
p_\mathrm{g}=&\, -\frac{\dot{H}}{4\pi G}\left[1 - 
\frac{\alpha_+}{\gamma_0}\left(\frac{GH^2\beta}{\pi\alpha_+}\right)^{1-\beta} 
 - 
\left(\frac{\alpha_+}{\gamma_0}\right)\left(\frac{\alpha_+}{\alpha_-}\right)^{
\beta}\left(\frac{GH^2\beta}{\pi\alpha_+}\right)^{1+\beta}\right] 
 - \rho_\mathrm{g} \,.
\label{energy and pressure-late time}
\end{align} 
Hence,  the effective equation-of-state parameter for the generalized 
holographic dark energy  $w_\mathrm{g}\equiv 
p_\mathrm{g}/\rho_\mathrm{g}$ becomes
\begin{align}
w_\mathrm{g}
= -1 - \left(\frac{2\dot{H}}{3H^2}\right)
\left[\frac{1 - 
\frac{\alpha_+}{\gamma_0}\left(\frac{GH^2\beta}{\pi\alpha_+}\right)^{1-\beta} 
 - 
\left(\frac{\alpha_+}{\gamma_0}\right)\left(\frac{\alpha_+}{\alpha_-}\right)^{
\beta}\left(\frac{GH^2\beta}{\pi\alpha_+}\right)^{1+\beta}}
{1 - 
\frac{\alpha_+}{\gamma_0(2-\beta)}\left(\frac{GH^2\beta}{\pi\alpha_+}\right)^{
1-\beta}}\right] \,.
\label{eos}
\end{align} 
Finally, one can express the holographic dark energy density parameter 
as~\cite{Nojiri:2022dkr} 
\begin{align}
\Omega_\mathrm{g}(z) = 1 - 
\left[\frac{\alpha_+}{\gamma_0(2-\beta)}\right]^{\frac{1}{2-\beta}}
\left[\frac{GH_0^2\beta}{\pi\alpha_+}~\Omega_{m0}(1+z)^3\right]^{\frac{1-\beta}{
2-\beta}} \,.
\label{fractional DE late time}
\end{align}

In summary, the  generalized entropy gives rise to a generalized 
holographic dark energy scenario, with interesting cosmological 
predictions~\cite{Nojiri:2022dkr,Odintsov:2022qnn, Odintsov:2023vpj, 
Jizba:2023ygi, 
Nojiri:2023bom, Brevik:2024nzf, Saha:2024ugn, Saha:2024fup, Elizalde:2024jte, 
Jawad:2024twt, Cimdiker:2025vfn, Odintsov:2025sew, Tariq:2025wiy, 
Tyagi:2025zov, 
Mushtaq:2025thw, Brevik:2025kya, Adhikary:2025khr, Elizalde:2025sat}.

\section{Entropic gravity and cosmology}
 \label{Entropicgravity}

The idea that gravity may not be a fundamental interaction, but rather an 
emergent phenomenon related to thermodynamic and informational principles, 
has gained considerable attention over the past years. Building on the  
connection between horizon thermodynamics, entropy bounds, and holography, this 
perspective suggests that gravitational dynamics could arise from the 
statistical behavior of microscopic degrees of freedom associated with 
spacetime 
itself. In this framework, spacetime geometry and gravitational interaction are 
no longer taken as primary ingredients, but instead they emerge from more 
fundamental entropic considerations tied to information storage on holographic 
screens.

The entropic approach to gravity provides a unifying conceptual framework in 
which black-hole thermodynamics, horizon entropy, and the holographic principle 
  coexist. By associating entropy and temperature with causal horizons 
and invoking general thermodynamic relations, one can reproduce the standard 
gravitational laws without assuming them a priori. This paradigm offers not 
only 
an alternative interpretation of known results, but also an interesting ground 
for exploring extensions of gravity and cosmology that may reveal aspects of 
quantum gravitational physics inaccessible to conventional formulations.

In this section we review how these ideas materialize at different 
levels. We begin with the entropic derivation of Newtonian gravity, 
illustrating 
how an effective gravitational force can arise from entropy gradients. We  
proceed to discuss the emergence of Einstein equations from thermodynamic 
arguments and then extend the analysis to cosmological settings, where 
entropic considerations lead to modified Friedmann equations and novel 
cosmological dynamics. This progression will set the stage for the discussion 
of 
entropic cosmology and inflation, where entropy-driven mechanisms play a 
central 
role in the early- and late-time evolution of the Universe.

\subsection{Entropic origin of Newtonian gravity}
\label{subsec:entropic_newton}

  Newtonian gravity is a very successful theory. After one postulates  
force and   inertia, one writes $F=ma$, and then one specifies a 
particular 
interaction
law, namely  $F=G Mm/R^{2}$. What Verlinde proposed in~\cite{Verlinde:2010hp} 
is to reverse the above logical arrow, namely
to treat the force law not as a microscopic input, but as a macroscopic 
bookkeeping
device that becomes unavoidable once we accept two intertwined ideas: (i) that 
space
is not fundamental but \emph{emergent}, and (ii) that the microscopic 
description
stores information in a holographic manner, i.e. in terms of degrees of freedom
associated with surfaces rather than volumes. In such a setting, the most 
basic objects are not coordinates and fields, but rather entropy, 
temperature, and energy,
namely quantities that remain meaningful even when ``space'' is not yet given. 
In other words,  gravity is interpreted as an emergent \emph{entropic force}, 
produced by an entropy gradient associated with the 
displacement of matter.

Before proceeding to the description of  gravity, one needs a clear 
definition of what is meant by an entropic force. In many-body systems with a 
huge number of microstates, a macroscopic force may appear simply because the 
system statistically prefers configurations with larger entropy. In that sense, 
the ``force'' is not carried by a fundamental mediator, but rather it is an 
effective description of the system's tendency to move
towards higher entropy. In this language, the mechanical work performed over a 
small displacement $\Delta x$ is written as
\begin{equation}
F\,\Delta x = T\,\Delta S,
\label{eq:entropic_force_def}
\end{equation}
where $T$ is the relevant temperature and $\Delta S$ the entropy change caused 
by the
displacement. This is the minimal (and essentially model-independent) statement 
of an
entropic force. 

In the emergent-space picture, this formula   becomes an organizing principle. 
\emph{If} the information associated with the presence and position of matter 
changes when matter is displaced, and \emph{if} this information is processed 
by 
microscopic degrees of freedom that can be assigned a temperature, then a force 
is not an extra assumption anymore, but rather it is a thermodynamic necessity. 
The only remaining 
task is to identify (a) the entropy change, and (b) the origin of the 
temperature.

The key step to proceed is to adopt a holographic perspective on information 
storage. Rather than assuming that information is attached to points in a 
pre-existing continuum, one assumes that the information relevant for the 
emergent description is encoded on \emph{screens} (surfaces) that separate 
regions, similarly to what  horizons do in gravitational settings. In this 
spirit, Verlinde 
introduced a postulate inspired by Bekenstein's
reasoning~\cite{Verlinde:2010hp,Verlinde:2016toy}, namely when a particle of 
mass $m$ 
approaches a screen, there is an associated
change in the entropy of the screen, and the minimal entropy jump occurs when 
the particle moves by about one Compton wavelength. Hence, the postulate 
is written as~\cite{Verlinde:2010hp}
\begin{equation}
\Delta S = 2\pi k_{B}
\qquad \text{when} \qquad
\Delta x=\frac{\hbar}{mc},
\label{eq:verlinde_postulate}
\end{equation}
or, in differential (linearized) form 
\begin{equation}
\Delta S = 2\pi k_{B}\,\frac{mc}{\hbar}\,\Delta x.
\label{eq:verlinde_postulate_linear}
\end{equation}
The proportionality to $m$ is natural since both mass and entropy are additive, 
namely splitting the particle into smaller constituents should split the 
entropy 
change accordingly. 

At first sight, it may look odd that $\hbar$ appears in what will become a 
classical law. But this is precisely part of the logic of the argument, i.e. 
$\hbar$ enters as a dimensional necessity in the postulate, yet it will 
disappear from the final classical force laws. In other words, the quantum 
scale 
is used as a ruler for the minimal displacement that changes the information 
content, but the macroscopic force emerges without explicit quantum
remnants.

Once Eqs.~\eqref{eq:entropic_force_def} and~\eqref{eq:verlinde_postulate_linear} 
are 
accepted, one still needs a temperature $T$.  At this stage Verlinde invokes 
the tight conceptual relation between acceleration and temperature, since the 
Unruh effect tells us that an accelerated observer associates a temperature 
with 
the vacuum. In the present logic, however, one does not interpret this as 
``acceleration produces temperature'', but rather as a useful
thermodynamic identification for the temperature needed to sustain an 
acceleration. Thus, one uses
\begin{equation}
k_{B}T = \frac{1}{2\pi}\,\frac{\hbar a}{c},
\label{eq:unruh_temp}
\end{equation}
with $a$ the acceleration. 

Combining Eqs.~\eqref{eq:entropic_force_def},
\eqref{eq:verlinde_postulate_linear}, and
\eqref{eq:unruh_temp}, one immediately obtains~\cite{Verlinde:2010hp}
\begin{equation}
F\,\Delta x
= T\,\Delta S
= \left(\frac{\hbar a}{2\pi c k_{B}}\right)
\left(2\pi k_{B}\frac{mc}{\hbar}\Delta x\right)
= ma\,\Delta x,
\end{equation}
and therefore
\begin{equation}
F=ma.
\label{eq:newton_second_from_entropy}
\end{equation}
We mention that this is not a   coincidence, rather it encodes a conceptual 
shift. Inertia, in this view, is not a fundamental axiom of mechanics but the 
macroscopic statement that, in the absence of entropy gradients (and hence in 
the absence of entropic forces), motion does not ``prefer'' any direction. The 
traditional law of inertia is thus reinterpreted as a statistical statement 
about information flow and coarse graining.

The previous steps reveal that once a temperature and an entropy gradient exist,
one obtains  $F=ma$. However,  gravity is more specific, namely  it is a 
particular 
force law that depends on the source mass $M$ and the distance $R$. Thus, the 
next step is to determine what sets the temperature of the screen. If the 
screen carries information, then it also carries degrees of freedom, and 
therefore if those degrees of freedom share energy, then a temperature emerges. 
In other words, the temperature is not postulated as an external bath, but 
rather it is induced by the energy content enclosed by the screen.

We consider  a closed holographic screen, taken for simplicity as a 
sphere of area
$A=4\pi R^{2}$. The holographic principle suggests that the number of 
fundamental bits
(or degrees of freedom) on the screen is proportional to $A$. Verlinde writes 
this as
\begin{equation}
N=\frac{A c^{3}}{G\hbar},
\label{eq:number_of_bits}
\end{equation}
introducing a constant $G$ which will later be identified with Newton's 
constant, however at this stage  it is simply the proportionality constant 
relating ``how much information'' fits on a surface to its area.  
Additionally, one  makes a minimal thermodynamic assumption, namely that the 
energy is distributed over the bits such that an equipartition-like relation 
holds, i.e.
\begin{equation}
E=\frac{1}{2}N k_{B}T,
\label{eq:equipartition}
\end{equation}
and the energy enclosed by the screen is identified with the mass $M$ through
\begin{equation}
E = Mc^{2}.
\label{eq:mass_energy}
\end{equation}
Thus, Eqs.~\eqref{eq:number_of_bits}-\eqref{eq:mass_energy}  
determine the temperature as
\begin{equation}
T=\frac{2Mc^{2}}{N k_{B}}
=\frac{2Mc^{2}}{k_{B}}\frac{G\hbar}{A c^{3}}
=\frac{2 G M \hbar}{A c k_{B}}.
\label{eq:screen_temperature}
\end{equation}
Moreover, using the entropic-force relation~\eqref{eq:entropic_force_def}, and 
the
entropy variation~\eqref{eq:verlinde_postulate_linear} for a test particle of 
mass $m$
located near the screen, we obtain
\begin{equation}
F
= T \frac{\Delta S}{\Delta x}
= \left(\frac{2 G M \hbar}{A c k_{B}}\right)
\left(2\pi k_{B}\frac{mc}{\hbar}\right)
= \frac{4\pi G M m}{A}.
\end{equation}
Finally, with $A=4\pi R^{2}$ we arrive at
\begin{equation}
F = G\frac{Mm}{R^{2}}.
\label{eq:newton_gravity_from_entropy}
\end{equation}
This expression is just Newton's law of gravitation, but reinterpreted 
as the entropic response of an information-carrying screen whose temperature is 
set by the enclosed energy. In particular, the
familiar $1/R^{2}$ scaling is not inserted by hand, but it is a geometric 
echo of holography, namely if information scales with area and energy 
per degree of freedom sets a temperature, then the resulting entropic force 
automatically inherits the area factor,
hence the inverse-square law.

One can summarize the minimal inputs of the derivation as three statements:
(i) there is an entropy change in the emergent (holographic) direction when 
matter is displaced, in the linear form~\eqref{eq:verlinde_postulate_linear};
(ii) the number of degrees of freedom scales with 
area, i.e. \eqref{eq:number_of_bits};
(iii) the degrees of freedom admit a thermodynamic description such that energy 
sets a
temperature, for instance through~\eqref{eq:equipartition}. From these, the 
Newtonian
force law follows essentially unavoidably. 
Concerning the validity of   these inputs, Verlinde 
notes that equipartition is perhaps the most fragile-looking assumption, since 
generic interacting systems need not share energy democratically. Yet the 
derivation really needs something weaker, namely that the relevant energy 
change 
associated with an entropy
shift scales with the energy per unit area, so that the final result is 
controlled by $E/A$
rather than by microscopic details. In that sense, equipartition functions as 
just a representative of a broader class of statements that require energy 
to spread over available degrees of freedom.
 Likewise, the explicit appearance of $\hbar$ 
in~\eqref{eq:verlinde_postulate_linear} should
not be overinterpreted. Since it is eliminated from the final 
classical force law, it can be viewed as an auxiliary scale introduced to write 
the entropy gradient in a dimensionally consistent manner. The true physical 
content is the existence of an entropy gradient proportional to $m$ and to the 
displacement in the emergent direction.  

We close this subsection by mentioning that the derivation above does not claim 
to be a microscopic theory of quantum gravity, rather it is, by construction, a 
macroscopic argument. However, it is precisely this ``thermodynamic''
character that makes it relevant in an entropy-focused discussion of cosmology, 
namely  it invites us to think of gravity as the effective dynamics of 
information, rather than the other way around. In this viewpoint, Newtonian 
gravity is not a fundamental force living on space, but  it is a statistical 
law that accompanies the emergence of space itself.

\subsection{Emergence of Einstein equations from entropic considerations}
\label{subsec:entropic_einstein}

The Newtonian discussion of the previous subsection is intentionally modest. It 
is embedded in a world where time is absolute, gravity is weak, and geometry is 
not yet forced upon us. On the other hand, General Relativity  is 
precisely the statement that gravity  is geometry,
and that the dynamics of this geometry is governed by the Einstein equations.
Verlinde's claim is that the same entropic logic can be pushed into the 
relativistic
domain, provided one upgrades the relevant notions in the only consistent way. 
In particular, the entropy gradients must now be understood in a 
redshifted setting, the temperature must be the gravitationally redshifted 
local temperature on a screen, and the appropriate relativistic notion of mass 
enclosed by a surface must be the Komar 
mass~\cite{Verlinde:2010hp,Verlinde:2016toy}. In this way, the Einstein 
equations 
appear not as an independent
postulate, but as the field equations required for the thermodynamic arguments
encoded by holographic screens to be self-consistent.

In a static spacetime one has a preferred time-like Killing vector $\xi^{a}$, 
and the
associated redshift potential can be written as
\begin{equation}
\phi=\frac{1}{2}\ln\!\bigl(-\xi^{a}\xi_{a}\bigr),
\label{eq:redshift_potential}
\end{equation}
so that $e^{\phi}$ plays the role of the redshift factor relative to a 
reference 
point
(often taken at infinity). The natural relativistic analogue of the Newtonian
equipotential surfaces is then a foliation by surfaces of constant $\phi$.
Thus, on such a screen the redshift is uniform, therefore the 
notion of
``time'' with respect to which the microscopic information is processed is 
coherent over
the whole screen. In other words, constant-redshift screens are the relativistic
holographic screens that best mimic the quasi-static reasoning used in the 
Newtonian derivation.

The entropic-gravity logic again needs two ingredients, namely a density of 
microscopic 
degrees of freedom on the screen, and a temperature assigned to these degrees 
of freedom. The bit density is taken to scale with the area element, i.e.
\begin{equation}
dN=\frac{dA}{G\hbar},
\label{eq:bit_density_rel}
\end{equation}
which is the direct relativistic counterpart of the holographic scaling used 
earlier. However, the temperature must now incorporate gravitational redshift. 
A 
convenient and physically
transparent choice is to define the screen temperature by a redshifted 
Unruh-like
expression, namely
\begin{equation}
T=\frac{\hbar}{2\pi}\,e^{\phi}\,N^{b}\nabla_{b}\phi,
\label{eq:screen_temperature_rel}
\end{equation}
where $N^{b}$ is the outward-pointing unit normal to the screen, orthogonal 
also 
to $\xi^{b}$. The factor $e^{\phi}$ ensures that $T$ is measured with respect 
to the same reference time used to define the global energy (and hence the 
mass) 
in the static geometry.  

At this stage, these identifications should be read in the same spirit as in 
the 
polymer analogy. One does  not declare that spacetime \emph{is} a heat 
bath in any microscopic sense, but one  is asserting that if a macroscopic 
description exists in which the relevant information is coarse-grained on 
screens, then temperature is the correct macroscopic parameter that captures 
how 
the energy is distributed over those degrees of freedom.

Assume now that the screen encloses a static matter configuration with total 
mass $M$. The relativistic analogue of the Newtonian step ``energy is evenly 
spread over the bits'' is to adopt an equipartition relation on the screen, 
namely
\begin{equation}
M=\frac{1}{2}\int_{\mathcal{S}} T\, dN.
\label{eq:equipartition_komar_start}
\end{equation}
Substituting~\eqref{eq:bit_density_rel} and~\eqref{eq:screen_temperature_rel} 
into~\eqref{eq:equipartition_komar_start} yields
\begin{equation}
M=\frac{1}{4\pi G}\int_{\mathcal{S}} e^{\phi}\,\nabla\phi\cdot dA,
\label{eq:komar_mass_surface}
\end{equation}
where, importantly, $\hbar$ drops out, as it should in a purely classical 
relation.

The expression~\eqref{eq:komar_mass_surface} is not an arbitrary construct, it 
is precisely the Komar mass associated with the time-like Killing vector 
$\xi^{a}$ in a static spacetime. Thus, from the entropic/holographic viewpoint, 
Komar's definition is reinterpreted as the natural relativistic generalization 
of Gauss's law that arises from counting bits and distributing energy on a 
redshifted screen~\cite{Verlinde:2010hp,Verlinde:2016toy}. We mention that  
this is a key conceptual turning point. In standard General Relativity, one 
typically starts from  the Einstein equations and then proves that the Komar 
integral defines a conserved notion of mass for stationary configurations. Here 
the flow of logic is reversed, namely  one arrives at the Komar expression from 
thermodynamic considerations, and then one asks what field equations are 
required so that this identification is universally consistent.

To proceed, one rewrites the Komar surface integral in a manifestly covariant 
form.
Using the Killing vector $\xi^{a}$, the Komar mass can be expressed as
\begin{equation}
M=\frac{1}{8\pi G}\int_{\mathcal{S}} dx^{a}\wedge dx^{b}\,
\epsilon_{abcd}\,\nabla^{c}\xi^{d}.
\label{eq:komar_mass_covariant}
\end{equation}
Applying Stokes' theorem, and employing the identity implied by the Killing 
equation 
\begin{equation}
\nabla^{a}\nabla_{a}\xi^{b}=-R^{b}{}_{a}\,\xi^{a},
\label{eq:killing_ricci_identity}
\end{equation}
one can relate the Komar mass to a volume integral involving the Ricci tensor.
On the matter side, the mass contained inside the screen can be expressed as an
appropriate contraction of the stress-energy tensor over the enclosed spacelike 
volume
$\Sigma$ bounded by $\mathcal{S}$. The outcome is the well-known integral 
relation
\begin{equation}
2\int_{\Sigma}\!\left(T_{ab}-\frac{1}{2}T g_{ab}\right)n^{a}\xi^{b}\,dV
=
\frac{1}{4\pi G}\int_{\Sigma}\!R_{ab}\,n^{a}\xi^{b}\,dV,
\label{eq:einstein_integral_relation}
\end{equation}
where $n^{a}$ is the unit normal to $\Sigma$, $T\equiv T^{a}{}_{a}$, and $dV$ 
the
induced volume element on $\Sigma$. 

Equation~\eqref{eq:einstein_integral_relation} is already   suggestive. 
It implies
that, under the thermodynamic identifications leading to the Komar mass, the
geometric quantity that naturally appears is $R_{ab}$, while the matter 
quantity 
is the
trace-reversed combination $T_{ab}-\frac{1}{2}T g_{ab}$. The appearance of this
combination is not accidental, it is precisely the one whose covariant 
divergence vanishes once the Bianchi identities are used together with 
$\nabla^{a}T_{ab}=0$, thereby matching
the conservation structure on both sides. In this sense, the entropic 
derivation 
is not merely dimensional analysis, but it is sensitive to the consistency 
requirements that any relativistic field equation must satisfy.

At first glance, Eq.~\eqref{eq:einstein_integral_relation} seems weaker than 
the 
full Einstein equations, because it involves contractions with both $n^{a}$ and 
the specific Killing vector $\xi^{a}$. Indeed, in a given static background one 
only accesses a restricted set of components. The crucial step is therefore 
conceptual rather than algebraic, namely one demands that the 
thermodynamic/gravitational identification holds not only for one
special screen, or one preferred global Killing field, but \emph{locally} and 
for  arbitrary  screens.

The above logic parallels the gravity-thermodynamics conjecture discussed in 
Sec.~\ref{Spacetimethermodynamics}, where local Rindler horizons and the 
Clausius relation 
are used to infer Einstein's equations. However, in the present case it is 
adapted to time-like holographic screens. One considers an arbitrarily small 
neighborhood of an event and arbitrarily short time scales. By the equivalence 
principle, the geometry in such a neighborhood is approximately Minkowskian, 
and 
one can introduce approximate local Killing vectors (corresponding to 
boosts and translations in the tangent space). One then imposes the physically 
robust requirement that when matter crosses a small patch of the
screen, the Komar integral must jump by precisely the amount corresponding to 
the
crossing energy (or mass) measured in the chosen local frame. Repeating the 
steps that lead to~\eqref{eq:einstein_integral_relation} for this family of 
local Killing vectors and for arbitrary choices of screens is sufficient to 
upgrade the contracted integral relation to a local tensor equation. The result 
is the Einstein field equations (up to the usual cosmological constant, which 
enters as an integration constant compatible with the Bianchi 
identities)~\cite{Verlinde:2010hp}, namely
\begin{equation}
R_{ab}-\frac{1}{2}R g_{ab} + \Lambda g_{ab}=8\pi G\,T_{ab}.
\label{eq:einstein_equations_entropic}
\end{equation}
As one can see, Eq.~\eqref{eq:einstein_equations_entropic} is not introduced as 
the 
axiomatic law of spacetime, rather it is the unique covariant completion 
that guarantees that the screen thermodynamics encoded in 
Eq.~\eqref{eq:komar_mass_surface} is universally
well-defined and compatible with energy-momentum conservation.

It is worth emphasizing here that the above argument is macroscopic, and it 
does not specify the microscopic constituents whose degrees of freedom are 
being 
counted on the screen, nor the dynamical rules by which they process 
information. This  is exactly the point of the entropic perspective.
Just as thermodynamics does not require one to know the detailed Hamiltonian of 
a gas in order to write $dE=T\,dS-p\,dV$, entropic gravity seeks the 
gravitational field equations as the effective, coarse-grained relations that 
must hold once holographic information bounds and thermodynamic consistency are 
accepted. In this sense, the Einstein equations appear as the statement that 
the redshift potential $\phi$ (and thus the emergent geometry) is precisely the 
field that keeps track of how information is coarse-grained in spacetime. 
Gravity is then the macroscopic response of this   system to the 
presence and motion of energy, while the curvature tensor encodes the local 
``information geometry'' required for the thermodynamic description to close. 

In summary, the approach of entropic gravity reveals a different way to 
describe gravitational dynamics. Nevertheless, it also naturally motivates 
why cosmology, where horizons, redshifts and information bounds are present, is 
an ideal framework to test entropic considerations, as we explore in the
subsequent subsections.

\subsection{Friedmann equations}

The entropic interpretation of gravity acquires particular significance when 
applied to cosmology, where gravity does not merely act as a force between 
localized bodies but governs the global dynamical evolution of spacetime 
itself. 
In this context, the FRW Universe provides a 
natural framework for extending the entropic-force paradigm to   
time-dependent geometries.
Remarkably, by suitably generalizing Verlinde’s arguments to cosmological 
settings, one can recover the Friedmann equations directly from thermodynamic 
and information-theoretic considerations, without postulating Einstein’s 
equations a priori~\cite{Cai:2010hk,Padmanabhan:2010qr,Shu:2010nv}.

We consider a spatially homogeneous and isotropic FRW Universe with line element
\begin{equation}
ds^{2}=-dt^{2}+a^{2}(t)\left(dr^{2}+r^{2}d\Omega^{2}\right),
\end{equation}
where $a(t)$ denotes the scale factor.
Following the entropic framework, one introduces a compact spatial region 
$\mathcal{V}$ bounded by a spherical holographic screen $\partial\mathcal{V}$ 
of 
physical radius $\tilde r = a r$.
The screen is assumed to encode the information associated with the degrees of 
freedom contained in the enclosed volume, in accordance with the holographic 
principle.

The number of fundamental bits on the screen is taken to be proportional to its 
area according to (\ref{eq:number_of_bits})
while the total energy associated with these degrees of freedom obeys the 
equipartition law  (\ref{eq:equipartition}).
As in the nonrelativistic derivation, the energy $E$ is identified with the 
mass-energy contained within $\mathcal{V}$ through $E=Mc^{2}$.
However, in a cosmological setting the matter content is described by a perfect 
fluid with energy density $\rho$ and pressure $p$, and therefore the relevant 
notion of gravitational mass is not simply $\rho V$, but rather the active 
gravitational (Tolman-Komar) mass.

The cosmological dynamics enters through the acceleration of the physical 
radius 
$\tilde r$,     given by 
\begin{equation}
a_{r}=-\ddot{\tilde r}=-\ddot a\, r,
\end{equation}
which, although not a proper acceleration for comoving observers, characterizes 
the relative acceleration induced by the matter content enclosed by the screen.
In the entropic framework, this acceleration is associated with a temperature 
via an Unruh-like relation, namely
\begin{equation}
T=\frac{\hbar}{2\pi k_{B} c}\, a_{r},
\end{equation}
that has to be interpreted as the temperature required to sustain the 
acceleration rather 
than one generated by it.
Combining the above relations leads, at first, to the Newtonian cosmology 
equation
\begin{equation}
\ddot a=-\frac{4\pi G}{3}\rho\, a,
\end{equation}
which already illustrates that cosmic expansion dynamics can emerge from 
holographic and thermodynamic reasoning alone.

In order to obtain the relativistic Friedmann equations, one must replace the 
total mass $M$ by the active gravitational mass, i.e.
\begin{equation}
\mathcal{M}=2\int_{\mathcal{V}} dV\left(T_{\mu\nu}-\frac{1}{2}T 
g_{\mu\nu}\right)u^{\mu}u^{\nu},
\end{equation}
which properly accounts for the contribution of pressure to gravitation.
Hence, this refinement finally yields the acceleration equation governing the 
evolution of the scale factor, 
namely~\cite{Cai:2010hk,Padmanabhan:2010qr,Shu:2010nv}
\begin{equation}
\label{entropic_accel}
\frac{\ddot a}{a}=-\frac{4\pi G}{3}\left(\rho+3p\right),
\end{equation}
that coincides exactly with the relativistic acceleration equation of standard 
FRW cosmology.
Upon multiplying Eq.~(\ref{entropic_accel}) by $\dot a a$ and using the 
continuity 
equation
$
\dot\rho+3H(\rho+p)=0,
$
one obtains, after integration, the first Friedmann equation, namely
\begin{equation}
\label{entropic_friedmann}
H^{2}+\frac{k}{a^{2}}=\frac{8\pi G}{3}\rho,
\end{equation}
where the integration constant $k$ naturally acquires the interpretation of the 
spatial curvature.

Finally, we mention that the above construction can be extended in a 
straightforward manner to arbitrary spacetime dimensions $d\geq4$.
In this case, the number of bits on the holographic screen generalizes 
appropriately, and the active gravitational mass acquires a dimension-dependent 
form.
The resulting cosmological equations are
\begin{equation}
\label{entropic_accel_d}
\frac{\ddot a}{a}=
-\frac{8\pi G}{(d-1)(d-2)}
\left[(d-3)\rho+(d-1)p\right],
\end{equation}
and
\begin{equation}
\label{entropic_friedmann_d}
H^{2}+\frac{k}{a^{2}}=
\frac{16\pi G}{(d-1)(d-2)}\,\rho,
\end{equation}
which are precisely the Friedmann equations of a $d$-dimensional FRW Universe.

In summary, through the entropic derivation of cosmological dynamics, the 
Friedmann equations are not viewed as fundamental gravitational field 
equations, 
but they emerge as effective relations encoding the thermodynamics of 
information stored on cosmological horizons. Hence, in this perspective the 
cosmic expansion reflects an entropic response of spacetime to the distribution 
and flow of energy, reinforcing the view that gravity and cosmology may 
ultimately be emergent, macroscopic phenomena related to the microscopic 
degrees 
of freedom. The cosmological applications of entropic gravity have been studied 
in detail, both at early and at late times and initiated a large body of 
subsequent work  exploring observational constraints, thermodynamic 
consistency, 
inflationary extensions, and generalized entropy formulations 
\cite{Cai:2010hk,Li:2010cj,Zhang:2010hi,Easson:2010av,Cai:2010zw,Sheykhi:2010wm,
Wei:2010wwa, 
Ling:2010zc,Wei:2010am,Sheykhi:2010yq,Cai:2010kp,Li:2010bc,Zhao:2010vt,
Gao:2010ong,Vancea:2010vf,Lee:2010xv,
Chang-Young:2010sam,Setare:2010xx,Mureika:2010wk,Nicolini:2010nb,Lee:2010za,
Fursaev:2010ix,Klinkhamer:2010qa,Neto:2010ds,Kobakhidze:2010mn,Bastos:2010au,
Porcelli:2010dz,
Casadio:2010fs,Myung:2010wz,Liu:2010zn,Koivisto:2010tb,Gogberashvili:2010em,
Nozari:2011et,
Neto:2011wg,Sahlmann:2011sk,
Chaichian:2011xc,Klinkhamer:2011un,Mann:2011rh,Chang-Young:2011pwc,
Sheykhi:2011vi,Chaichian:2011hc,
Qiu:2011zr,Sheykhi:2011sqb,Sheykhi:2012zza,
Padmanabhan:2012gx,Cai:2012ip,Yang:2012wn,Komatsu:2012zh,
Klinkhamer:2012vg,Abreu:2012msk,
Shalyt-Margolin:2012asj,Sheykhi:2012vf,Gregory:2012an,Wang:2012gc,
Sheykhi:2013oac,
Komatsu:2013qia,Sheykhi:2013ffa,Tu:2013gna,Ling:2013qoa,
Eune:2013ima,Nozari:2013obh,Haranas:2013fna,Chen:2013lqa,Pazy:2013yxj,
Mehdipour:2013cza,
Abreu:2013rxe,
Abreu:2013zea,Wen:2013sya,
Tu:2013fpa,Ai:2013jha,Hashemi:2013oia,Sheykhi:2013tqa,
Ai:2013cvr,Chang-Young:2013gwa,Komatsu:2014lsa,Basilakos:2014tha,
Mehdipour:2014oxa,Yang:2014kna,HamidMehdipour:2014iwa,
Garcia-Islas:2014gfa,
Komatsu:2014vna,Komatsu:2015nkb,Dabrowski:2015tia,vanPutten:2015wma,
Wang:2015cna,Mathew:2015yaa,Hashemi:2015vsa,Yuan:2016pkz,Li:2016vdq,
Hadi:2016zed,Carroll:2016lku,Nunes:2016two,Abreu:2017fhw,
Feng:2017vvw,Wibisono:2017dkt,Abreu:2017puf,Sefiedgar:2017sbw,Komatsu:2017gtf,
Komatsu:2018meb,Bhattacharya:2018wfr,
Plastino:2018krc,Keppens:2018cnu,Xiao:2018dnh,Diaz-Saldana:2018gxx,
Kibaroglu:2018mnx,Kibaroglu:2019dwd,El-Nabulsi:2019odr,Kibaroglu:2019odt,
Plastino:2020mjn,Ipek:2020mvd,Ma:2020aab,Schimmoller:2020kvg,
Perez-Cuellar:2021otp,Senay:2021tjg,Sadeghnezhad:2021ekw,
Tu:2021wwn,Naeem:2022jdq,Schlatter:2022rzs,
Komatsu:2022bik,Odintsov:2023vpj,Naeem:2023ipg,Komatsu:2023xzs,Sung:2023jir,
Schlatter:2024hrv,Jusufi:2024tmb,Schlatter:2024tcu,Tyagi:2024cqp,
Perez-Cuellar:2024lau,
Prasanthan:2024xsl,Chen:2024crb,Okcu:2024llu,Komatsu:2024lov,Chagoya:2024tqv,
Hadi:2025zoe,
Kibaroglu:2025hzc,Rezazadeh:2025htb,Okcu:2025sby,
Nojiri:2025fiu,Kibaroglu:2025nzp,
Sahakian:2025irl,He:2025uff,Aldam-Tajima:2026zwq,Kastner:2026atu,
Nojiri:2026wkb,
Nojiri:2026ish,Chishtie:2026xde}.
Although the original proposal 
remains phenomenological and its microscopic foundation is still debated, it 
represents one of the most direct attempts to connect horizon entropy with the 
observed accelerated expansion of the Universe. More generally, it illustrates 
how thermodynamic concepts may influence cosmological dynamics not only through 
modified entropy-area relations, as discussed in previous sections, but also 
through the possibility that gravity itself emerges from underlying statistical 
degrees of freedom.

\subsection{Entropic cosmology and late-time acceleration}

Motivated by  Verlinde's proposal,  Easson, Frampton and Smoot  proposed 
one of the first cosmological realizations of entropic gravity, showing that 
entropy associated with the cosmological horizon can modify the late-time 
dynamics of the Universe and potentially drive cosmic acceleration without 
introducing an explicit dark-energy 
component~\cite{Easson:2010av,Easson:2010xf}.

As we mentioned above, the basic argument relies on the existence of entropy 
and temperature associated with a cosmological horizon. For a horizon of radius 
$R_H$, one assigns the Bekenstein-Hawking entropy
\begin{equation}
S_H
=\frac{\pi R_H^2}{G},
\end{equation}
and the corresponding horizon temperature
\begin{equation}
T_H=\frac{1}{2\pi R_H}.
\end{equation}
A displacement of the horizon then changes its entropy, generating an effective 
entropic force through
\begin{equation}
F=-T_H\frac{dS_H}{dr}.
\end{equation}
For a Hubble horizon $R_H=H^{-1}$ one obtains an outward acceleration 
proportional to the Hubble parameter,
\begin{equation}
a_H \sim H.
\end{equation}
This contribution can be interpreted as an additional repulsive effect 
associated with the entropy stored on the cosmic horizon.

Incorporating this effect into the cosmological dynamics leads to a modified 
acceleration equation of the form
\begin{equation}
\frac{\ddot a}{a}=
-\frac{4\pi G}{3}(\rho+3p)
+C_H H^2
+C_{\dot H}\dot H,
\label{EntropicAccelerationEquation}
\end{equation}
where $C_H$ and $C_{\dot H}$ are dimensionless constants encoding the strength 
of the entropic contribution. The additional terms arise from surface 
contributions associated with the horizon and become increasingly important at 
late times. For suitable parameter values, they can generate accelerated 
expansion even in the absence of a cosmological constant.

Combining Eq.~(\ref{EntropicAccelerationEquation}) with the Friedmann 
constraint 
equation yields an effective dark-energy sector whose origin is entirely 
entropic. The resulting cosmological evolution reproduces the standard 
matter-dominated era at early times while naturally approaching an accelerated 
phase at low redshift. In this sense, the acceleration of the Universe is 
interpreted not as the consequence of an exotic fluid, but as a manifestation 
of 
horizon thermodynamics~\cite{Easson:2010av}.

\subsection{Entropic cosmology and inflation}

 The cosmological implications of  Verlinde's entropic-force  
scenario have been extensively investigated, ranging from late-time 
acceleration
to modifications of early-universe dynamics. At the same time, concerns were
raised regarding the status and uniqueness of gravity in a purely emergent
framework. This motivated alternative formulations in which Einstein gravity
remains fundamental, but acquires an additional holographic boundary structure
that induces entropic forces in the bulk~\cite{Easson:2010av,Easson:2010xf}. 
Remarkably, such constructions were shown to
admit accelerated expansion both at early and late cosmological epochs,
including realizations of inflation without invoking a fundamental inflaton
field, although they also face important observational and conceptual
challenges~\cite{Danielsson:2010uy}. In what follows, we focus on a particularly
transparent realization of entropic inflation based on a double-screen
holographic setup, and examine its implications at both the background and
perturbative levels.

\subsubsection{A double-screen model of entropic cosmology}
\label{dbscrentr}

In order to investigate thermal inflation within entropic cosmology, we focus
on a framework involving two holographic screens~\cite{Cai:2010zw}. Before
introducing this extension, it is useful to briefly recall the essential
features of the standard one-screen realization of entropic cosmology, also
known as the   Easson-Frampton-Smoot (EFS) 
scenario~\cite{Easson:2010av,Easson:2010xf}.

\paragraph{One-screen entropic cosmology.}

In the one-screen setup, gravity is described by a bulk Einstein-Hilbert action
supplemented by boundary contributions encoding holographic degrees of freedom.
Variation of the total action leads to modified Einstein equations, where an
additional boundary-induced term effectively accounts for energy and momentum
exchange between the bulk and the holographic screen. Assuming a thermodynamic
description of the boundary, the number of degrees of freedom scales with the
screen area, yielding the standard holographic entropy
$S_{BH}=A/(4G)=\pi r_b^2/G$.

Variation of the boundary entropy gives rise to an entropic 
force~\cite{Verlinde:2010hp}, which, combined with the Unruh relation, induces 
an
effective acceleration $a_e=2\pi T_b$. When applied to a flat FRW Universe, the
holographic screen is assumed to lie close to the Hubble horizon,
$r_b=(\beta H)^{-1}$, with temperature $T_b=\beta H/(2\pi)$. As a result, the
cosmological acceleration equation is modified as
\begin{equation}
\frac{\ddot a}{a}=-\frac{4\pi G}{3}(\rho+3p)+\beta^2H^2,
\end{equation}
where the last term reflects the contribution of the entropic force. Quantum and
string-inspired corrections to the horizon entropy further introduce higher
powers of $H$, leading to additional terms proportional to $H^4$ and 
beyond~\cite{Easson:2010xf}.

Although qualitatively appealing, the one-screen model faces difficulties in
reproducing the standard radiation and matter-dominated eras for $\beta\sim
\mathcal{O}(1)$. This motivates the consideration of an additional holographic
screen.

\paragraph{Double-screen entropic cosmology.}

In the double-screen scenario~\cite{Cai:2010zw}, the outer holographic screen is
again associated with the Hubble horizon, while an inner screen is introduced at
the Schwarzschild radius corresponding to the total energy content of the
Universe. The two screens possess distinct temperatures, $T_b$ and $T_S$,
leading to competing entropic forces. The resulting net acceleration is
determined by their temperature difference,
\begin{equation}
a_e=2\pi (T_b-T_S),
\end{equation}
and modifies the cosmological acceleration equation to
\begin{equation}
\frac{\ddot a}{a}=-\frac{4\pi G}{3}(\rho+3p)
+\beta^2H^2\left(1-\frac{3\beta^2H^2}{16\pi G\rho}\right).
\end{equation}

Including quantum-corrected entropy expressions on both screens further
generalizes the acceleration equation through higher-order terms in $H$ and
$\rho$. The cosmological evolution is then governed by a modified Friedmann
equation together with a non-standard continuity equation that allows for energy
exchange between the bulk and the holographic boundaries. Importantly, when the
two screens reach thermal equilibrium, $T_b=T_S$, and for a specific choice of
$\beta$, the standard Friedmann equations are exactly recovered.

At early times, the system is naturally driven away from equilibrium, making
the double-screen framework particularly suitable for describing inflationary
dynamics. In the following we   focus exclusively on the 
early-Universe
regime, and do not consider late-time effects such as the evaporation of the
inner screen.

\subsubsection{Thermal inflation in the early Universe}

Having outlined the structure and dynamics of double-screen entropic cosmology,
we now turn our attention to the behavior of the Universe at very early times,
with particular emphasis on the realization of an inflationary phase. As we
will show, inflation arises rather naturally in this framework, without the
need to introduce additional scalar degrees of freedom.

At sufficiently early epochs the Universe is expected to be radiation
dominated, and therefore throughout this paragraph we assume an equation of
state of the form $p=\rho/3$. Focusing on this regime, one can solve the
cosmological equations  at leading
order, retaining only the first quantum correction to the entropy. This leads
to an approximate expression for the Hubble parameter at high energy 
densities~\cite{Cai:2010zw,Cai:2010kp}
\begin{eqnarray}\label{HFRW}
H^2 = \frac{8\pi G}{3}\left[\rho+\frac{8g}{69}G^2\rho^2+\cdots\right],
\end{eqnarray}
where the effective coefficient $g\equiv g_H-4g_S$ encodes the combined quantum
corrections from the two holographic screens. Clearly, when $g=0$ the standard
Friedmann equation is recovered. However, a qualitatively new behavior emerges
for $g>0$, since at sufficiently high energy scales the Hubble parameter becomes
proportional to $\rho$, rather than $\sqrt{\rho}$. This modification plays a
crucial role in enabling an inflationary phase driven purely by holographic
effects.

Let us now examine this inflationary realization in more detail. For positive
$g$, the $\rho^2$ contribution in Eq.~(\ref{HFRW}) inevitably dominates at early
times, leading to an accelerated expansion even though the Universe is filled
with radiation. Owing to its thermal origin, this phase is naturally referred
to as thermal inflation. This mechanism differs substantially from earlier
entropic inflationary scenarios based on a single holographic screen, where
thermal effects are either neglected, rendering inflation 
marginal~\cite{Easson:2010xf}, eternal~\cite{Qiu:2011zr}, or making inflation 
altogether 
unattainable~\cite{Li:2010bc}. As the Universe expands, the energy density 
gradually
decreases, and once it falls below the critical value
$\rho_C\simeq 69/(8gG^2)$, the linear $\rho$ term takes over, signaling the end
of inflation and the onset of the standard post-inflationary evolution.

Solving the cosmological equations in this regime, one finds that the energy
density evolves approximately as~\cite{Cai:2010kp}
\begin{eqnarray}\label{inflation1}
\rho\simeq\sqrt{\rho_C^2-\frac{512\pi^3t}{27g^{5/2}G^{9/2}}}.
\end{eqnarray}
In this description, the initial Big Bang singularity is formally pushed to
$t\to-\infty$. The physically relevant entropic thermal inflation begins at a
time $-\infty<t_i\ll0$, with the constraint
$t_i\gtrsim -g^{3/2}G^{1/2}$ ensuring that the Hubble scale remains below the
Planck scale, and it terminates at $t_C=0$, when $\rho$ reaches $\rho_C$.

At early stages of this evolution ($t\ll0$), the Hubble parameter behaves as
\begin{eqnarray}
\label{Htapprox}
H(t)\simeq 24.25\times\frac{(-t)^{1/2}}{(gG)^{3/4}},
\end{eqnarray}
which allows for a direct estimation of the slow-roll parameter
\begin{eqnarray}
\label{epsilondef}
\epsilon\equiv-\frac{\dot H}{H^2}
\simeq 2.06\times10^{-2}\frac{(gG)^{3/4}}{(-t)^{3/2}}.
\end{eqnarray}
For sufficiently early times, namely $t\ll-\sqrt{gG}$, one has $\epsilon\ll1$,
confirming that the expansion is indeed quasi-de Sitter. The total number of
e-folds during the observable inflationary period, starting at $t_i$ and ending
at $t_C=0$, can then be estimated as
\begin{eqnarray}
\label{efolds}
{\cal N}\equiv\int_{t_i}^{t_C}H(t)\,dt
\simeq 16.17\times\frac{(-t_i)^{3/2}}{(gG)^{3/4}}.
\end{eqnarray}
We stress that this expression is only approximate, since the asymptotic 
form~(\ref{Htapprox}) ceases to be valid close to the end of inflation.

Finally, we note that phenomenological considerations suggest values
$g\sim\mathcal{O}(10^{16})$~\cite{Cai:2010kp}, as implied by the requirement 
that
the inner holographic screen evaporates within the current age of the Universe,
allowing for a consistent late-time cosmological acceleration.

\subsubsection{Primordial perturbations in entropic cosmology}
\label{primperturb}

In the previous paragraph we established that double-screen entropic cosmology
can naturally accommodate an early thermal-inflationary stage at the background
level. Nevertheless, the phenomenological viability of any inflationary
mechanism is ultimately assessed through its predictions for primordial
perturbations, since these can be directly confronted with CMB and LSS data.
Therefore, we now extend the discussion to the fluctuation sector and examine
the origin and properties of the curvature perturbations generated during
thermal inflation.

Contrary to conventional cold slow-roll inflation, the Universe in the present
setup remains radiation-filled even at very early times. Hence, thermal
fluctuations dominate over vacuum quantum fluctuations, as anticipated by
thermal field theory~\cite{Kraemmer:2003gd}. In double-screen entropic
cosmology there are, in principle, two independent thermal sources, namely  
ordinary
radiation fluctuations in the bulk, and holographic fluctuations associated
with the boundary degrees of freedom on the screens. Neglecting possible
cross-correlations and noting that the inner-screen fluctuations remain
sub-Hubble during the relevant epoch, we focus on the bulk radiation and the
outer screen. Working in the longitudinal gauge 
\begin{eqnarray}
ds^2=a(\tau)^2\left[(1+2\Phi)d\tau^2-(1-2\Phi)dx^idx^i\right],
\end{eqnarray}
the perturbed Einstein constraint relates $\Phi$ to $\delta\rho$, and the
resulting power spectrum at Hubble crossing takes the generic thermal 
form~\cite{Cai:2009rd}
\begin{eqnarray}\label{PPhi}
P_{\Phi}(k)\equiv\frac{k^3}{2\pi^2}\langle\Phi_k^2\rangle
=\frac{8G^2\langle\delta\rho^2\rangle}{H^4}\Big|_{t_*(k)}~,
\end{eqnarray}
with
\begin{eqnarray}\label{deltarho2}
\langle\delta\rho^2\rangle=C_V\frac{T^2}{R^6},
\end{eqnarray}
where $C_V$ is the heat capacity and $R$ the physical correlation length.

For normal radiation one has $C_V^r=g_rR_r^3T_r^3$ with $R_r=c_s/H$, and near
thermal equilibrium $T_r\simeq \beta H/2\pi$. This yields a scale-invariant
contribution
\begin{eqnarray}\label{radspect1}
P_{\Phi}^r=\frac{g_r\beta^5}{4\pi^5c_s^3}G^2H^4~,
\end{eqnarray}
which is, however, parametrically suppressed. On the other hand, the holographic
degrees of freedom on the outer screen satisfy area scaling. Using
$\langle E\rangle\sim r_bT_b/G$ one finds
\begin{eqnarray}\label{CVb}
C_V^b=c_v\frac{r_b^2}{G},
\end{eqnarray}
with $c_v=\mathcal{O}(1)$ and $R\simeq r_b$. Repeating the above steps leads to
\begin{eqnarray}
P_{\Phi}^b=\frac{2c_v\beta^6}{\pi^2}GH^2~,
\end{eqnarray}
which dominates over $P_{\Phi}^r$ once normalized to the observed amplitude,
implying that the main scalar fluctuations are seeded by the outer screen
rather than by the bulk radiation.

Passing to the comoving curvature perturbation $\zeta$, and using that on
super-Hubble scales $\zeta\simeq\Phi/\epsilon$, we obtain
\begin{eqnarray}
P_{\zeta}\simeq\frac{16c_v}{\epsilon^2\pi^2}GH^2~.
\end{eqnarray}
Consequently, the spectral tilt is
\begin{eqnarray}\label{spectrond}
n_S-1\equiv\frac{d\ln P_\zeta}{d\ln k}=-2\epsilon-2\eta,
\qquad \eta\equiv\frac{\dot\epsilon}{H\epsilon},
\end{eqnarray}
and, combining with the background relations of thermal inflation, one arrives
at the simple estimate
\begin{eqnarray}\label{nSN}
n_S\simeq 1-\frac{8}{3{\cal N}},
\end{eqnarray}
namely a nearly scale-invariant spectrum with a red tilt that becomes closer to
unity as the e-folding number    ${\cal N}$ increases.

Finally, tensor modes remain essentially standard,
$P_T=16GH^2/\pi$~\cite{Sasaki:1995aw}, since the holographic sector does not
modify the tensor perturbation equation. Thus, the tensor-to-scalar ratio reads
\begin{eqnarray}\label{te-to-sc}
r\equiv\frac{P_T}{P_\zeta}=\frac{\epsilon^2\pi}{c_v},
\end{eqnarray}
which is doubly suppressed by $\epsilon$ (but can be mildly enhanced if $c_v$
is tuned small). Finally, using the approximate relation $\epsilon\simeq 
1/(3{\cal N})$
one finds~\cite{Cai:2010kp}
\begin{eqnarray}\label{rN}
r\approx \frac{\pi}{9c_v{\cal N}^2},
\qquad
r\approx \frac{\pi(1-n_S)^2}{64c_v},
\end{eqnarray}
showing that for realistic ${\cal N}$ the model naturally predicts a small $r$,
consistent with current observational bounds.

\subsubsection{Non-Gaussianities}
\label{non-gauss}

Beyond the power spectrum, primordial non-Gaussianities provide a sensitive
probe of the underlying inflationary mechanism and can place strong
observational constraints on early-universe scenarios. In the context of
thermal inflation within double-screen entropic cosmology, the dominant source
of scalar perturbations originates from holographic fluctuations on the outer
screen, and thus the corresponding non-Gaussian signal is determined by their
higher-order correlators. The non-linearity parameter is defined as
\begin{eqnarray}
 f_{NL}=\frac{5\langle\zeta_k^3\rangle}{18k^{3/2}\langle\zeta_k^2\rangle^2},
\end{eqnarray}
and can be evaluated by computing the three-point function of holographic
energy-density fluctuations at Hubble crossing. Using equilibrium statistical
mechanics for the boundary degrees of freedom, one finds
$\langle\delta\rho_b^3\rangle\propto T_b^3/(G R_b^7)$, which, together with the
relation $\zeta\simeq\Phi/\epsilon$, yields~\cite{Cai:2010kp}
\begin{eqnarray}
 f_{NL}\simeq \frac{5\epsilon}{36\sqrt{2}\pi c_v},
\end{eqnarray}
after expressing the correlation length and temperature in terms of the Hubble
scale. Hence, the resulting non-Gaussianity is scale-invariant and suppressed
by the slow-roll parameter, reflecting the inflationary nature of the
background. Nevertheless, a potentially observable $f_{NL}$ can arise if the
holographic coefficient $c_v$ is sufficiently small, indicating that a more
microscopic understanding of holographic degrees of freedom is crucial for a
fully quantitative assessment.

In summary, in this subsection we have examined thermal inflation within a
double-screen entropic cosmology, showing that an inflationary phase naturally
arises from the relaxation of the system toward thermal equilibrium. As we saw,
although both bulk radiation and holographic boundary fluctuations generate
nearly scale-invariant curvature perturbations, the dominant contribution
originates from the outer holographic screen, leading to a red-tilted spectrum
consistent with observations and modified consistency relations compared to
standard slow-roll inflation. The predicted non-Gaussianities are generally
small and slow-roll suppressed, although they may become observationally
relevant depending on the microscopic properties of the holographic degrees of
freedom. While uncertainties in holographic thermodynamics prevent sharp
constraints at present, the framework offers distinctive and potentially
testable signatures, and moreover provides a unified entropic description of
early-time inflation and late-time cosmic acceleration, linking naturally with
holographic dark-energy scenarios.

\section{Conclusions}
 \label{Conclusions}

In this review we have explored the role of entropy as a central organizing 
principle in gravitational physics and cosmology, with the aim of providing a 
coherent and conceptually transparent account of its diverse manifestations. 
What emerges from this broad survey is that entropy is a concept that repeatedly 
appears at the interface between gravity, quantum theory, and spacetime 
structure. From black-hole thermodynamics and entropy bounds to spacetime 
thermodynamics, holography, and cosmological applications, entropy serves as a 
unifying notion that encodes deep information about the degrees of freedom 
underlying gravitational phenomena.

We began by reviewing entropy in gravitational systems, focusing on black-hole 
thermodynamics and the Bekenstein-Hawking area law. The realization that black 
holes behave as genuine thermodynamic systems, characterized by a temperature 
and an entropy proportional to the horizon area, has led to profound 
implications. The area scaling of black-hole entropy stands in sharp contrast to 
the volume scaling familiar from conventional thermodynamics, strongly 
suggesting that gravitational degrees of freedom are organized in a 
fundamentally non-local or holographic manner. This observation is reinforced by 
entropy bounds and by the holographic principle, which impose strict limits on 
the information content of spacetime regions and hint at a lower-dimensional 
description of quantum gravity.

A recurring theme throughout the review is the remarkable robustness of the area 
law. Despite the very different physical mechanisms involved, a wide range of 
approaches, including quantum field theory, entanglement entropy, string theory, 
loop quantum gravity, and holography, reproduce the area law at leading order. 
At the same time, essentially all of these frameworks predict subleading 
corrections, such as logarithmic, power-law, square-root, exponential, or 
non-extensive terms. The universality of both the leading area contribution and 
the qualitative structure of its corrections strongly suggests that entropy 
captures a fundamental aspect of spacetime microphysics, largely insensitive to 
the specific details of the underlying theory.

A substantial part of this review was devoted to clarifying the physical origin 
and interpretation of these corrections. We emphasized that subleading entropy 
terms arise from a variety of sources, including quantum entanglement across 
horizons, ultraviolet regularization, quantum fluctuations of geometry, 
discreteness effects, minimal-length scenarios, and modifications of the 
statistical framework itself. Although these corrections are typically small for 
macroscopic horizons, they provide important insight into the microscopic 
degrees of freedom of spacetime and may become significant in regimes where the 
horizon area is not parametrically large compared to the Planck scale.

One of the main conceptual goals of this review has been to disentangle the 
distinct ways in which entropy is employed in cosmology. We have shown that, 
despite their common inspiration from horizon thermodynamics and information 
theory, these approaches play fundamentally different logical roles.

The first major class of applications falls under the umbrella of 
\emph{spacetime thermodynamics}. In this framework, gravitational dynamics are 
interpreted as emergent relations arising from thermodynamic principles applied 
to horizons. The seminal result of Jacobson, demonstrating that the Einstein 
field equations can be derived from the Clausius relation, provides a powerful 
indication that gravity itself may be an emergent phenomenon rather than a 
fundamental interaction. When this perspective is extended to cosmology, the 
Friedmann equations can be obtained from the first law of thermodynamics 
applied 
to the apparent horizon of an FRW Universe. Within this approach, entropy is 
not 
a passive quantity but an active ingredient, since  modifying the entropy-area 
relation directly modifies the cosmological field equations and hence the 
evolution of the Universe.

This observation has motivated a large class of entropic cosmological models 
based on generalized entropy expressions inspired by quantum gravity, modified 
statistics, or horizon microstructure. Tsallis, R\'{e}nyi, Sharma-Mittal, 
Kaniadakis, Barrow, Luciano-Saridakis, and loop-quantum-gravity-motivated 
entropies all lead to modified Friedmann equations with rich phenomenology, 
including accelerated expansion, effective dark-energy components, and novel 
inflationary dynamics. In this context, entropy acts as a bridge between 
microscopic physics and macroscopic cosmological behavior, providing a 
phenomenological window into possible quantum-gravitational effects.

The second major class of entropy applications in cosmology is conceptually 
distinct and is based on holographic considerations. In holographic dark-energy 
models, entropy does not alter the gravitational field equations themselves. 
Instead, entropy bounds are used to constrain the energy content of the 
Universe, leading to an upper limit on the vacuum energy determined by an 
infrared cutoff scale. The resulting dark-energy density depends on global 
properties of spacetime, such as the size of a cosmological horizon, rather 
than 
on local dynamics. Although holographic dark-energy models are inspired by the 
same foundational ideas that underlie spacetime thermodynamics, their logical 
role is different, namely entropy acts as a guiding principle rather than as a 
dynamical variable. 

We have emphasized that maintaining a clear conceptual distinction between these 
two uses of entropy is essential for a consistent interpretation of entropic 
cosmological models. Conflating spacetime thermodynamics with holographic dark 
energy can obscure the physical meaning of entropy and lead to 
misinterpretations of model assumptions and predictions. At the same time, the 
parallel development of these approaches highlights the versatility of entropy 
as a tool for exploring gravitational and cosmological phenomena.

Beyond these two main frameworks, we also reviewed entropic-gravity scenarios, 
in which gravity itself is interpreted as an entropic or emergent force. While 
these ideas remain more speculative and face significant conceptual and 
phenomenological challenges, they provide an additional perspective on the deep 
connections between information, thermodynamics, and spacetime dynamics, 
particularly in cosmological settings.

In summary, the results surveyed in this review suggest that entropy plays a 
structural role in gravitational physics that goes far beyond its original 
thermodynamic meaning. Whether entropy should ultimately be regarded as a 
fundamental quantity, an emergent effective description, or a phenomenological 
proxy for unknown microscopic degrees of freedom remains an open question. 
Nevertheless, the repeated appearance of entropy in diverse and seemingly 
unrelated contexts strongly indicates that it captures an essential aspect of 
the quantum nature of spacetime.

Looking to the future, several directions appear particularly promising. On
the theoretical side, further progress in quantum gravity, holography, and
non-perturbative approaches may clarify the microscopic origin of generalized
entropy expressions and their relation to spacetime dynamics. A major open
challenge is to determine whether the numerous generalized entropy proposals
represent different effective descriptions of a common underlying framework or
instead encode genuinely distinct aspects of quantum spacetime. Closely
related to this issue is the need for a deeper understanding of the roles
played by quantum entanglement, information, and the emergence of geometry,
which are increasingly recognized as fundamental ingredients of gravitational
physics.

On the phenomenological side, increasingly precise cosmological observations,
including future measurements of the cosmic microwave background, large-scale
structure, gravitational waves and black-hole environments, will provide
progressively more stringent tests of entropy-inspired models. Such
observations may help constrain deviations from the Bekenstein-Hawking area
law, discriminate among competing generalized entropy frameworks and assess
the viability of entropic explanations for inflation, dark energy, and the
late-time evolution of the Universe.

Equally important is the continued effort to strengthen the conceptual
foundations of entropy-based approaches. In particular, establishing a clear
logical distinction between spacetime thermodynamics, holographic dark-energy
models and entropic-gravity scenarios will remain essential for developing a
coherent theoretical framework. Understanding which features are universal
consequences of horizon thermodynamics and which instead depend on specific
microscopic assumptions represents one of the central challenges for future
research.

Ultimately, entropy should not be viewed as providing a final answer to the
problem of gravity and cosmology, but rather as a powerful guiding principle
that continues to illuminate the deep and subtle connections between
thermodynamics, information and the structure of the Universe.

\raggedleft
\bibliographystyle{apsrev4-1}
\bibliography{bibliographycomplete}

\end{document}